\documentclass[useAMS,usenatbib]{mn2e}
\usepackage{amssymb}
\usepackage[english]{babel}
\usepackage{color}
\usepackage{dblfloatfix}
\usepackage{fontawesome}
\usepackage{graphics}
\usepackage{graphicx}
\usepackage{natbib}
\usepackage{hyperref}
\usepackage[capitalise]{cleveref}
\usepackage{pifont}
\usepackage{subfiles}
\usepackage{orcidlink}
\usepackage{wasysym}

\defcitealias{Etherington2022}{E22}

\def\simlt{\lower.5ex\hbox{$\; \buildrel < \over \sim \;$}}
\def\simgt{\lower.5ex\hbox{$\; \buildrel > \over \sim \;$}}

\newcommand*\tens[1]{\ensuremath{\mathsf{#1}}}

\definecolor{purple}{rgb}{0.525,0,0.525}

\title[Detection of an Ultramassive Black Hole]
{Abell 1201: Detection of an Ultramassive Black Hole in a Strong Gravitational Lens}
\author[Nightingale et al.]
{\parbox{\textwidth}{James.\ W.\ Nightingale$^{1}$\orcidlink{0000-0002-8987-7401}\thanks{e-mail: james.w.nightingale@durham.ac.uk},
Russell J.\ Smith$^{1}$\orcidlink{0000-0001-5998-2297},
Qiuhan He$^{1}$\orcidlink{0000-0003-3672-9365},
Conor M.\ O'Riordan$^{2}$\orcidlink{0000-0003-2227-1998},
Jacob A. Kegerreis$^{3}$\orcidlink{0000-0001-5383-236X},
Aristeidis Amvrosiadis$^{1}$\orcidlink{0000-0002-4465-1564},
Alastair C.\ Edge$^{1}$\orcidlink{0000-0002-3398-6916},
Amy Etherington$^{1}$,
Richard G.\ Hayes$^{1}$,
Ash Kelly$^{1}$\orcidlink{0000-0003-3850-4469},
John R.\ Lucey$^{1}$\orcidlink{0000-0002-9748-961X},
Richard J.\ Massey$^{1}$\orcidlink{0000-0002-6085-3780}\\
}\\
$^{1}$Centre for Extragalactic Astronomy, Department of Physics, Durham University, South Road, Durham, DH1 3LE, UK\\
$^{2}$Max Planck Institute for Astrophysics, Karl-Schwarzschild-Strasse 1, 85748 Garching bei München, Germany\\
$^{3}$NASA Ames Research Center, Moffett Field, CA 94035, USA\\
}

\makeatletter
\newcommand{\github}[1]{%
   \href{#1}{\faGithubSquare}%
}
\makeatother

\begin{document}

\bibliographystyle{mn2e}
\bibpunct{(}{)}{;}{a}{}{;}
\date{\today}
\pagerange{\pageref{firstpage}--\pageref{lastpage}} 
\pubyear{2018}
\maketitle
\label{firstpage}

\begin{abstract}

Supermassive black holes (SMBHs) are a key catalyst of galaxy formation and evolution, leading to an observed correlation between SMBH mass $M_{\rm BH}$ and host galaxy velocity dispersion $\sigma_{\rm e}$. Outside the local Universe, measurements of $M_{\rm BH}$ are usually only possible for SMBHs in an active state: limiting sample size and introducing selection biases. Gravitational lensing makes it possible to measure the mass of non-active SMBHs. We present models of the $z=0.169$ galaxy-scale strong lens Abell~1201. 
A cD galaxy in a galaxy cluster, it has sufficient `external shear' that a magnified image of a $z = 0.451$ background galaxy is projected just $\sim 1$ kpc from the galaxy centre. Using multi-band Hubble Space Telescope imaging and the lens modeling software \texttt{PyAutoLens} we reconstruct the distribution of mass along this line of sight. Bayesian model comparison favours a point mass with $M_{\rm BH} = 3.27 \pm 2.12\times10^{10}$\,M$_{\rm \odot}$ (3$\sigma$ confidence limit); an ultramassive black hole. One model gives a comparable Bayesian evidence without a SMBH, however we argue this model is nonphysical given its base assumptions. This model still provides an upper limit of $M_{\rm BH} \leq 5.3 \times 10^{10}$\,M$_{\rm \odot}$, because a SMBH above this mass deforms the lensed image $\sim 1$ kpc from Abell 1201's centre. This builds on previous work using central images to place upper limits on $M_{\rm BH}$, but is the first to also place a lower limit and without a central image being observed.
The success of this method suggests that surveys during the next decade could measure thousands more SMBH masses, and any redshift evolution of the $M_{\rm BH}$--$\sigma_{\rm e}$ relation. Results are available at \url{https://github.com/Jammy2211/autolens_abell_1201}.

\end{abstract}


\section{Introduction}\label{Intro}

\begin{figure*}
\centering
\includegraphics[width=0.32\textwidth]{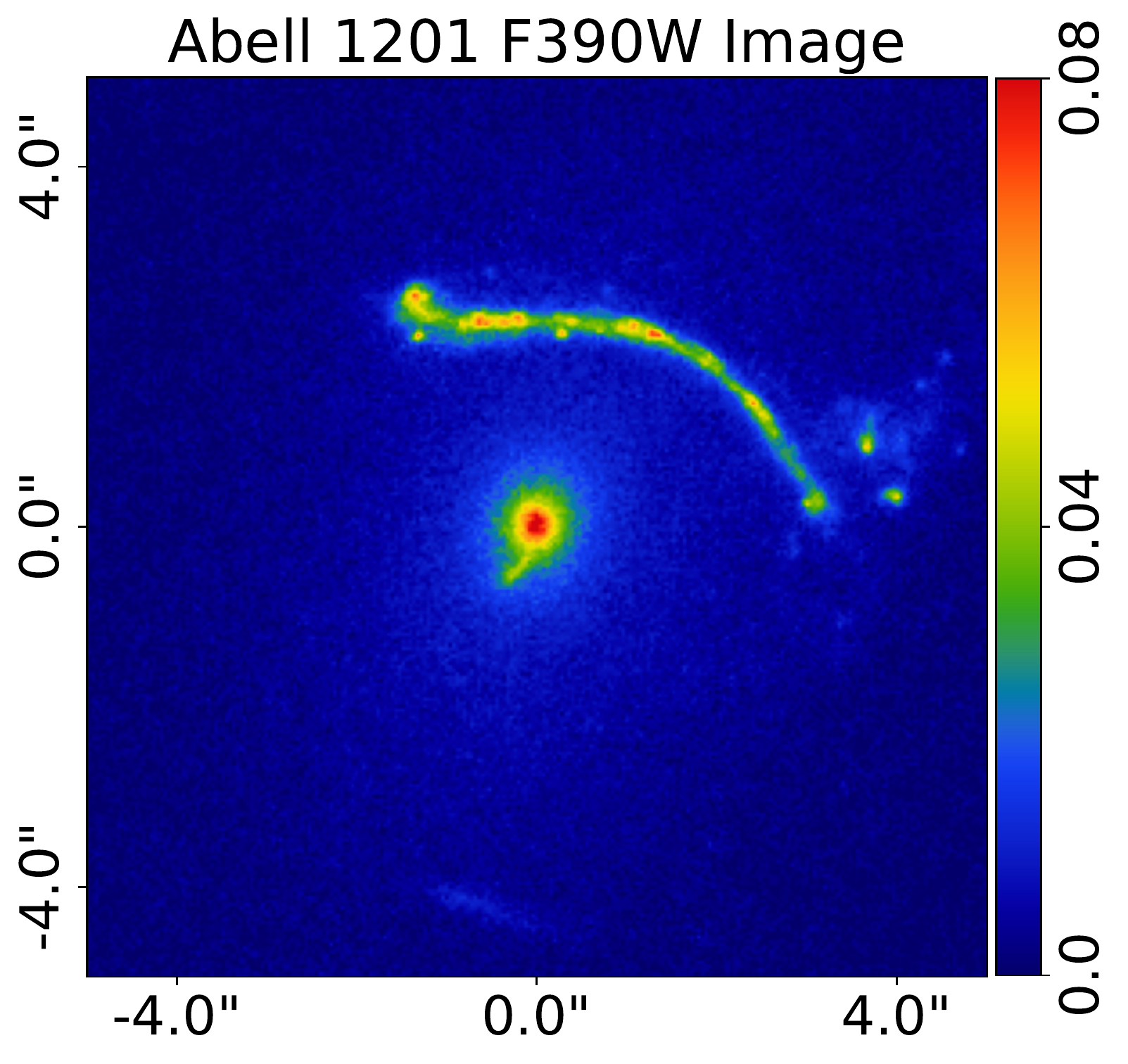}
\includegraphics[width=0.32\textwidth]{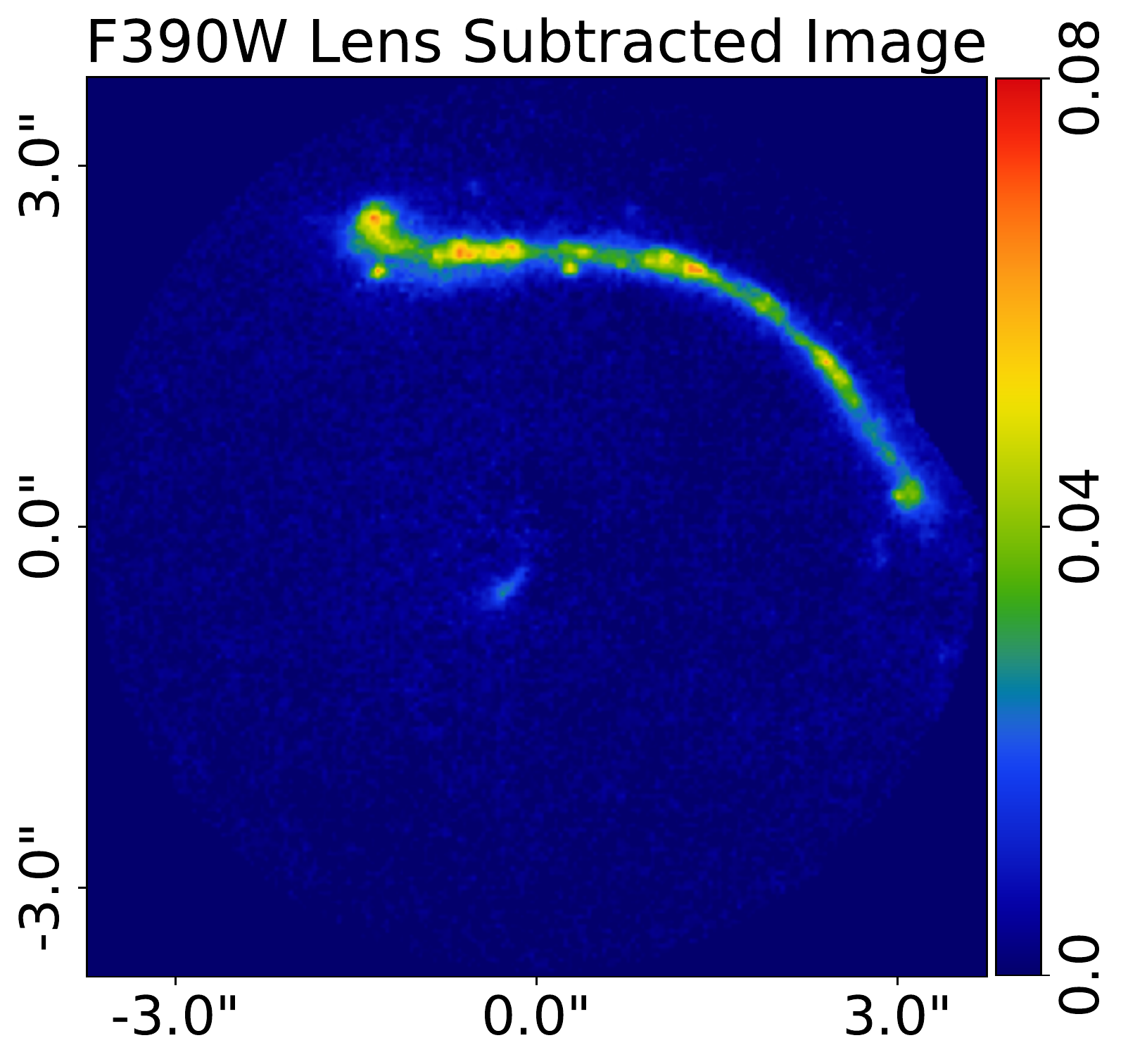}
\includegraphics[width=0.32\textwidth]{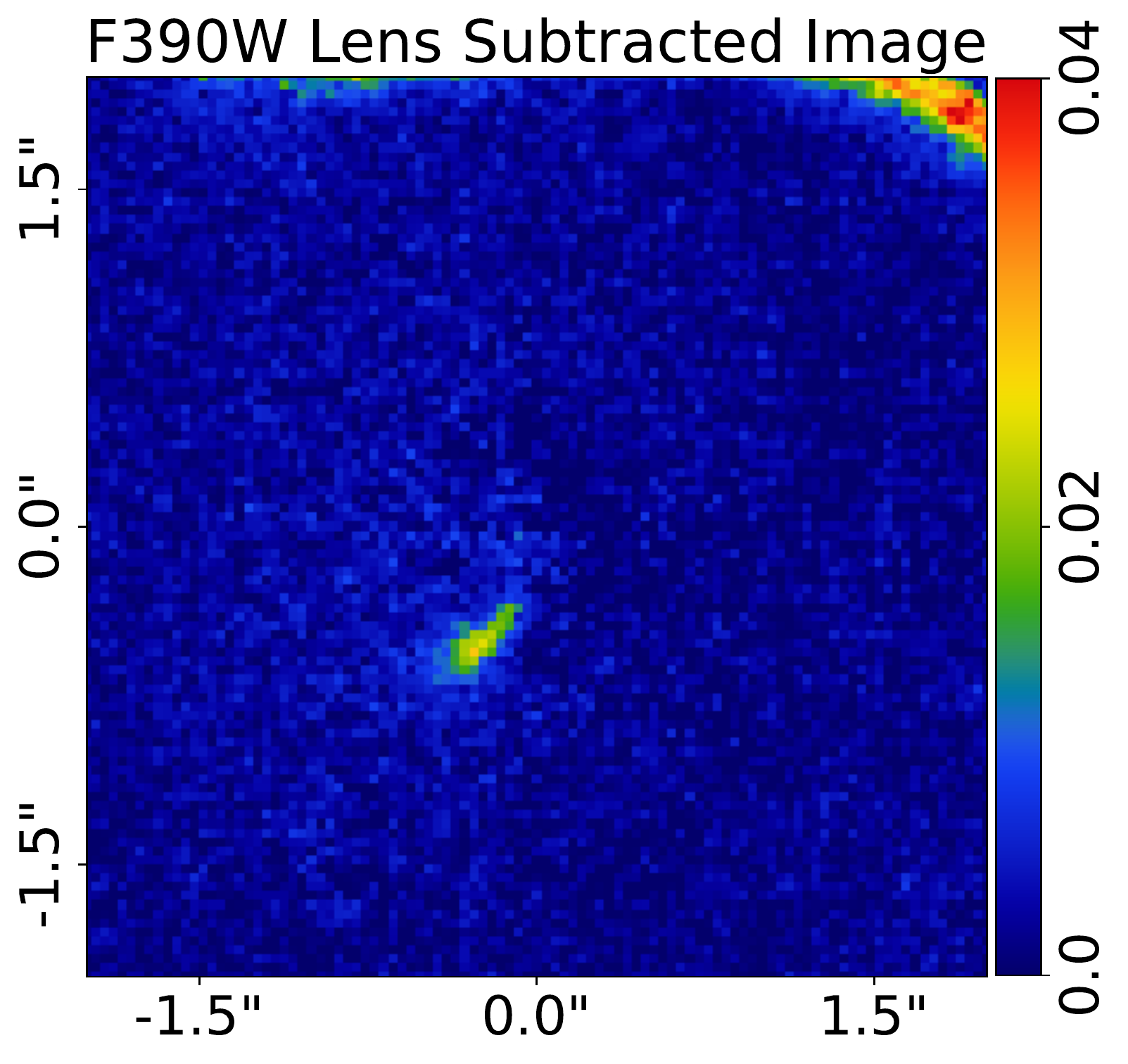}
\includegraphics[width=0.32\textwidth]{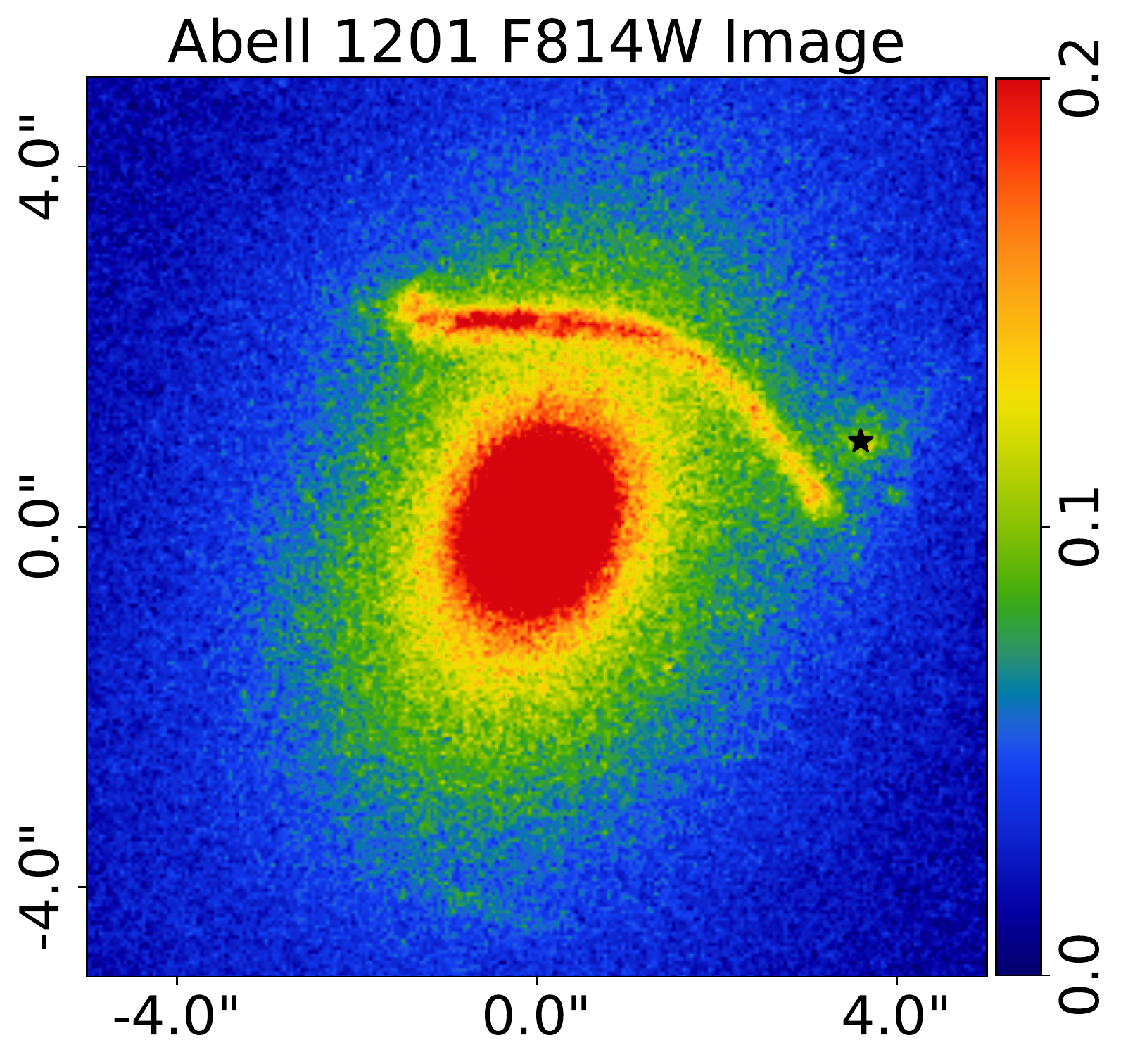}
\includegraphics[width=0.32\textwidth]{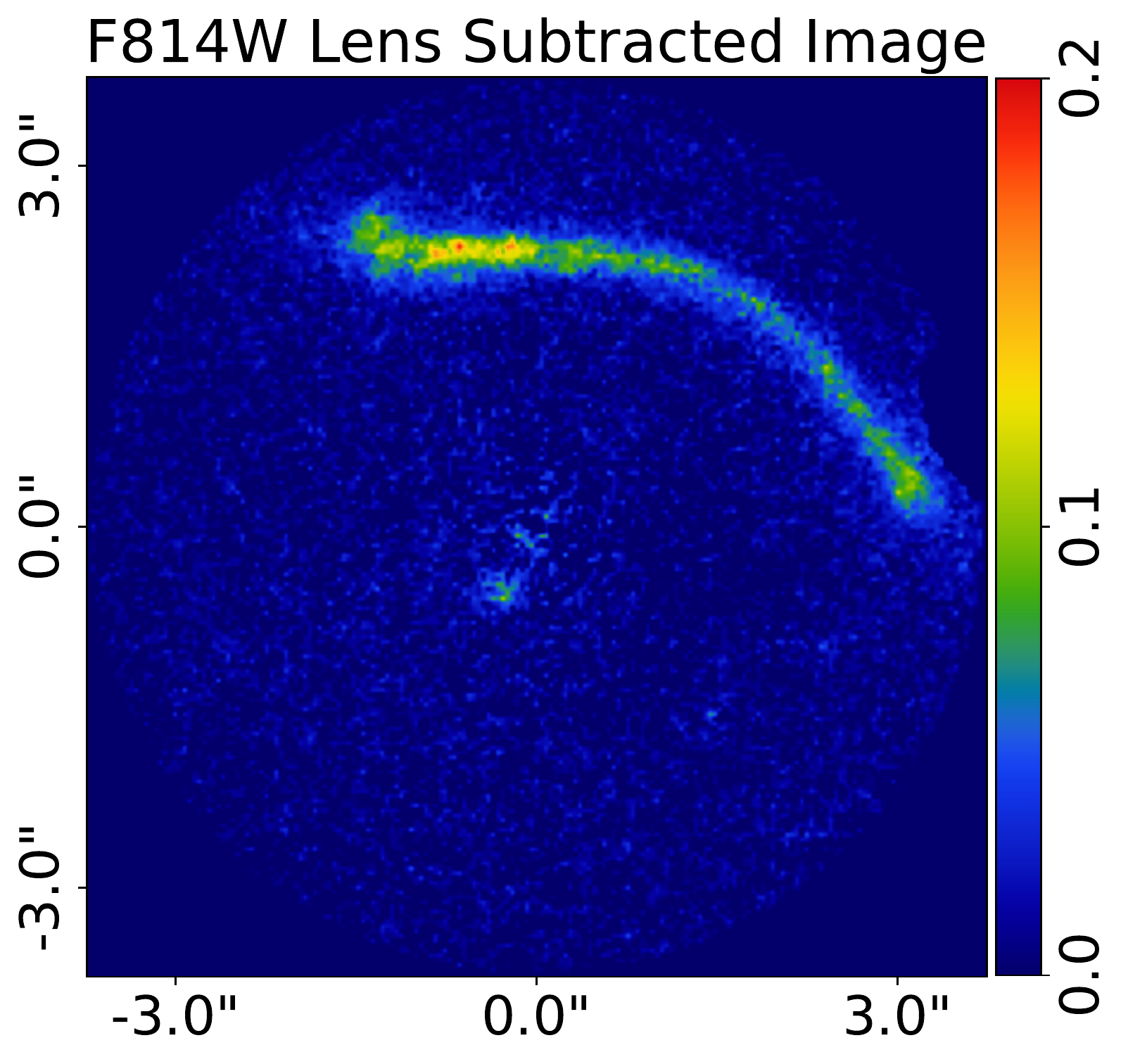}
\includegraphics[width=0.32\textwidth]{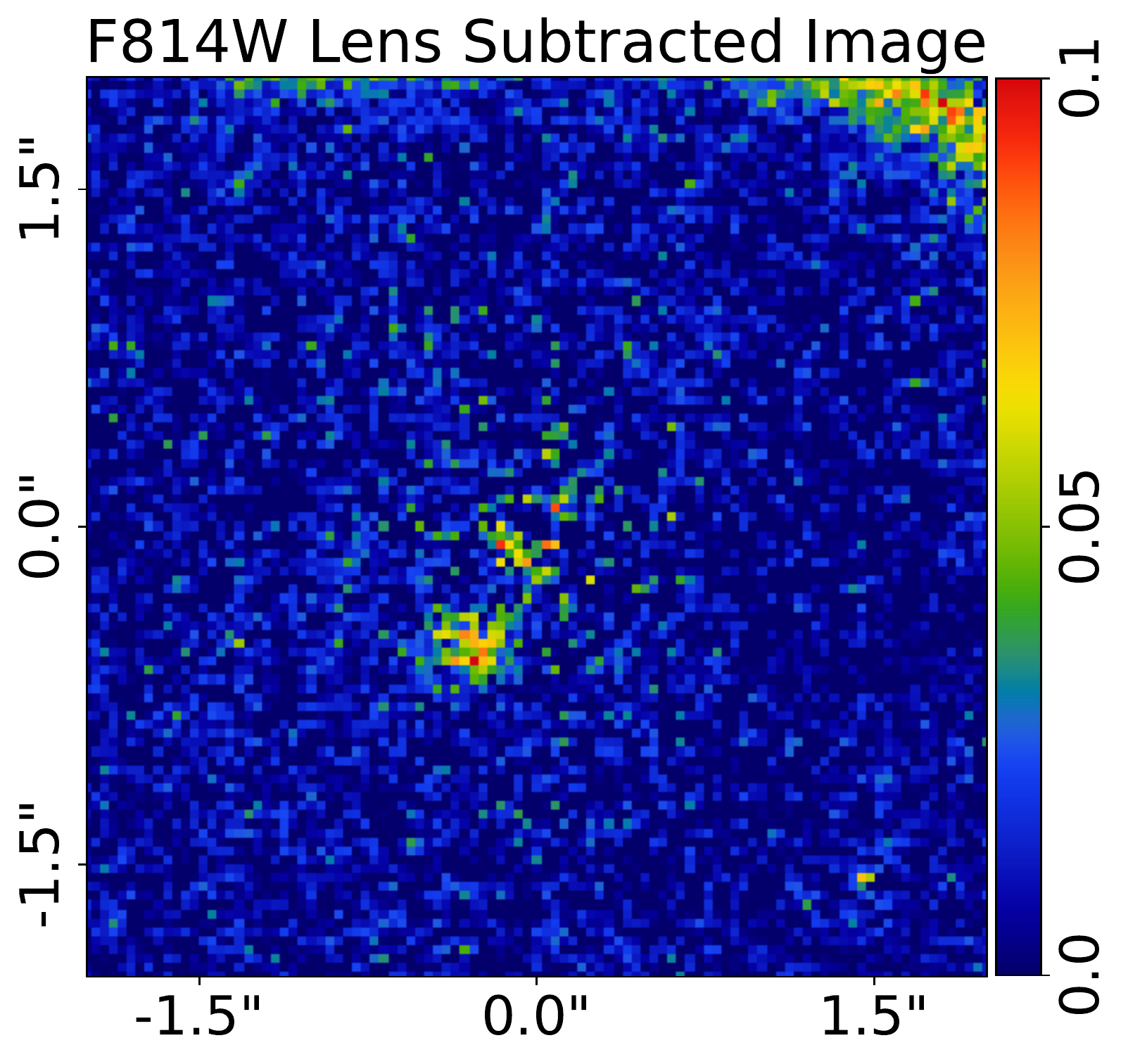}
\caption{The observed images (left column), masked and lens subtracted images (middle column) and images zoomed in on the central regions containing the counter image (right column) of Abell 1201. The top row show the HST optical image taken using the F390W filter, the bottom row shows an image taken at near infrared wavelengths using the F814W filter, which are both in units of electrons per second. The lens subtractions are performed using the highest likelihood model found for each image, however their visual appearance does not change significantly for other high likelihood models. The counter image of the giant arc can be clearly seen at both wavelengths but has much higher contrast and more clumpy structure in the bluer F390W nwaveband. In the F814W image, residuals from the lens light subtraction around the coordinates (0, 0)\,arcsec are seen; these are not a central image of the source galaxy, which would be brighter in F390W. The black star marks a line-of-sight galaxy at $z = 0.273$ which is included in certain lens models.}
\label{figure:Data}
\end{figure*}

Supermassive black holes (SMBHs) have emerged as an integral part of models of galaxy formation and evolution, owing to the tight correlation observed between SMBH mass, $M_{\rm BH}$, and host galaxy bulge velocity dispersion, bulge mass and other galaxy properties \citep{Kormendy2013, Graham2012, Bosch2016}. It is posited that a SMBH resides at the centre of every galaxy and that galaxies and SMBHs coevolve with one another from their initial formation in the early Universe \citep{Heckman2014, Smith2019}. The mass of an individual SMBH, $M_{\rm BH}$, can be measured via spatially resolved dynamics of nearby tracers such as stars and gas \citep{Davis2017a, Thater2019}. This technique has provided over $100$ measurements of $M_{\rm BH}$ which show tight correlations with other galaxy properties such as bulge luminosity or velocity dispersion \citep{Kormendy1995, Ferrarese2000, Gebhardt2000, Graham2001a}. The need for spectroscopy at high spatial resolution that resolves the SMBH's sphere of influence restricts this approach to nearby galaxies, preventing the study of how these relations evolve with redshift. Spectral fitting of active galactic nuclei \citep{Peterson2004, McLure2004, Shen2013} and reverberation mapping techniques can provide measurements of $M_{\rm BH}$ in higher redshift galaxy populations which therefore enable evolutionary studies. However, these observations necessitate that the galaxy's SMBH is actively accreting, bringing in potential selection effects. A method that can measure $M_{\rm BH}$ for non-active galaxies outside the local Universe would be highly complementary to these existing approaches. Analysing the strong gravitational lensing of background sources, acting in some specific (and perhaps rare) circumstances and configurations, might provide such a technique.

In this paper, we present a re-examination of the strong-lensing brightest cluster galaxy (BCG) in Abell 1201. A tangential gravitational arc was first identified in shallow {\it Hubble Space Telescope} (HST) WFPC2 images of this system, by \cite{Edge2003}. Compared to most cluster lenses, the arc is unusual in being formed at small projected radius from the BCG ($\sim2$\,arcsec; $\sim$5\,kpc). \cite{Edge2003} found that a high ellipticity and/or strong external shear was necessary to match the arc shape. Integral-field spectroscopic data later revealed a faint counter-image to the main arc, projected even closer to the lens centre ($\sim$0.3\,arcsec; $\sim$1\,kpc) \citep{Smith2017a}. Using a simplified position-based model of the lensing configuration, \citep{Smith2017a} argued that an additional mass of $\sim$10$^{10}$\,M$_\odot$ at small radius was necessary to reproduce the counter-image as observed. The spatially-resolved stellar kinematics support this conclusion \citep{Smith2017}. The authors concluded that the necessary central mass could be a SMBH, but with the limited imaging data available, and the rudimentary lensing analysis employed, a degeneracy with the inner stellar mass distribution of the lens could not be excluded. 

\begin{figure*}
\centering
\includegraphics[width=0.24\textwidth]{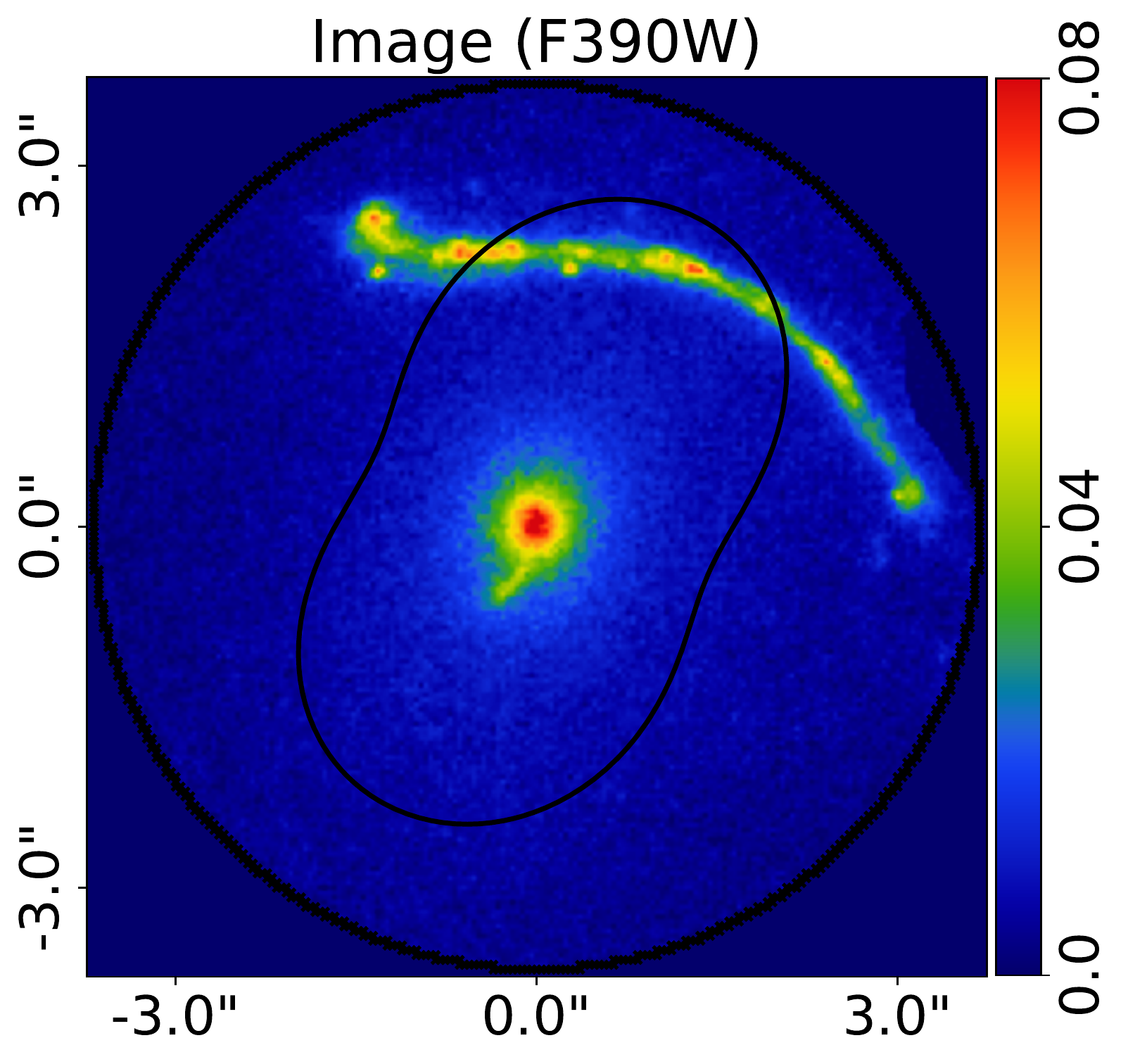}
\includegraphics[width=0.24\textwidth]{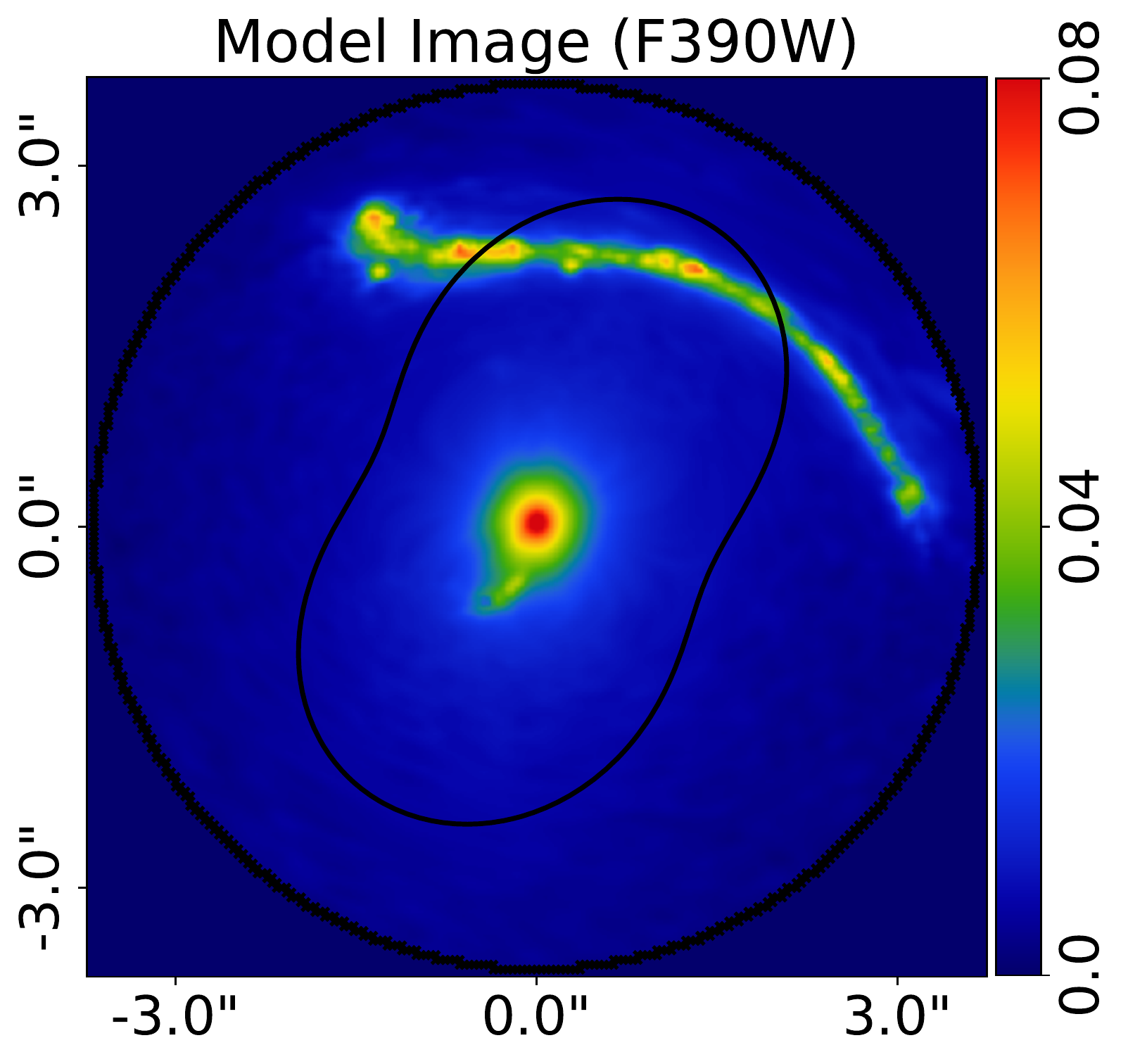}
\includegraphics[width=0.24\textwidth]{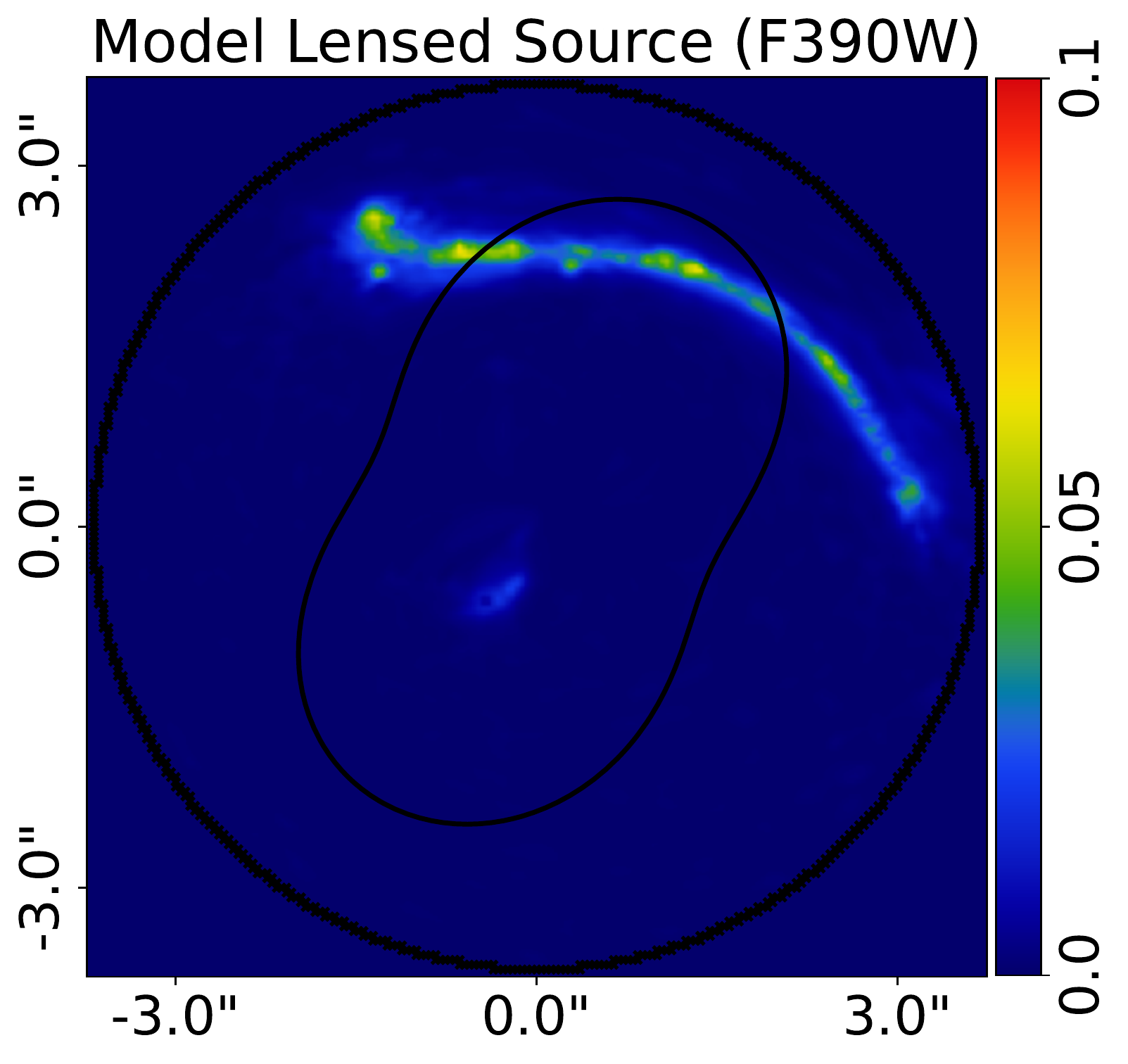}
\includegraphics[width=0.24\textwidth]{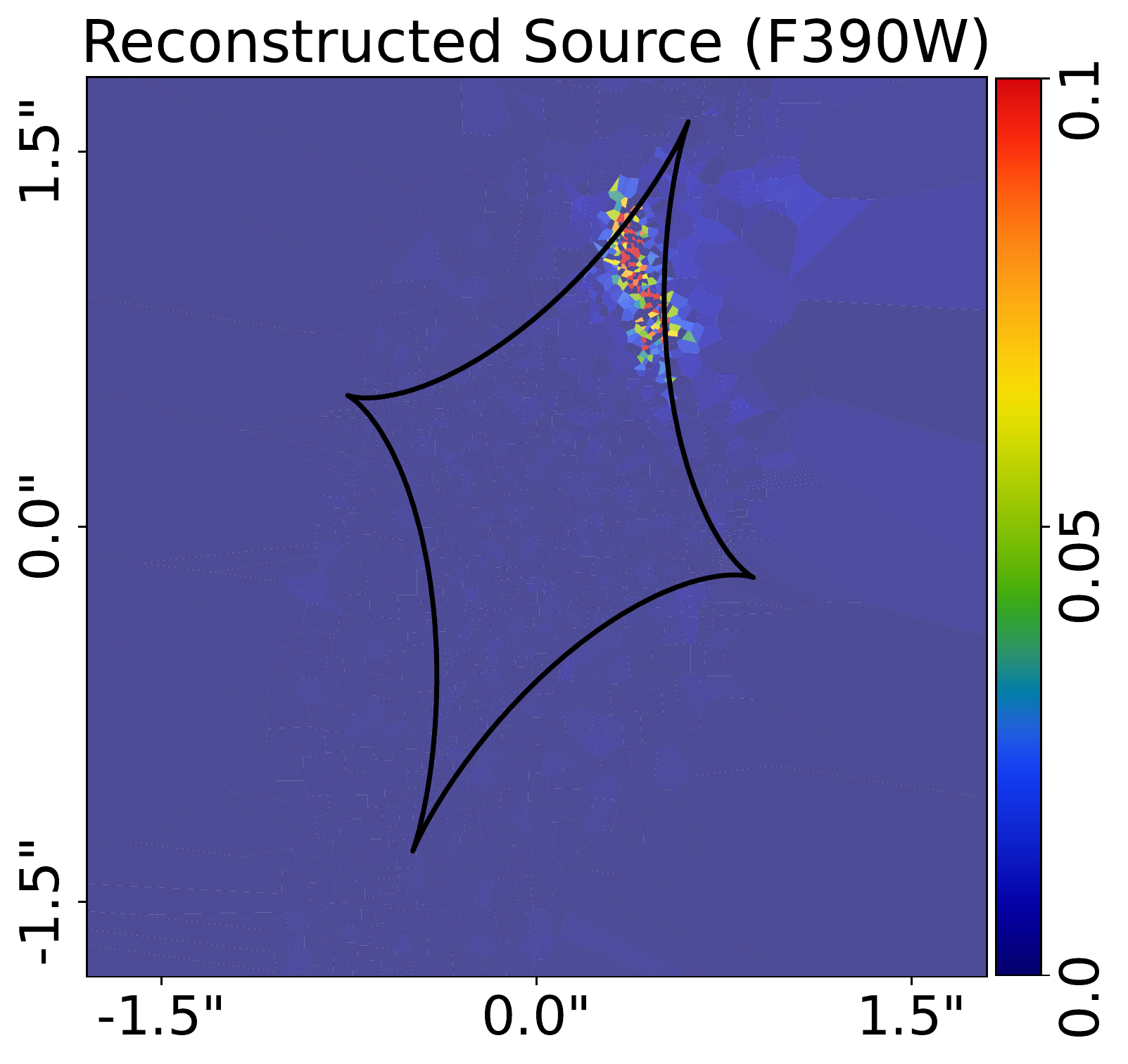}
\includegraphics[width=0.24\textwidth]{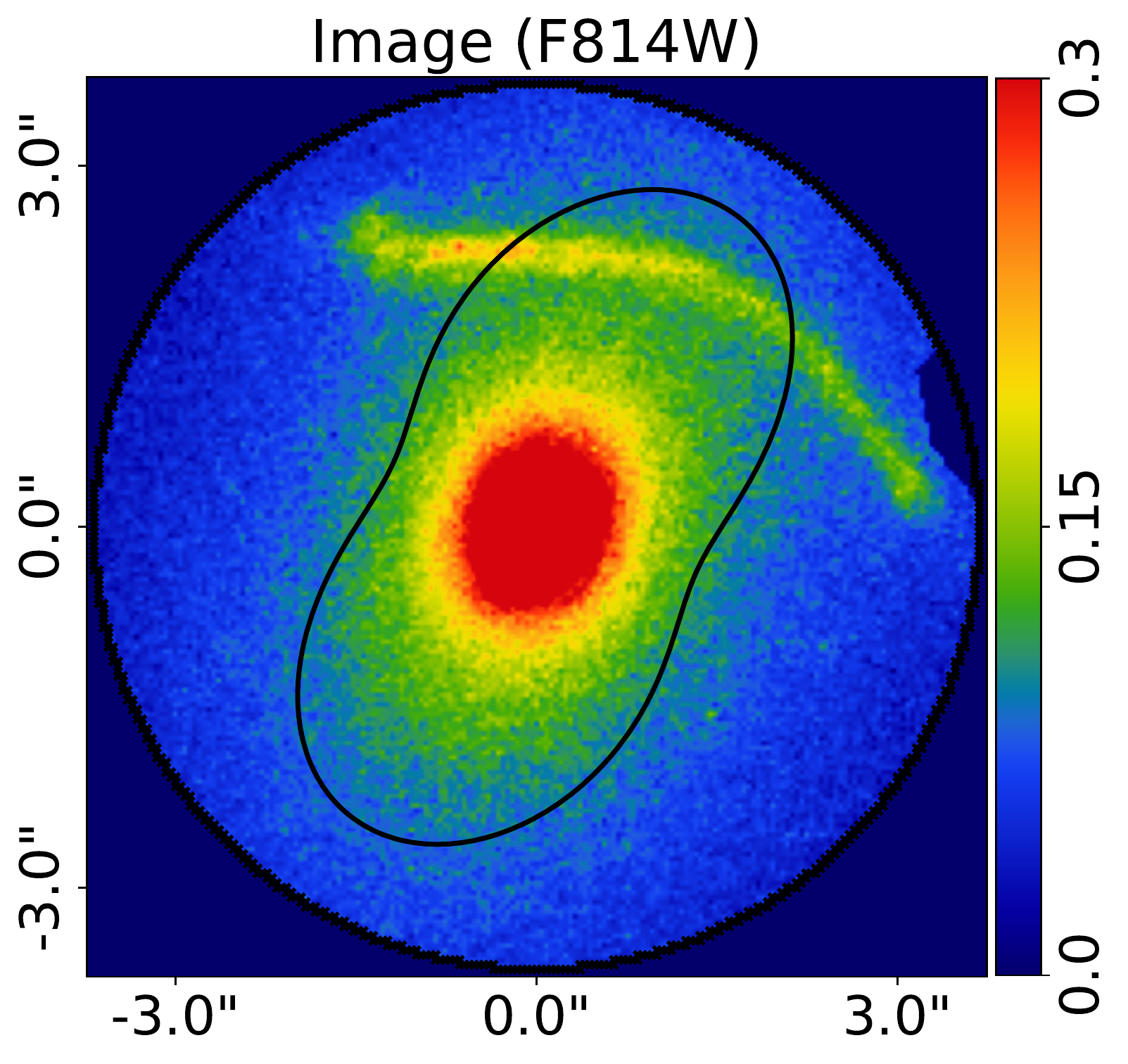}
\includegraphics[width=0.24\textwidth]{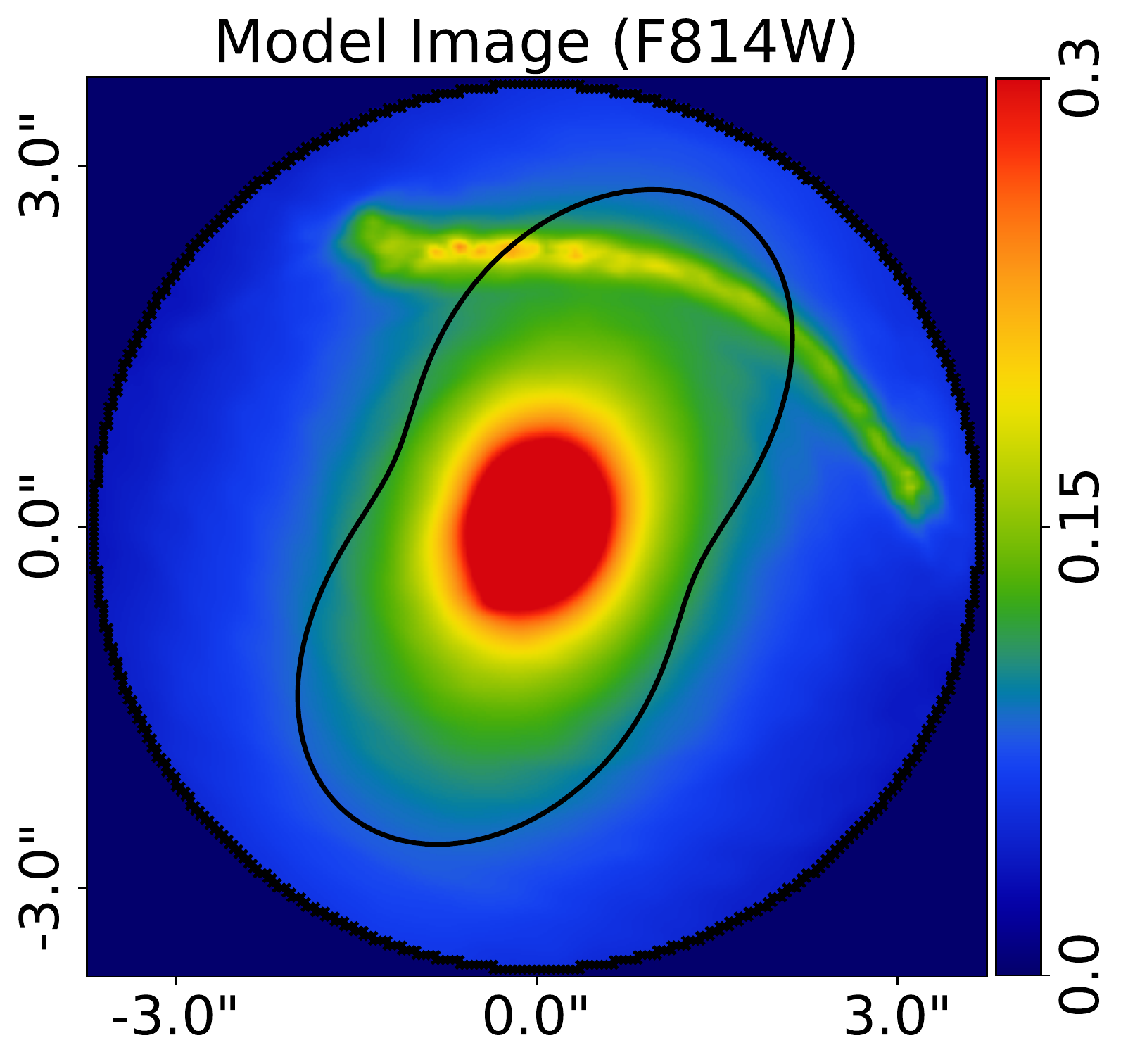}
\includegraphics[width=0.24\textwidth]{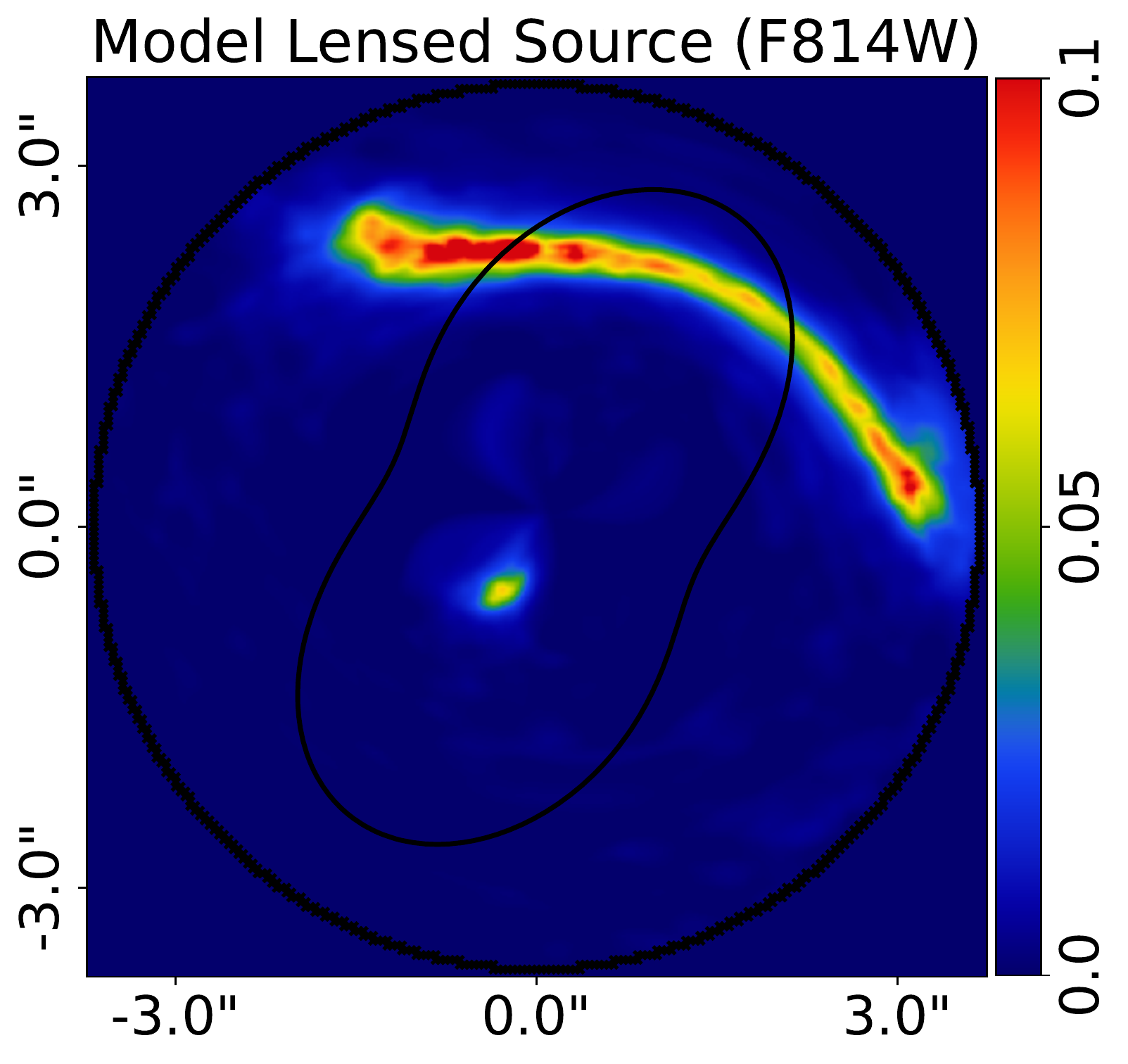}
\includegraphics[width=0.24\textwidth]{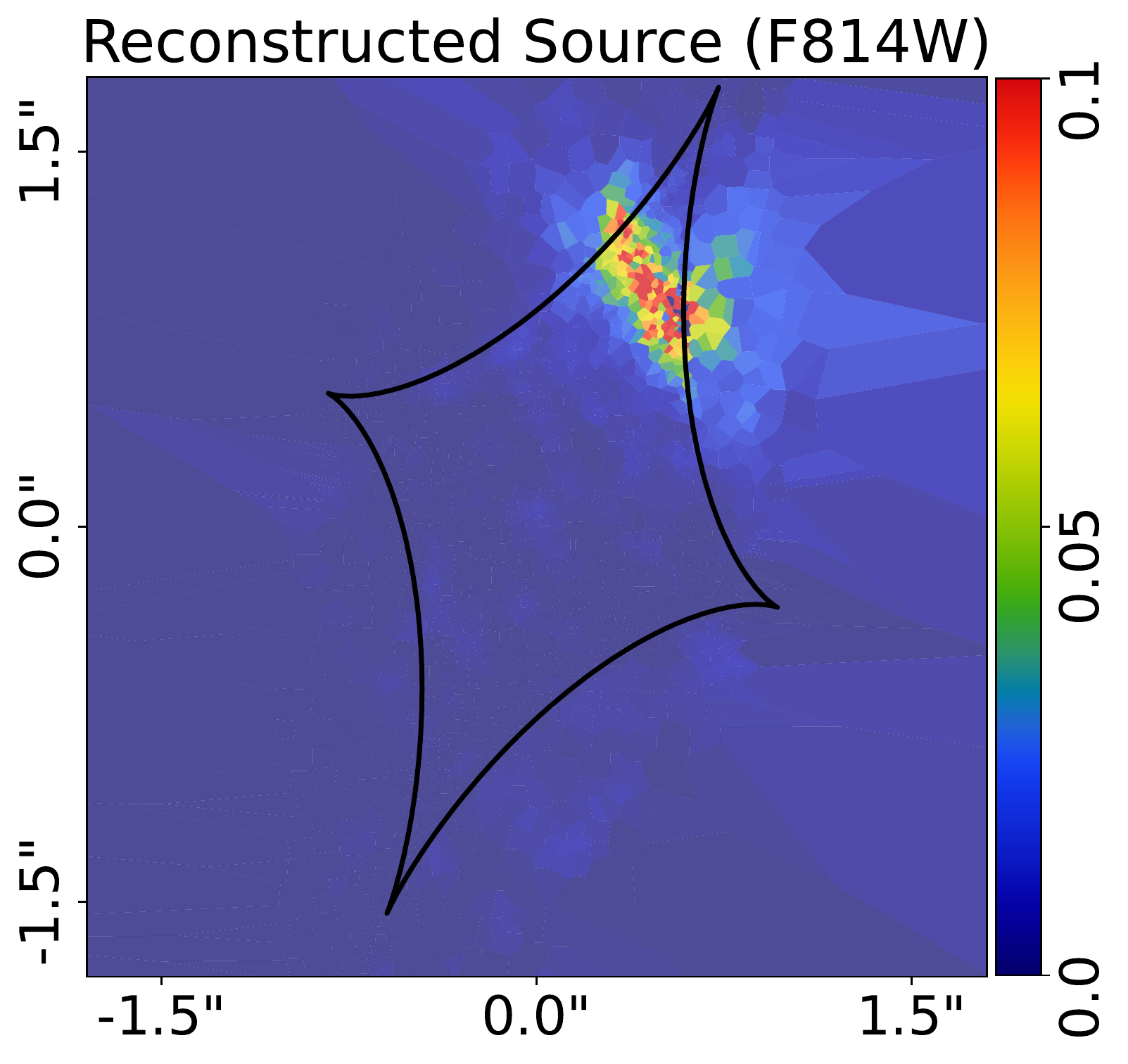}
\caption{Fits to HST imaging of Abell 1201 via {\tt PyAutoLens}. The observed data (left column), the image-plane model images of the lens and source galaxies (left-centre column), the lensed source only (right-centre column) and source-plane pixelized source reconstruction (right column) are shown. The top row shows fits to the F390W and bottom row the F814W wavebands respectively. All images are in units of electrons per second. The lens model is the maximum likelihood model inferred at the end of the first SLaM (see \cref{SLAM}) pipeline run, which produces a lens subtracted image. The black lines show the mass model's tangential critical curve for all panels in the image plane (central columns) and the tangential caustic for panels in the source plane (right hand column).}
\label{figure:ModelFit}
\end{figure*}


Here, we analyse new HST WFC3/UVIS imaging of higher spatial resolution and greater signal-to-noise ratio, using advanced lens modelling techniques, to reassess the evidence for a lensing-detected SMBH in Abell 1201. We show that the detailed structure observed in the 
counter image constrains the inner mass distribution of the lens, and allows us to place constraints on the central SMBH. We perform a Bayesian model comparison of a variety of lens models that include and omit a point-mass representing a SMBH. The majority of models favour the inclusion of a SMBH and produce consistent estimates of $M_{\rm BH}$, with some dependence on the form and flexibility of the assumed lens galaxy mass model. This work marks the second observation of a strong lens that provides constraints on the SMBH at the centre of its lens galaxy, following the work of \citet{Winn2003} who detected the ``central'' image of a lensed source via radio observations. Our study is the first where a measurement of $M_{\rm BH}$ is inferred via strong lensing (as opposed to an upper limit) and does so without the rare observation of a central image. 

Over the next decade, of order one-hundred thousand strong lenses will be discovered by cosmological surveys such as Euclid, LSST and SKA \citep{Collett2015}, a three orders of magnitude increase over the hundreds of systems that are currently known \citep{Bolton2008a, Shu2016, Sonnenfeld2013b, Bolton2012, Shu2016}. This will naturally lead to the discovery of more exotic and peculiar strong lens systems \citep{OrbandeXivry2008}, whose rare lensing configurations may provide constraints on $M_{\rm BH}$. We conclude with a discussion of whether galaxy-scale strong lensing can become a viable technique to measure large samples of SMBH masses in the future\footnote{For other lensing-related techniques, see also \citet{EventHor} for a measurement of the SMBH at the centre of M87 by mapping the lensed shadow of surrounding gas, \citet{Banik2019} for a discussion of using strong lensing to detect intermediate mass black holes, \citet{Chen2018} and \citet{Mahler2022} for discussions of searching for wandering SMBHs in strong lensing galaxy clusters and \citet{Hezaveh2015, Tamura2015, Wong2017, Quinn2016} for studies based around strong lens central images.}

This paper is structured as follows.
In Section~\ref{Data}, we describe the Hubble Space Telescope imaging of Abell 1201.
In Section~\ref{Method}, we describe the {\tt PyAutoLens} method and model fits performed in this work.
In Section~\ref{Results}, we present the results of model fits using a variety of lens models.
In Section~\ref{Discussion}, we discuss the implications of our measurements, and we give a summary in \S\ref{Summary}.
{We assume a Planck 2015 cosmology throughout \citep{PlanckCollaboration2015a}}.
Text files and images of every model-fit performed in this work are available at \url{https://github.com/Jammy2211/autolens_abell_1201}. Full \texttt{dynesty} chains of every fit are available at \url{https://doi.org/10.5281/zenodo.7695438}.

\section{Data}\label{Data}

We acquired the new {\it HST} imaging of Abell 1201 in Programme 14886 using the Ultraviolet and VISible channel on the Wide Field Camera 3 (WFC3/UVIS). A total of five exposures with a total integration time of 7150\,s were taken in the F390W bandpass, tracing the clumpy rest-frame ultraviolet emission from star-forming regions in the source galaxy. This filter probes wavelengths shorter than the 4000-\AA\ break at the redshift of Abell 1201; hence the foreground light of the lens is suppressed and the contrast of the source enhanced. Additionally, we acquired three exposures in F814W, totalling 1009\,s, to trace the distribution of stellar mass in the BCG. The observatory-provided reduced single-exposure images were registered and combined using {\sc astrodrizzle}, projecting onto an output pixel scale of 0.04\,arcsec. An accurate estimate of the point-spread function (PSF) is required for the lens modelling. To this end, we employed  the empirical PSF provided by STScI \footnote{\url{
https://www.stsci.edu/hst/instrumentation/wfc3/data-analysis/psf}}, as appropriate to the position of the target in each individual exposure, and propagated the PSF images through the same stacking process as for the real observation. The final combined images of Abell 1201 in the two bandpasses are shown in \cref{figure:Data}.

\cref{figure:Data} shows the 
F390W and F814W imaging, 
alongside lens-subtracted versions which highlight the lensed source galaxy. There is a giant arc 2.0--3.0\,arcsec away from the lens galaxy on one side of the lens with a counter image just $\sim$0.3\,arcsec ($\sim 0.9\,$kpc) from the lens galaxy centre. The lens itself is a cD galaxy residing in the central regions of a galaxy cluster, in contrast to most galaxy-scale (e.g. Einstein radius $< 5.0$\,arcsec) strong lens systems which are massive elliptical field galaxies and not in a cluster environment. 

The cluster Abell 1201 has also been investigated. X-ray analysis reveals an offset gas core 500 kpc northwest of the lens \citep{Ma2012}, which is interpreted as a tail of gas stripped from the offset core. The gas has different a different density, entropy and temperature than gas in the surrounding area, providing evidence indicative of a minor merger at second core passage. Alignment between the mass distribution of Abell 1201's BCG mass distribution (inferred via lens modeling performed by \citealt{Edge2003}) and the offset core is also noted, which could be the result of a sloshing mechanism.        
\section{Method}\label{Method}

\subsection{Overview}

We use version \texttt{2022.03.30.1} of the lens modeling software {\tt PyAutoLens}\github{https://github.com/Jammy2211/PyAutoLens} \citep{Nightingale2021}. {\tt PyAutoLens} fits the lens galaxy's light and mass and the source galaxy simultaneously. The method assumes a model for the lens's foreground light (e.g. one or more Sersic profiles), which is convolved with the instrumental PSF and subtracted from the observed image. A mass model (e.g. an isothermal mass distribution) ray-traces image-pixels from the image-plane to the source-plane and a pixelized source reconstruction, using an adaptive Voronoi mesh, is performed. \cref{figure:ModelFit} provides an overview of a {\tt PyAutoLens} lens model, where models of Abell 1201 for the image-plane lens galaxy emission and lensed source are shown alongside the source-plane source reconstruction.

By fitting the source's extended surface brightness distribution, {\tt PyAutoLens} considers light rays emanating from different parts of the source, therefore constraining different regions of the lens galaxy's potential. If a small-mass clump is near the lensed source's emission, it may cause observable distortions to one or more of its multiple images. This technique has provided detections of three non-luminous dark matter substructures \citep{Vegetti2010, Vegetti2012, Hezaveh2016, Nightingale2022} in strong lenses, where their presence is inferred by how they perturb the appearance of the lensed images. 

\begin{figure}
\centering
\includegraphics[width=0.235\textwidth]{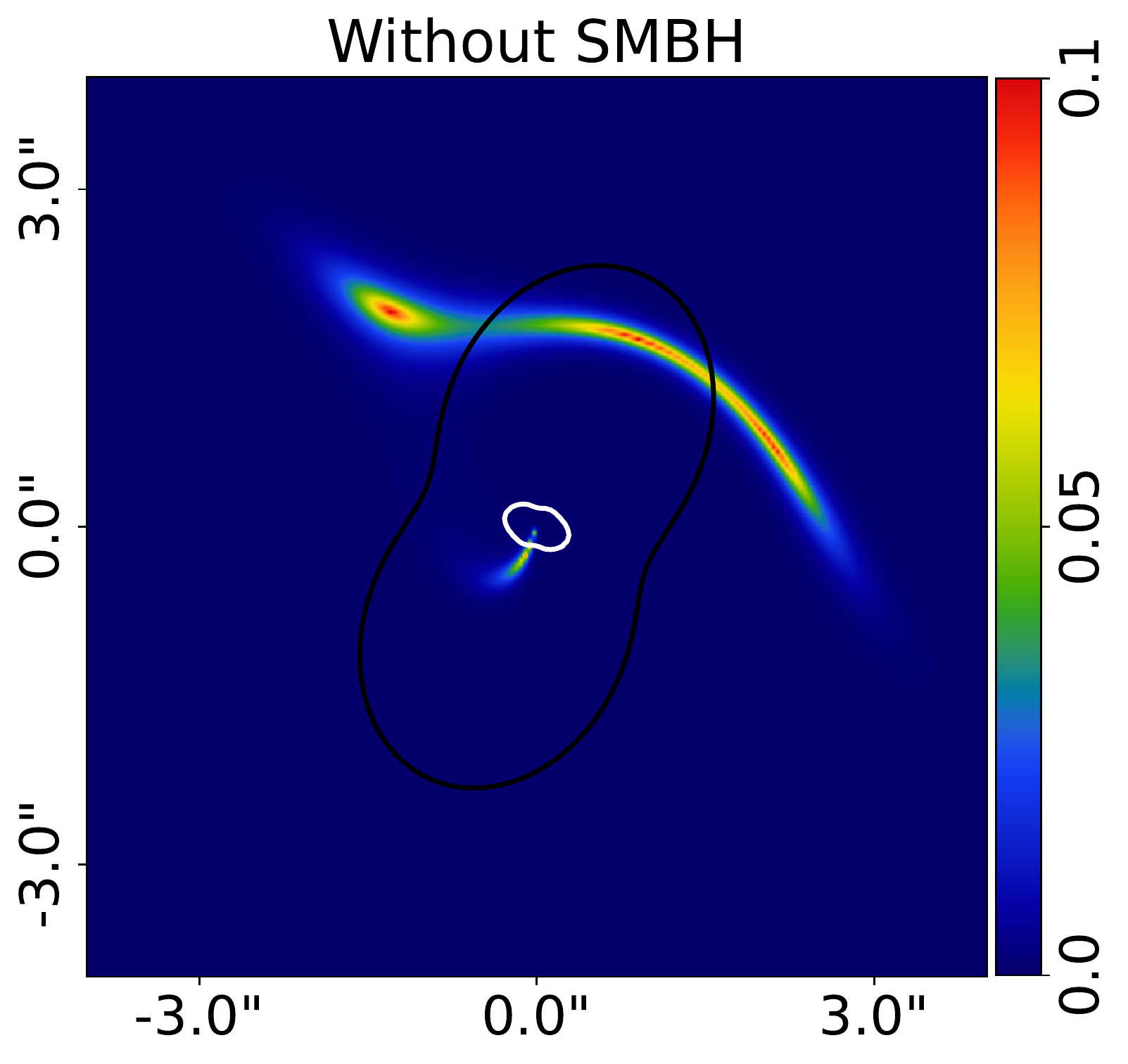}
\includegraphics[width=0.235\textwidth]{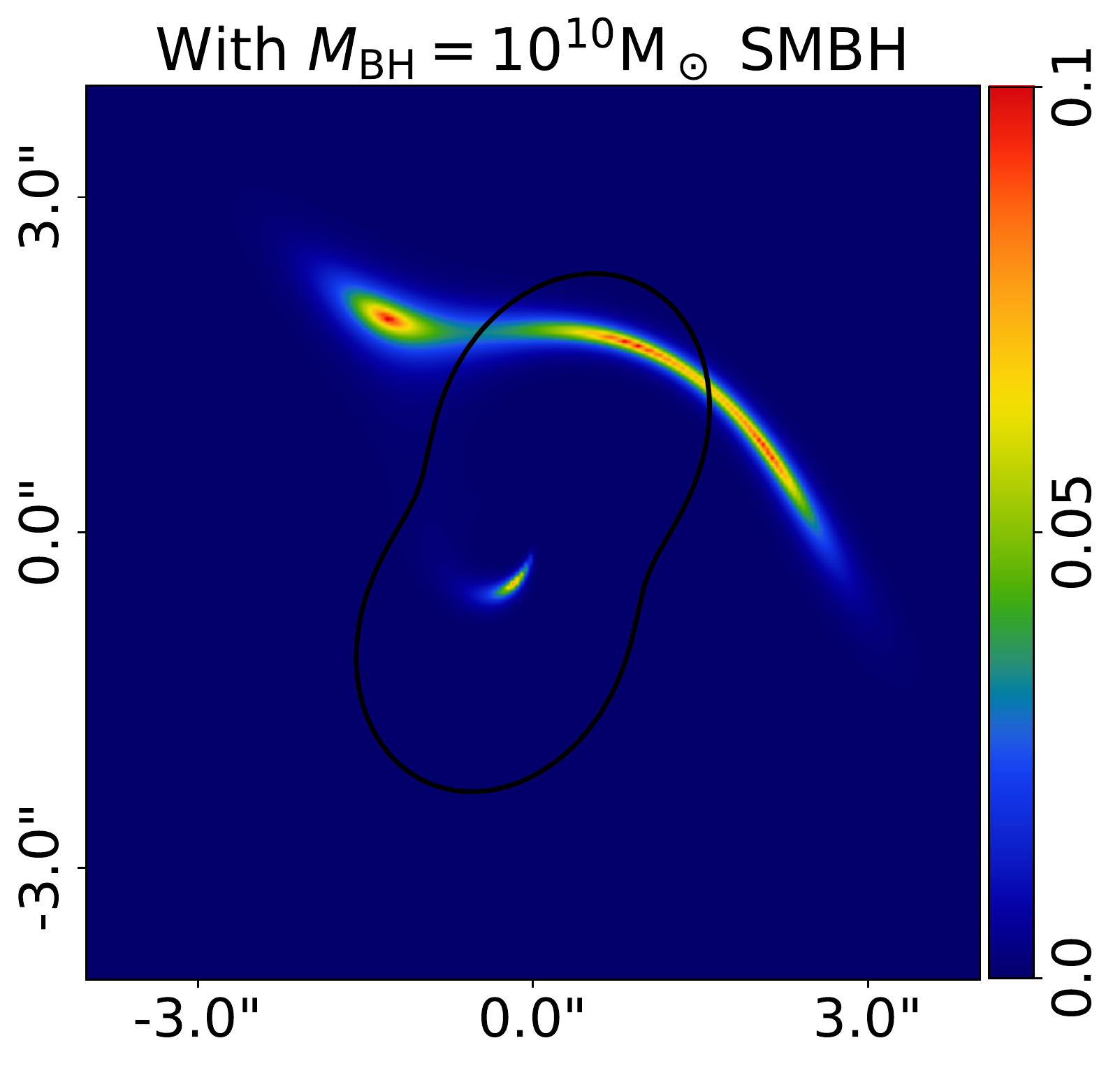}
\includegraphics[width=0.235\textwidth]{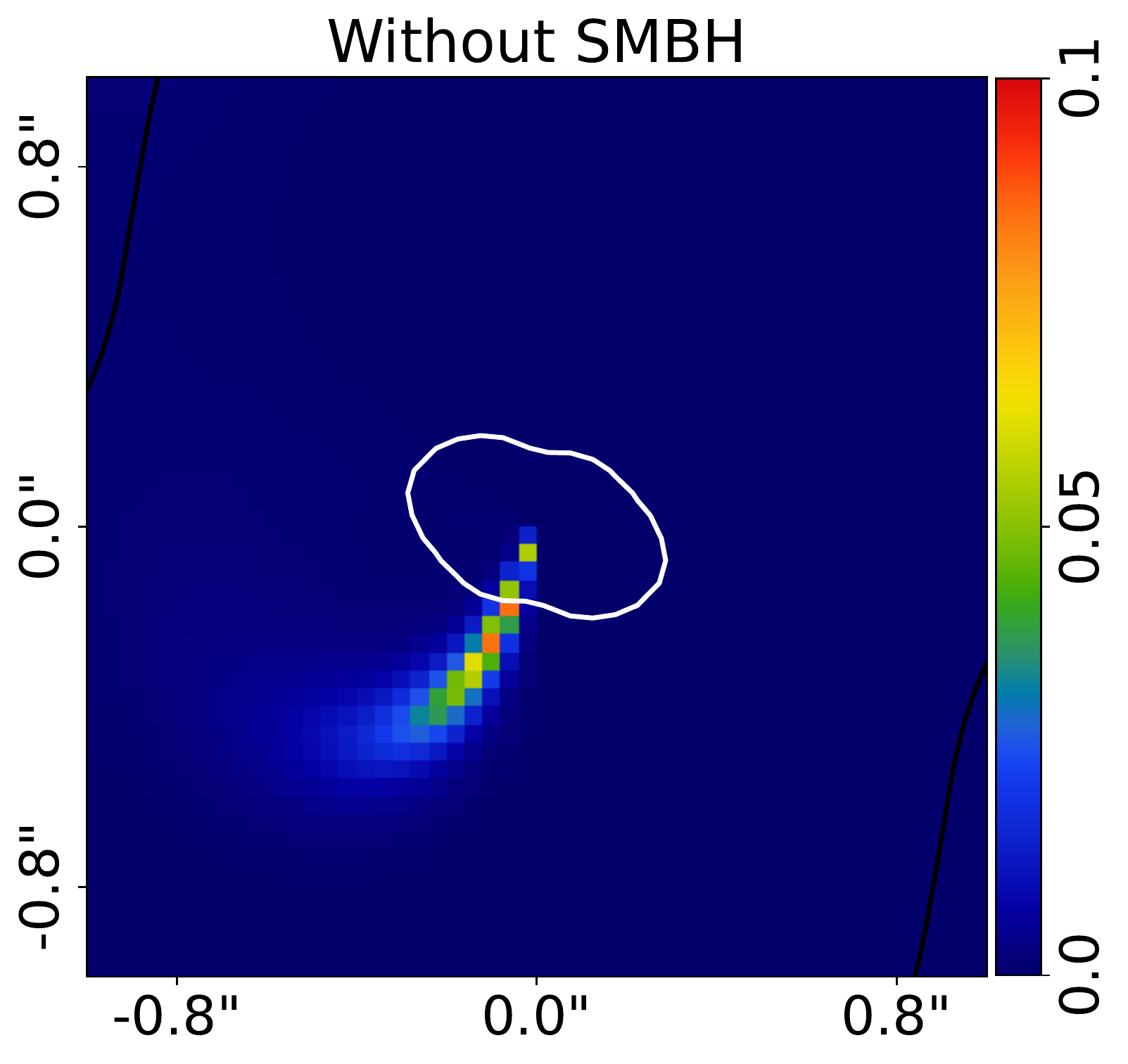}
\includegraphics[width=0.235\textwidth]{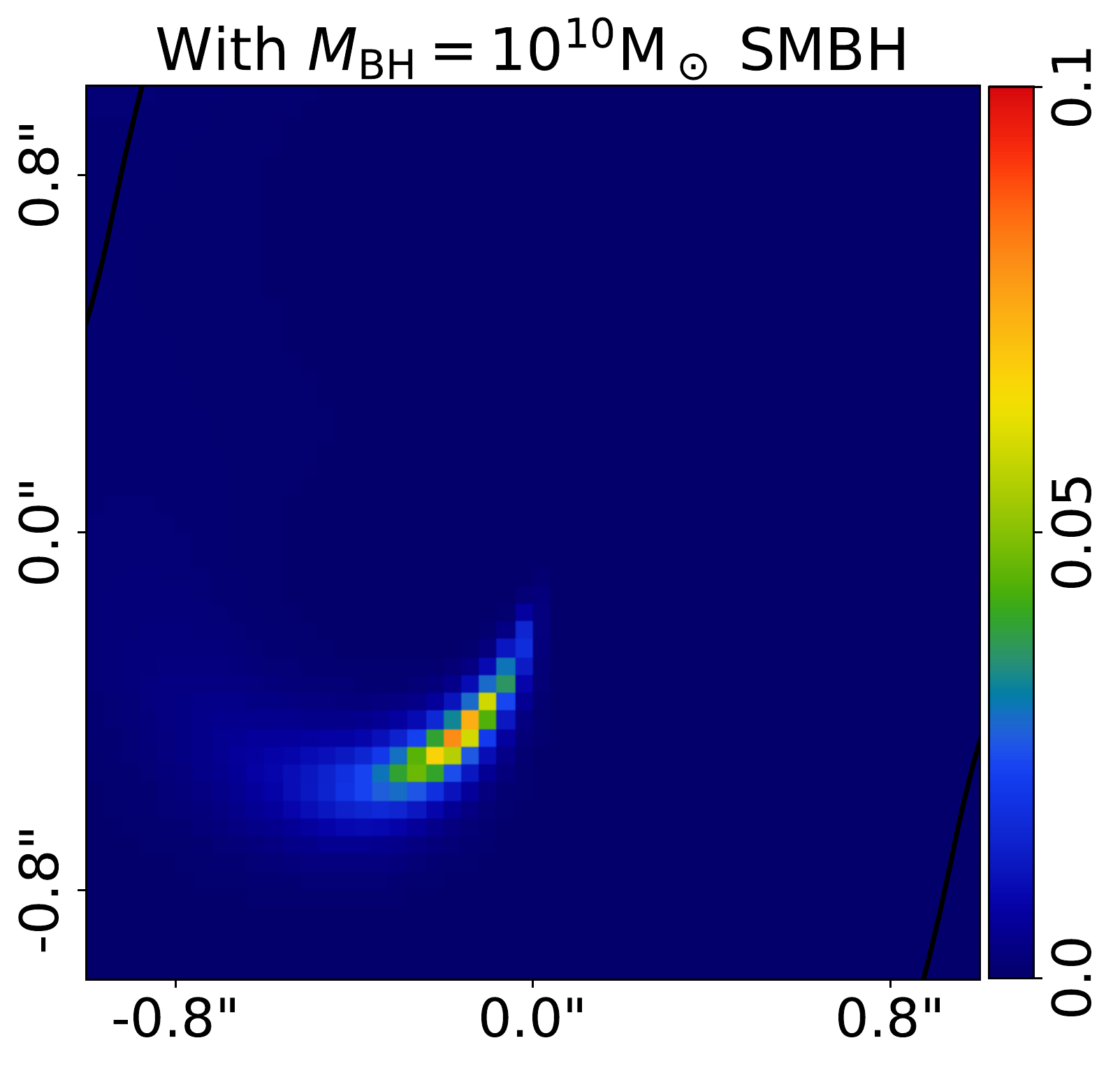}
\caption{
Illustration of how Abell 1201's lens configuration is sensitive to the lens galaxy's SMBH. Both images on the top row are simulated using the same lens mass model (a power-law with external shear) and source galaxy light model (an elliptical Sersic). In the left panel a SMBH is not included, whereas in the right panel a $M_{\rm BH} = 10^{10}$\,M$_{\rm \odot}$ SMBH is included at $(0.0, 0.0)$, which is marked with a black cross. The bottom row shows a zoom in on the counter image. The SMBH changes the location, appearance and brightness of the counter image but does not lead to visible changes in the giant arc. The tangential critical curve is shown by a black line and radial critical curve a white line. The latter does not form for sufficiently steep mass profiles \citep{Kochanek2004a}, including the model with a SMBH shown here. 
} 
\label{figure:SMBHDemo}
\end{figure}

\begin{table}
\tiny
\resizebox{\linewidth}{!}{
\begin{tabular}{ l | l | l } 
\multicolumn{1}{p{1.3cm}|}{\centering Number of \\ Sersics} 
& \multicolumn{1}{p{1.8cm}|}{Aligned Elliptical Components} 
& \multicolumn{1}{p{1.8cm}|}{Aligned Centres} 
\\ \hline
 1 & N/A & N/A \\[2pt]
2 & \checkmark & \checkmark   \\[2pt]
2 & \ding{55} & \checkmark \\[2pt]
2 & \ding{55} & \ding{55}  \\[2pt]
3 & \ding{55} & \ding{55} \\[2pt]
\end{tabular}
}
\caption{The five models for the lens's light that are fitted and compared in the Light pipeline. The lens light model assumes either one, two or three Sersic profiles and makes different assumptions as to whether their $(x,y)$ centre and elliptical components ($\epsilon_{\rm 1}$, $\epsilon_{\rm 2}$) are aligned. A tick mark indicates that this assumption is used in the model, for example the second row is a model where both the elliptical components and centres are aligned.}
\label{table:LightModels}
\end{table}

\begin{table*}
\tiny
\resizebox{\linewidth}{!}{
\begin{tabular}{ l | l | l | l l l l l l} 
\multicolumn{1}{p{1.8cm}|}{\centering \textbf{Model}} 
& \multicolumn{1}{p{1.35cm}|}{\centering \textbf{Component}} 
& \multicolumn{1}{p{1.1cm}|}{\centering \textbf{Represents}} 
& \multicolumn{1}{p{1.5cm}}{\textbf{Parameters}} 
& \multicolumn{1}{p{1.5cm}}{} 
\\ \hline
& & & & & & \\[-4pt]
\textbf{Point Mass} & Mass & Black Hole & $\theta^{\rm smbh}_{\rm E}$: Einstein Radius $(\arcsec)$ \\[2pt]
\hline
& & & & & & \\[-4pt]
\textbf{Sersic} & Light + & Stellar Matter & ($x$,$y$): centre $(\arcsec)$ & ($\epsilon_{\rm 1}$, $\epsilon_{\rm 2}$): elliptical components  \\[2pt]
               &  Mass     & (Bulge, Disk, & $I$: Intensity & $R$: Effective Radius $(\arcsec)$\\[2pt]
              & & Envelope) & $n$: Sersic index &  \\[2pt]
               & & & $\Psi$: Mass-to-Light Ratio (e$^{\rm -}$\,s$^{\rm -1}$) & $\Gamma$: Radial Gradient & \\[2pt]
               \hline
& & & & & & \\[-4pt]
\textbf{Elliptical NFW} & Mass & Dark Matter & ($x^{\rm dark}$,$y^{\rm dark}$): centre $(\arcsec)$ & ($\epsilon_{\rm 1}^{\rm dark}$, $\epsilon_{\rm 2}^{\rm dark}$): elliptical components  & \\[2pt]
            &   &   & $M_{\rm 200}^{\rm dark}$: Mass at 200 ($M_{\rm \odot}$) \\[2pt]
\hline
& & & & & & \\[-4pt]
\textbf{Shear} & Mass & Line-of-sight & ($\epsilon_{\rm 1}^{\rm ext}$, $\epsilon_{\rm 2}^{\rm ext}$): elliptical components  \\[2pt] 
\hline
& & & & & & \\[-4pt]
\textbf{Elliptical} & Mass & Total (Stellar  & ($x^{\rm mass}$,$y^{\rm mass}$): centre $(\arcsec)$ & ($\epsilon^{\rm mass}_{\rm 1}$, $\epsilon^{\rm mass}_{\rm 2}$): elliptical components  \\[2pt]
\textbf{Power-Law (PL)} &      & + Dark Matter) & $\theta^{\rm mass}_{\rm  E}$: Einstein Radius $(\arcsec)$ & $\gamma^{\rm mass}$: density slope \\[2pt]
\hline
& & & & & & \\[-4pt]
\textbf{Broken} & Mass & Total (Stellar  & ($x^{\rm mass}$,$y^{\rm mass}$): centre $(\arcsec)$ & ($\epsilon^{\rm mass}_{\rm 1}$, $\epsilon^{\rm mass}_{\rm 2}$): elliptical components  \\[2pt]
 \textbf{Power-Law (BPL)} &      & + Dark Matter) & $\theta^{\rm mass}_{\rm E}$: Einstein Radius $(\arcsec)$ & $t^{\rm mass}_{\rm 1}$: inner density slope \\[2pt]
         &      &                & $r^{\rm mass}_{\rm B}$: Break Radius $(\arcsec)$ & $t^{\rm mass}_{\rm 2}$: outer density slope \\[2pt]
\end{tabular}
}
\caption{The light and mass profiles used in this work. Column 1 gives the model name. Column 2 whether it models the lens's light, mass or both. Column 3 states what component of mass it represents. Column 4 gives its associated parameters and units.}
\label{table:Models}
\end{table*}

This work uses the same technique, albeit we are in this case investigating whether the perturbing 
effects of the 
central SMBH 
are detected in the lensed source emission. This is why the proximity of Abell 1201's counter-image to the lens galaxy's centre, and therefore SMBH, is so important. A high mass SMBH will induce a local perturbation to the counter image's appearance that does not produce a significant change in the appearance of the other multiple images of source in the giant arc. This is shown in \cref{figure:SMBHDemo}, where two simulated lenses based on our models of Abell 1201 are shown. In the right panel, a $M_{\rm BH} = 10^{10}$\,M$_{\rm \odot}$ SMBH is added to the lens model, which changes the location, appearance and brightness of the counter image without producing a visible change in the giant arc. Our results are therefore not based on whether the source forms a central image \citep{Winn2003, Rusin2005} \footnote{By central image, we are referring to the hypothetical third or fifth image that would form directly over the centre of the lens galaxy, provided its mass distribution were sufficiently cored. We therefore do not consider the counter image located ~0.3 arcsec to the southwest of the lens galaxy a central image, and will always refer to it as the counter-image.}.


At the heart of the {\tt PyAutoLens} model fitting process is the computation of the likelihood function.  We provide a brief description of this calculation in \cref{Coordinates} - \cref{Source} below. Furthermore, to assist readers less familiar with strong lens modelling, we provide Jupyter notebooks providing a visual step-by-step guide, including URL links to previous literature and explanations of technical aspects of the linear algebra and Bayesian inference. The notebooks can be found at the following link: \url{https://github.com/Jammy2211/autolens_likelihood_function}.

Recent works using {\tt PyAutoLens} include modeling strong lenses simulated using stellar dynamics models \citep{Cao2021} and via a cosmological simulation \cite{He2023}, an automated analysis of $59$ lenses \citep{Etherington2022, Etherington2022a} and studies of dark matter substructure \citep{He2022a, He2022b, Amorisco2022}. 

\subsection{Coordinate System}\label{Coordinates}

Light and mass profile quantities are computed using elliptical coordinates $\xi = \sqrt{{x}^2 + y^2/q^2}$, with minor to major axis-ratio $q$ and position angle $\phi$ defined counter clockwise from the positive x-axis. For model-fitting, these are parameterized as two components of ellipticity
\begin{equation}
\epsilon_{1} =\frac{1-q}{1+q} \sin 2\phi, \,\,
\epsilon_{2} =\frac{1-q}{1+q} \cos 2\phi.    
\label{eq: ellip}
\end{equation}
To convert parameters from arc-second units to stellar masses we require the critical surface mass density
\begin{equation}
    \Sigma_\mathrm{crit}=\frac{{\rm c}^2}{4{\rm \pi} {\rm G}}\frac{D_{\rm s}}{D_{\rm l} D_{\rm ls}},
    \label{eq: sigma crit}
\end{equation}
where $D_{\rm l}$, $D_{\rm s}$, and $D_{\rm ls}$ are respectively the angular diameter distances to the lens, to the source, and from the lens to the source, and ${\rm c}$ is the speed of light.


\subsection{Lens Light Models}\label{LensLightModel}

The lens light profile intensities $I$ are computed using one or more elliptical Sersic profiles \citep{Sersic1968}
\begin{equation}
\label{eqn:Sersic}
I_{\rm  Ser} (\xi_{\rm l}) = I \exp \bigg\{ -k \bigg[ \bigg( \frac{\xi}{R} \bigg)^{\frac{1}{n}} - 1 \bigg] \bigg\} ,
\end{equation}
which has seven free parameters: $(x,\,y)$, the light centre, $(\epsilon_{\rm 1},\,\epsilon_{\rm 2})$ the elliptical components, $I$, the intensity at the effective radius $R$ and $n$, the Sersic index. $k$ is a function of $n$ \citep{Trujillo2004}. These parameters are given superscripts depending on which component of the lens galaxy they are modeling, for example the sersic index of the bulge component is $n^{\rm bulge}$. Models with multiple light profiles are evaluated by summing each individual component’s intensities. Up to three light profiles are fitted to the lens galaxy representing a bulge, bulge + disk or bulge + disk + envelope, with their superscript matching these terms. 
The Sersic profile 
intensities are computed using an adaptive oversampling routine which computes all values to a fractional accuracy of $99.99\%$.

Bayesian model comparison is used to determine the light model complexity, from the five models listed in \cref{table:LightModels}. These models assume one, two or three Sersic profiles and make different assumptions for how their centres and elliptical components are aligned. Model comparison is performed separately for the F390W and F814W images. \cref{LensModels} provides the priors of every Sersic profile parameter assumed in this study.

\subsection{Lens Mass Models}

This work fits a variety of lens galaxy mass models, which are summarised in \cref{table:Models}. We fit \textit{decomposed} mass models, where the light profile(s) that represent the lens's light are translated to stellar density profiles (via a mass-to-light profile) to perform ray-tracing \citep{Nightingale2019}.

The lens's light and stellar mass are modeled as a sum of Sersic profiles, where the Sersic profile given by \cref{eqn:Sersic} is used to give the light matter surface density profile
\begin{equation}
\label{eqn:Sersickap}
\kappa_{Ser} (\xi) = \Psi \bigg[\frac{q  \xi}{R}\bigg]^{\Gamma} I_{Ser} (\xi_l) \, \, ,
\end{equation}
where $\Psi$ gives the mass-to-light ratio in electrons per seconds (the units of the HST imaging) and $\Gamma$ folds a radial dependence into the conversion of mass to light. A constant mass-to-light ratio is given for $\Gamma = 0$. If there are multiple light profile components (e.g. a bulge and disk) they assume independent values of $\Psi$ and $\Gamma$. Deflection angles for this profile are computed via an adapted implementation of the method of \citet{Oguri2021}, which decomposes the convergence profile into multiple cored steep elliptical profiles and efficiently computes the deflection angles from each.

Observationally, early-type galaxies are observed to exhibit steep internal gradients in some spectral features associated with dwarf stars. If these features are truly driven by variations in the initial mass function, as advocated by \citet{VanDokkum2017, LaBarbera2019}, then substantial stellar mass-to-light ratio gradients are expected \citep{Ferreras2019}. Some evidence for such trends have indeed been reported by previous lensing studies using decomposed mass models (e.g. \citealt{Oldham2018}). We therefore fit a stellar mass model which allows for different mass-to-light ratios and radial gradients in each stellar component (bulge, disk and envelope). This ensures that we do not incorrectly favour the inclusion of a SMBH, as could otherwise occur if there is no other way for the lens model to increase the amount of mass centrally.

The dark matter component is given by an elliptical Navarro-Frenk-White (NFW) profile. Parameters associated with the lens's dark matter have superscript `dark'. The NFW represents the universal density profile predicted for dark matter halos by cosmological N-body simulations \citep{Zhao1996, Navarro1997} and with a volume mass density given by
\begin{equation}
    \rho = \frac{\rho_{\rm s}^{\rm dark}}{(r/r_{\rm s}^{\rm dark}) (1 + r/r_{\rm s}^{\rm dark})^2}.
\end{equation}
The halo normalization is given by $\rho_{\rm s}^{\rm dark}$ and the scale radius by $r_{\rm s}^{\rm dark}$. The dark matter normalization is parameterized using the mass at 200 times the critical density of the Universe, $M_{\rm 200}^{\rm dark}$, as a free parameter. The scale radius is set via $M_{\rm 200}^{\rm dark}$ using the mean of the mass-concentration relation of \citet{Ludlow2016}, which uses the lens and source redshifts to convert this to units of solar masses. 

The dark matter model has five free parameters: ($x^{\rm dark}$,\,$y^{\rm dark}$), the centre, ($\epsilon_{\rm 1}^{\rm dark}$,\,$\epsilon_{\rm 2}^{\rm dark}$), the elliptical components and; the mass, $M_{\rm 200}^{\rm dark}$. In \cref{ResultSIE} we fit an elliptical NFW using a parameterization which also varies the concentration of the NFW, to test models which can increase the dark matter central density. The deflection angles of the elliptical NFW are computed via the same method used for the Sersic profile \citep{Oguri2021}.

An external shear field is included and parameterized as two elliptical components $(\epsilon_{\rm 1}^{\rm ext},\,\epsilon_{\rm 2}^{\rm ext})$, where parameters associated with the lens's external shear have superscript `ext'. The shear magnitude $\gamma^{\rm  ext}$ and the orientation of the semi-major axis $\theta^{\rm  ext}$, measured counter-clockwise from north, are given by
\begin{equation}
    \label{eq:shear}
    \gamma^{\rm ext} = \sqrt{\epsilon_{\rm 1}^{\rm ext^{2}}+\epsilon_{\rm 2}^{\rm ext^{2}}}, \, \,
    \tan{2\phi^{\rm ext}} = \frac{\epsilon_{\rm 2}^{\rm ext}}{\epsilon_{\rm 1}^{\rm ext}}.
\end{equation}
Deflection angles are computed analytically.

To test for the presence of a SMBH via Bayesian model comparison, every model is fitted with and without a point-mass, whose parameters have superscript `smbh'. This model includes a single free parameter, the Einstein radius $\theta^{\rm smbh}_{\rm Ein}$, which is related to mass as 
\begin{equation}
\label{eqn:PointMass}
M_{\rm BH} = \Sigma_\mathrm{crit} \, \pi \, (\theta^{\rm smbh}_{\rm Ein})^2 \, .
\end{equation}
$\theta^{\rm smbh}_{\rm Ein}$ is in units of arc-seconds and $M_{\rm BH}$ in stellar masses. Point mass deflection angles are computed analytically. The SMBH $(x^{\rm smbh},\,y^{\rm smbh})$ centre is aligned with the highest Sersic index light profile (e.g.\ the bulge) for decomposed mass models.

In \cref{ResultSIE}, we fit \textit{total} mass models that represent all the mass (e.g. stellar plus dark) in a single profile, either the elliptical power-law (PL) \citep{Tessore2016} or the elliptical broken power-law (BPL) introduced by \citet{Oriordan2019, Oriordan2020, Oriordan2021}. Parameters associated with the total mass model have superscript `mass'. For these models the SMBH $(x^{\rm smbh},\,y^{\rm smbh})$ centre is aligned with the centre of the PL or BPL mass profile. The results of fitting this model are summarized in the main paper.
 
We fit a number of additional lens mass models which make different assumptions, in order to verify that none change any of this paper's main results. An additional galaxy is present towards the right of the giant arc, as shown in the first panel of \cref{figure:Data}. In \cref{MassClump}, we include this galaxy in the lens mass model. In \cref{MassCentreFree}, we fit models where the SMBH position is free to vary. In \cref{Radial}, we fit mass models with a shallower inner density profile, which form a large radial critical curve.

\subsection{Source Reconstruction}\label{Source}

After subtracting the foreground lens emission and ray-tracing the coordinates to the source-plane via the mass model, the source is reconstructed in the source-plane using an adaptive Voronoi mesh which accounts for irregular or asymmetric source morphologies (see \cref{figure:ModelFit}). Our results use the \texttt{PyAutoLens} pixelization \texttt{VoronoiBrightnessImage}, which adapts the centres of the Voronoi pixels to the reconstructed source morphology, such that more resolution is dedicated to its brighter central regions (see \citealt{Nightingale2018}).

The reconstruction computes the linear superposition of PSF-smeared source pixel images that best fits the observed image. This uses the matrix $f_{\rm  ij}$, which maps the $j$th pixel of each lensed image to each source pixel $i$. When constructing $f_{\rm  ij}$ we apply image-plane subgridding of degree $4 \times 4$, meaning that $16 \times j$ sub-pixels are fractionally mapped to source pixels with a weighting of $\frac{1}{16}$, removing aliasing effects \citep{Nightingale2015}.

Following the formalism of \citep[][WD03 hereafter]{Warren2003}, we define the data vector $\vec{D}_{i} = \sum_{\rm  j=1}^{J}f_{ij}(d_{j} - b_{j})/\sigma_{j}^2$ and curvature matrix $\tens{F}_{ik} = \sum_{\rm  j=1}^{J}f_{ij}f_{kj}/\sigma_{j}^2$, where $d_{j}$ are the observed image flux values with statistical uncertainties $\sigma{\rm _j}$, and $ b_{j}$ are the model lens light values. The source pixel surface brightnesses values are given by $s = \tens{F}^{-1} \vec{D}$, which are solved via a linear inversion that minimizes
\begin{equation}
\label{eqn:ChiSquared}
\chi^2 = \sum_{\rm  j=1}^{J} \bigg[ \frac{(\sum_{\rm  i=1}^{I} s_{i} f_{ij}) + b_{j} - d_{j}}{\sigma_{j}} \bigg] \, .
\end{equation}
The term $\sum_{\rm  i=1}^{I} s_{i} f_{ij}$ maps the reconstructed source back to the image-plane for comparison with the observed data. 

This matrix inversion is ill-posed, therefore to avoid over-fitting noise the solution is regularized using a linear regularization matrix $\tens{H}$ (see WD03). The matrix $\tens{H}$ applies a prior on the source reconstruction, penalizing solutions where the difference in reconstructed flux of neighboring Voronoi source pixels is large. We use the \texttt{PyAutoLens} regularization scheme \texttt{AdaptiveBrightness}, which adapts the degree of smoothing to the reconstructed source's luminous emission (see \citealt{Nightingale2018}). The degree of smoothing is chosen objectively using the Bayesian formalism introduced by \citet{Suyu2006}. The likelihood function used in this work is taken from \citep{Dye2008a} and is given by
\begin{eqnarray}
-2 \,{  \mathrm{ln}} \, \mathcal{L} &=& \chi^2 + s^{T}\tens{H}s
+{ \mathrm{ln}} \, \left[ { \mathrm{det}} (\tens{F}+\tens{H})\right]
-{ \mathrm{ln}} \, \left[ { \mathrm{det}} (\tens{H})\right]
\nonumber \\
& &
+ \sum_{\rm  j=1}^{J}
{ \mathrm{ln}} \left[2\pi (\sigma{_j})^2 \right]  \, .
\label{eqn:evidence2}
\end{eqnarray}

The step-by-step Jupyter notebooks linked to above describes how the different terms in this likelihood function compare and ranks different source reconstructions, allowing one to objectively determine the lens model that provides the best fit to the data in a Bayesian context.

\subsection{Data Preparation}

In both the F390W and F814W wavebands there is emission from nearby interloper galaxies towards the right of the giant arc, which can be most clearly seen in the upper left panel of \cref{figure:Data}. Including this emission would negatively impact our analysis, therefore we remove it beforehand. Our lens analysis assumes a circular mask of radius 3.7\,arcsec, whereby all image-pixels outside this circular region are not included in the fitting procedure. The central panels of \cref{figure:Data} show that this mask removes the majority of foreground emission, however a small fraction is still within this circle. We therefore subtract it using a graphical user interface, replacing it with background noise in the image and increasing the RMS noise-map values of these pixels to ensure they do not contribute to the likelihood function. We also consider lens models which include this galaxy in the ray-tracing (see \cref{MassClump}).

\subsection{Light Model Waveband}\label{F390WModel}

The wavelength at which the lens galaxy's emission is observed is important for tracing its stellar mass distribution. The F390W image of Abell 1201 observes the lens galaxy at rest-frame ultra-violet wavelengths, possibly probing younger stellar populations with lower mass-to-light ratios. The F814W image observes rest-frame near infrared (NIR) emission and probes more aged and reddened stellar populations which make up a greater fraction of the stellar mass. This can be seen in \cref{figure:Data}, where only the central regions of the lens are visible in the F390W image compared to the F814W image. The F390W image is therefore less appropriate for constraining the stellar mass component of the lens model. 

Therefore, to fit the decomposed mass model to the F390W image we use the maximum likelihood Sersic light model parameters of the F814W fits that are chosen after the lens light Bayesian model comparison (see \cref{LensLightModel}). The mass-to-light ratio and gradient parameters of each Sersic remain free to vary, ensuring high flexibility in the model's stellar mass distribution. Fits are performed using a lens light subtracted image for the F390W image which is output midway through the analysis. To ease the comparison between fits to the F390W and F814W images, we follow the same approach with the F814W image, using the same fixed maximum likelihood Sersic parameters and fitting a lens subtracted image output midway through the analysis.

\begin{figure*}
\centering
\includegraphics[width=0.24\textwidth]{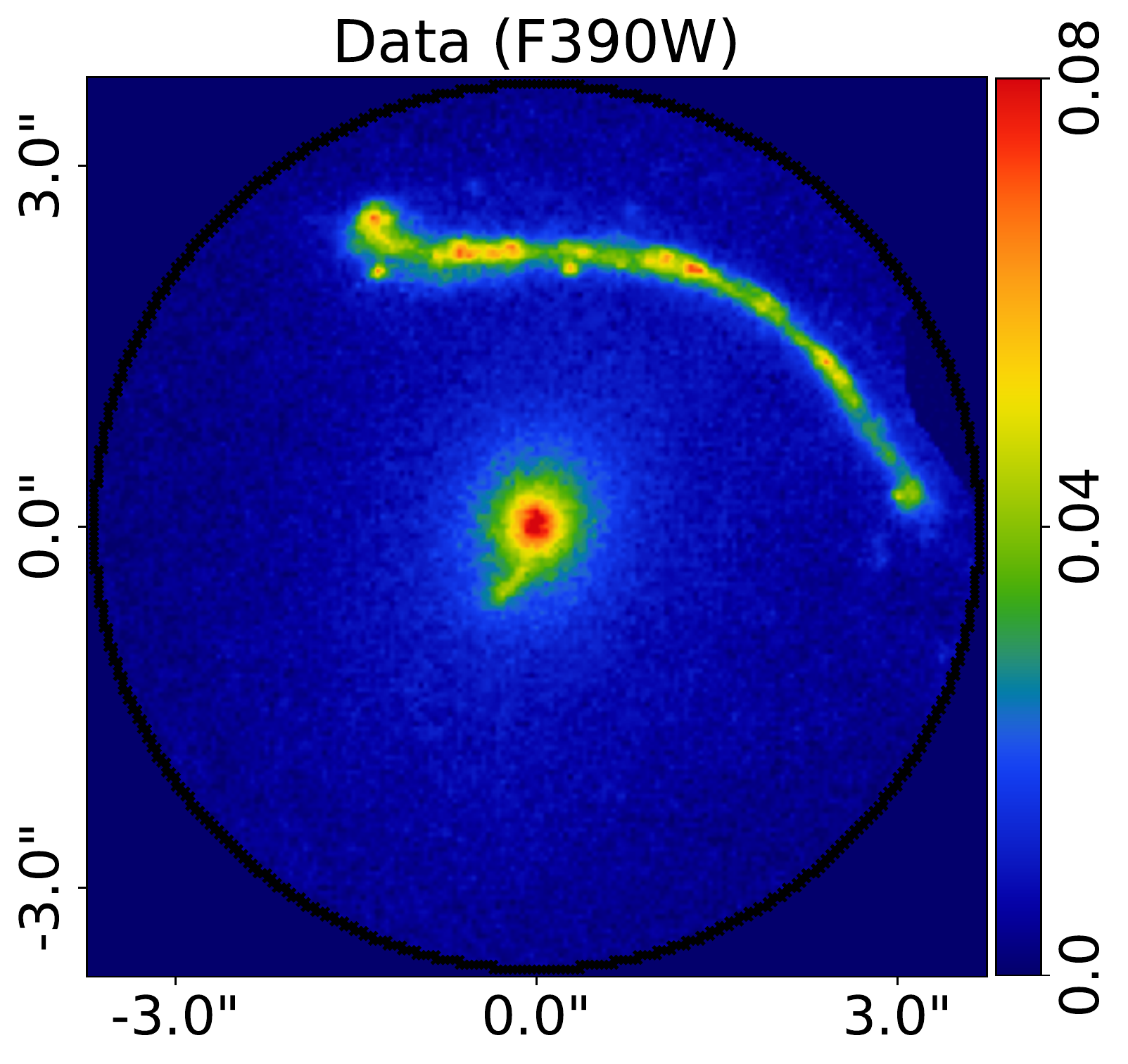}
\includegraphics[width=0.24\textwidth]{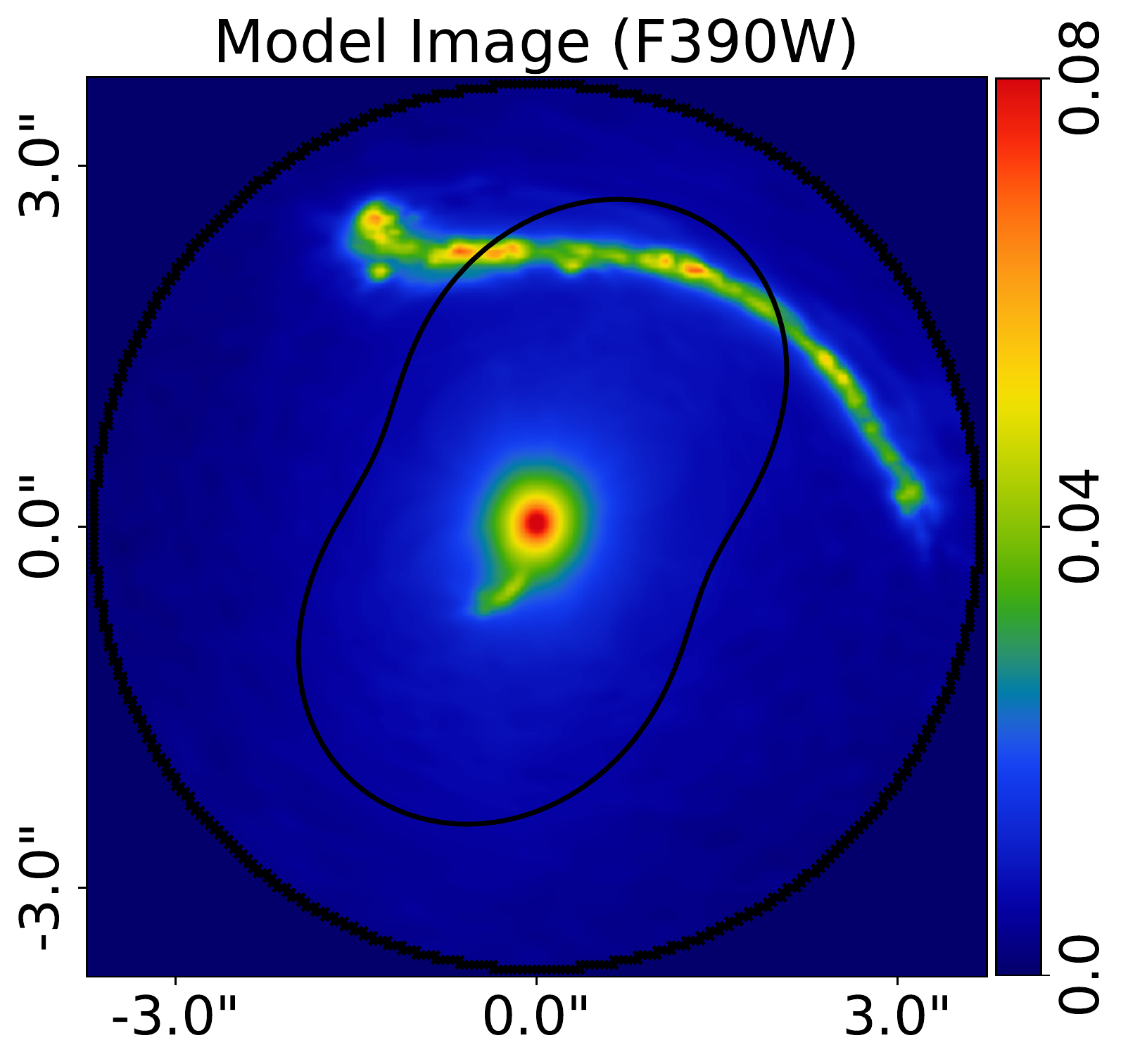}
\includegraphics[width=0.24\textwidth]{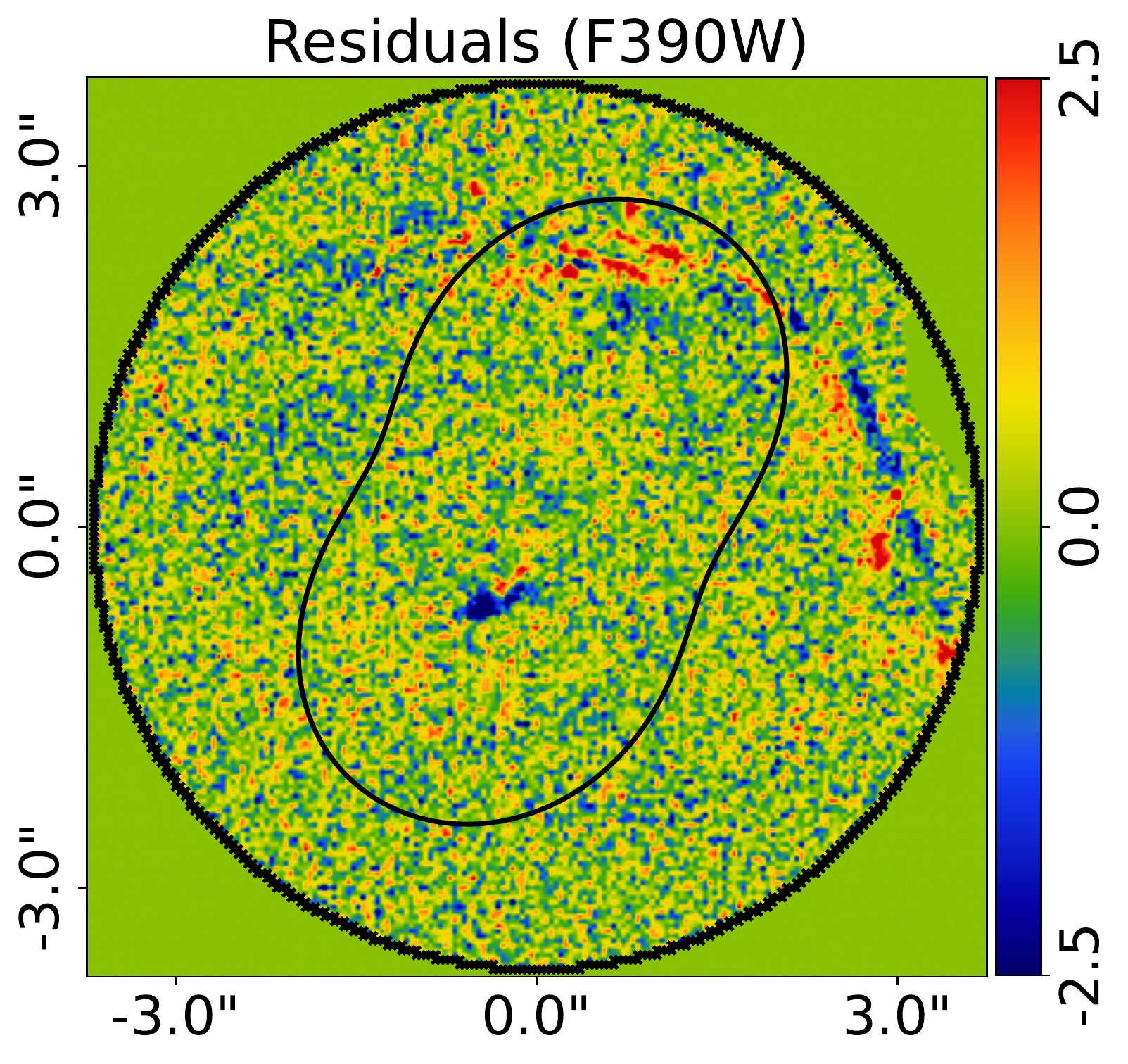}
\includegraphics[width=0.24\textwidth]{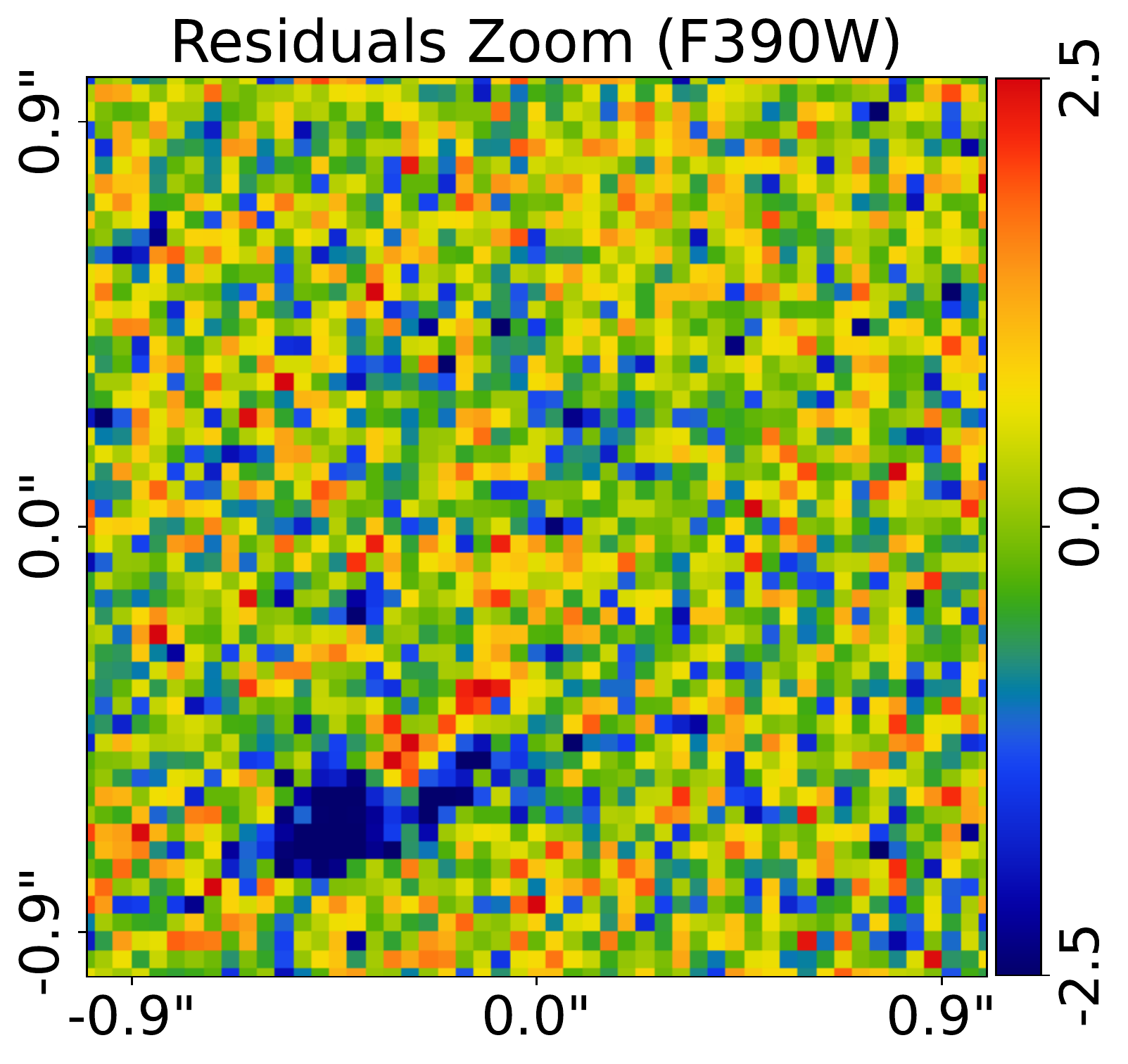}
\includegraphics[width=0.24\textwidth]{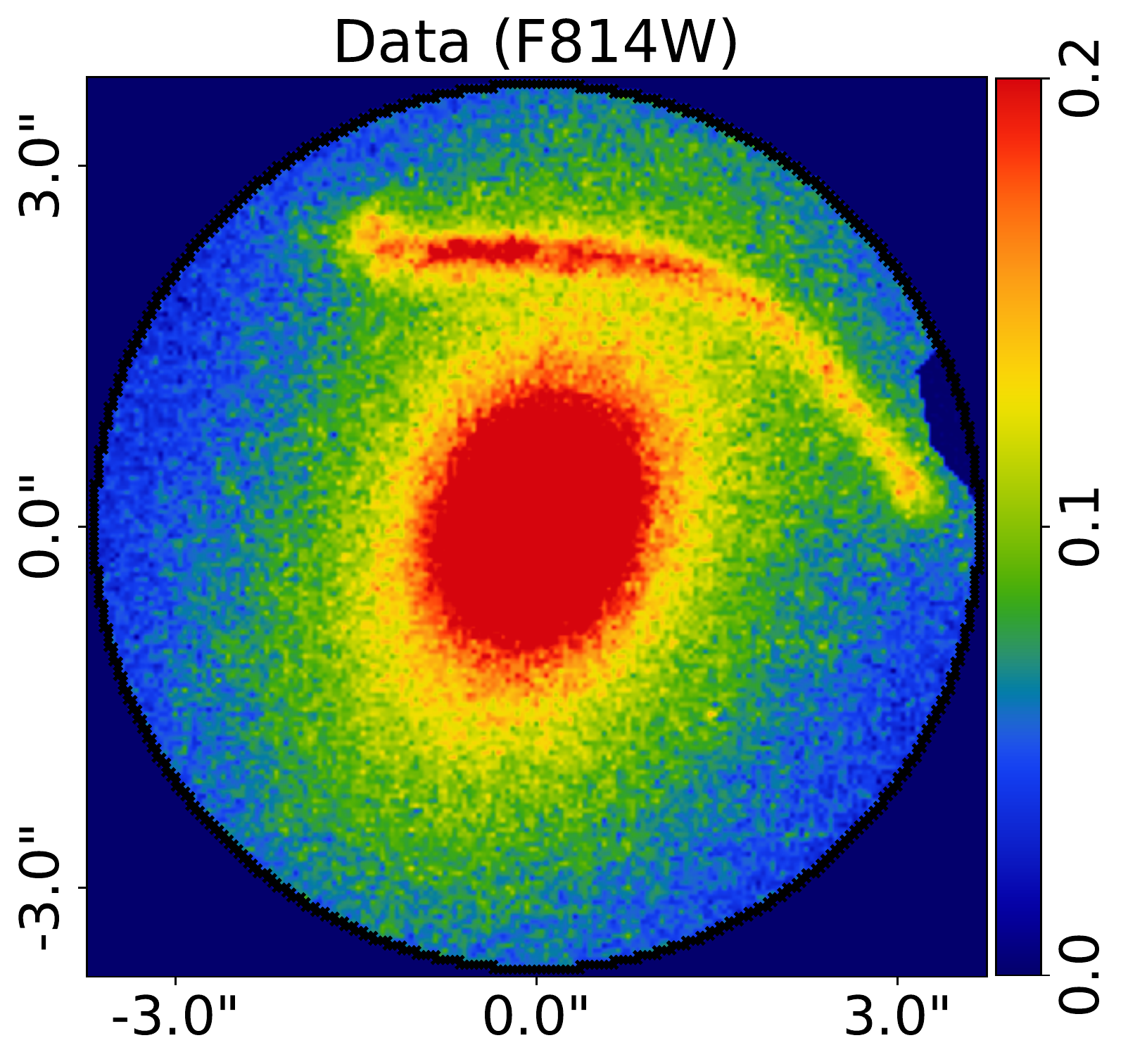}
\includegraphics[width=0.24\textwidth]{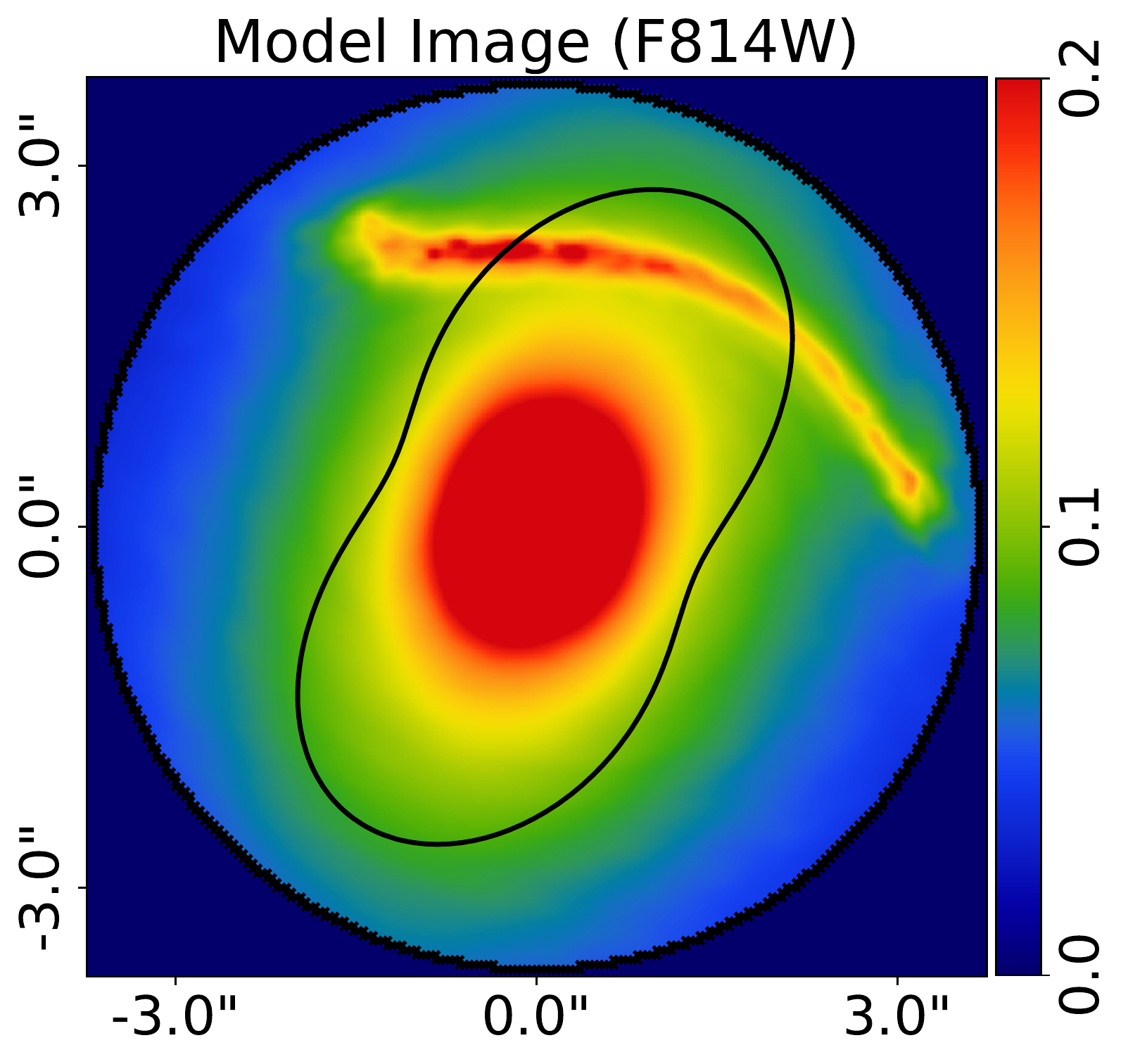}
\includegraphics[width=0.24\textwidth]{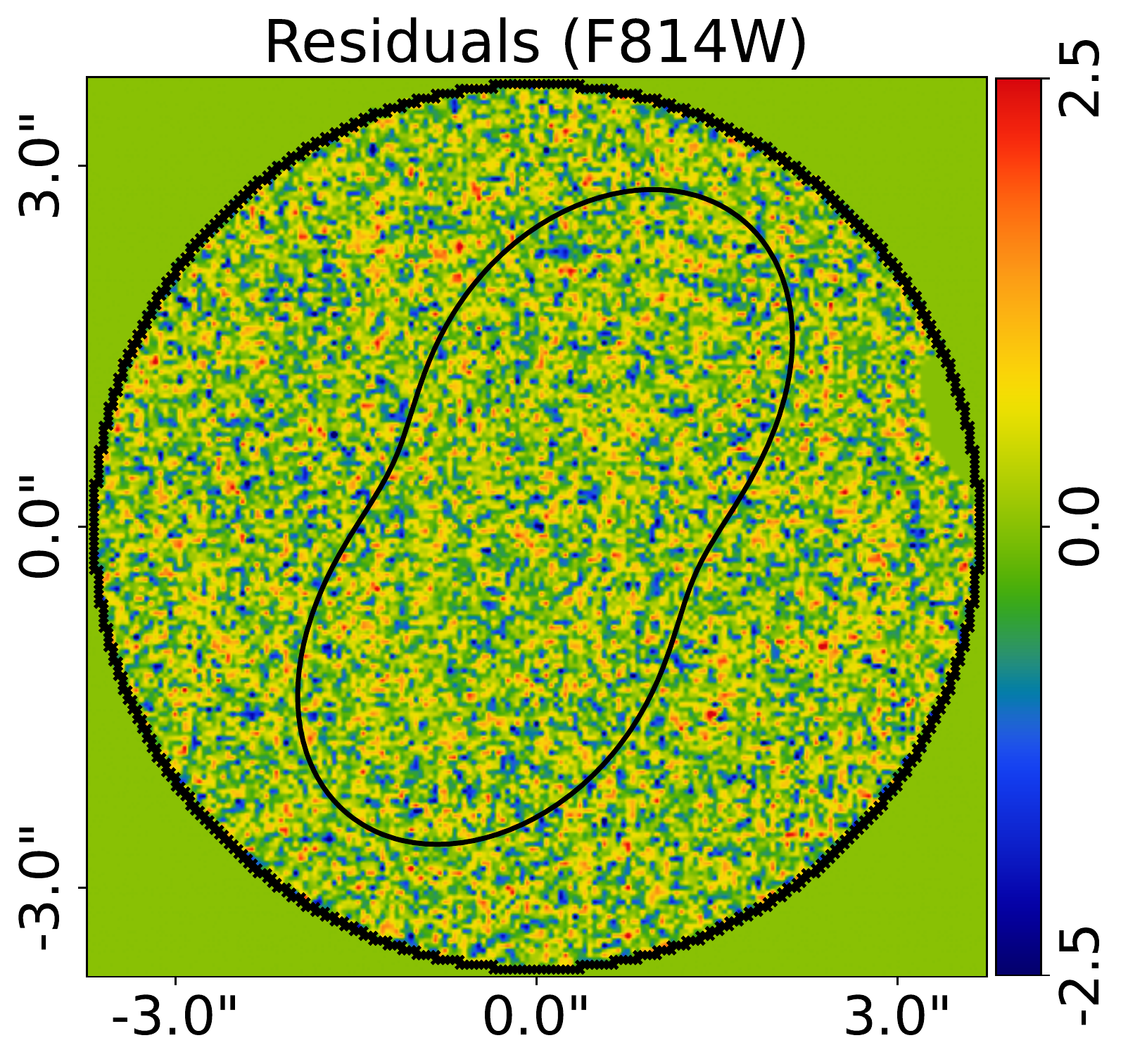}
\includegraphics[width=0.24\textwidth]{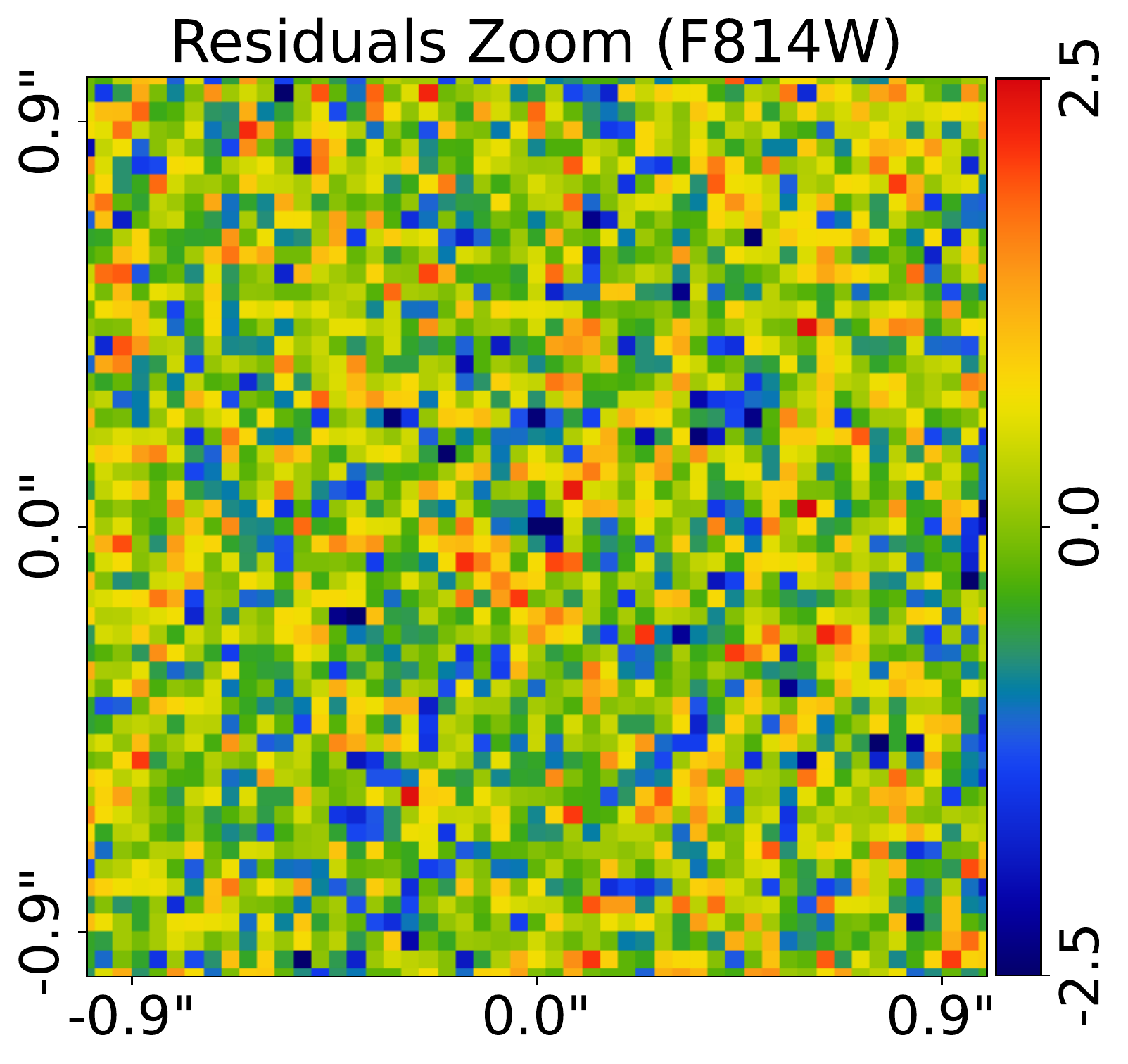}
\caption{Lens light subtractions of HST imaging of Abell 1201. The observed images (left column), image-plane model images of the lens and source galaxies (left-centre column), normalized residuals (right-centre column) and a zoom-in of these residuals near the counter image (right column) are shown. The top row shows the F390W and bottom row the F814W wavebands respectively. The lens model is the maximum likelihood model inferred at the end of the first SLaM pipeline run, which produces a lens subtracted image. For the F390W data, a double Sersic lens light model with offset centres and elliptical components is shown, whereas for the F814W data a triple Sersic model is used. The magenta circle indicates residuals that are due to the lens light subtraction. The counter image is fitted poorly in the F390W image, because the mass model (which is an isothermal mass model with shear) does not enable an accurate reconstruction of the lensed source's structure.} 
\label{figure:LightFit2}
\end{figure*}

\subsection{SLaM Pipelines}\label{SLAM}

The models of lens mass, lens light and source light are complex and their parameter spaces highly dimensional. Without human intervention or careful set up, a model-fitting algorithm (e.g. a Markov chain Monte Carlo search) may converge very slowly to the global maximum likelihood solution. 
\texttt{PyAutoLens} therefore applies `non-linear search chaining' to break the search into a sequence of tractable operations.
Using the probabilistic programming language {\tt PyAutoFit} \github{https://github.com/rhayes777/PyAutoFit}, we fit a series of parametric lens models that approximate the form of the desired model, with growing complexity. A fit to the simplest model provides information to initialise a fit to the next model. The final search is started around the global maximum likelihood and with priors reflecting the likelihood surface. Each fit in this chain uses the nested sampler \citep[{\tt dynesty}][]{Speagle2020} \github{https://github.com/joshspeagle/dynesty}.
The models used to perform this analysis extend the Source, Light and Mass (SLaM) pipelines described by \citet[][hereafter E22]{Etherington2022}, \citet{Cao2021} and \citet{He2023}. They are available at \url{https://github.com/Jammy2211/autolens_workspace}.

The first pipeline, called the Source pipeline, initializes the pixelized source model by inferring a robust lens light subtraction (using a double Sersic model) and total mass model (using a PL with $\gamma=2$ plus shear). The highest likelihood lens model and source reconstruction at this stage of the pipeline are shown in \cref{figure:ModelFit}. They give an accurate foreground lens subtraction and reconstruction of the lensed source's light. 

The Light pipeline follows, which uses fixed values of the mass and source parameters corresponding to the maximum likelihood model of the Source pipeline. The lens's mass is therefore again fitted using a total mass model such that the lens light model does not yet contribute to the ray-tracing. The only free parameters in this pipeline are those of the lens light and all five of the models listed in \cref{table:LightModels} are fitted independently, enabling Bayesian model comparison. The results of the Light pipeline, including the models chosen for all subsequent model fits, are presented in \cref{LightModels}.

The final pipeline is the Mass pipeline, which in \citetalias{Etherington2022} directly follows the Light Pipeline, fitting PL mass profiles representing the total mass distribution. In this work, we do not immediately start the Mass pipeline after the Light pipeline, due to the complications of fitting the stellar component of the decomposed models to the F390W imaging data discussed previously (see \cref{F390WModel}). Instead, the lens light models favoured by model comparison are used to output lens-subtracted F390W and F814W images. An analysis of these images is then performed from scratch, starting a new SLaM pipeline fit that uses a scaled down Source pipeline, which removes models that fit the lens light, and which omits the Light pipeline completely (see \citetalias{Etherington2022}). 

When this analysis reaches the Mass pipeline, it fits the decomposed models (models assuming two or three Sersic profiles for the lens light and stellar mass) and the total mass models (the PL and BPL), whose results are described in \cref{ResultSIE}. Every mass model is fitted twice, with and without a point mass representing a SMBH. The Bayesian model comparison of these mass models is the main component of this work's results.

As described in \citetalias{Etherington2022}, the SLaM pipelines use prior passing to initialize the regions of parameter space that \texttt{dynesty} will search in later pipelines, based on the results of earlier pipelines. \cref{LensModels} gives a description of the priors used in this work. We also use the likelihood cap analysis described in \citetalias{Etherington2022} to estimate errors on lens model parameters.

\subsection{Bayesian Evidences}\label{BayesEvi}

The Bayesian evidence, $\mathcal{Z}$, of every lens model we fit is estimated by \texttt{dynesty} and is given by equation 2 of \citet{Speagle2020}. The Bayesian evidence is the integral over all parameters in the model and therefore naturally includes a penalty term for including too much complexity in a model – if a model has more free parameters it is penalized for this complexity. The evidence is computed via sampling of Eq. (9). Our analysis therefore incorporates the principle of Occam's razor, whereby more complex models are only favoured if they improve the fit enough to justify their additional complexity compared a simpler model. To compare models, we use the difference in log evidence, $\Delta \ln \mathcal{Z}$. An increase of $\Delta \ln \mathcal{Z} = 4.5$ for one model over another corresponds to odds of 90:1 in favour of that model. For comparisons of lens models with and without a SMBH this corresponds to a $3\sigma$ detection of the SMBH. An increase of $\Delta \ln \mathcal{Z} = 11$ corresponds to a $5\sigma$ detection. 

However, there are sources of uncertainty in the evidence estimate that means taking these numbers at face value is problematic. For example, there is an error on the evidence estimated by \texttt{dynesty}, with identical runs of a lens model showing variations of $\ln \mathcal{Z} \sim 5$ (due to stochasticity in the \texttt{dynesty} sampling process). Adjusting the priors on the lens model parameters or reparameterizing the model also change its value, with tests showing variations up to $\ln \mathcal{Z} \sim 5$. Accordingly, we consider values of $\Delta \ln \mathcal{Z} > 10 $ sufficient to favour more complex models over simpler ones, including the detection of a SMBH.   
\section{Results}\label{Results}

\begin{figure*}
\centering
\includegraphics[width=0.195\textwidth]{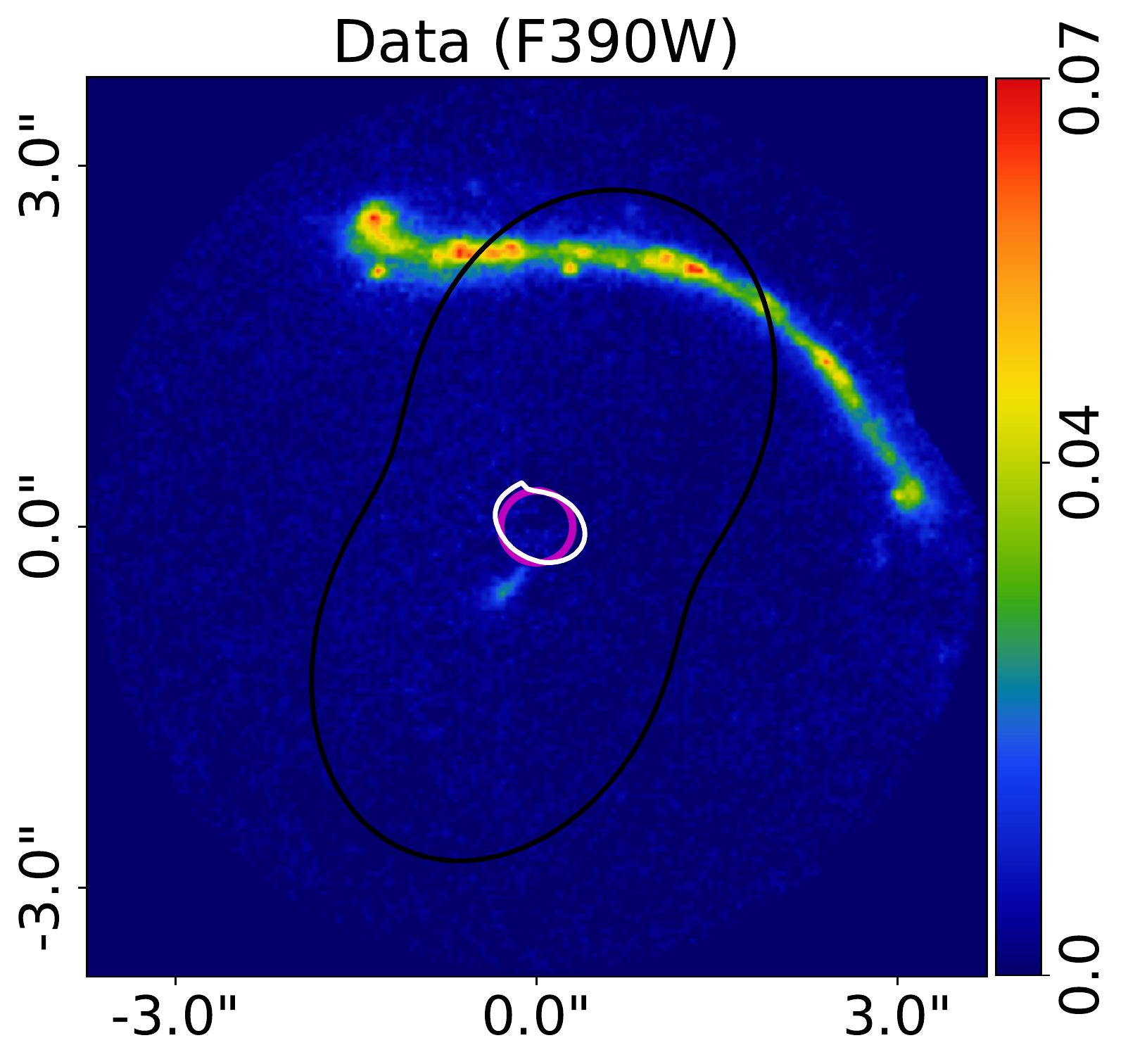}
\includegraphics[width=0.195\textwidth]{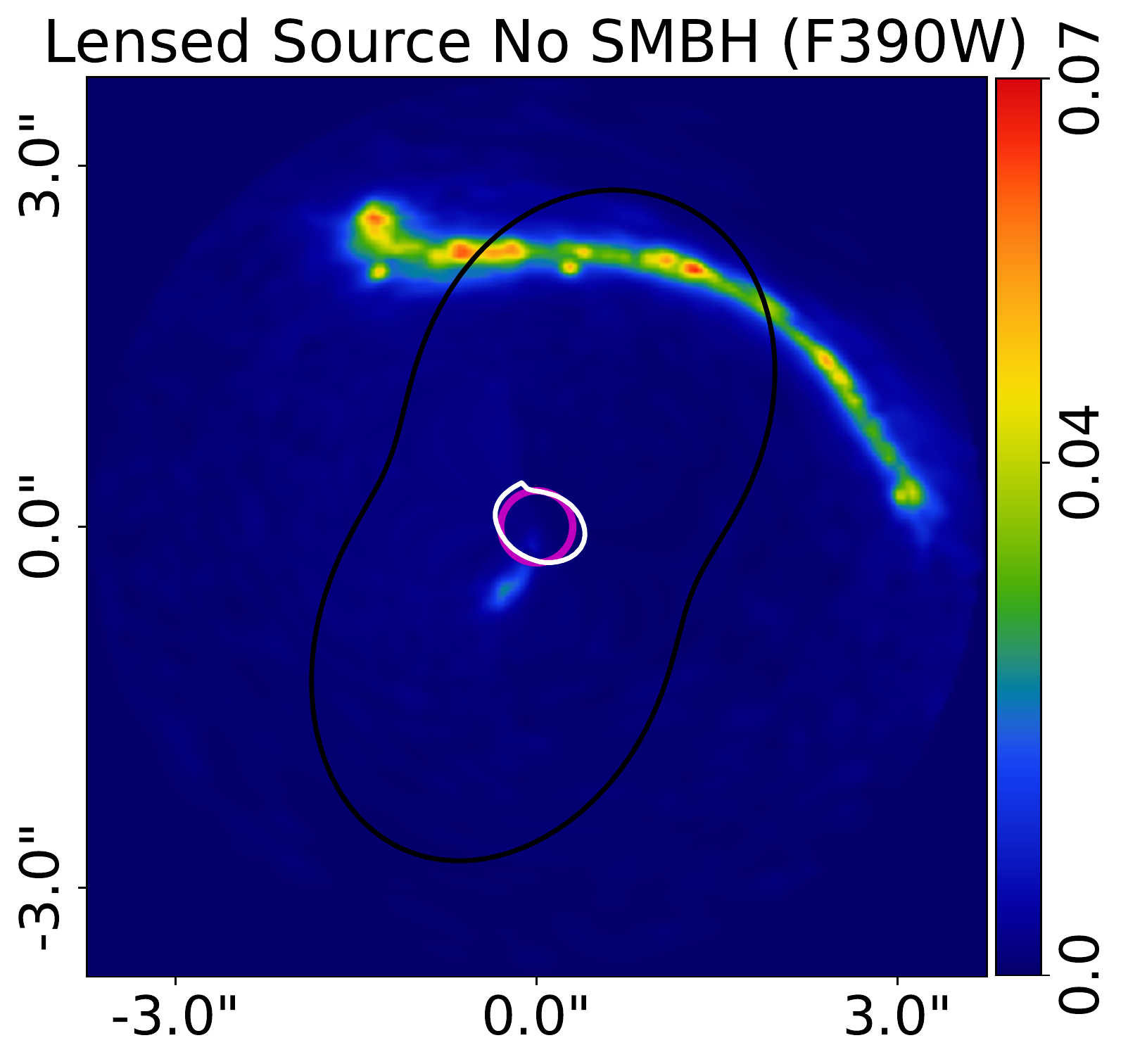}
\includegraphics[width=0.195\textwidth]{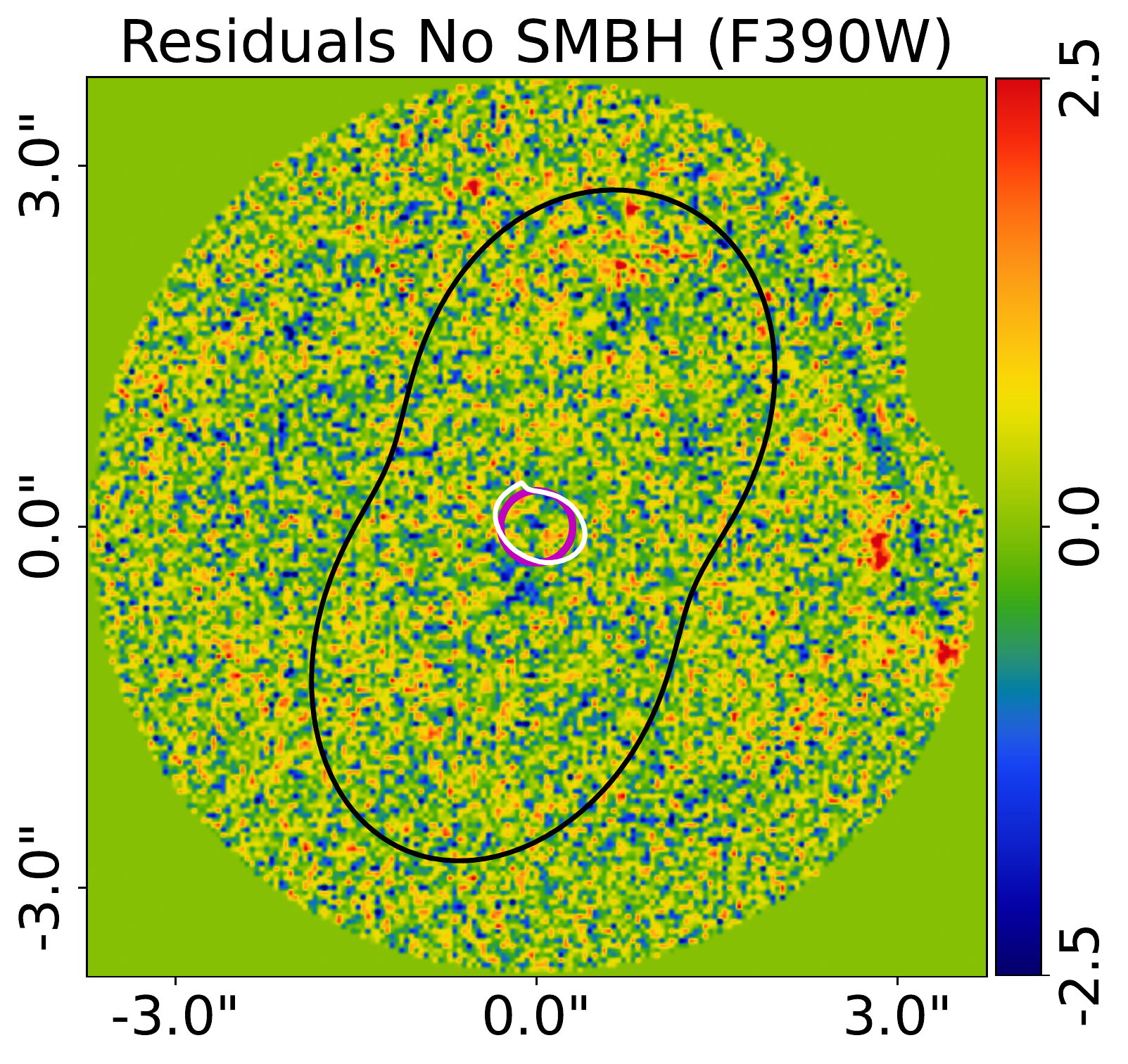}
\includegraphics[width=0.195\textwidth]{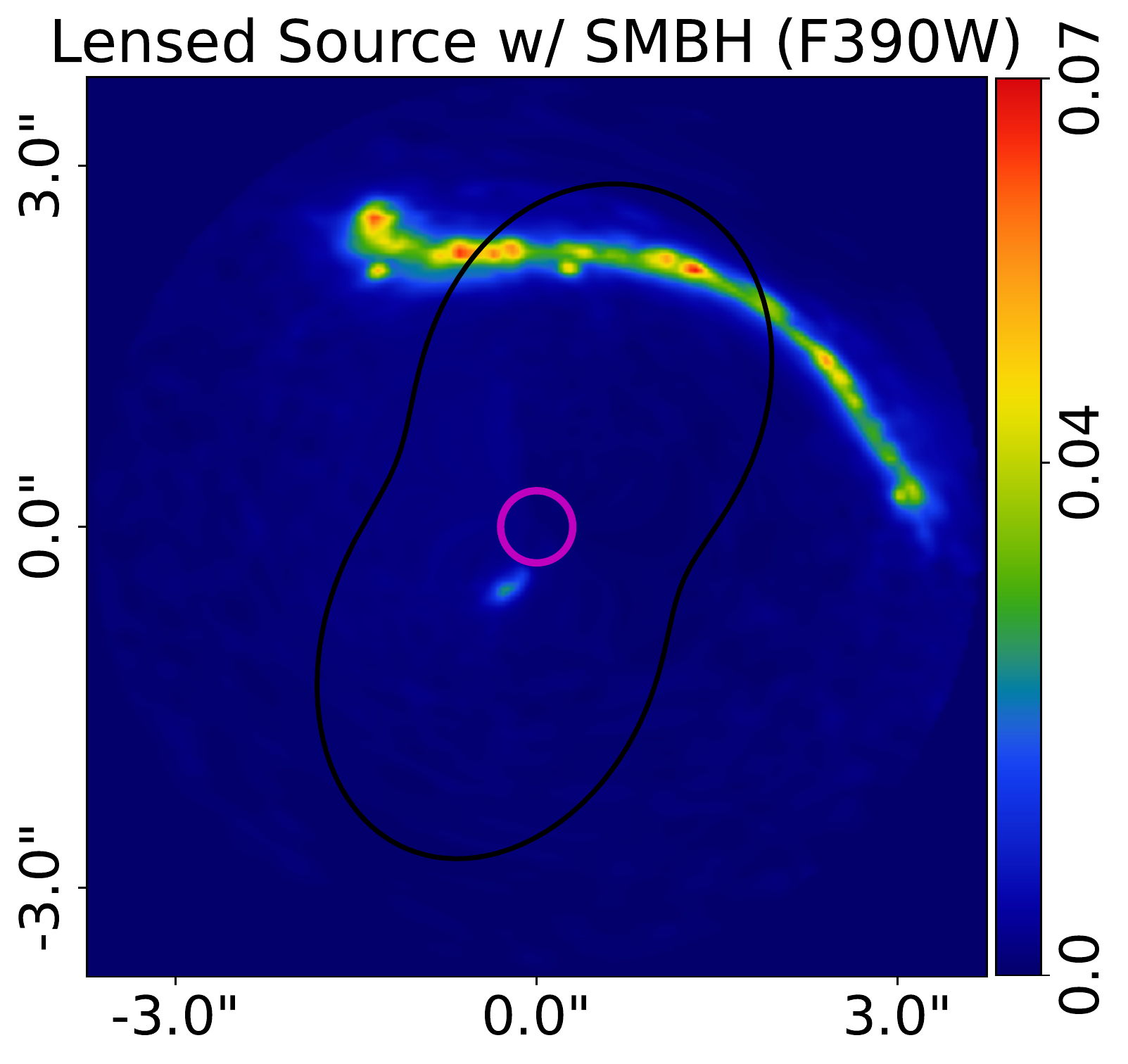}
\includegraphics[width=0.195\textwidth]{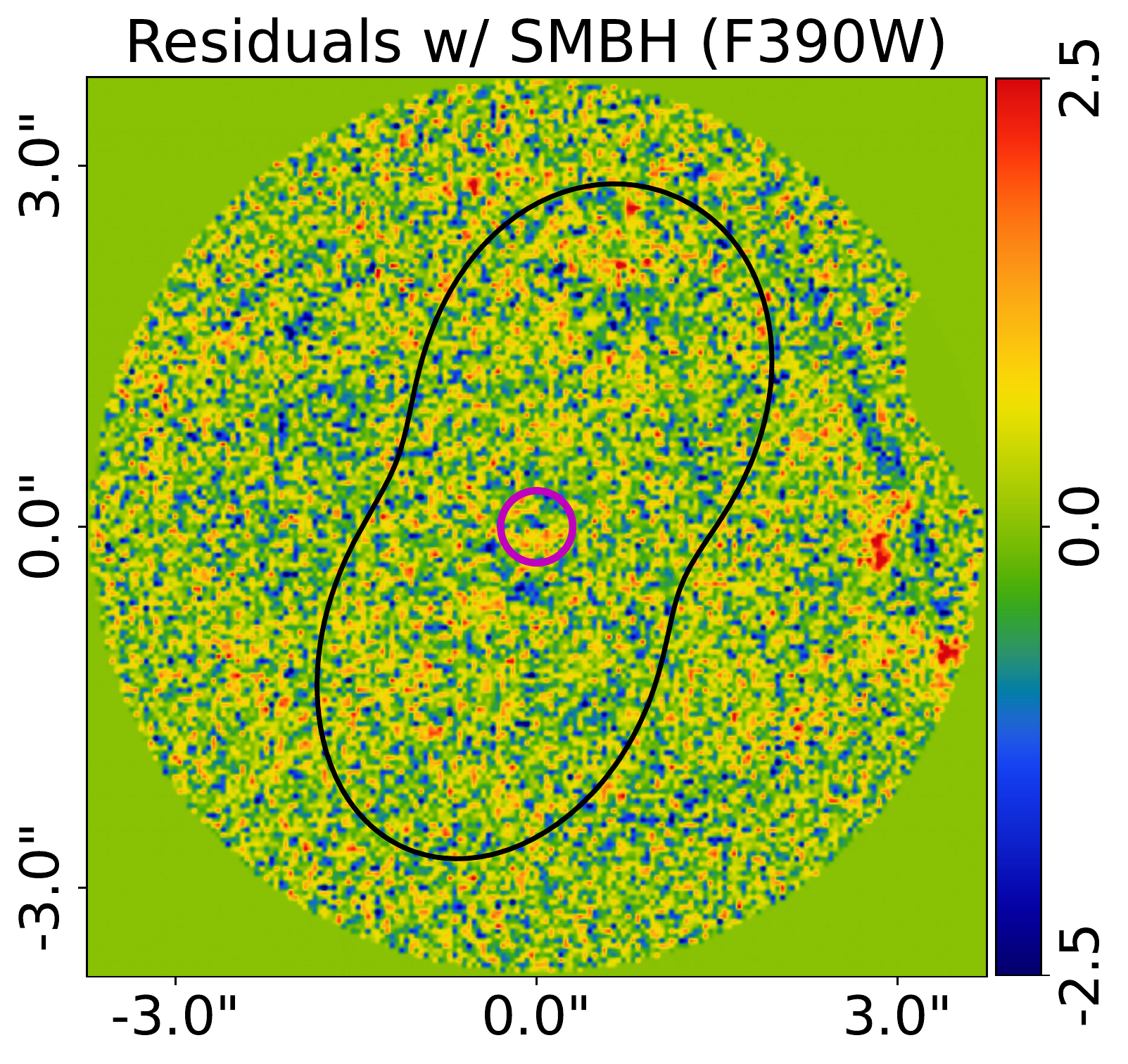}
\includegraphics[width=0.195\textwidth]{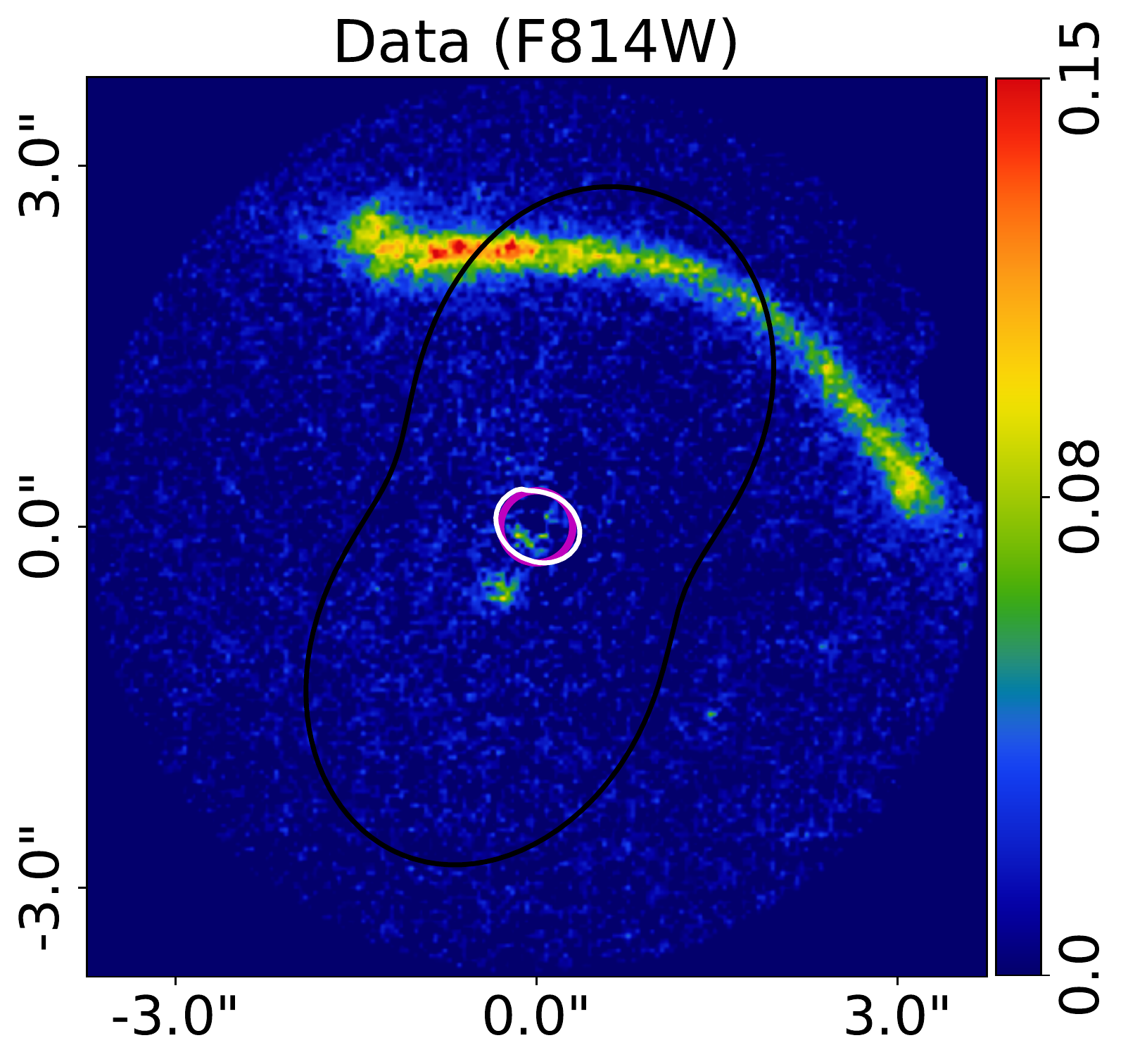}
\includegraphics[width=0.195\textwidth]{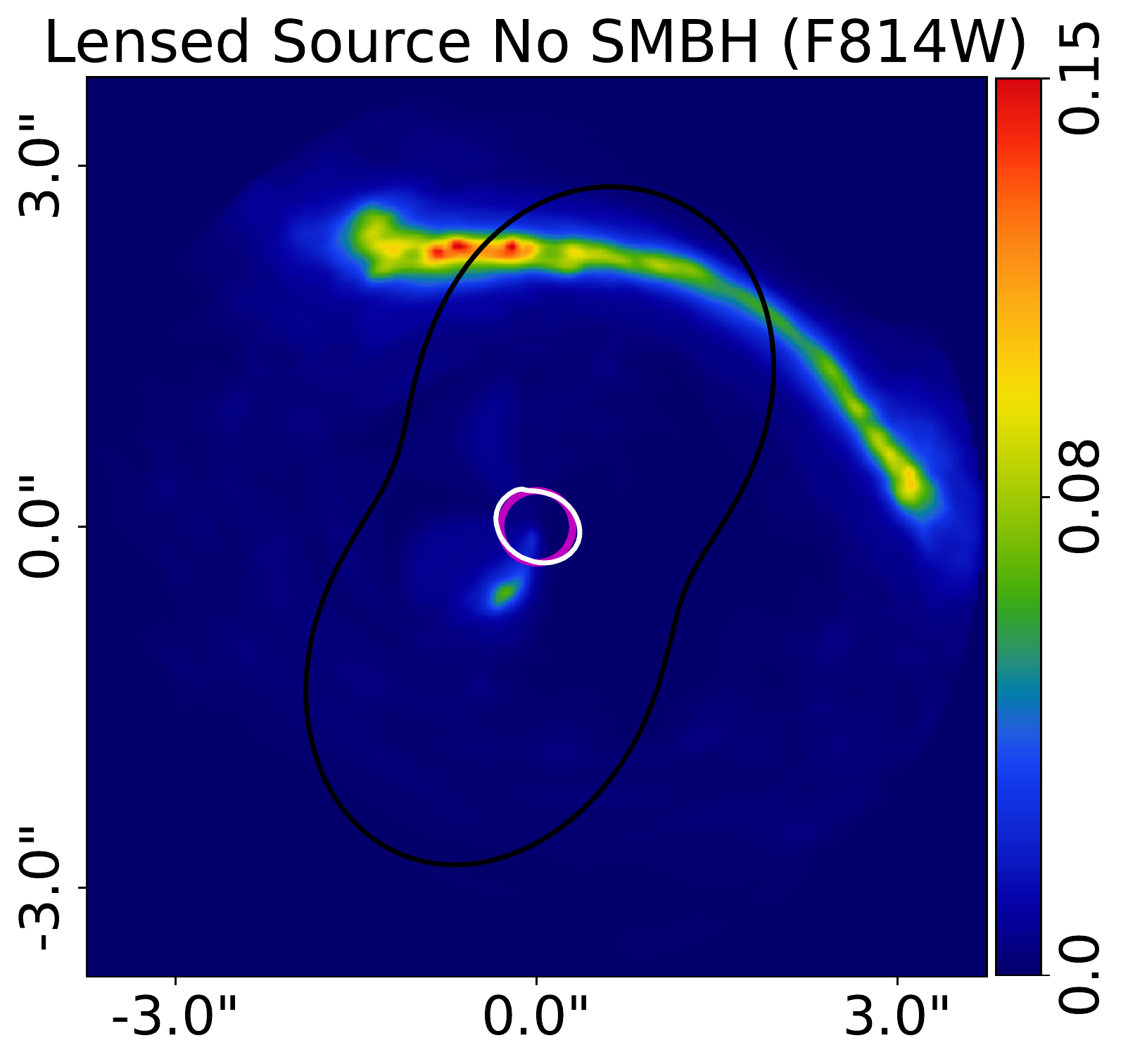}
\includegraphics[width=0.195\textwidth]{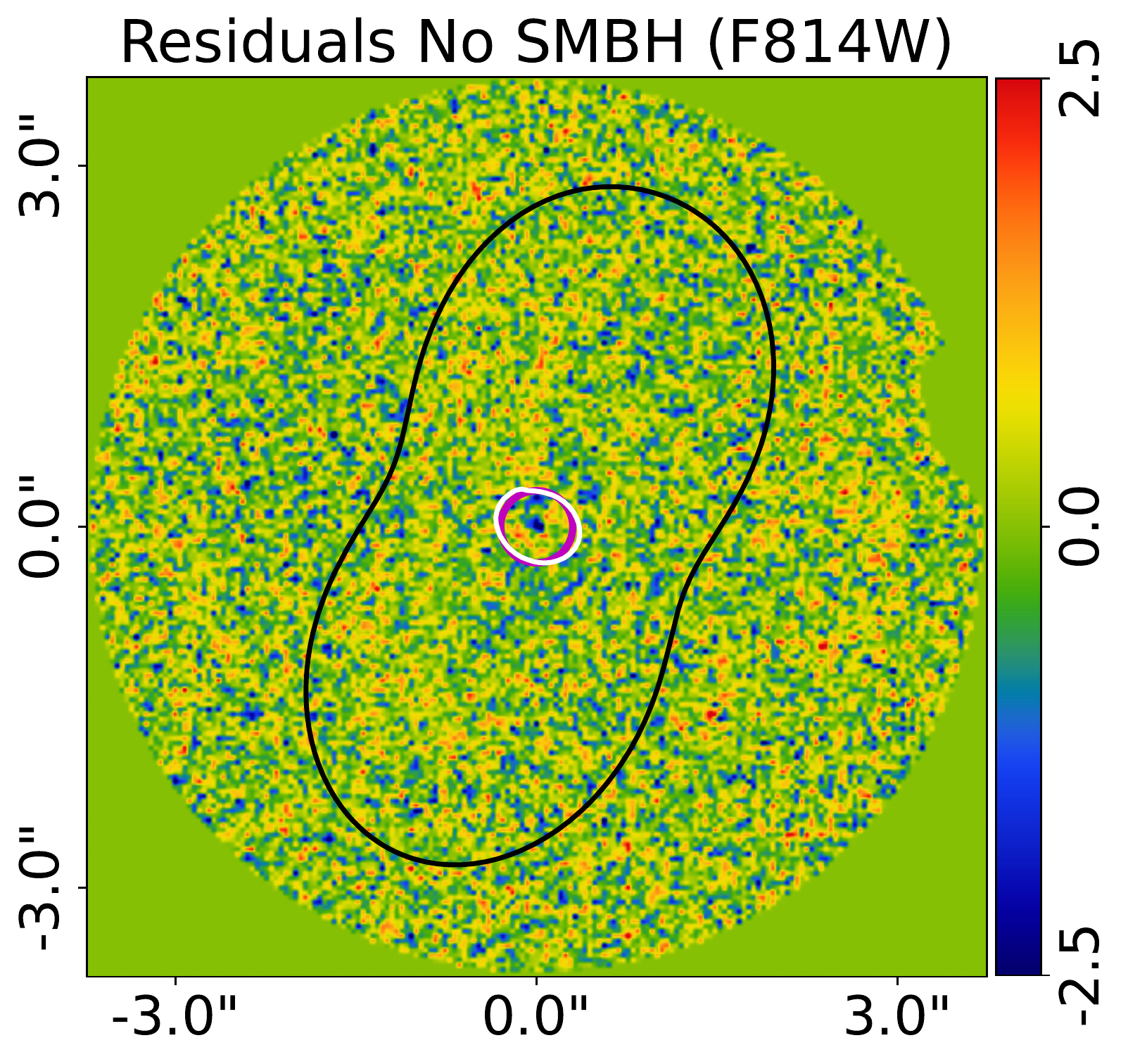}
\includegraphics[width=0.195\textwidth]{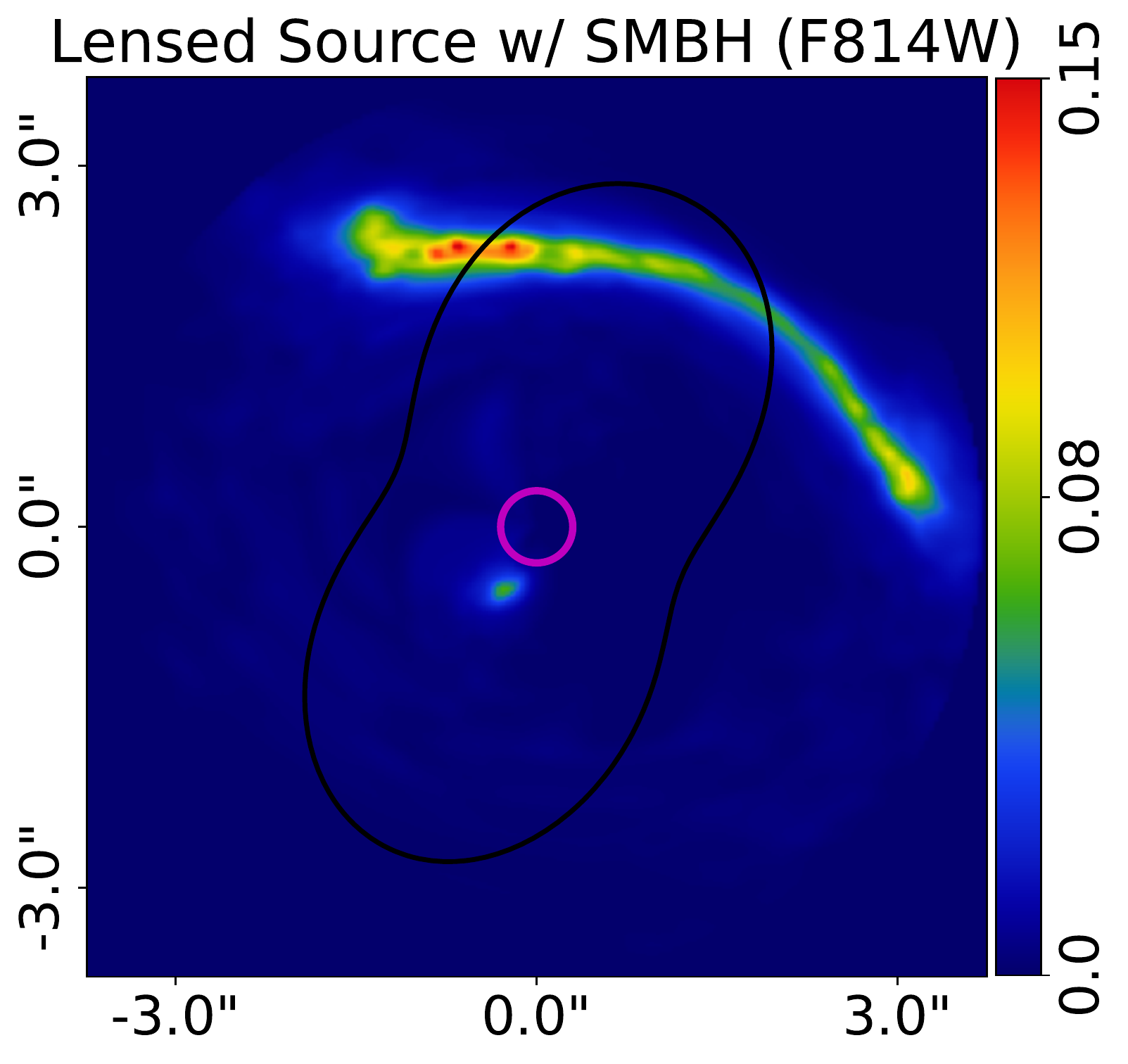}
\includegraphics[width=0.195\textwidth]{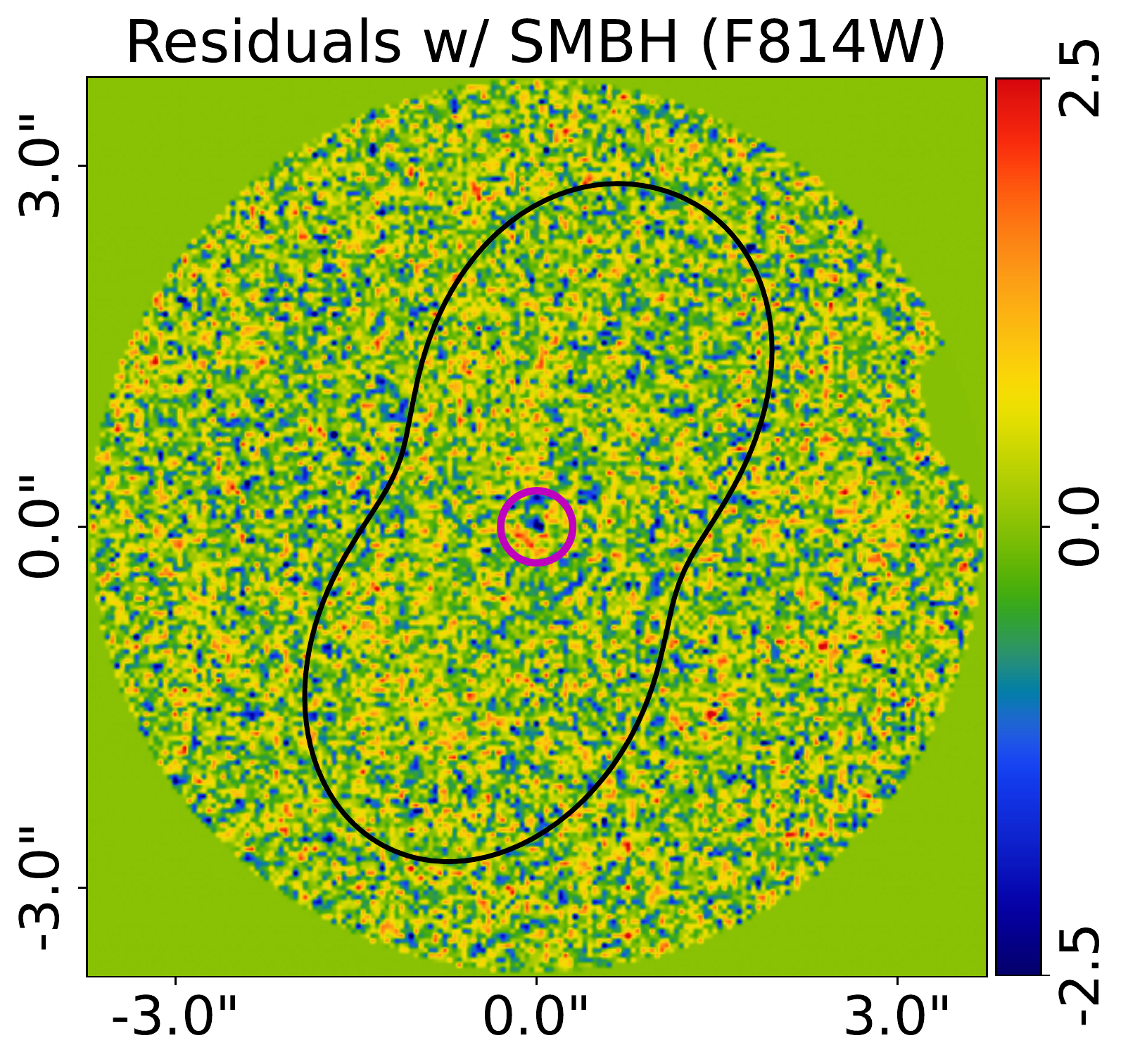}
\caption{The observed image (left panel), model lensed source and normalized residuals for decomposed model-fits without a SMBH (left-centre and centre panels) and with a SMBH (right-centre and right panel). The top row shows fits to the F390W image and bottom row the F814W image. Minimal residuals are seen in the central regions, indicating the lens light model and subtraction are accurate (the lens light models are visualized in \cref{figure:Light2D}). The magenta circle marks regions of the data where the brightest regions of the lens light were observed and subtracted. The tangential caustic is shown by a black line and radial critical curve and caustic a white line; the latter does not form for models including a SMBH. Source reconstructions with and without a SMBH successfully reproduce the giant arc and counter image, although residuals are present in both indicating their detailed structure is not fitted accurately. Figures \ref{figure:LightDarkF390Wx3} and \ref{figure:LightDarkF814Wx3} show zoom-ins of the counter image to better illustrate how these different models reconstruct the data.} 
\label{figure:LightDarkF390Wx3Global}
\end{figure*}

\begin{table*}
\tiny
\resizebox{\linewidth}{!}{
\begin{tabular}{ l l l l l l l l l } 
\multicolumn{1}{p{1.1cm}|}{\centering \textbf{Model}} 
& \multicolumn{1}{p{1.2cm}|}{\centering \textbf{Component}} 
& \multicolumn{1}{p{1.1cm}}{$x$ (\arcsec)} 
& \multicolumn{1}{p{1.1cm}}{$y$ (\arcsec)} 
& \multicolumn{1}{p{1.1cm}}{$\epsilon_{\rm 1}$} 
& \multicolumn{1}{p{1.1cm}}{$\epsilon_{\rm 2}$} 
& \multicolumn{1}{p{1.3cm}}{$I$ (e$^{\rm -}$\,s$^{\rm -1}$)} 
& \multicolumn{1}{p{1.1cm}}{$R_{\rm eff}$ (\arcsec)} 
& \multicolumn{1}{p{1.1cm}}{$n$} 
\\ \hline
& & & & & & & & \\[-4pt]

Sersic x1 & Bulge & 
$ 0.001^{+0.003}_{-0.003}$ &
$ 0.008^{+0.003}_{-0.003}$ &
$ 0.087^{+0.005}_{-0.005}$ & 
$-0.086^{+0.005}_{-0.005}$ & 
$ 0.010^{+0.002}_{-0.002}$ & 
$ 7.18^{+1.28}_{-1.08}$ & 
$ 3.78^{+0.18}_{-0.18}$ 
\\[-6pt] 

\\ \hline
& & & & & & & & \\[-4pt]

Sersic x2 & Bulge & 
$-0.009^{+0.003}_{-0.004}$ & 
$ 0.003^{+0.003}_{-0.003}$ & 
$ 0.030^{+0.011}_{-0.012}$ & 
$-0.062^{+0.010}_{-0.001}$ & 
$ 0.26^{+0.01}_{-0.02}$ & 
$ 0.46^{+0.04}_{-0.04}$ & 
$ 1.25^{+0.08}_{-0.07}$  \\[6pt] 
& Disk & 
$ 0.069^{+0.015}_{-0.012}$ &
$ 0.032^{+0.018}_{-0.016}$ &
$ 0.16^{+0.015}_{-0.012}$ & 
$-0.14^{+0.011}_{-0.013}$ &
$ 0.030^{+0.005}_{-0.006}$ & 
$ 5.14^{+0.94}_{-0.63}$ & 
$ 1.31^{+0.26}_{-0.12}$ \\[-6pt]

\\ \hline
& & & & & & & & \\[-4pt]

Sersic x3 & Bulge & 
$-0.005^{+0.004}_{-0.005}$ & 
$-0.002^{+0.004}_{-0.004}$ & 
$ 0.047^{+0.010}_{-0.013}$ & 
$-0.046^{+0.015}_{-0.013}$ & 
$ 0.22^{+0.012}_{-0.014}$ & 
$ 0.46^{+0.03}_{-0.03}$ & 
$ 1.28^{+0.06}_{-0.06}$ \\[6pt] 
  & Disk & 
$ 0.12^{+0.06}_{-0.04}$ & 
$-0.048^{+0.045}_{-0.0657}$ & 
$ 0.22^{+0.042}_{-0.030}$ & 
$-0.11^{+0.03}_{-0.02}$ & 
$ 0.025^{+0.004}_{-0.006}$ & 
$ 4.63^{+1.42}_{-1.30}$ & 
$ 1.16^{+0.25}_{-0.28}$  \\[6pt] 
  & Envelope & 
$-0.071^{+0.079}_{-0.06}$ &
$ 0.033^{+0.08}_{-0.10}$ & 
$ 0.032^{+0.079}_{-0.083}$ & 
$-0.27^{+0.061}_{-0.068}$ & 
$ 0.0024^{+0.0042}_{-0.0018}$ & 
$12.15^{+17.13}_{-7.71}$ & 
$ 2.37^{+0.90}_{-0.64}$ \\[-2pt]

\end{tabular}
}
\caption{
The inferred model parameters of the lens light models with one, two and three Sersic profiles fitted to the F814W image in the Light pipeline. The two Sersic model does not assume alignment in its geometric parameters. Errors are given at 3$\sigma$ confidence intervals.
}
\label{table:ModelsLight}
\end{table*}

We now present the results of lens modeling of Abell 1201. We first examine the preferred choice of lens light models, inferred using an isothermal mass model that omits a SMBH. Then we present results using the more complex stellar plus dark matter decomposed mass models, which may also include a SMBH. We discuss additional mass models which assume a total mass profile. In each case we examine the reconstruction of the near-centre counter image that is highly sensitive to the central mass distribution and SMBH, as well as the quantitative Bayesian evidence, $\mathcal{Z}$.

\subsection{Lens Light Model}

The choice of lens light model via Bayesian model comparison is described in \cref{LightModels} and summarized as follows:

\begin{itemize}
 \item All light models with two or three Sersic profiles are favoured over models with one Sersic, producing Bayesian evidence increases of $\Delta \ln \mathcal{Z} > 300$.
 \item The two Sersic models whose centres, position angles, and axis ratios are unaligned produce values of $\Delta \ln \mathcal{Z} > 100$ compared to two Sersic models which assume alignment.
 \item For the F814W image, the three-Sersic model marginally gives the highest evidence overall, where $\Delta \ln \mathcal{Z} = 12$ compared with the double Sersic with unaligned geometric parameters. We use this image to create the lens light subtracted image that mass models are fitted to.
 \item The triple Sersic could not be constrained in the F390W band, owing to its observed lower rest-frame wavelength. We therefore use the double Sersic with aligned parameters to create the F390W lens light subtracted image.
\end{itemize}

\begin{figure*}
\centering
\includegraphics[width=0.241\textwidth]{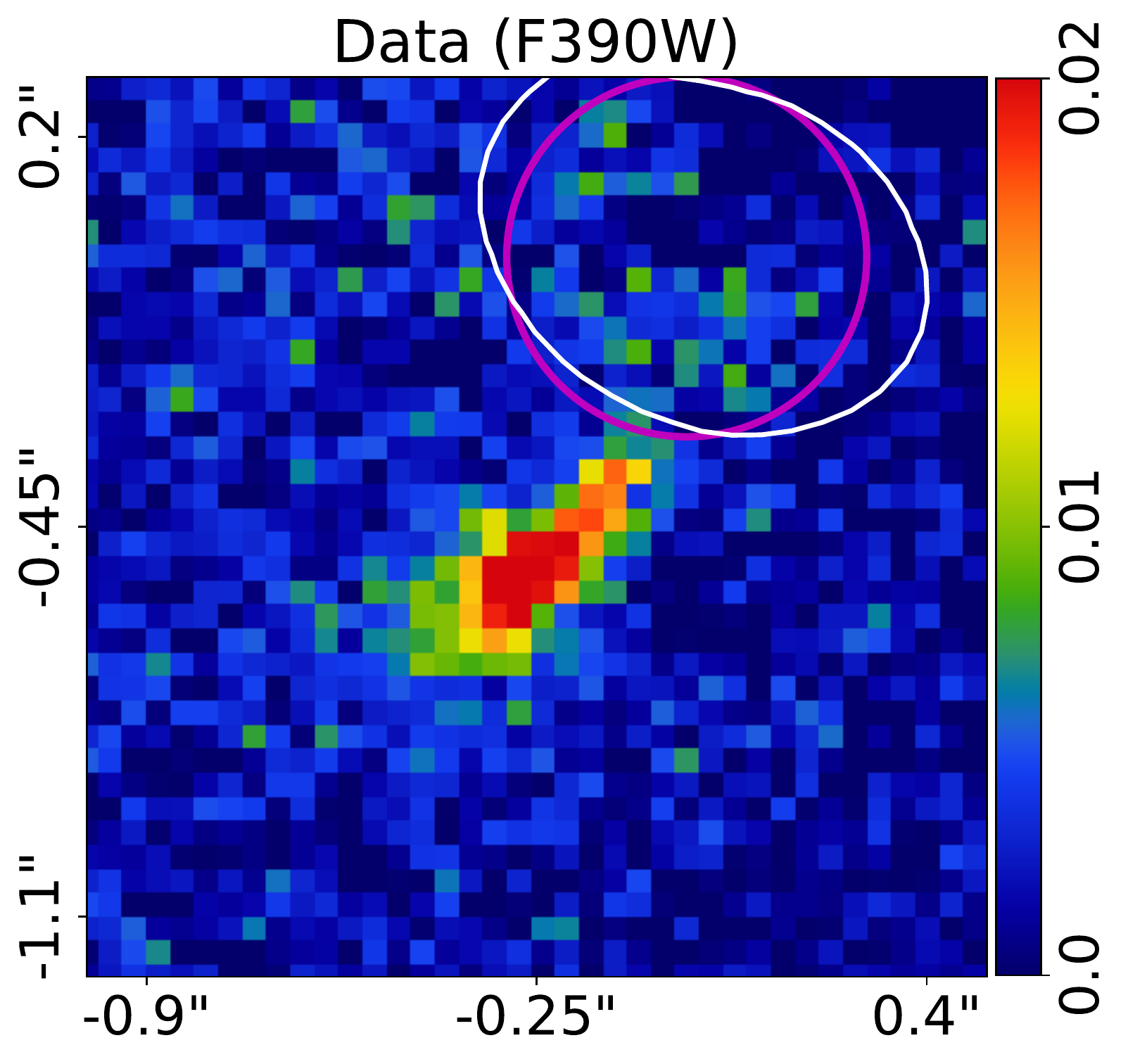}
\includegraphics[width=0.241\textwidth]{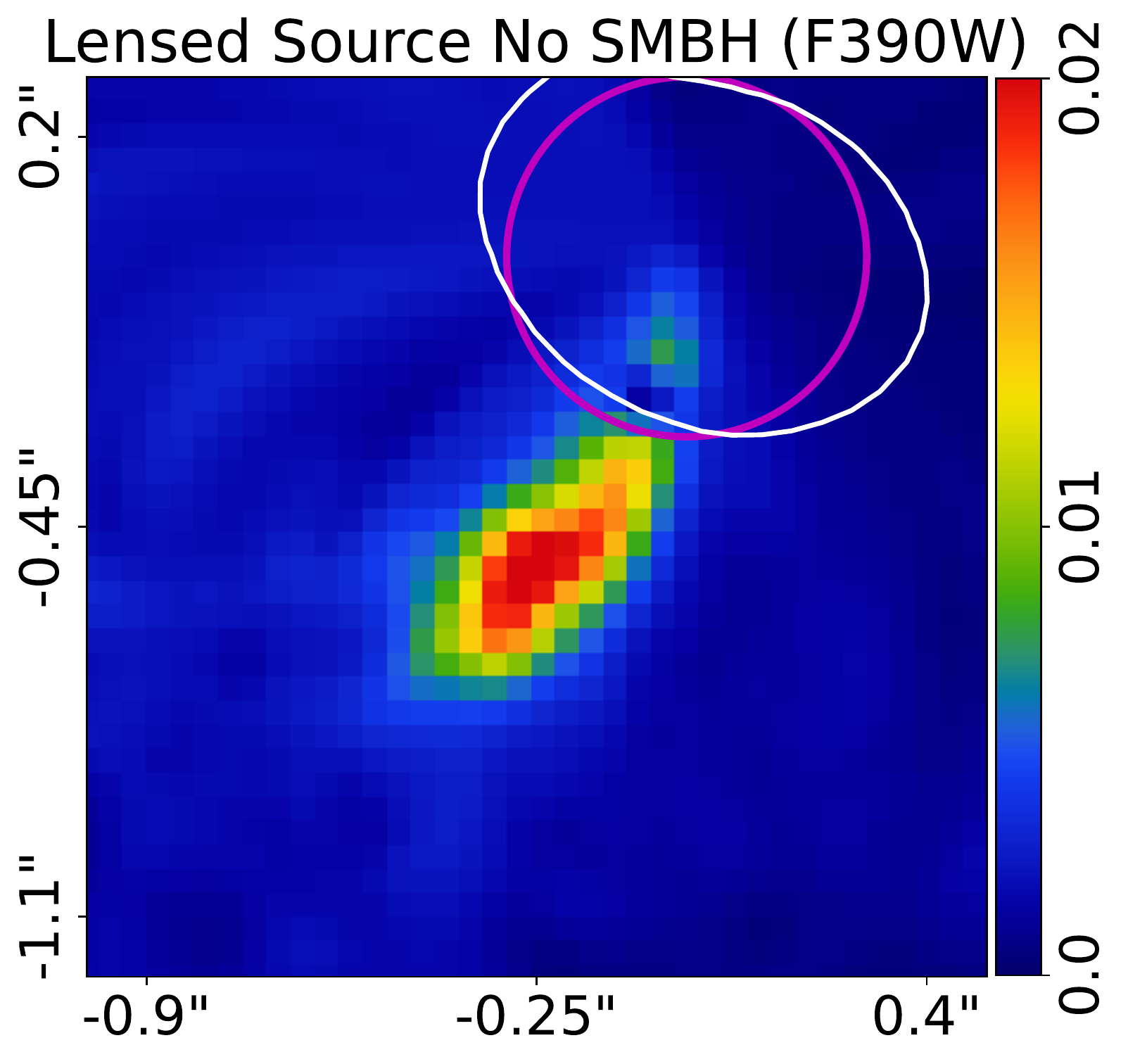}
\includegraphics[width=0.241\textwidth]{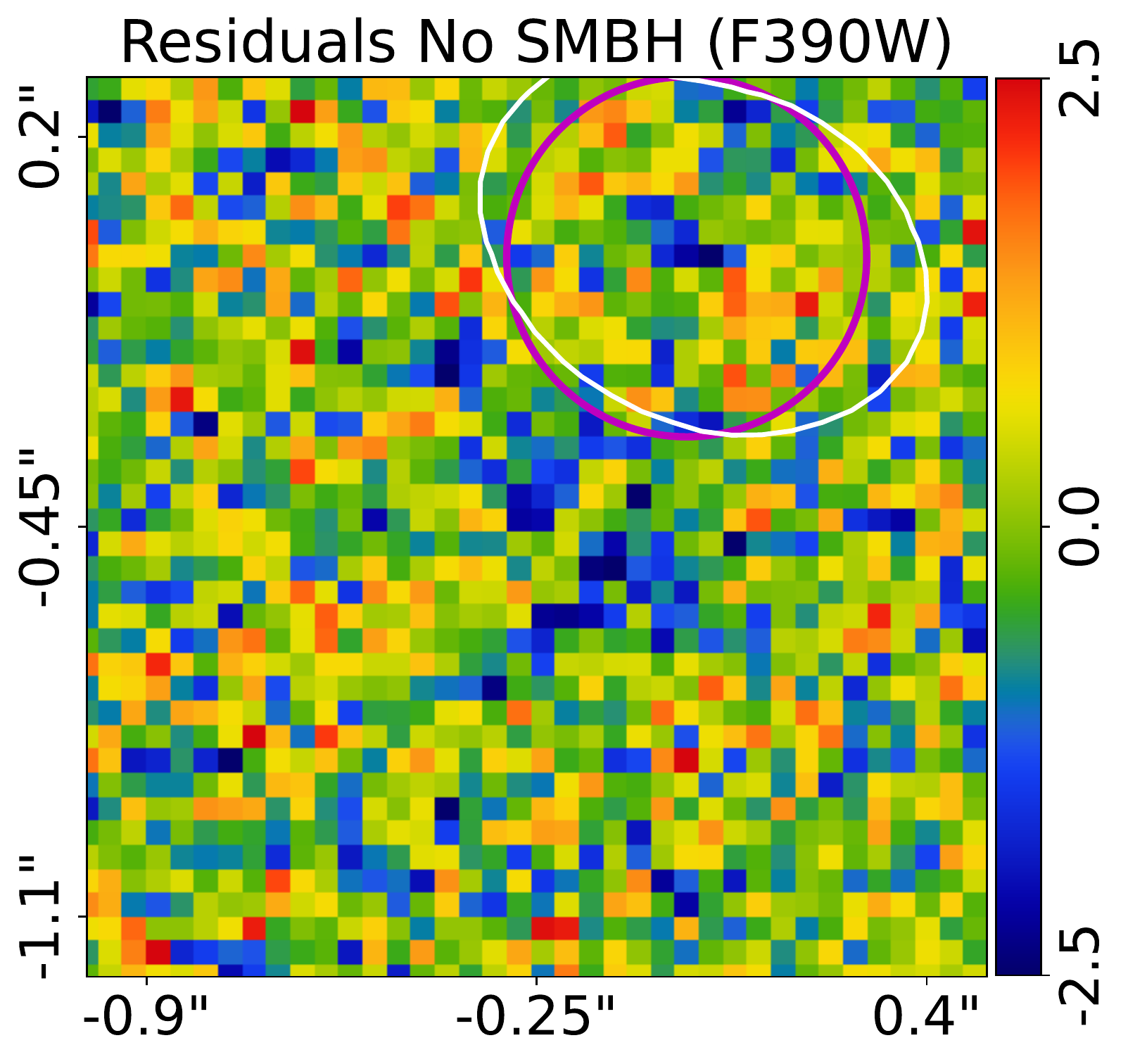}
\includegraphics[width=0.241\textwidth]{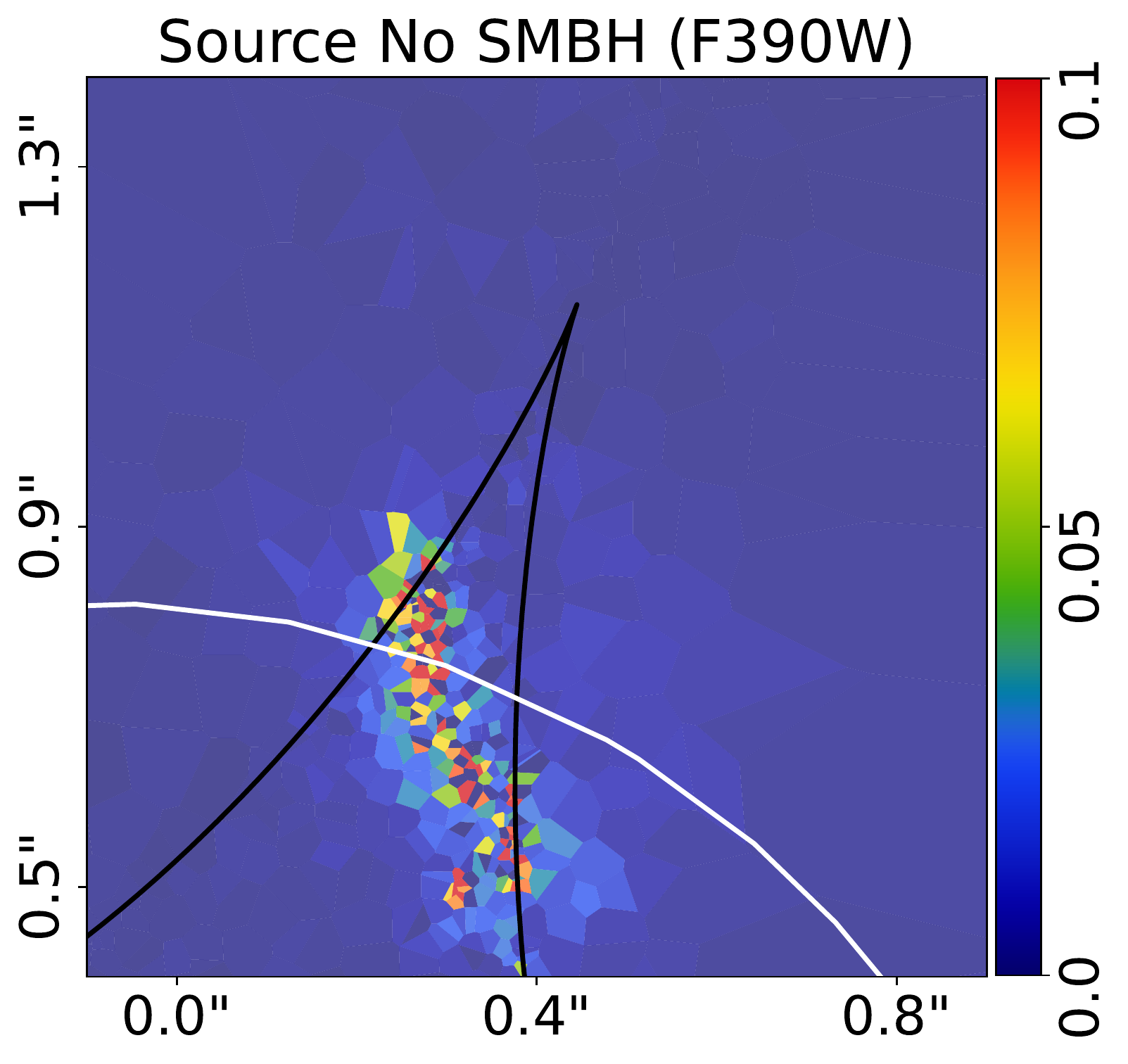}
\includegraphics[width=0.241\textwidth]{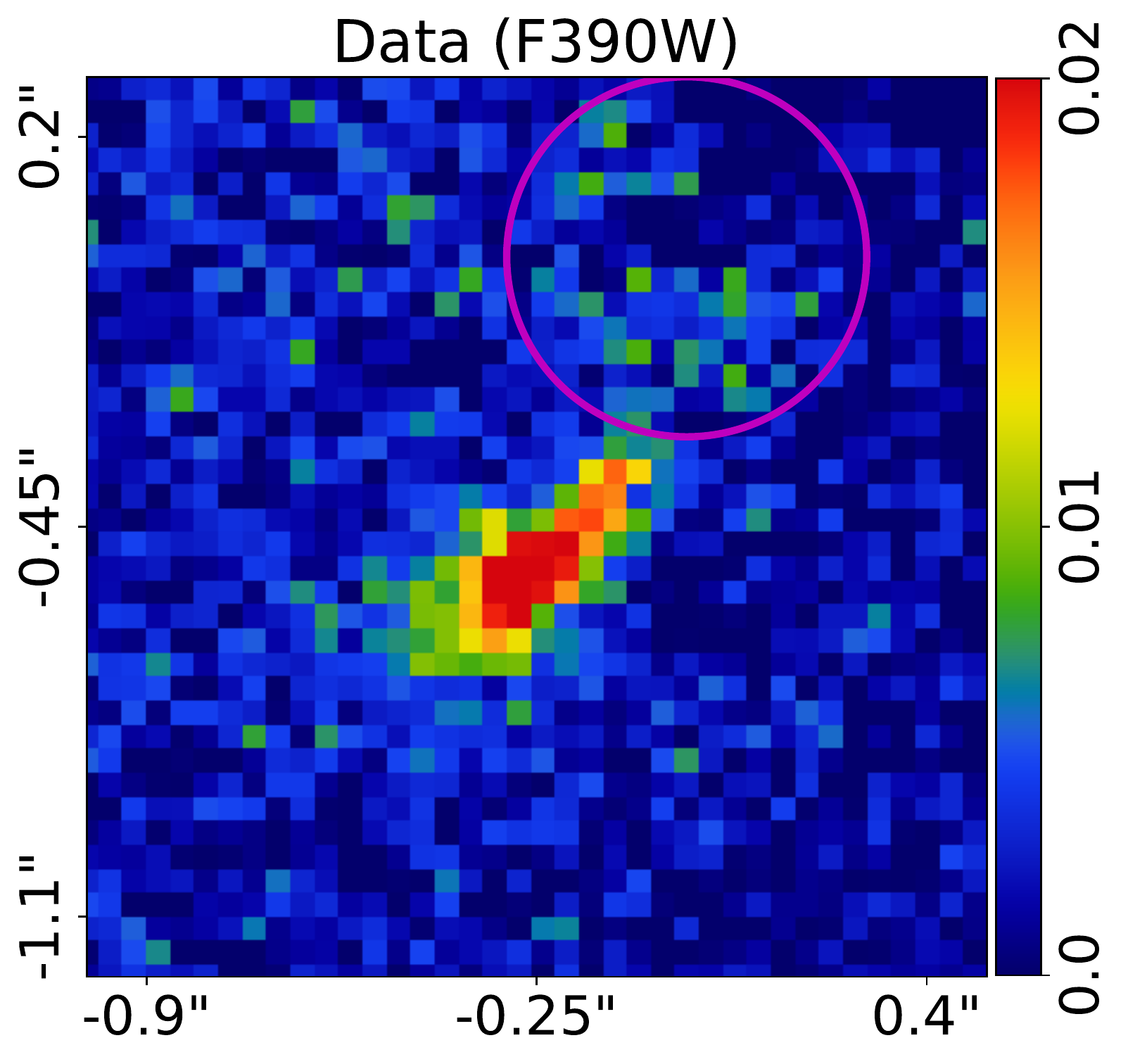}
\includegraphics[width=0.241\textwidth]{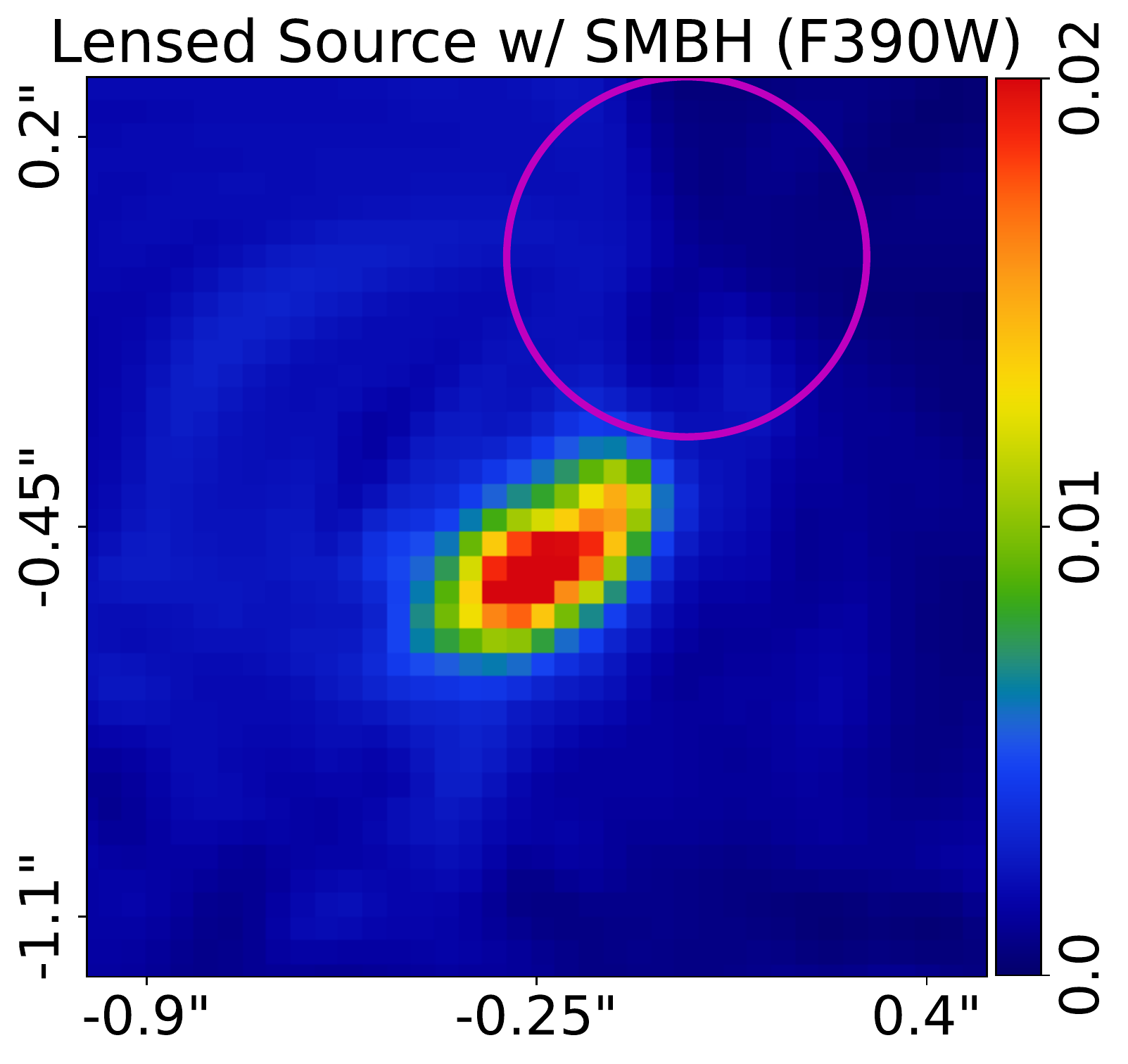}
\includegraphics[width=0.241\textwidth]{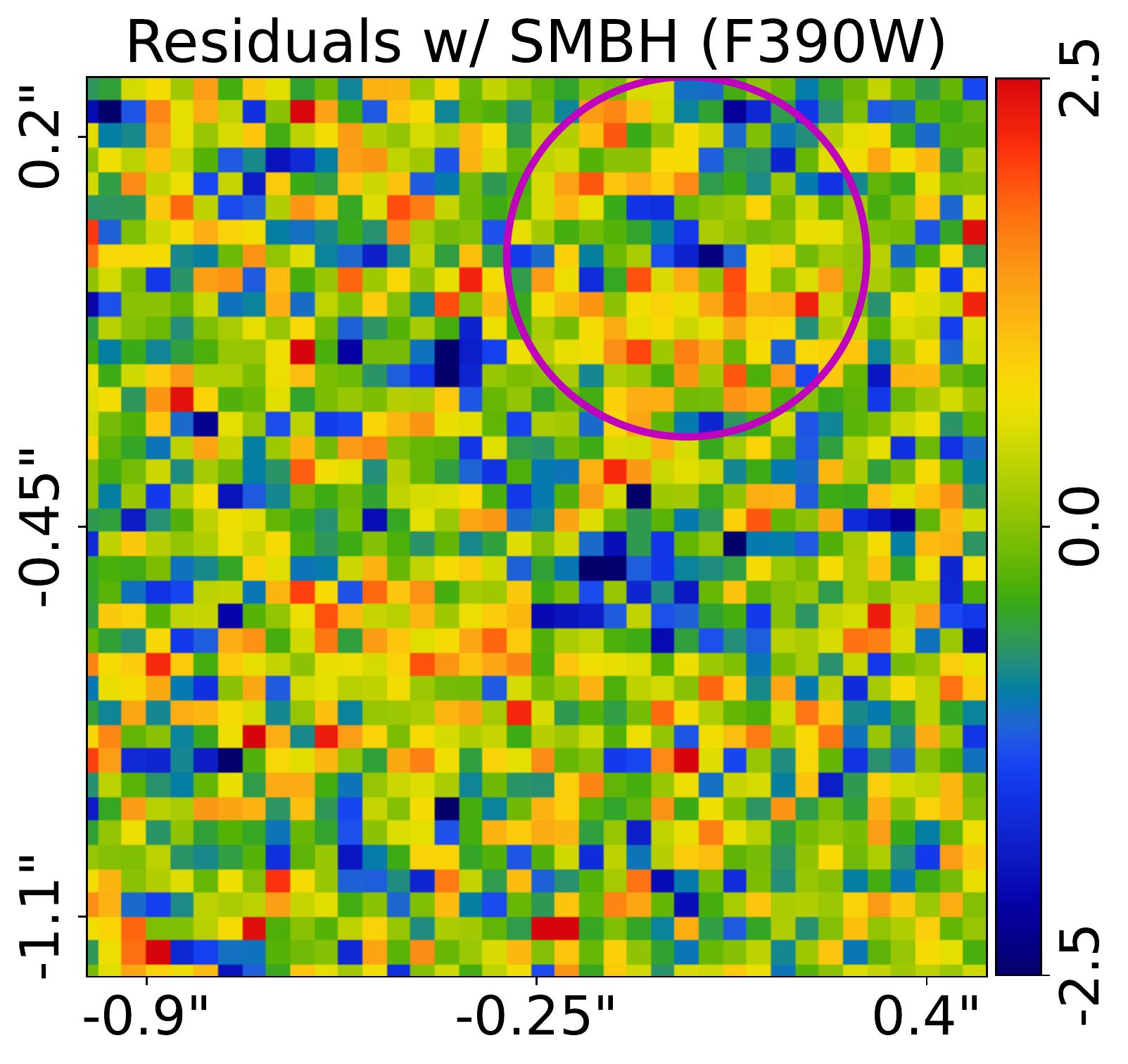}
\includegraphics[width=0.241\textwidth]{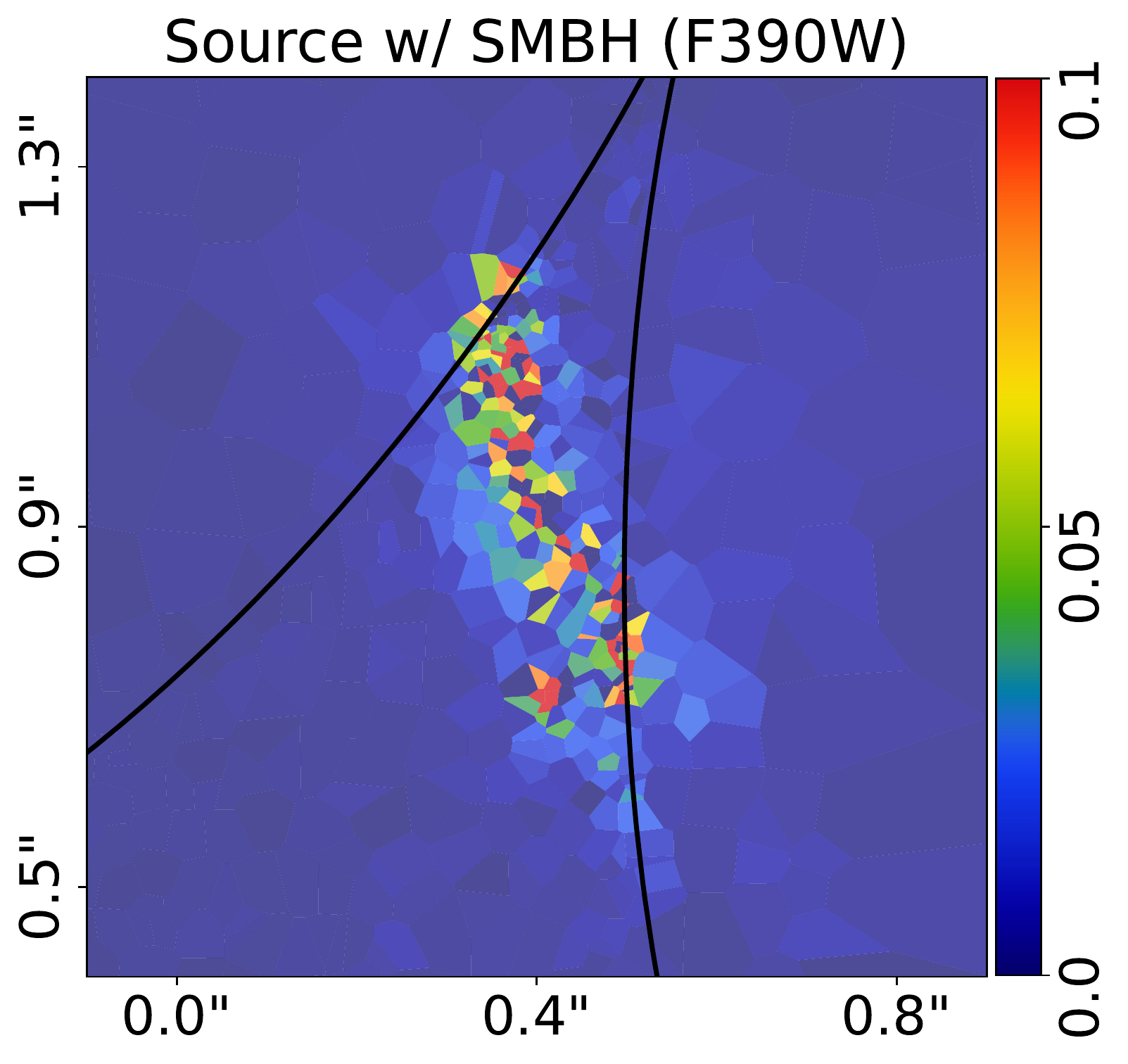}
\caption{
Zoom-ins of the observed counter image in the F390W data (left panel), the model lensed source (left-centre panel), the normalized residuals (right-centre panel) and the source reconstruction (right panel). The top and bottom rows show triple Sersic plus NFW decomposed model-fits without and with a SMBH respectively. All models include an external shear. The magenta circle marks regions of the data where the brightest regions of the lens light were observed and subtracted. Models which omit a SMBH form extraneous light in the reconstructed counter image (which is seen just inside the magenta circle), which is not present in the data. The tangential caustic is shown by a black line and the radial critical curve and caustic are shown with a white line; the latter does not form for models including a SMBH.
} 
\label{figure:LightDarkF390Wx3}
\end{figure*}

\cref{figure:LightFit2} shows the highest evidence lens light model fits to the F390W and F814W images. A good fit to the lens galaxy's emission and a clean subtraction is seen. \cref{table:ModelsLight} gives a subset of inferred parameters for fits to the F814W data and the full results of lens light model comparison are presented in \cref{LightModels}.

A small magenta circle is plotted on this figure and subsequent figures to indicate where the centre of the lens galaxy is. Within this magenta circle faint correlated residuals due to a slightly imperfect lens light subtraction can be seen. These are more visible in the F814W image, which is expected given the lens stellar emission is much brighter. The residuals appear as a dipole-like feature, which is commonly seen for lens light subtractions of HST imaging of strong lenses (e.g. \citealt{Etherington2022}). 

We considered whether these residuals might be a central image of the lensed source galaxy, but in this case the feature would be much brighter in the F390W image. Dust absorption could lower the F390W emission, however HST F606W observations of Abell 1201 also show no central emission \citet{Smith2017a}, making dust absorption unlikely. In \cref{Radial} we fit mass models with priors manually tuned to include a centrally cored mass profile, which for the F814W (or the F390W) data do not reconstruct this central emission. Lens modeling therefore confirms it is not a central image.

\subsection{Decomposed Mass Models}\label{Decomp}

\begin{table}
\resizebox{\linewidth}{!}{
\begin{tabular}{ l | l | l | l } 
\multicolumn{1}{p{1.1cm}|}{Filter} 
& \multicolumn{1}{p{1.3cm}|}{Number of Sersics} 
& \multicolumn{1}{p{1.3cm}|}{Includes SMBH?} 
& \multicolumn{1}{p{1.5cm}|}{$\ln \mathcal{Z}$}  
\\ \hline
F390W & 2 & \ding{55} & 125637.18  \\[1pt]
F390W & 2 & \checkmark & 125669.13  \\[0pt]
\hline
F390W & 3 & \ding{55} & 125598.48  \\[1pt]
F390W & 3 & \checkmark & \textbf{125699.06} \\[0pt]
\hline
F814W & 2 & \ding{55} & 78330.51  \\[1pt]
F814W & 2 & \checkmark & 78328.12  \\[0pt]
\hline
F814W & 3 & \ding{55} & 78329.19   \\[1pt]
F814W & 3 & \checkmark & 78332.19 \\[0pt]
\end{tabular}
}
\caption{The Bayesian Evidence, $\ln \mathcal{Z}$, of each model-fit performed by the Mass pipelines using decomposed mass models that assume two and three Sersic profiles, an elliptical NFW and external shear. Fits to both the F390W and F814W images are shown, where the F390W fits assume the Sersic parameters of the F814W image for the stellar mass. The favoured model given our criteria of $\Delta \ln \mathcal{Z} > 10$ is shown in bold. For the F390W image, all models with a SMBH produce $\Delta \ln \mathcal{Z}$ values of at least $30$ above models without a SMBH.}
\label{table:SMBHMCDecomp}
\end{table}

\begin{figure*}
\centering
\includegraphics[width=0.241\textwidth]{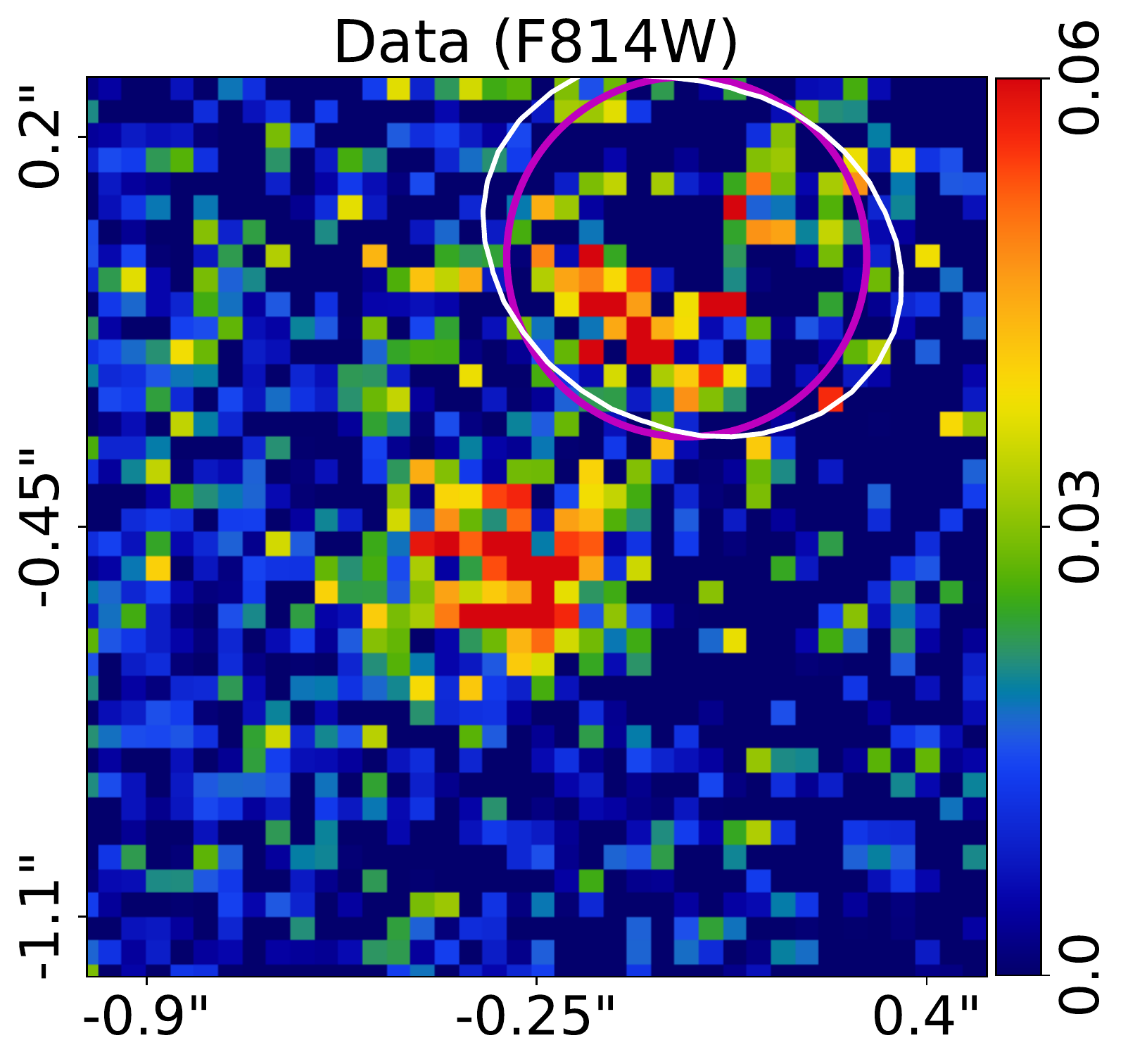}
\includegraphics[width=0.241\textwidth]{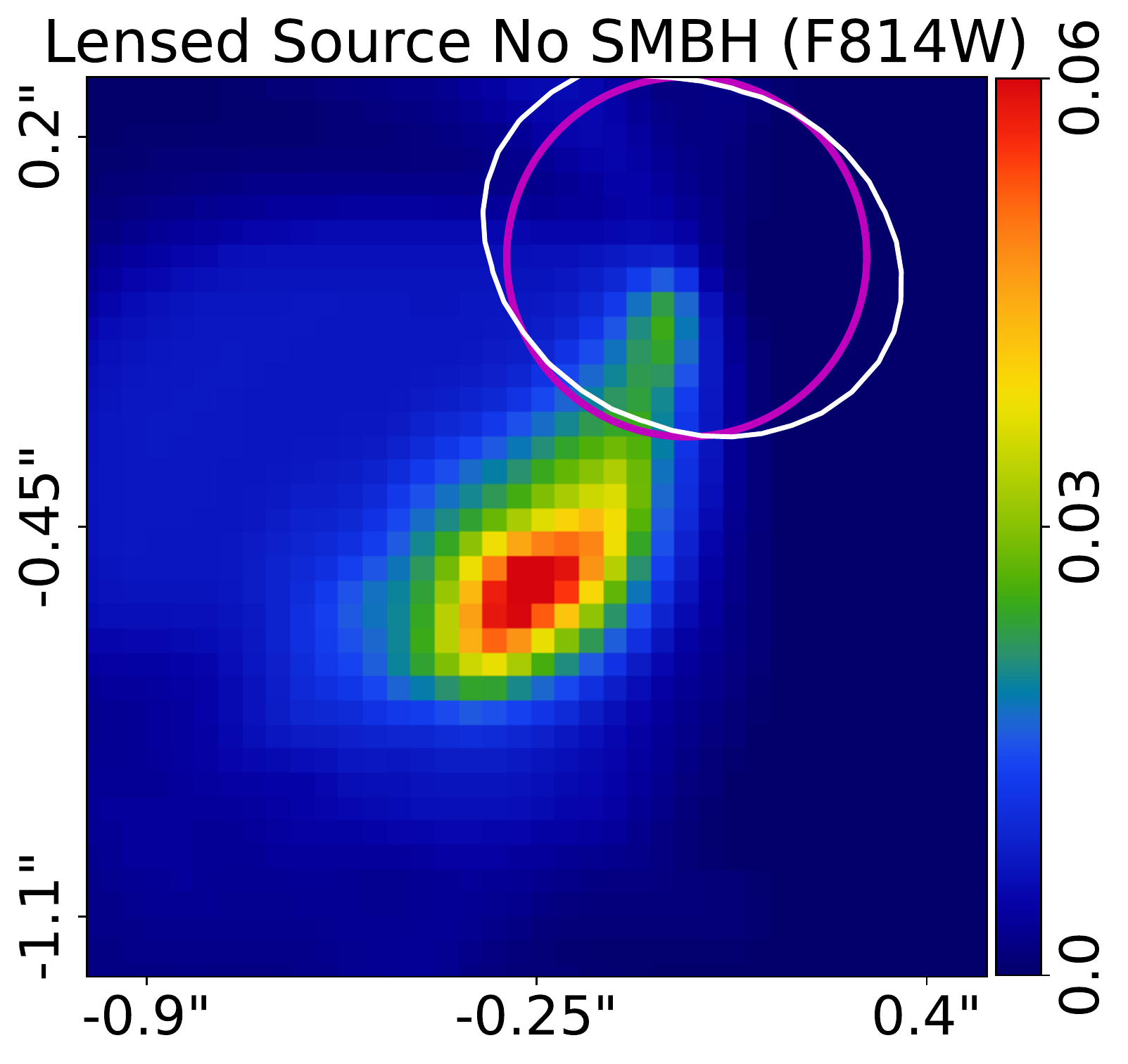}
\includegraphics[width=0.241\textwidth]{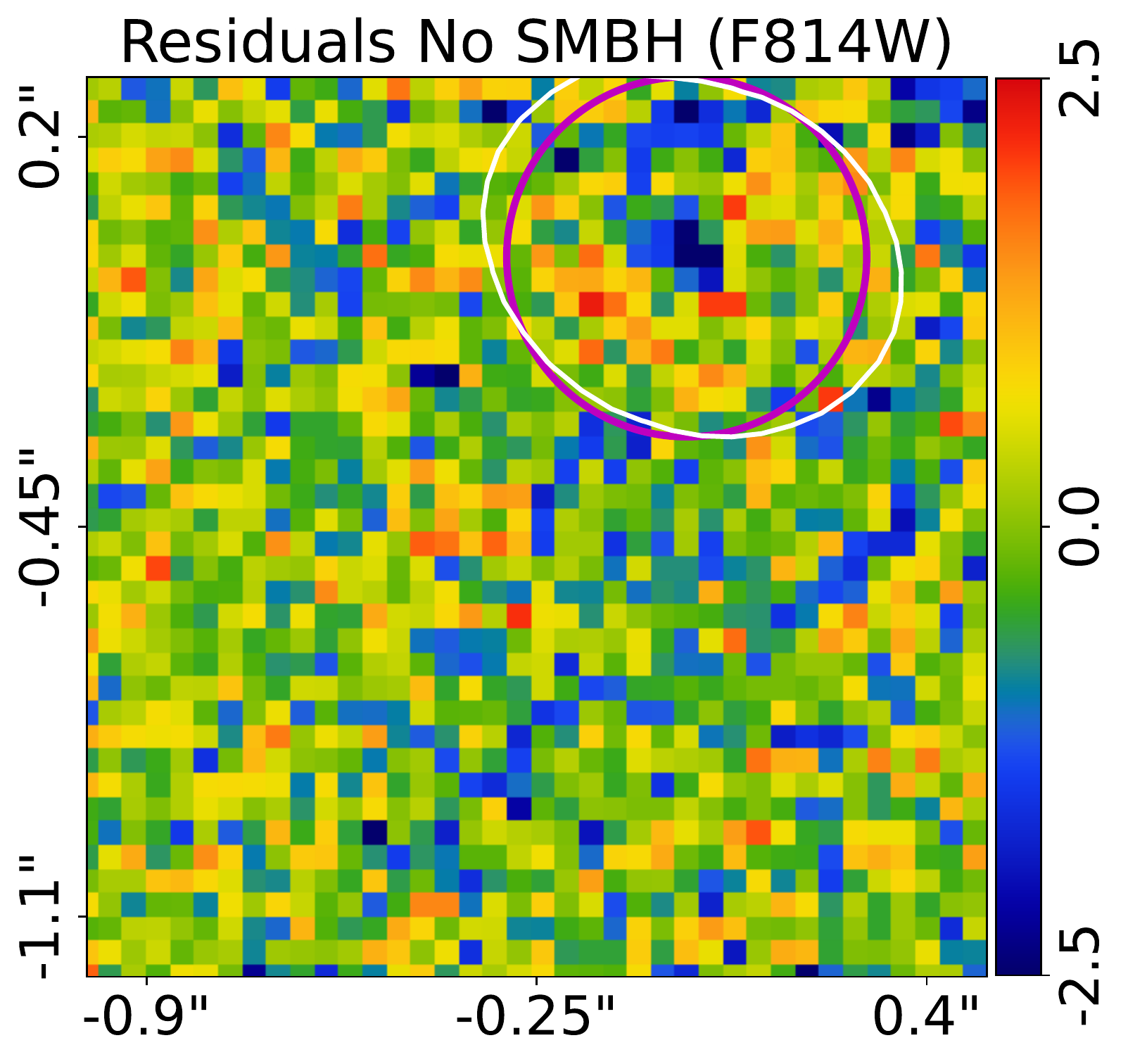}
\includegraphics[width=0.241\textwidth]{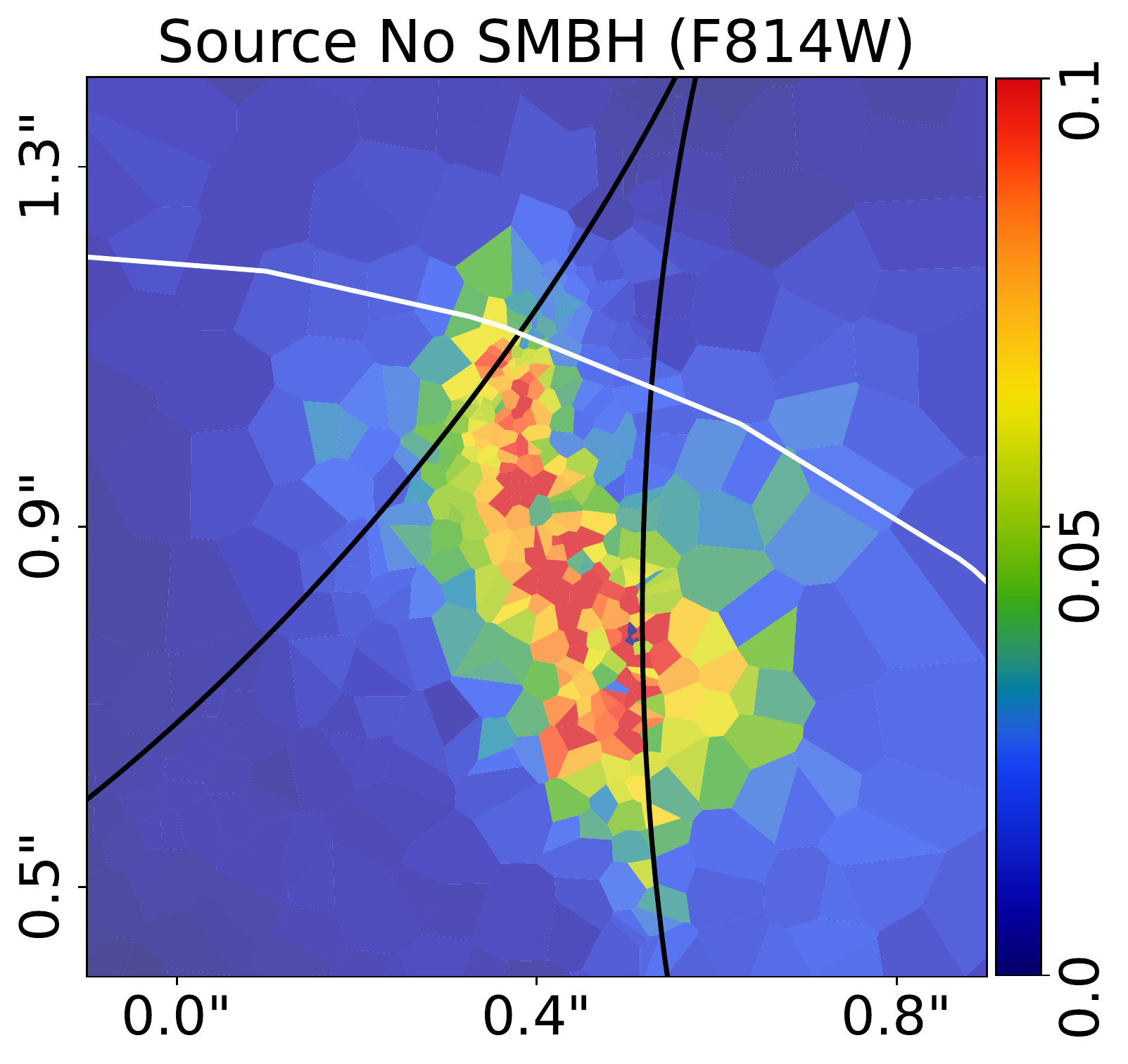}
\includegraphics[width=0.241\textwidth]{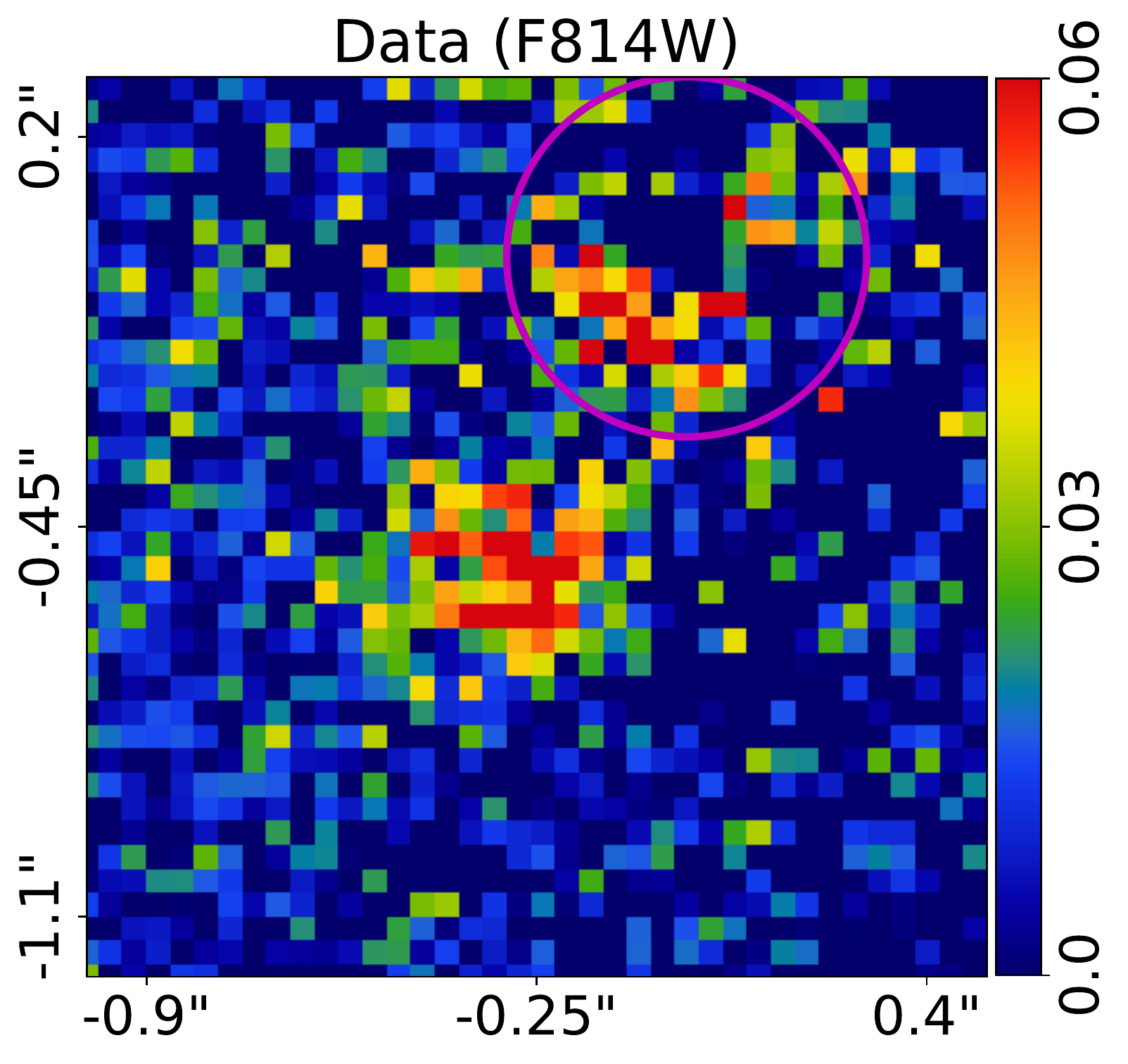}
\includegraphics[width=0.241\textwidth]{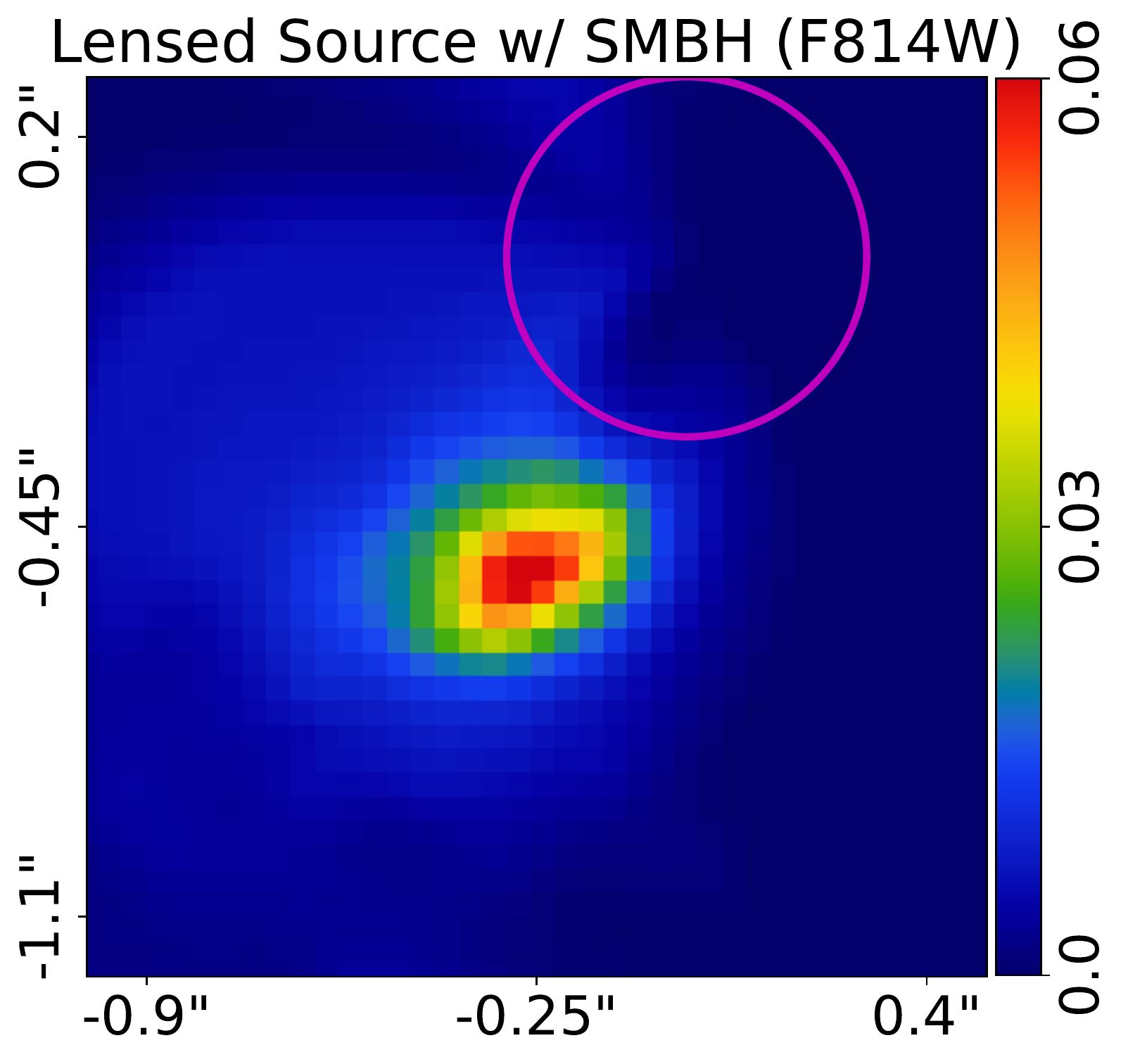}
\includegraphics[width=0.241\textwidth]{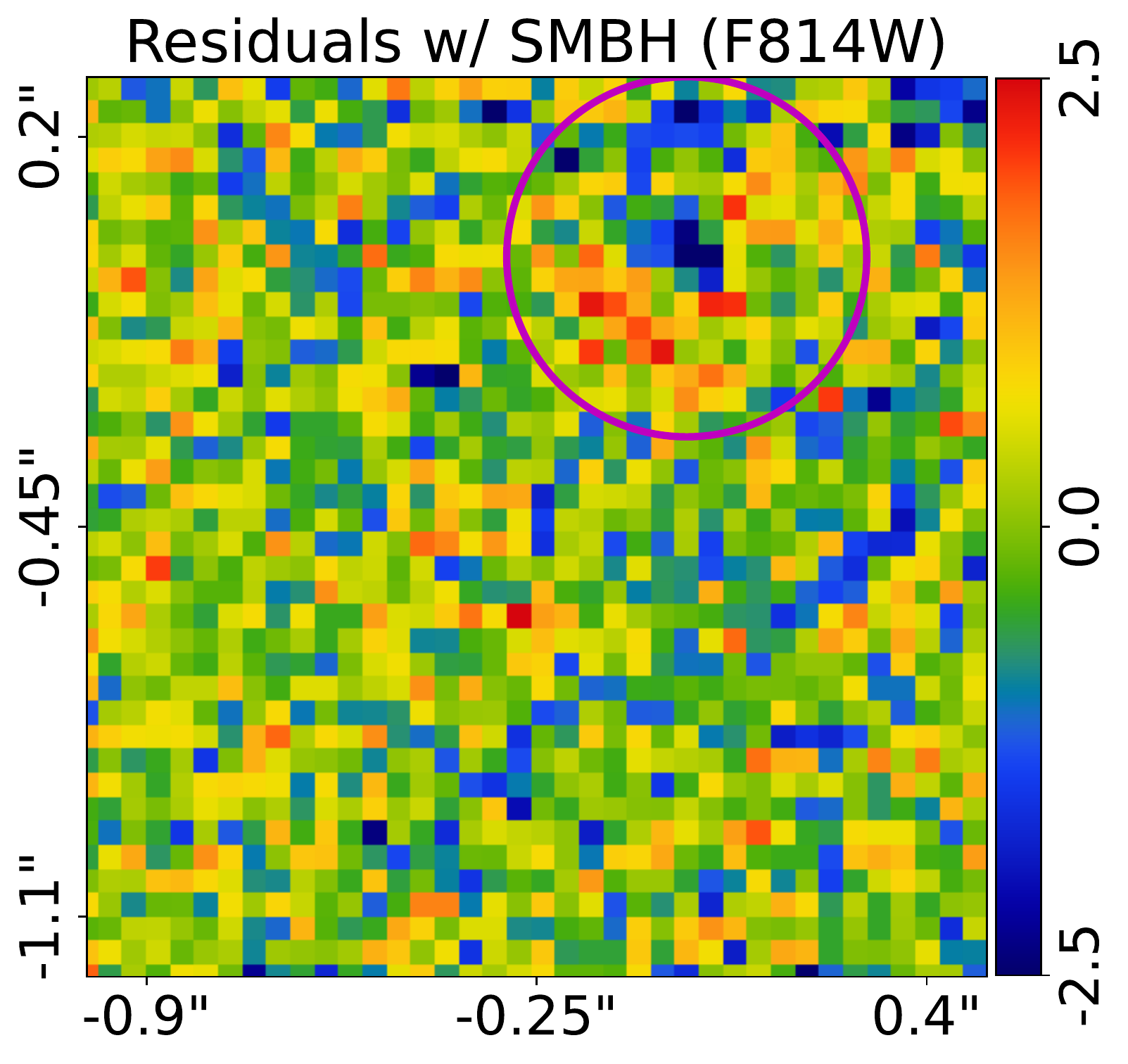}
\includegraphics[width=0.241\textwidth]{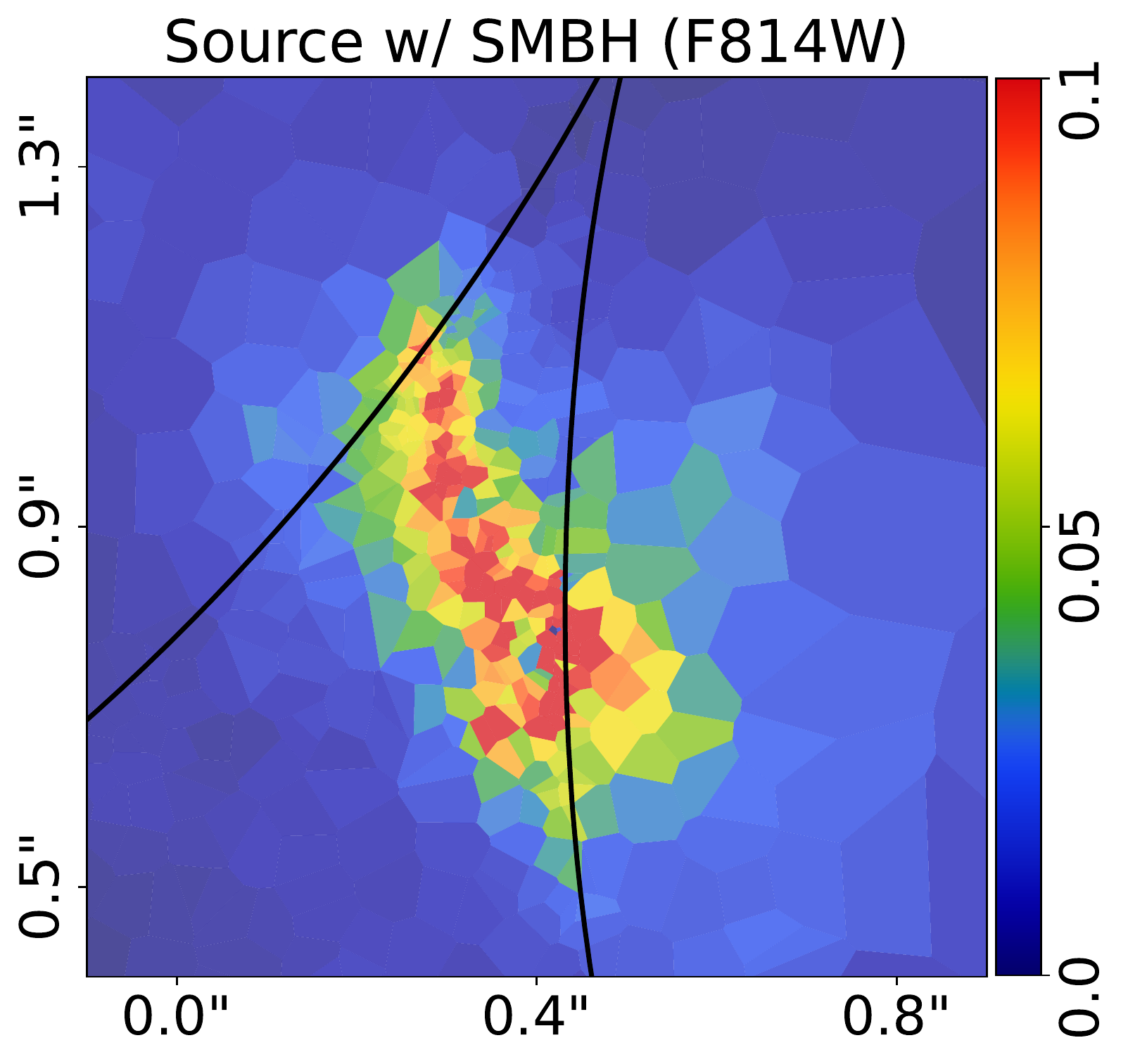}
\caption{
The same as \cref{figure:LightDarkF390Wx3} but for the F814W data.
} 
\label{figure:LightDarkF814Wx3}
\end{figure*}

\begin{table*}
\tiny
\resizebox{\linewidth}{!}{
\begin{tabular}{ l l l l l l l l l l} 
\multicolumn{1}{p{1.8cm}|}{\centering \textbf{Model}} 
& \multicolumn{1}{p{1.0cm}}{$\Psi^{\rm{bulge}}$ (e$^{\rm -}$\,s$^{\rm -1}$)} 
& \multicolumn{1}{p{1.0cm}}{$\Psi^{\rm{disk}}$ (e$^{\rm -}$\,s$^{\rm -1}$)} 
& \multicolumn{1}{p{1.0cm}}{$\Psi^{\rm{envelope}}$ (e$^{\rm -}$\,s$^{\rm -1}$)} 
& \multicolumn{1}{p{1.0cm}}{$\Gamma^{\rm{bulge}}$} 
& \multicolumn{1}{p{1.0cm}}{$\Gamma^{\rm{disk}}$} 
& \multicolumn{1}{p{1.0cm}}{$\Gamma^{\rm{envelope}}$} 
& \multicolumn{1}{p{1.0cm}}{$\epsilon_{\rm 1}^{\rm{ext}}$} 
& \multicolumn{1}{p{1.0cm}}{$\epsilon_{\rm 2}^{\rm{ext}}$} 
& \multicolumn{1}{p{1.0cm}}{$\theta_{\rm Ein}^{\rm{smbh}}$(\arcsec)} 
\\ \hline
& & & & & & \\[-4pt]

x3 Sersic & 
$1.47^{+0.69}_{-0.55}$ & 
$1.14^{+0.44}_{-0.28}$ & 
$2.32^{+1.05}_{-0.95}$ & 
$0.57^{+0.21}_{-0.32}$ & 
$0.34^{+0.19}_{-0.18}$ & 
$0.12^{+0.08}_{-0.12}$ & 
$-0.097^{+0.042}_{-0.044}$ & 
$0.19^{+0.05}_{-0.07}$  \\[-6pt]

\\ \hline
& & & & & & \\[-2pt]

x2 Sersic & 
$1.65^{+0.69}_{-0.61}$ & 
$0.95^{+0.74}_{-0.30}$ & 
& 
$0.46^{+0.35}_{-0.34}$ & 
$0.29^{+0.16}_{-0.14}$ & 
&
$-0.084^{+0.061}_{-0.057}$ & 
$0.13^{+0.06}_{-0.06}$ \\[-6pt]

\\ \hline
& & & & & & \\[-2pt]

x3 Sersic + SMBH & 
$0.54^{+0.26}_{-0.38}$ & 
$1.45^{+0.55}_{-0.33}$ & 
$1.69^{+0.63}_{-1.05}$ & 
$0.50^{+0.37}_{-0.25}$ & 
$0.41^{+0.14}_{-0.11}$ & 
$0.33^{+0.20}_{-0.12}$ & 
$-0.14^{+0.02}_{-0.02}$ & 
$0.22^{+0.02}_{-0.04}$ & 
$0.48^{+0.07}_{-0.11}$ \\[-6pt]

\\ \hline
& & & & & & \\[-2pt]
x2 Sersic + SMBH & 
$0.87^{+0.61}_{-0.80}$ & 
$1.25^{+0.57}_{-0.46}$ & 
& 
$0.15^{+0.65}_{-1.12}$ & 
$0.48^{+0.12}_{-0.18}$ & 
&
$-0.14^{+0.04}_{-0.03}$ & 
$0.20^{+0.03}_{-0.04}$ & 
$0.42^{+0.09}_{-0.17}$ \\[-2pt]
\end{tabular}
}
\caption{
The inferred stellar mass, shear and SMBH model parameters of the decomposed mass models with two and three Sersic components fitted to the F390W image in the Mass pipeline. Errors are given at 3$\sigma$ confidence intervals.
}
\label{table:ModelsDecompStellar}
\end{table*}

We now present results using decomposed mass models that separately model Abell 1201's stellar and dark matter. Based on the lens light model comparison, we fit models assuming both two and three Sersic profiles (where parameters for the F390W fits use those inferred from fits to the F814W, see \cref{F390WModel}). We fit both models independently to both the F390W and F814W images. Visualization in this section shows the triple Sersic fits, \cref{MassFits} shows figures for the double Sersic fits. 

\cref{table:SMBHMCDecomp} shows the values of $\ln \mathcal{Z}$ for decomposed models with and without a SMBH. Values of $\Delta \ln \mathcal{Z} > 30$ are seen for all model fits to the F390W image with a SMBH compared to those without. The highest overall value of $\ln \mathcal{Z}$ is the triple Sersic decomposed mass model with a SMBH, which is a $\ln \mathcal{Z}$ value more than 60 greater than that for any decomposed model without a SMBH. For the F814W images all models produce nearly consistent values of $\ln \mathcal{Z}$ with or without a SMBH, indicating that the higher S/N of the F390W data or the source's different structure is enabling the SMBH detection. 

\cref{figure:LightDarkF390Wx3Global} shows the reconstructed lensed sources and normalized residuals for fits to the F390W and F814W images with and without a SMBH. All models reproduce the giant arc and counter image. Residuals are seen around the giant arc in the F390W image indicating missing complexity in the mass model. These residuals are seen across all mass models compared in this work (including fits which include the mass of the line-of-sight galaxy to the right of the giant arc, see \cref{MassClump}). We therefore do not anticipate they impact our inference on the SMBH. The reconstructed counter images for the models with and without a SMBH are visibly distinct and they produce different residuals, albeit this is difficult to discern from \cref{figure:LightDarkF390Wx3Global} due to the large arc-second scales over which the image is plotted.

\begin{table*}
\tiny
\resizebox{\linewidth}{!}{
\begin{tabular}{ l l l l l l} 
\multicolumn{1}{p{1.8cm}|}{\centering \textbf{Model}} 
& \multicolumn{1}{p{1.2cm}}{$x^{\rm{dark}}$(\arcsec)} 
& \multicolumn{1}{p{1.2cm}}{$y^{\rm{dark}}$(\arcsec)} 
& \multicolumn{1}{p{1.2cm}}{$\epsilon_{\rm 1}^{\rm{dark}}$} 
& \multicolumn{1}{p{1.2cm}}{$\epsilon_{\rm 2}^{\rm{dark}}$} 
& \multicolumn{1}{p{2.5cm}}{$M_{\rm 200}^{\rm{dark}}$ ($M_{\rm \odot} \times 10^{14}$)} 
\\ \hline
& & & & & \\[-7pt]

x3 Sersic & 
$0.048^{+0.108}_{-0.126}$ & 
$0.050^{+0.100}_{-0.121}$ & 
$0.022^{+0.110}_{-0.090}$ & 
$0.13^{+0.12}_{-0.15}$ & 
$3.43^{+1.24}_{-7.76}$  \\[-7pt]

\\ \hline
& & & & & \\[-7pt]

x2 Sersic & 
$0.031^{+0.059}_{-0.058}$ & 
$0.11^{+0.17}_{-0.12}$ & 
$0.032^{+0.084}_{-0.106}$ & 
$ 0.017^{+0.119}_{-0.097}$ &  
$5.55^{+3.29}_{-2.19}$ \\[-7pt]

\\ \hline
& & & & & \\[-7pt]

x3 Sersic + SMBH & 
$0.13^{+0.08}_{-0.17}$ & 
$0.17^{+0.03}_{-0.10}$ & 
$-0.036^{+0.066}_{-0.115}$ & 
$ 0.19^{+0.12}_{-0.16}$ & 
$1.43^{+6.48}_{-6.28}$ \\[-7pt]

\\ \hline
& & & & & \\[-7pt]
x2 Sersic + SMBH & 
$-0.22^{+0.12}_{-0.12}$ & 
$0.008^{+0.084}_{-0.088}$ & 
$-0.015^{+0.086}_{-0.159}$ & 
$ 0.095^{+0.094}_{-0.067}$ 
& $1.69^{+8.31}_{-4.37}$  \\[-2pt]
\end{tabular}
}
\caption{
The inferred dark matter model parameters of the decomposed mass models with two and three Sersic components fitted to the F390W image in the Mass pipeline. Errors are given at 3$\sigma$ confidence intervals.
}
\label{table:ModelsDecompDark}
\end{table*}

\cref{figure:LightDarkF390Wx3} therefore shows zoom-ins around the counter image for the F390W image, where models without and with a SMBH are shown on the top and bottom rows respectively. The model without a SMBH places extraneous flux in the reconstructed counter image, which is not present when the SMBH is included (this flux can be seen within the radial critical curve shown by a white line and the magenta circle, and is not related to the lens light residuals). \cref{figure:LightDarkF814Wx3} shows zoom-ins for the F814W image where the same extraneous flux is seen for the model not including a SMBH. The same behaviour is seen in \cref{MassFits} for the decomposed models which fits two Sersics instead of three. The inclusion of the SMBH therefore allows the counter image to be reconstructed more accurately in both wavebands, removing central luminous emission that is not observed in the data. Removing this extraneous flux increases $\ln \mathcal{Z}$ for the F390W image data only, implying that the F814W data are too low S/N for changes in the counter image reconstruction to improve the fit in a Bayesian sense.

\begin{figure}
\centering
\includegraphics[width=0.47\textwidth]{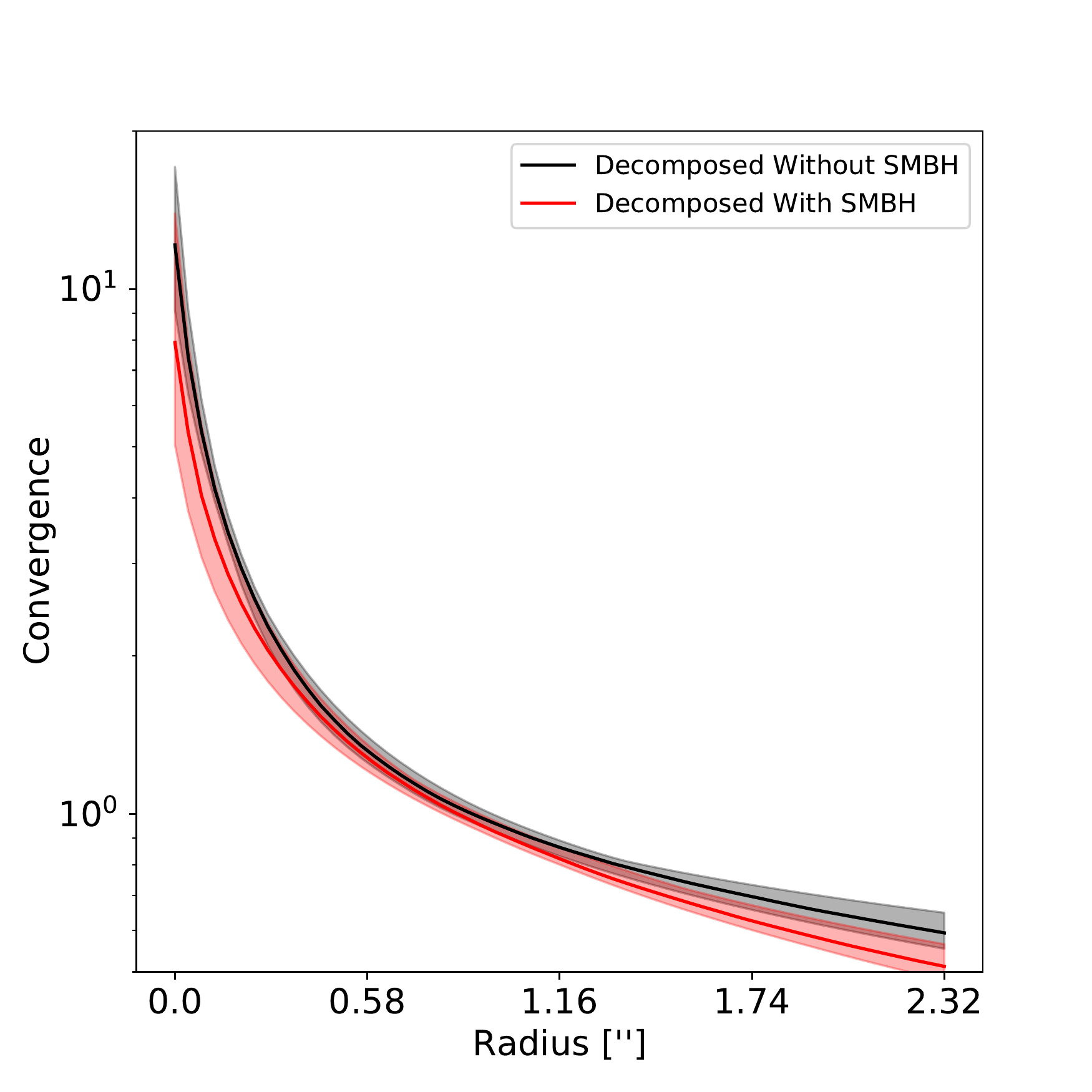}
\caption{
The convergence as a function of radius inferred using the F390W image for the decomposed mass models which assume three Sersic profiles, without a SMBH (black) and with a SMBH (red). All models include an external shear. Each line is computed using coordinates that extend radially outwards from the centre of the mass profile and are aligned with its major axis. Shaded regions for each mass model's convergence are shown, corresponding to the inferred $3\sigma$ confidence intervals. The 1D convergence of the SMBH is not included, to make comparison of each mass model's convergence straightforward. 
} 
\label{figure:LightDark1Dx3}
\end{figure}

The inferred 1D convergence profiles for the decomposed models with three Sersic profiles are shown in \cref{figure:LightDark1Dx3}. When a SMBH is included the inferred mass model convergence is shallower. Increasing the central density of the lens galaxy's mass model therefore produces a similar lensing effect to including a SMBH and is an alternative way to improve the counter image fit. We will expand on this further when we discuss the total mass model fits.

\begin{figure*}
\centering
\includegraphics[width=0.98\textwidth]{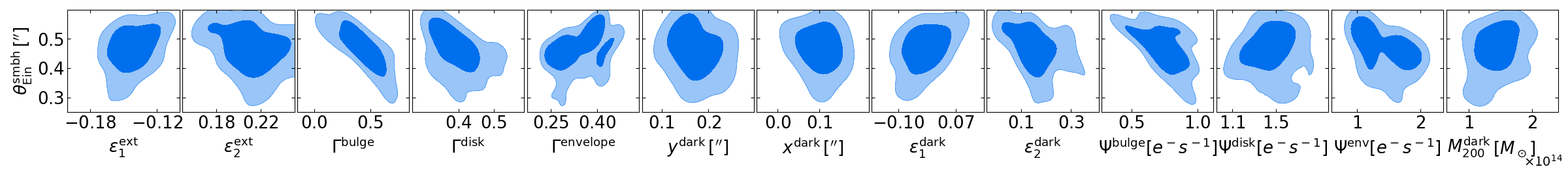}
\caption{
The 2D probability density functions (PDF) of fits to the F390W image of Abell 1201 using the triple Sersic decomposed mass model. Marginalized 2D contours are shown for every lens mass model parameter paired with the SMBH normalization $\theta^{\rm smbh}_{\rm Ein}$ which is related to $M_{\rm BH}$, see \cref{eqn:PointMass}. The inner and outer contours cover the 1 and 2$\sigma$ confidence intervals respectively.
}
\label{figure:LightDarkPDFx3}
\end{figure*}

\cref{figure:LightDarkPDFx3} shows the 2D probability density functions (PDF) of the mass model parameters and the SMBH normalization $\theta^{\rm smbh}_{\rm Ein}$ for fits to the F390W image using the triple Sersic decomposed models. $\theta^{\rm smbh}_{\rm Ein}$ depends on the parameters controlling the mass distribution (e.g., $\Gamma^{\rm bulge}$, $\Gamma^{\rm disk}$, $\Gamma^{\rm env}$). 

\cref{table:ModelsDecompStellar} and \cref{table:ModelsDecompDark} give the inferred parameter estimates of the decomposed models. We can compare our inferred dark matter halo mass to the virial mass estimate of \citet{Rines2013} (from the infall caustic method), which for an NFW dark matter halo gives $M_{\rm 200} = 3.9 \pm 0.1 \times 10^{14} M_{\rm \odot}$. \cref{table:ModelsDecompDark} shows our estimates of $M_{\rm 200}$ range between $M_{\rm 200} = 0 - 5 \times 10^{14} M_{\rm \odot}$. Both models with a SMBH are consistent with \citet{Rines2013}. Our lens model is therefore inferring a realistic dark matter host halo.  

\subsection{Total Mass Models}

The results of fitting total mass models which represent the stars and dark matter with a single projected mass distribution are given in \cref{ResultSIE}. For the power-law (PL) mass model, which has reduced flexibility in adjusting its central density, the inferred $\ln \mathcal{Z}$ values without a SMBH are over 100 below models including a SMBH (PL or decomposed). When the PL includes a SMBH, $\ln \mathcal{Z}$ increases to within $\sim 10$ of the decomposed models with a SMBH. The PL fits therefore strongly favour a SMBH. The counter image reconstructions also reflect those seen above, whereby PL models without a SMBH show extraneous flux which is removed when a SMBH is included. 

For the broken power-law (BPL), which has more flexibility in adjusting its central density, the model without a SMBH infers $\ln \mathcal{Z} = 125699.90$ for the F390W data. This is within $10$ of the highest evidence decomposed and PL models with a SMBH. This model also reconstructs the counter image without extraneous flux. In a Bayesian sense, the BPL model without a SMBH is therefore as likely as any model fitted in this work with a SMBH, calling into question whether a SMBH is necessary in the lens model.  

The inferred BPL model increases its central density above any decomposed model inferred in \cref{Decomp}. In \cref{ResultSIE}, we therefore verify that the decomposed model parameterization can attain the same central density as the BPL. We show that it does when the bulge's radial gradient parameter is increased to $\Gamma^{\rm bulge} = 0.9$. The reason we do not infer this model is because this model is lower likelihood than models inferred above, where $\Gamma^{\rm bulge} = \sim 0.5$, indicating that increasing the stellar mass density produces a different lensing effect to including a SMBH. In \cref{ResultSIE} we also fit models where the dark matter concentration is free to vary, such that it can reach the same central density as the BPL. These models again do not produce solutions with as high an evidence as those presented above.

For the high density BPL model to fit the data as well as the mass models with a SMBH, its ($x^{\rm mass}$,\,$y^{\rm mass}$) centre assumes values that are $\geq 100$\,pc offset from the centre of the bulge's luminous emission. In \cref{ResultSIE}, we show that if the BPL model centre is aligned with the luminous bulge it produces a much lower $\ln \mathcal{Z}$. The BPL model is built-on the assumption that it can simultaneously represent both the stellar and dark matter mass distributions \citep{Oriordan2019}. Therefore, on the grounds that a $100$\,pc offset between the light and total mass distribution is non-physical and breaks the underlying assumption on which the BPL is built, we favour models including a SMBH which do not require this offset. 

\subsection{Alternative Models}

We verify that the inclusion of a SMBH is still favoured for a number of alternative lens galaxy mass models. In \cref{MassClump}, we include the ray-tracing effects of the line-of-sight galaxy to the north-east of the giant arc, by modelling it as a singular isothermal sphere. In \cref{MassCentreFree}, we fit lens models which allow the centre of the SMBH to vary as a free parameter. In \cref{Radial}, we explore a family of solutions where the lens mass model has a shallow (or cored) inner density, therefore forming a larger radial critical curve than those inferred in the main paper. For all alternative models, a SMBH is favoured with the same or greater significance than shown for the models above.

\subsection{SMBH Mass}

\begin{figure}
\centering
\includegraphics[width=0.48\textwidth]{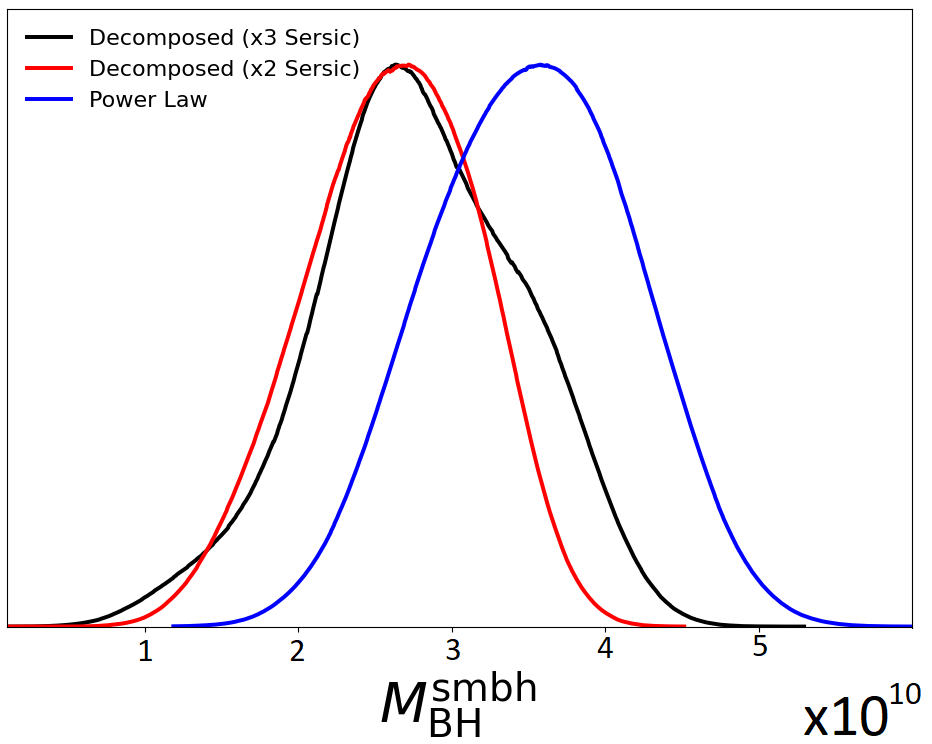}
\caption{
The 1D probability density functions (PDF) of the SMBH mass $M_{\rm BH}$ for fits to the F390W image of Abell 1201. Inferred values of $M_{\rm BH}$ are shown for the decomposed mass model with three Sersics, two Sersics and the power-law total mass model. The broken power-law fitted in \cref{ResultSIE} and discarded due to its nonphysical 100pc offset.
}
\label{figure:SMBHPDF}
\end{figure}

The 1D PDFs for $M_{\rm BH}$ for the decomposed two and three Sersic models and PL total mass model are shown in \cref{figure:SMBHPDF}. At $3\sigma$ confidence, the SMBH mass inferred for fits to the F390W image for different mass models (excluding the BPL due to its nonphysical offset centre) are: 

\begin{itemize}
    \item $M_{\rm BH} = 2.22^{+1.41}_{-1.06}\times10^{10}$\,M$_{\rm \odot}$ for the triple Sersic decomposed model.
    \item $M_{\rm BH} = 2.91^{+1.17}_{-0.91}\times10^{10}$\,M$_{\rm \odot}$ for double Sersic fits.
    \item $M_{\rm BH} = 3.95^{+1.44}_{-1.47}\times10^{10}$\,M$_{\rm \odot}$ for the PL model.
\end{itemize}

To estimate a final value of $M_{\rm BH}$ we simply estimate the value which spans the full range of measurements, producing $M_{\rm BH} = 3.27 \pm 2.12\times10^{10}$\,M$_{\rm \odot}$ at $3\sigma$ confidence.

\subsection{Upper Limit Analysis}

\begin{figure}
\centering
\includegraphics[width=0.235\textwidth]{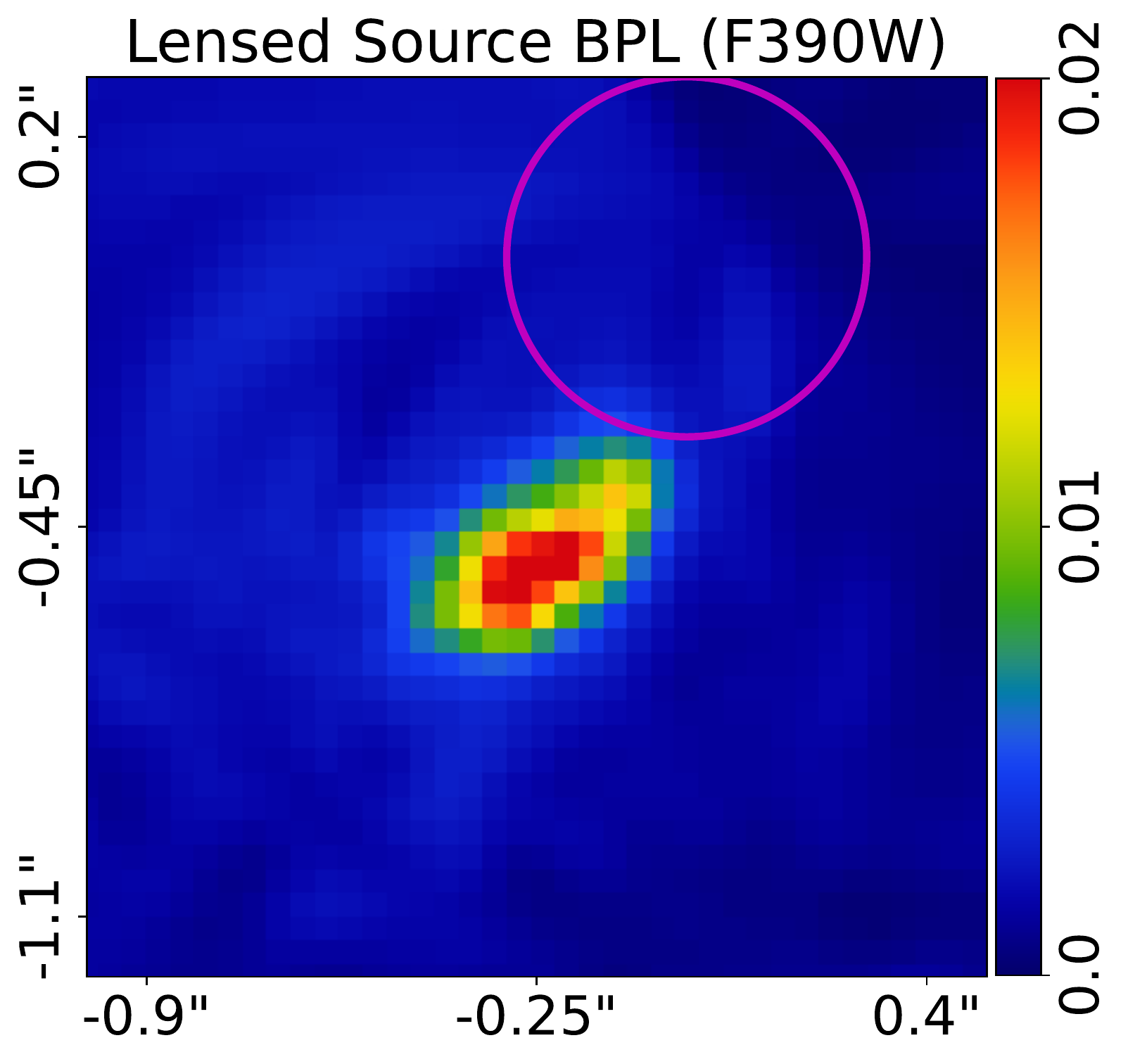}
\includegraphics[width=0.235\textwidth]{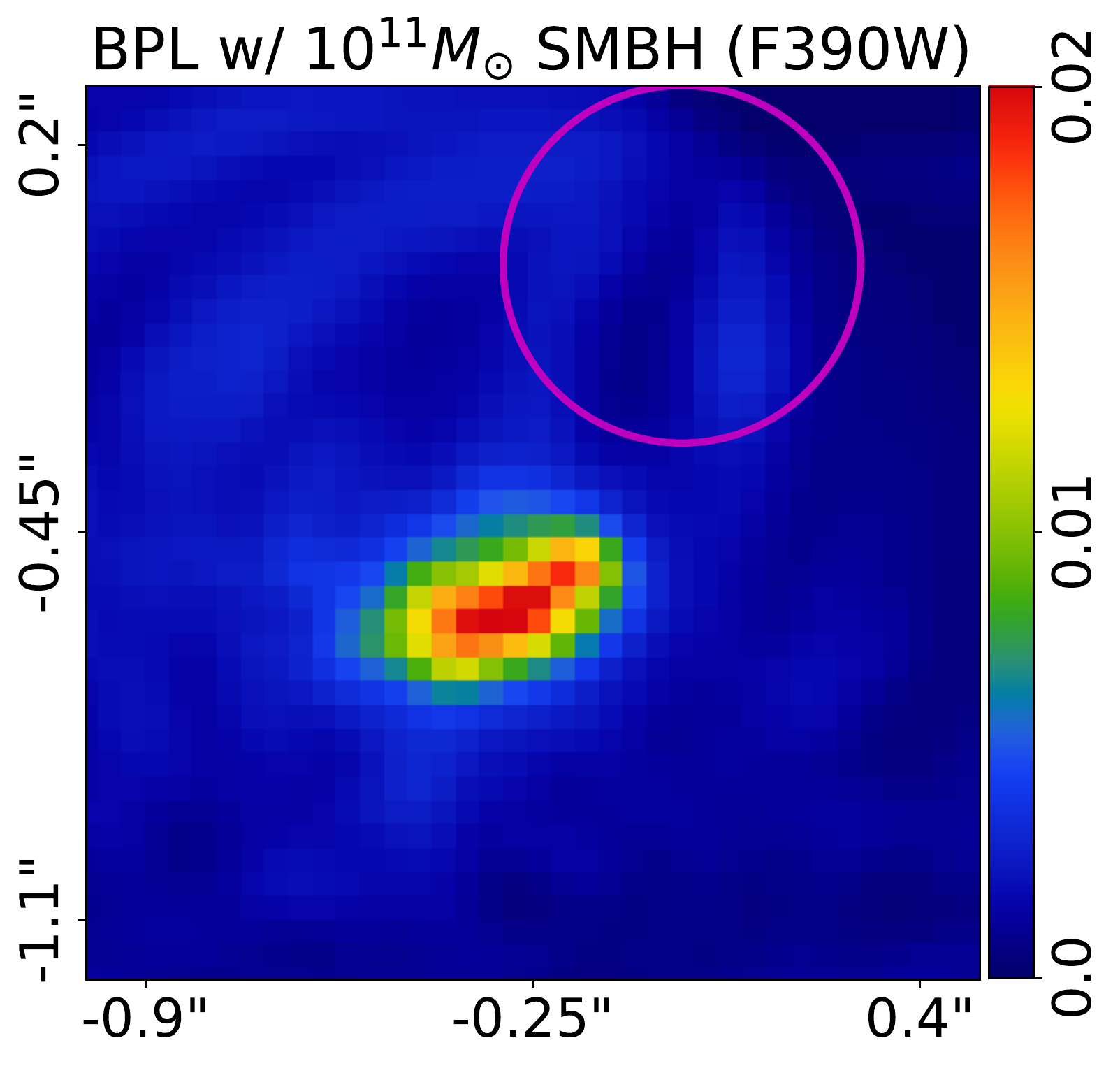}
\caption{
The reconstructed counter image for a BPL models without a SMBH (left panel) and including a $M_{\rm BH} = 1.0 \times 10^{11}$\,M$_{\rm \odot}$ SMBH for fits to the F390W image. The fit including a $M_{\rm BH} = 10^{11}$\,M$_{\rm \odot}$ SMBH displaces the reconstructed counter image such that it does not fit the data accurately.
}
\label{figure:WithSMBH}
\end{figure}

Although we have discarded the BPL model on the grounds of physical plausibility, it can still be used to place an upper limit on $M_{\rm BH}$, even with the offset centre. Once a SMBH of sufficiently high mass is included in the mass model, it deforms the counter image reconstruction in a way which cannot be compensated for by reducing the inner density of the mass model. To demonstrate this, \cref{figure:WithSMBH} shows the reconstructed counter images of a BPL model fit without a SMBH and with a SMBH whose mass is fixed to $M_{\rm BH} = 10^{11}$\,M$_{\rm \odot}$. The SMBH displaces the counter image, producing a reconstruction that is not consistent with the observed data.

The value $M_{\rm BH} = 10^{11}$\,M$_{\rm \odot}$ was chosen to visually emphasise how a high mass SMBH disfigures the counter image. We can fit a grid of BPL plus SMBH models where $M_{\rm BH}$ is fixed to incrementally higher values between $1 - 10 \times 10^{10}$\,M$_{\rm \odot}$ to the F390W data. \cref{table:SMBHUpperLimit} shows the $\ln \mathcal{Z}$ values for each fit, where a drop of $\ln \mathcal{Z} = 20$ is seen above masses of $M_{\rm BH} = 5.3 \times 10^{10}$\,M$_{\rm \odot}$. The BPL model with a nonphysical offset centre therefore still places an upper limit of $M_{\rm BH} \leq 5.3 \times 10^{10}$\,M$_{\rm \odot}$.

Whilst in this study Abell 1201's counter image contains sufficient information to provide a measurement of $M_{\rm BH}$, in less fortuitous circumstances upper limits on $M_{\rm BH}$ will still be possible in many strong lenses.

\begin{table}
\tiny
\resizebox{\linewidth}{!}{
\begin{tabular}{ l | l | l} 
\multicolumn{1}{p{1.2cm}}{$\theta_{\rm Ein}^{\rm{smbh}}$ (\arcsec)} 
& \multicolumn{1}{p{2.0cm}}{$M_{\rm BH}$ ($M_{\rm \odot} \times 10^{10}$)} 
& \multicolumn{1}{p{1.0cm}|}{$\ln \mathcal{Z}$}  
\\ \hline
 None  & 0.0   & 125699.90 \\[2pt]
 0.2   & 0.513 & 125706.18 \\[2pt]
 0.3   & 1.145 & 125693.19 \\[2pt]
 0.4   & 2.030 & 125686.63 \\[2pt]
 0.5   & 3.168 & 125657.25 \\[2pt]
 0.6   & 4.557 & 125676.59  \\[2pt]
 0.625 & 4.945 & 125699.68  \\[2pt]
 0.65  & 5.349 & \textbf{125655.86}  \\[2pt]
 0.675 & 5.765 & 125676.04 \\[2pt]
 0.7   & 6.202 & 125636.15 \\[2pt]
 0.725 & 6.651 & 125624.91 \\[2pt]
 0.75  & 7.118 & 125648.55 \\[2pt]
 0.775 & 7.580 & 125656.45 \\[2pt]
 0.8   & 8.099 & 125617.61 \\[2pt] 
 0.9   & 10.248 & 125464.21 \\[2pt]  
\end{tabular}
}
\caption{
The Bayesian evidences, $\ln \mathcal{Z}$, of BPL mass model fits which include a SMBH with a fixed mass. A 1D grid of fits are shown, which iteratively increase the SMBH mass $M_{\rm BH}$. For SMBHs above masses of $M_{\rm BH} = 5.349 \times 10^{10}$\,M$_{\rm \odot}$ all $\ln \mathcal{Z}$ values are at least $20$ below the BPL model without a SMBH where $\ln \mathcal{Z} = 125699.90$. Therefore SMBHs above this mass are ruled out by the data, because they deform the reconstruction of the counter image (see \cref{figure:WithSMBH}).
}
\label{table:SMBHUpperLimit}
\end{table}

\section{Discussion}\label{Discussion}

\subsection{Super Massive Black Holes}

\subsubsection{$M_{\rm BH}$--$\sigma_{\rm e}$ relation}

\begin{figure}
\centering
\includegraphics[width=0.47\textwidth]{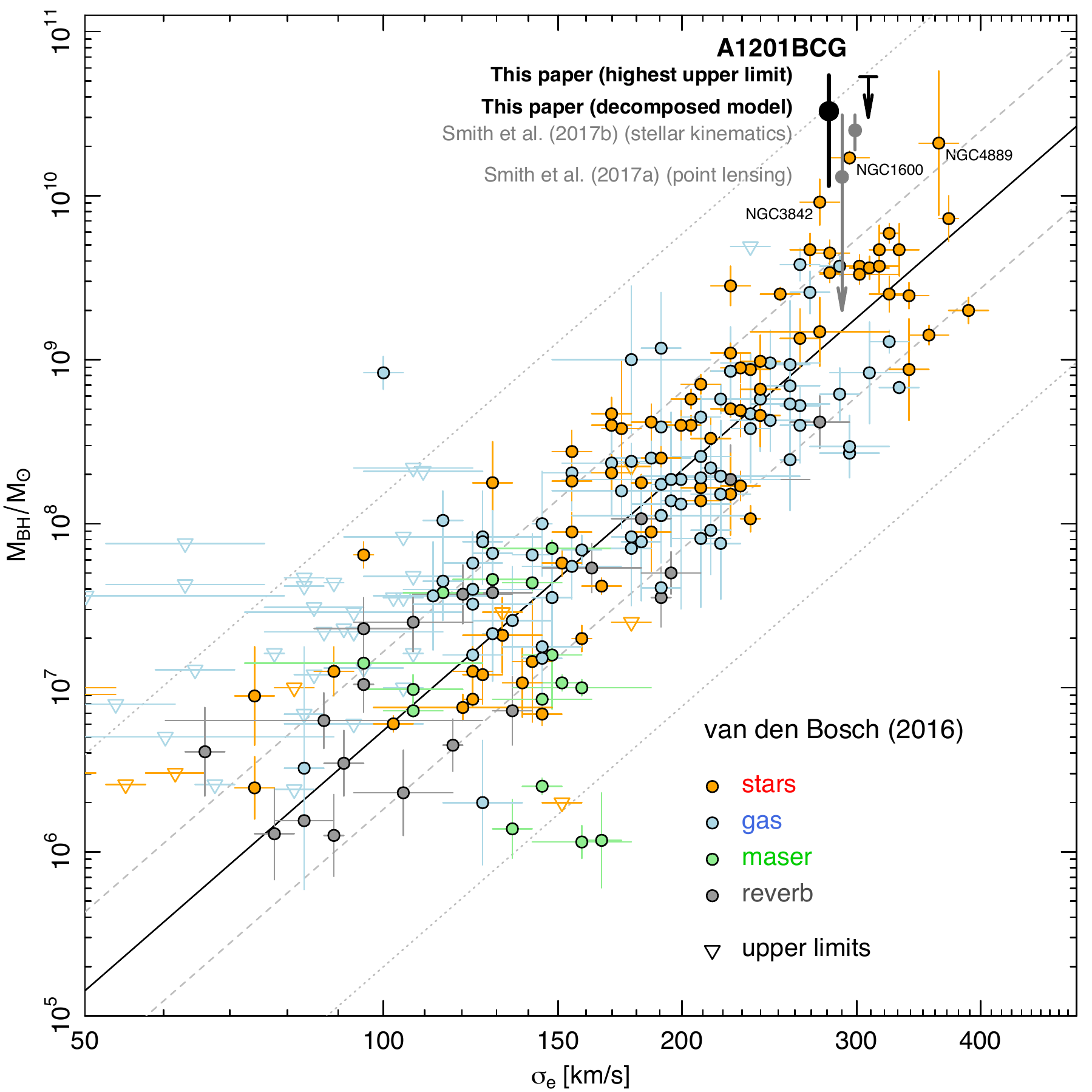}
\caption{
This work's measurements of Abell 1201's SMBH's mass in comparison to the black-hole mass versus velocity dispersion relation, from the compilation of \citet{Bosch2016}. Abell 1201's $\sigma_{\rm e}$ value is taken from \citet{Smith2017}. This work's measurement of $M_{\rm BH} = 3.27 \pm 2.12\times10^{10}$\,M$_{\rm \odot}$ is shown in black, which comes from averaging over all mass models. The upper limit of $M_{\rm BH} \leq 5.3 \times 10^{10}$\,M$_{\rm \odot}$ inferred for the broken power law mass model (without a SMBH) is shown for completeness, although we have argued this model is less trustworthy due to being nonphysical (see \cref{ResultSIE}). This figure is adapted from \citet{Smith2017a} and shows their inferred SMBH masses in grey, which come from independent analyses using either point-source based strong lens modeling \citep{Smith2017a} or stellar kinematics \citep{Smith2017}. Both works report that a SMBH with $M_{\rm BH} \geq 10^{10}$\,M$_{\rm \odot}$ fits the data, but neither work could break a degeneracy with models that assumed a radial gradient in the conversion of mass to light. Our inferred value of $M_{\rm BH}$ in Abell 1201 makes it one of the highest mass SMBH's measured. The grey dashed and dotted diagonal lines show 1$\sigma$ and 2$\sigma$ scatter of the mean $M_{\rm BH}$--$\sigma_{\rm e}$ relation, with Abell 1201's SMBH approximately a 2$\sigma$ positive outlier.
} 
\label{figure:SMBHRelation}
\end{figure}

\cref{figure:SMBHRelation} shows the inferred value of $M_{\rm BH} = 3.27 \pm 2.12  \times 10^{10}$\,M$_{\rm \odot}$ on the black-hole mass versus velocity dispersion relation.
This figure shows that Abell 1201 has one of the largest reported black hole masses measured so far, making it an ultramassive black hole \citep{Hlavacek-Larrondo2012}. Its mass is comparable to the SMBH of the brightest cluster galaxies NGC 3842 and NGC 4889 \citep{Mcconnell2013} and the field elliptical NGC 1600 \citep{Thomas2016}, all of which are measured via stellar orbit analysis. All three objects have similar values of $\sigma_{\rm e}$ to Abell 1201. 

The SMBH of Abell 1201 is a $\sim$ $2\sigma$ outlier above the scatter of the $M_{\rm BH}$--$\sigma_{\rm e}$ relation. Two other objects with similar $\sigma_{\rm e}$ values to Abell 1201, NGC3842 and NGC1600, are $\sim 1.5$-$2\sigma$ outliers above the mean relation. There are no corresponding outliers at $\sim 1.5\sigma$ below the mean relation, indicating that for $\sigma_{\rm e} > 250$km\,s$^{\rm -1}$ SMBH masses tend to be above the mean $M_{\rm BH}$--$\sigma_{\rm e}$ relation. Although there are too few objects to draw definitive conclusions,  such an upwards kink at high $\sigma_{\rm e}$ is a prediction of different physical processes. For example, binary SMBH scouring, which saturates $\sigma_{\rm e}$ whilst increasing $M_{\rm BH}$ \citep{Kormendy2013a, Thomas2014}, as well as AGN feedback processes \citep{Hlavacek-Larrondo2012}.  

\subsubsection{Stellar Core}

Massive ellipticals are often observed with a stellar core, quantified via the Nuker or cored Sersic models \citep{Hernquist1990, Trujillo2004, Dullo2013, Dullo2014}. BCGs like Abell 1201 may have extremely large and flat cores \citep{Postman2012}. It is posited that these cores form via SMBH scouring, whereby the dissipationless merging of two SMBHs in the centre of a galaxy preferentially ejects high mass stars via three-body interactions \citep{Faber1997, Merritt2006, Kormendy2009,Kormendy2013a, Thomas2014}. We fitted the core-Sersic model to Abell 1201's lens galaxy light during our initial analysis, however the model did not produce an improved fit to the data. Typical core sizes are $0.02$--$0.5$\,kpc \citep{Dullo2019}, therefore if Abell 1201 has a stellar core it may be we simply cannot resolve it, due to the data's resolution of $\sim 120$\,pc\,pixel$^{\rm -1}$. 

Aspects of the lens models which include a SMBH point towards a cored (or at least shallow) inner density. For example, the power-law mass model with a SMBH infers a slope $\gamma^{\rm mass} = 1.65^{+0.12}_{-0.12}$, which is much shallower than many massive elliptical strong lenses with near isothermal slopes of $\gamma^{\rm mass} \sim 2$ \citep{Koopmans2009}. Decomposed models including a SMBH give comparable inner densities. When fitting the core-Sersic model we only included it in the model for the lens galaxy's light. We did not fit it as part of a decomposed mass model and therefore did not try to constrain the stellar core via the ray-tracing and lensing analysis. Future studies hunting for SMBHs in strong lenses may benefit from doing this, because an improved model of the lens's central mass density could help break the degeneracy seen in this work with $M_{\rm BH}$.

\subsubsection{Outlook for Strong Lensing}

Abell 1201 is the second strong lens in which the central SMBH mass has been constrained. It is the first to do so without a central image, as well as the first to provide a measurement of $M_{\rm BH}$ as opposed to an upper limit. This raises a number of questions: what is so special about Abell 1201 that makes it sensitive to its SMBH? Can $M_{\rm BH}$ be measured in other known strong lenses? How common an occurrence will this be amongst the incoming samples of 100,000 strong lenses?

Abell 1201 is a unique strong lens in that its counter image is close to the lens centre and it is a cD galaxy in a galaxy cluster. The cluster potential exerts a large external shear (which is seen in our lens models) that brings the counter-image even closer to the lens centre \citep{Smith2017a}, an effect that is not present in most known galaxy-scale strong lenses, which are typically field galaxies. Thus, a very specific set of circumstances may make Abell 1201 sensitive to its SMBH, and a strategy to finding more systems is to target cD / BCG galaxies with instruments like Multi Unit Spectroscopic Explorer (MUSE). 

On the other hand, some known strong lenses in surveys like the Sloan Lens ACS Survey \citep{Bolton2008a} and Strong Lensing in the Legacy Survey \citep{Sonnenfeld2013b} may be sensitive to their central SMBH and appropriate lens modeling has simply not been performed. Certainly, every strong lens will provide an upper limit on $M_{\rm BH}$, the question is whether any are low enough to be informative for models of galaxy evolution. Whilst the multiple images of strong lenses are predominantly observed at radii well beyond Abell 1201's $1\,$kpc counter image, there are examples of strong lenses where the extended emission of the lensed source goes this close. For example, SLACS1250+0523, which was modeled by \citet{Nightingale2019}. In many surveys, for a candidate strong lens to be worthy of following up with higher resolution imaging, a visible counter image clearly distinct from the lens's emission is typically required. Systems like Abell 1201 may therefore be common in nature but rarely selected for followup. We leave it to future work to investigate what constraints known strong lenses can place on $M_{\rm BH}$. 


It has long been expected that strong lensing can constrain SMBH masses when a central third or fifth image is observed \citep{Rusin2000, Mao2000, Keeton2003, Hezaveh2015}. Such a system was presented by \citet{Winn2003}, who placed an upper limit of $M_{\rm BH} < 2 \times 10^8$\,M$_{\rm \odot}$. These systems require the inner density profile of the lens galaxy to be sufficiently cored that the central image is not demagnifed below the observing instrument's detection limit. Given that no other such observation has been made despite numerous attempts \citep{Jackson2015, Wong2017} this appears to be a rare occurrence. Lower limits on $M_{\rm BH}$ have been placed in systems where a central image is not detected \citep{Quinn2016}.

Abell 1201 demonstrates that a SMBH mass measurement is possible even when the lens's inner density is not cored. This offers hope that large samples of strong lenses can one day constrain the $M_{\rm BH}$--$\sigma_{\rm e}$ relation. This would enable the masses of non-active black holes to be measured at high redshifts, and would provide measurements on the high $\sigma_{\rm e}$ end of the relation where few ETGs are observed in the local Universe. With over 100000 strong lenses set to be observed in the next decade \citep{Collett2015} it is inevitable that more SMBH measurements via strong lensing will be made, however more work is necessary to determine how common an occurrence this will be, and in what types of strong lenses and at how high of a redshift such constraints are feasible. If the detectability of a strong lens's SMBH depends on a specific set of circumstances like Abell 1201, there will also be unavoidable selection effects that must be accounted for.
\section{Summary}\label{Summary}

We present an analysis of the galaxy-scale strong gravitational lens Abell 1201 using multi-waveband Hubble Space Telescope imaging. Abell 1201 is a particularly unique system for two reasons: (i) its counter image is just $1$\,kpc away from the lens galaxy centre; (ii) it is a cD galaxy located within a galaxy cluster. After extensive strong lens modeling, we show that reconstructions of Abell 1201's counter image provide constraints for mass models that include a central super massive black hole (SMBH). After performing Bayesian model comparison, we find that all but one lens model of Abell 1201 prefer the inclusion of a SMBH. By averaging over these models, we infer a value of $M_{\rm BH} = 3.27 \pm 2.12  \times 10^{10}$\,M$_{\rm \odot}$, in agreement with previous lensing and stellar dynamics models of the system \cite{Smith2017a, Smith2017}. This makes it one of the largest black hole masses measured to date and qualifies it as an ultramassive black hole. Its mass is comparable to other high velocity dispersion $\sigma_{\rm e}$ systems on the $M_{\rm BH}$--$\sigma_{\rm e}$ relation whose masses were measured via stellar orbit analysis.

There is one mass model without a SMBH which, in a Bayesian sense, is as plausible as any model including a SMBH. This model has a lot of flexibility in adjusting its central density and mimics the lensing effect of the SMBH by increasing its density to be extremely peaked; far more so than any other mass model. However, the model simultaneously requires that its mass centre is offset from the luminous centre of the bulge by $\geq 100$\,pc. This offset is not necessary when a SMBH is included in the lens model and we therefore rule-out this model as being nonphysical. Even allowing for this nonphysical offset centre, the model still provides an upper limit of  $M_{\rm BH} \leq 5.3 \times 10^{10}$\,M$_{\rm \odot}$, as including a SMBH above this mass completely deforms the counter image reconstruction. Therefore, even strong lens systems which are not as fortuitous as Abell 1201 in their configuration could provide meaningful constraints on SMBHs as upper limits.

Abell 1201 is the second strong lens to provide constraints on on its central SMBH mass, following the upper limit of $M_{\rm BH} \leq 2 \times 10^8 $\,M$_{\rm \odot}$ placed by \citet{Winn2003} in a strong lens whose central image was observed. Our work is therefore the first to not only place an upper limit but measure $M_{\rm BH}$ and it does so without the rare observation of a central image. This offers hope that many more strong lens systems can potentially constrain the mass of their central SMBH, although the unique properties of Abell 1201 may mean this remains a somewhat rare occurrence. Further investigation is necessary to draw firm conclusions, but with over one hundred thousand strong lens systems set to be discovered in the next decade there is hope that strong lensing can one day constrain the redshift evolution of the $M_{\rm BH}$--$\sigma_{\rm e}$ relation.

\section*{Software Citations}

This work uses the following software packages:

\begin{itemize}

\item
\href{https://github.com/astropy/astropy}{{Astropy}}
\citep{astropy1, astropy2}

\item
\href{https://bitbucket.org/bdiemer/colossus/src/master/}{{Colossus}}
\citep{colossus}

\item
\href{https://github.com/dfm/corner.py}{{Corner.py}}
\citep{corner}

\item
\href{https://github.com/joshspeagle/dynesty}{{Dynesty}}
\citep{Speagle2020}

\item
\href{https://github.com/matplotlib/matplotlib}{{Matplotlib}}
\citep{matplotlib}

\item
\href{numba` https://github.com/numba/numba}{{Numba}}
\citep{numba}

\item
\href{https://github.com/numpy/numpy}{{NumPy}}
\citep{numpy}

\item
\href{https://github.com/rhayes777/PyAutoFit}{{PyAutoFit}}
\citep{pyautofit}

\item
\href{https://github.com/Jammy2211/PyAutoGalaxy}{{PyAutoGalaxy}}
\citep{pyautogalaxy}

\item
\href{https://github.com/Jammy2211/PyAutoLens}{{PyAutoLens}}
\citep{Nightingale2015, Nightingale2018, Nightingale2021}

\item
\href{https://github.com/AshKelly/pyquad}{{Pyquad}}
\citep{pyquad}

\item
\href{https://www.python.org/}{{Python}}
\citep{python}

\item
\href{https://github.com/scikit-image/scikit-image}{{Scikit-image}}
\citep{scikit-image}

\item
\href{https://github.com/scikit-learn/scikit-learn}{{Scikit-learn}}
\citep{scikit-learn}

\item
\href{https://github.com/scipy/scipy}{{Scipy}}
\citep{scipy}

\item
\href{https://www.sqlite.org/index.html}{{SQLite}}
\citep{sqlite}

\end{itemize}

\section*{Data Availability}

Text files and images of every model-fit performed in this work are available at \url{https://github.com/Jammy2211/autolens_abell_1201}. Full \texttt{dynesty} chains of every fit are available at \url{https://zenodo.org/record/7695438}.

\section*{Acknowledgements}

JWN is supported by the UK Space Agency, through grant ST/N001494/1, and a Royal Society Short Industry Fellowship.
RJM is supported by a Royal Society University Research Fellowship and by the STFC via grant ST/T002565/1, and the UK Space Agency via grant ST/W002612/1.
JAK acknowledges support from a NASA Postdoctoral Program Fellowship.
AE is supported by STFC via grants ST/R504725/1 and ST/T506047/1.  
AA and QH acknowledge support from the European Research Council (ERC) through Advanced Investigator grant DMIDAS (GA 786910). 
This work used both the Cambridge Service for Data Driven Discovery (CSD3) and the DiRAC Data-Centric system, project code dp195, which are operated by the University of Cambridge and Durham University on behalf of the STFC DiRAC HPC Facility (www.dirac.ac.uk). These were funded by BIS capital grant ST/K00042X/1, STFC capital grants ST/P002307/1, ST/R002452/1, ST/H008519/1, ST/K00087X/1, STFC Operations grants ST/K003267/1, ST/K003267/1, and Durham University. DiRAC is part of the UK National E-Infrastructure.

\bibliography{library, citations, manual}            

\appendix
\section{Lens Profiles}\label{LensModels}

\subsection{Priors}

\begin{table*}
\tiny
\begin{tabular}{ l | l | l | l | l | l} 
\multicolumn{1}{p{1.5cm}|}{\centering \textbf{Model}} 
& \multicolumn{1}{p{0.9cm}|}{\centering \textbf{Parameter}} 
& \multicolumn{1}{p{1.1cm}|}{\centering \textbf{Prior}} 
& \multicolumn{1}{p{0.9cm}|}{\centering \textbf{Units}} 
& \multicolumn{1}{p{2.5cm}}{\textbf{Values (F390W)}} 
& \multicolumn{1}{p{2.5cm}}{\textbf{Values (F814W)}} 
\\ \hline
& & & & & \\[-4pt]
\textbf{Elliptical}     & $x^{\rm mass}$ & Gaussian & \arcsec & $\mu = 0.036$, $\sigma = 0.05$ & $\mu = 0.047$, $\sigma = 0.05$ \\[2pt]
\textbf{Power-Law (PL)} & $y^{\rm mass}$ & Gaussian & \arcsec & $\mu = 0.115$, $\sigma = 0.05$ & $\mu = 0.070$, $\sigma = 0.05$ \\[2pt]
                        & $\epsilon_{\rm 1}^{\rm mass}$ &  Gaussian & & $\mu = 0.094$, $\sigma = 0.2$ & $\mu = 0.201$, $\sigma = 0.2$ \\[2pt]
                        & $\epsilon_{\rm 1}^{\rm mass}$ & Gaussian & & $\mu = 0.048$, $\sigma = 0.2$ & $\mu = -0.016$, $\sigma = 0.2$ \\[2pt]
                        & $\theta^{\rm mass}_{\rm  E}$ & Gaussian & \arcsec & $\mu = 1.964$, $\sigma = 0.491$ & $\mu = 2.002$, $\sigma = 0.05$ \\[2pt]
                        & $\gamma^{\rm mass}$ & Uniform & & $l = 1.5$, $u = 3.0$ & $l = 1.5$, $u = 3.0$ \\[2pt]                        
\hline
& & & \\[-4pt]
\textbf{Broken}          & $x^{\rm mass}$ & Gaussian & \arcsec& $\mu = 0.036$, $\sigma = 0.05$ & $\mu = 0.047$, $\sigma = 0.05$ \\[2pt]
\textbf{Power-Law (BPL)} & $y^{\rm mass}$ & Gaussian & \arcsec& $\mu = 0.115$, $\sigma = 0.05$ & $\mu = 0.070$, $\sigma = 0.05$ \\[2pt]
                         & $\epsilon_{\rm 1}^{\rm mass}$ & Gaussian & & $\mu = 0.094$, $\sigma = 0.2$ & $\mu = 0.201$, $\sigma = 0.05$ \\[2pt]
                         & $\epsilon_{\rm 1}^{\rm mass}$ & Gaussian & & $\mu = 0.048$, $\sigma = 0.2$ & $\mu = -0.016$, $\sigma = 0.05$ \\[2pt]
                         & $\theta^{\rm mass}_{\rm  E}$ & Gaussian & \arcsec & $\mu = 1.964$, $\sigma = 0.491$ & $\mu = 2.002$, $\sigma = 0.05$ \\[2pt]
                         & $t^{\rm mass}_{\rm 1}$ & Uniform & &  $l = 0.3$, $u = 1.5$ & $l = 0.3$, $u = 1.5$ \\[2pt]  
                         & $t^{\rm mass}_{\rm 2}$ & Uniform & &  $l = 0.3$, $u = 2.0$ & $l = 0.3$, $u = 2.0$ \\[2pt]  
                         & $r^{\rm mass}_{\rm B}$ & Uniform & \arcsec & $l = 0.0$, $u = 1.0$ & $l = 0.0$, $u = 1.0$ \\[2pt]  
\hline
& & & \\[-4pt]
\textbf{Sersic x3}       & $\Psi^{\rm bulge}$ & Log Uniform & e$^{\rm -}$\,s$^{\rm -1}$ & $l = 0.200$, $u = 100.163$ & $l = 0.237$, $u = 118.31$\\[2pt]
                         & $\Psi^{\rm disk}$ & Log Uniform & e$^{\rm -}$\,s$^{\rm -1}$ & $l = 0.048$, $u = 24.500$ & $l = 0.105$, $u = 52.57$\\[2pt]
                         & $\Psi^{\rm env}$ & Log Uniform & e$^{\rm -}$\,s$^{\rm -1}$ & $l = 0.092$, $u = 46.200$ & $l = 0.315$, $u = 157.66$\\[2pt]   
\hline
& & & \\[-4pt]
\textbf{Sersic x2}       & $\Psi^{\rm bulge}$ & Log Uniform & e$^{\rm -}$\,s$^{\rm -1}$ & $l = 0.200$, $u = 98.872$ & $l = 0.233$, $u = 116.60$ \\[2pt]
                         & $\Psi^{\rm disk}$ & Log Uniform & e$^{\rm -}$\,s$^{\rm -1}$ & $l = 0.035$, $u = 17.704$ & $l = 0.079$, $u = 39.68$ \\[2pt]         
\hline
& & & \\[-4pt]
\textbf{All Sersics}     &$\Gamma$ & Uniform & & $l = -0.2$, $u = 1.0$ & $l = -0.2$, $u = 1.0$ \\[2pt]
\hline
& & & \\[-4pt]
\textbf{Elliptical NFW} & $x^{\rm dark}$ & Gaussian & \arcsec & $\mu = 0.0$, $\sigma = 0.1$ & $\mu = 0.0$, $\sigma = 0.1$ \\[2pt]
                        & $y^{\rm dark}$ & Gaussian & \arcsec & $\mu = 0.0$, $\sigma = 0.1$ & $\mu = 0.0$, $\sigma = 0.1$ \\[2pt]
                        & $\epsilon_{\rm 1}^{\rm dark}$ & Gaussian & & $\mu = 0.0$, $\sigma = 0.2$ & $\mu = 0.0$, $\sigma = 0.2$ \\[2pt]
                        & $\epsilon_{\rm 1}^{\rm dark}$ & Gaussian & & $\mu = 0.0$, $\sigma = 0.2$ & $\mu = 0.0$, $\sigma = 0.2$ \\[2pt]
                        & $M_{\rm 200}^{\rm dark}$ & Log Uniform & $M_{\rm \odot}$ & $l = 1 \times 10^{10}$, $u = 1 \times 10^{15}$ & $l = 1 \times 10^{10}$, $u = 1 \times 10^{15}$ \\[2pt]                        
                        
\hline
& & & \\[-4pt]                         
\textbf{Point Mass}     & $\theta^{\rm smbh}_{\rm E}$ & Uniform & \arcsec & $l = 0.0$ $u = 3.0$ & $l = 0.0$ $u = 3.0$ \\[2pt]
\hline
& & & \\[-4pt]                         
\textbf{Shear}          & $\epsilon_{\rm 1}^{\rm ext}$ & Gaussian & & $\mu = -0.141$, $\sigma = 0.05$ & $\mu = -0.105$, $\sigma = 0.05$\\[2pt]
                        & $\epsilon_{\rm 2}^{\rm ext}$ & Gaussian & & $\mu = 0.236$, $\sigma = 0.05$ & $\mu = 0.210$, $\sigma = 0.05$\\[2pt]                         
\end{tabular}
\caption{The priors on every parameter for the mass profiles used in this work, when they are fitted in the Mass pipeline and therefore from which our final parameter estimates and Bayesian evidences are based. Column 1 gives the model name. Column 2 gives the parameter. Column 3 gives the type of prior. Column 4 the values of that prior for fits to the F390W image and column 5 to the F814W image. For uniform and log uniform priors $l$ and $u$ give the lower and upper limits assumed. For Gaussian priors $\mu$ is the centre of the Gaussian and $\sigma$ its width.}
\label{table:ModelsAppendix}
\end{table*}

Non-linear search chaining (see \cref{SLAM}) updates the priors on the lens model parameters throughout the SLaM pipelines. \cref{table:ModelsAppendix} lists the priors assumed for every mass model parameter in the Mass pipeline which our Bayesian evidences are based. Details of the specific prior used for every lens model parameter in every model-fit are provided at \url{https://zenodo.org/record/7695438}, where the full sets of \texttt{dynesty} results are also provided.

The mass-to-light ratio of each stellar light model assumes log uniform priors, where the lower and upper limits correspond to values that give Einstein masses of $0.01$ and $5$ times the Einstein mass inferred for the total mass profile fitted previously. Radial gradients assume uniform priors between $-0.2$ and $1.0$. The NFW dark matter profile is parameterized with its normalization as the mass at two hundred times the critical density of the Universe, $M^{\rm dark}_{\rm 200}$, and assumes a log uniform prior between $10^{9}$\,M$_{\rm \odot}$ and $10^{15}$\,M$_{\rm \odot}$.

Identical prior passing is used in the Source and Light pipelines as in \citet{Etherington2022} and we also use the likelihood cap described in this work to infer errors on lens model parameters, with all errors quoted at a $3\sigma$ confidence interval unless stated otherwise.

Due to prior passing, the prior on the BPL centre is not $(0.0\,",\,0.0\,")$ but offset to $(0.036\,",\,0.115\,")$ for the F390W fit. An important aspect of our results is that we infer a BPL centre that is offset from the luminous bulge, which we argue is non-physical. We verify that manually setting this prior to be centred on $(0.0\,",\,0.0\,")$ does not infer an accurate model that is not offset from the bulge light (these results are included at \url{https://zenodo.org/record/7695438}). In fact, these fits infer much lower Bayesian evidences.
\section{Light Models}\label{LightModels}

\begin{figure*}
\centering
\includegraphics[width=0.32\textwidth]{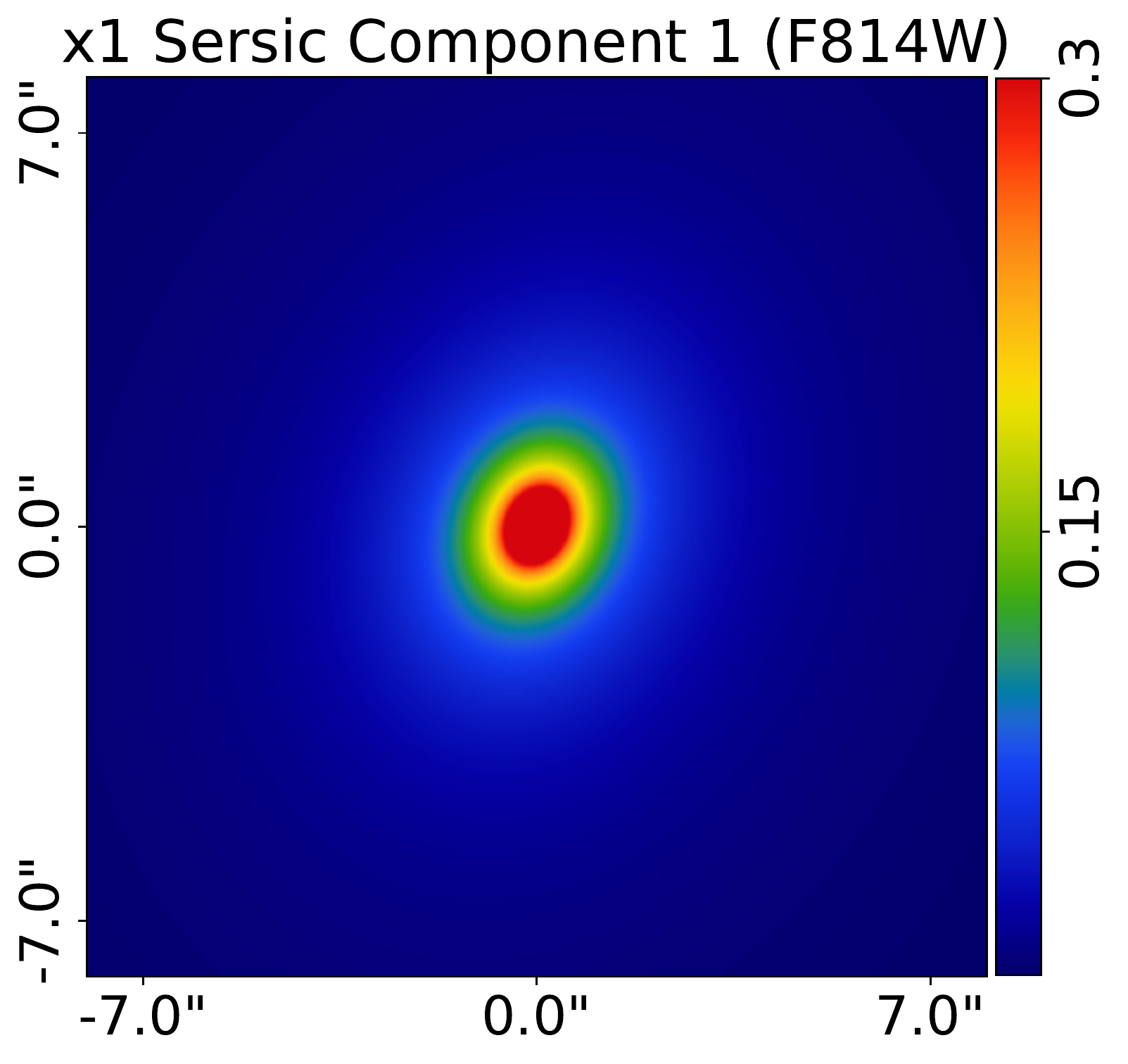}
\includegraphics[width=0.32\textwidth]{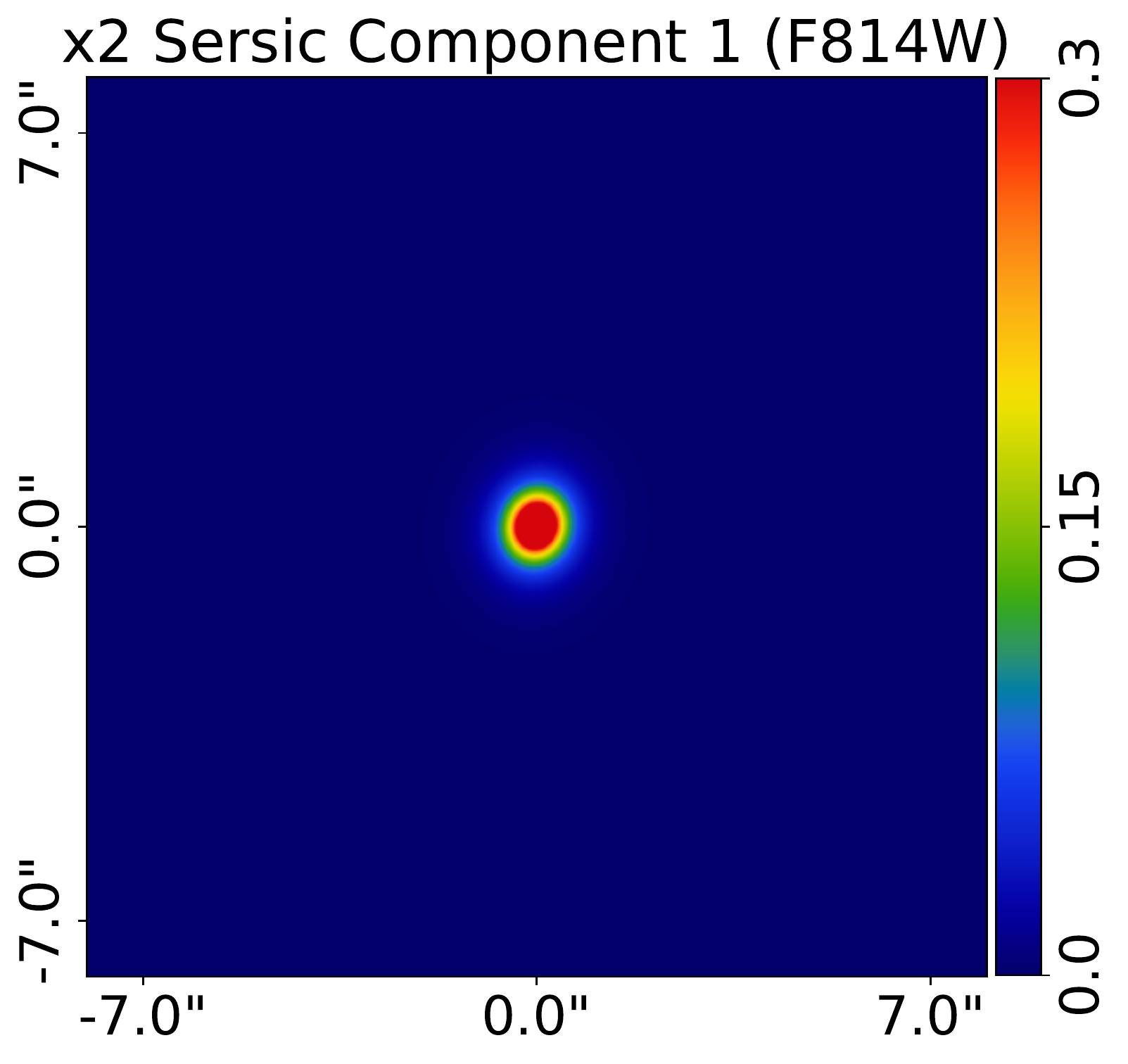}
\includegraphics[width=0.32\textwidth]{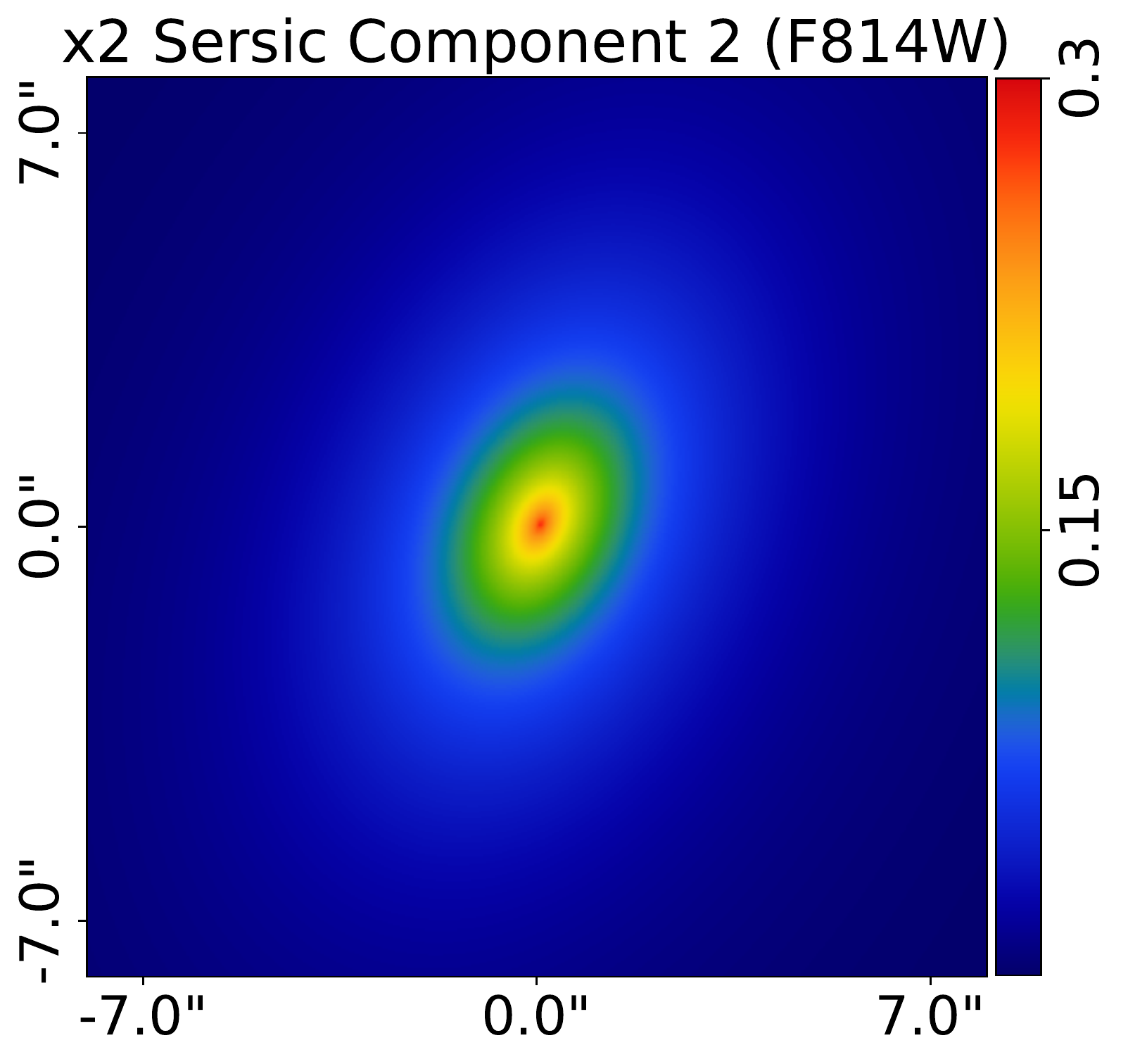}
\includegraphics[width=0.32\textwidth]{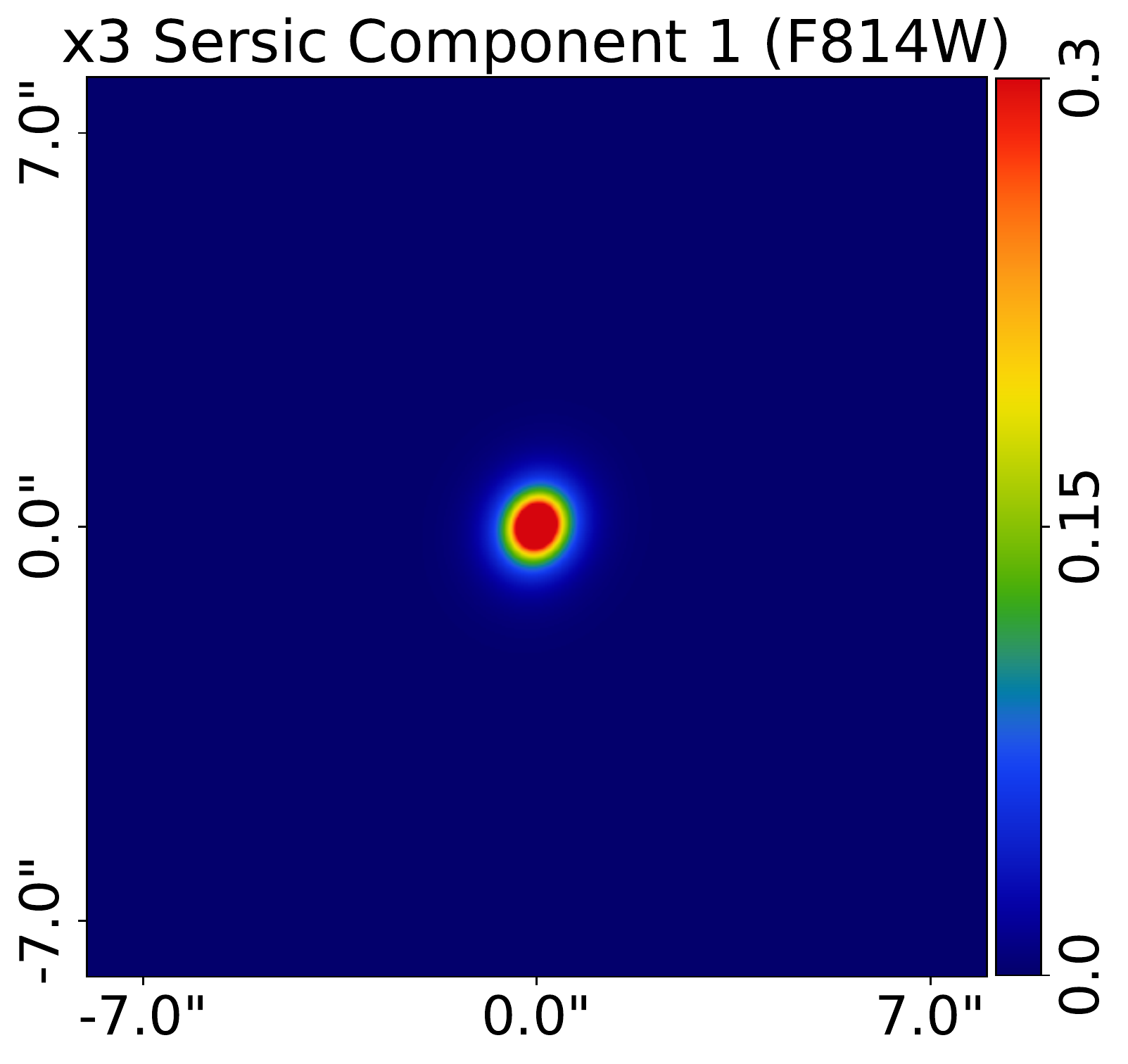}
\includegraphics[width=0.32\textwidth]{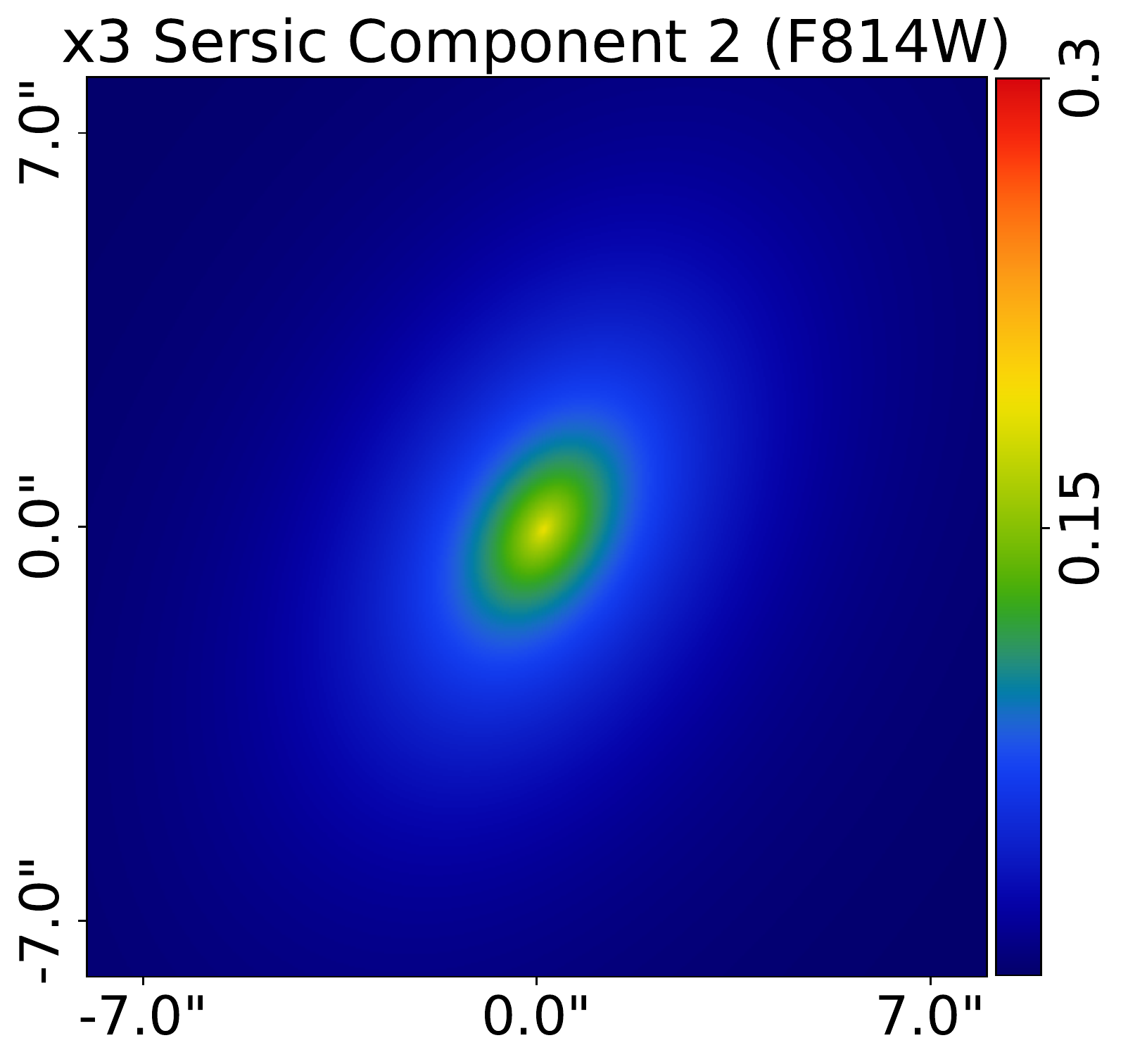}
\includegraphics[width=0.32\textwidth]{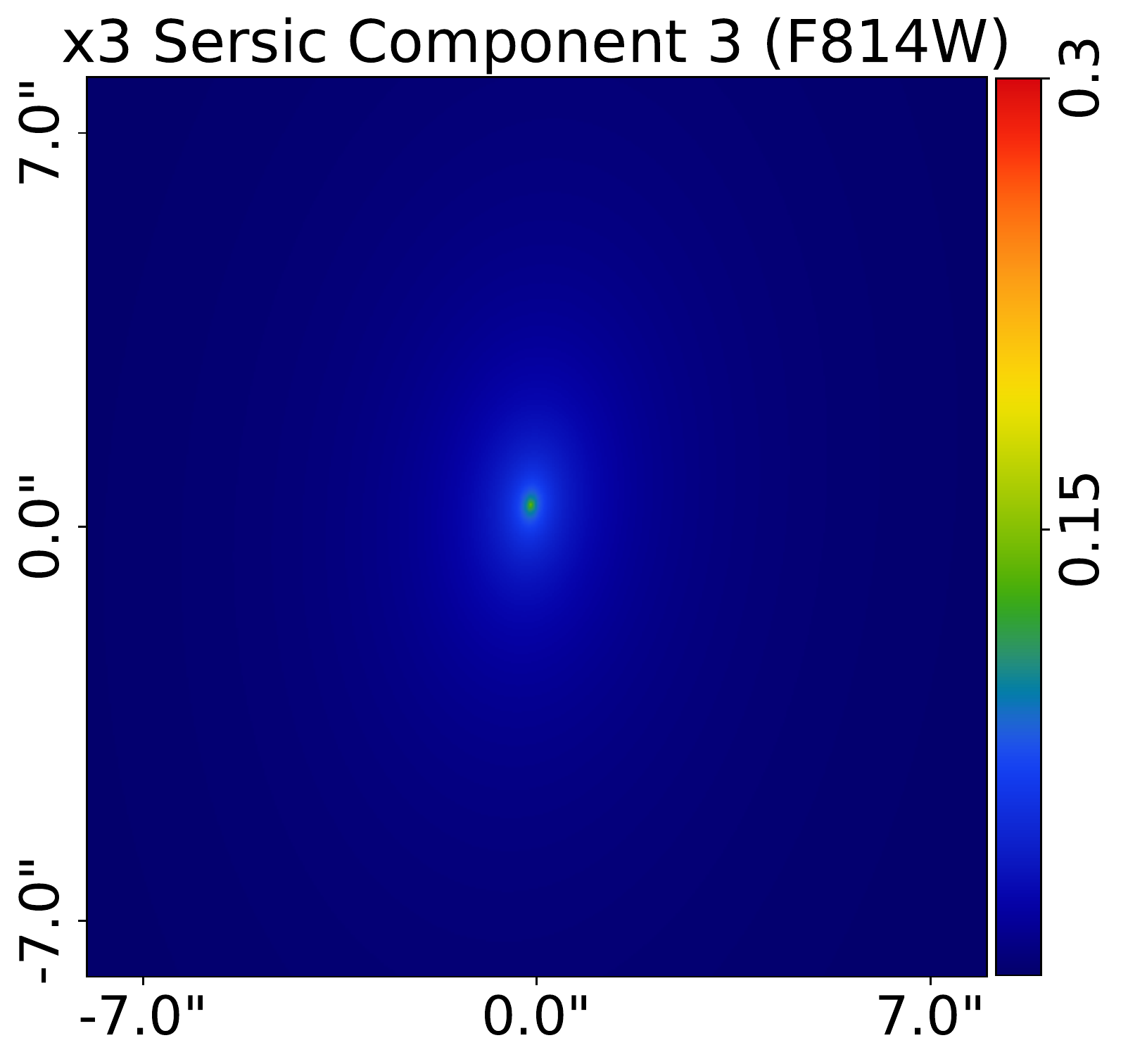}
\caption{Two dimensional projections of the individual light profiles for the following three lens galaxy light models: (i) a single Sersic profile (top left panel); (ii) a double Sersic profile where the centres and elliptical components are not aligned, representing a central bulge (top middle panel) and extended component (top right panel); (iii) a triple Sersic model where no geometric components are aligned, representing a bulge, an extended component and a third inner component (bottom row). All models are fitted with a fixed isothermal mass profile with external shear and a pixelized source reconstruction which changes for every light profile fitted. Each intensity plot corresponds to the maximum likelihood light model of a model-fit using the F814W image (the F390W image's blue wavelength makes it is less suited to tracing the lens galaxy's stellar mass).} 
\label{figure:Light2D}
\end{figure*}

\begin{figure*}
\centering
\includegraphics[width=0.19\textwidth]{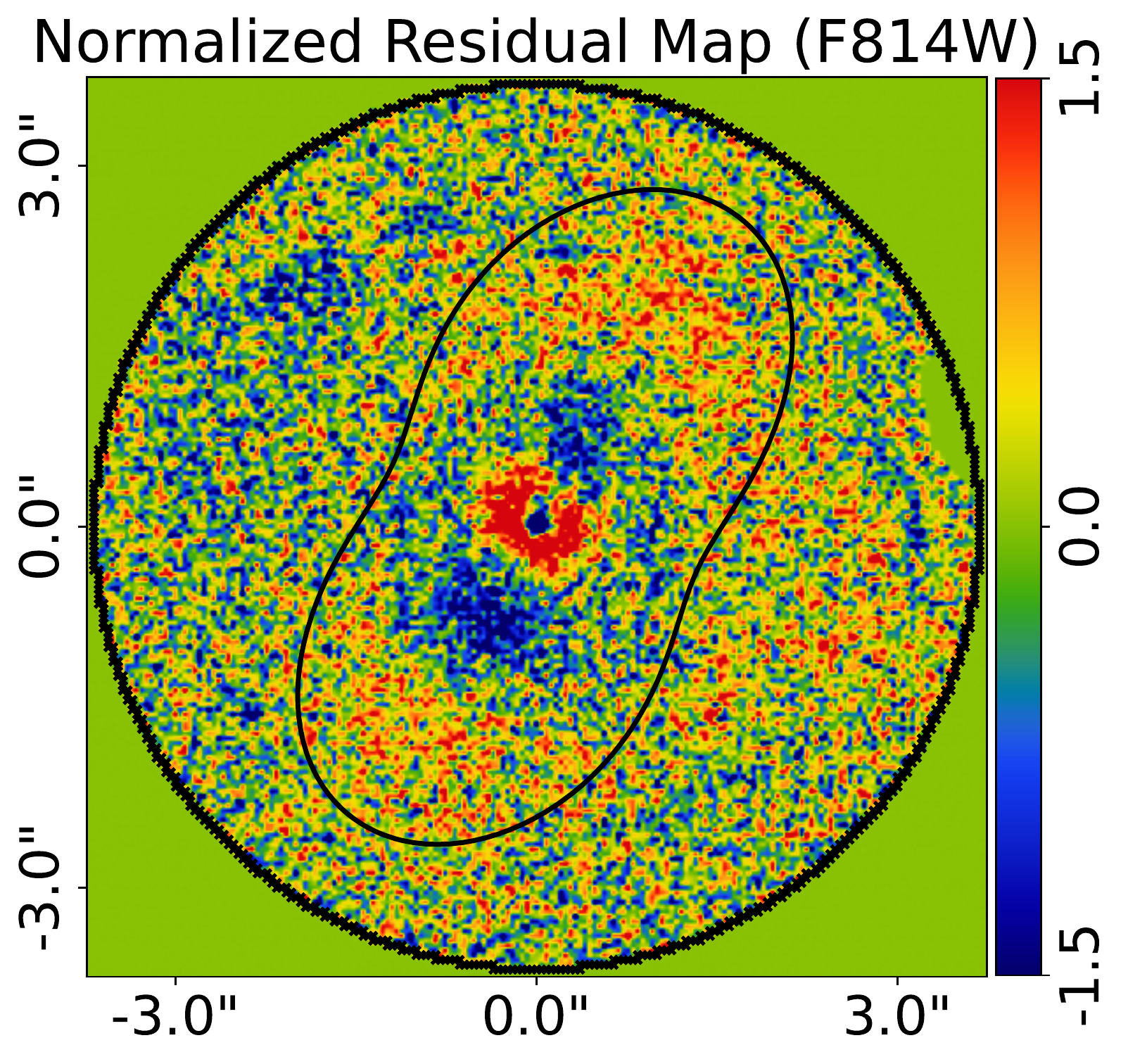}
\includegraphics[width=0.19\textwidth]{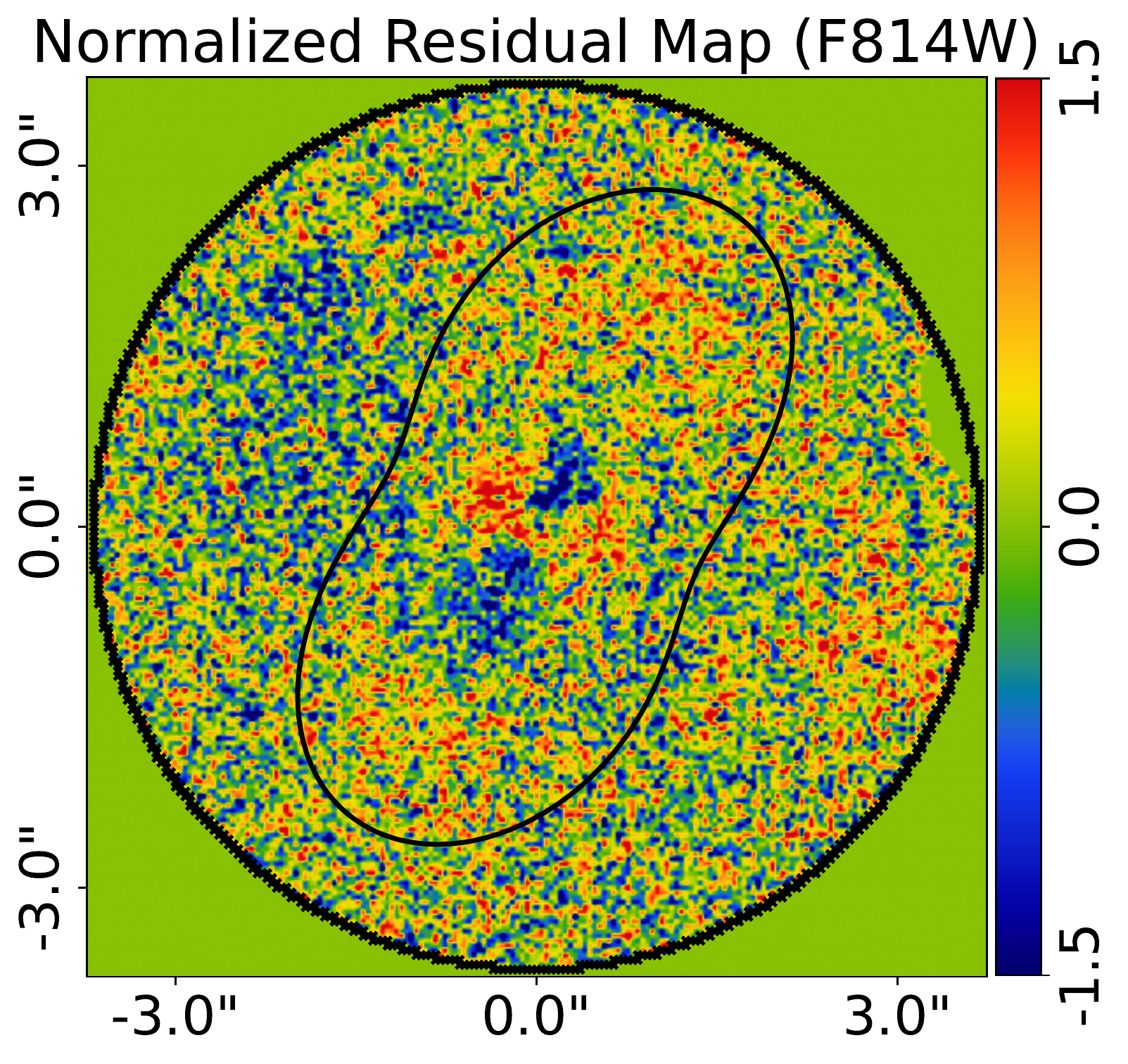}
\includegraphics[width=0.19\textwidth]{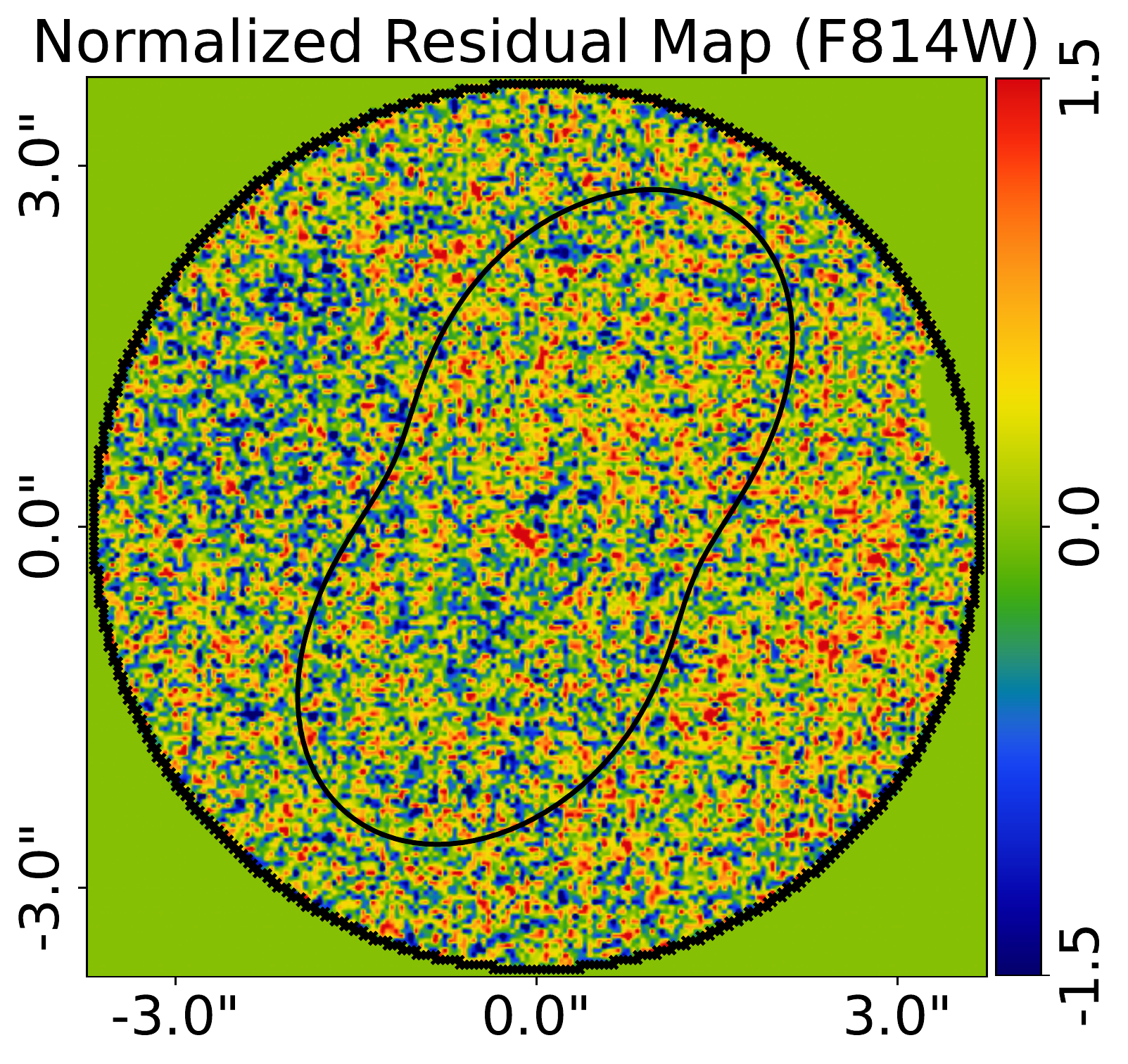}
\includegraphics[width=0.19\textwidth]{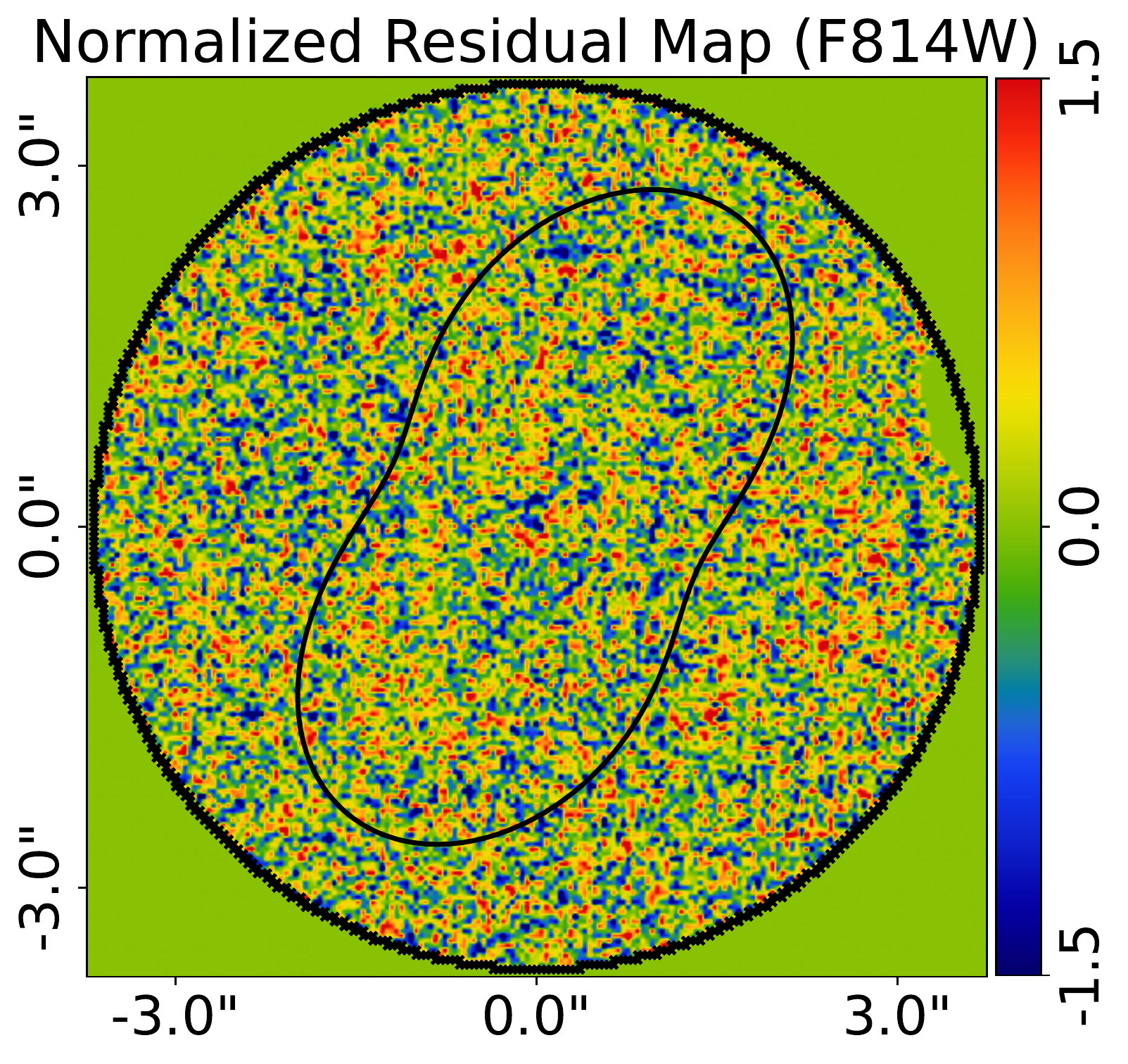}
\includegraphics[width=0.19\textwidth]{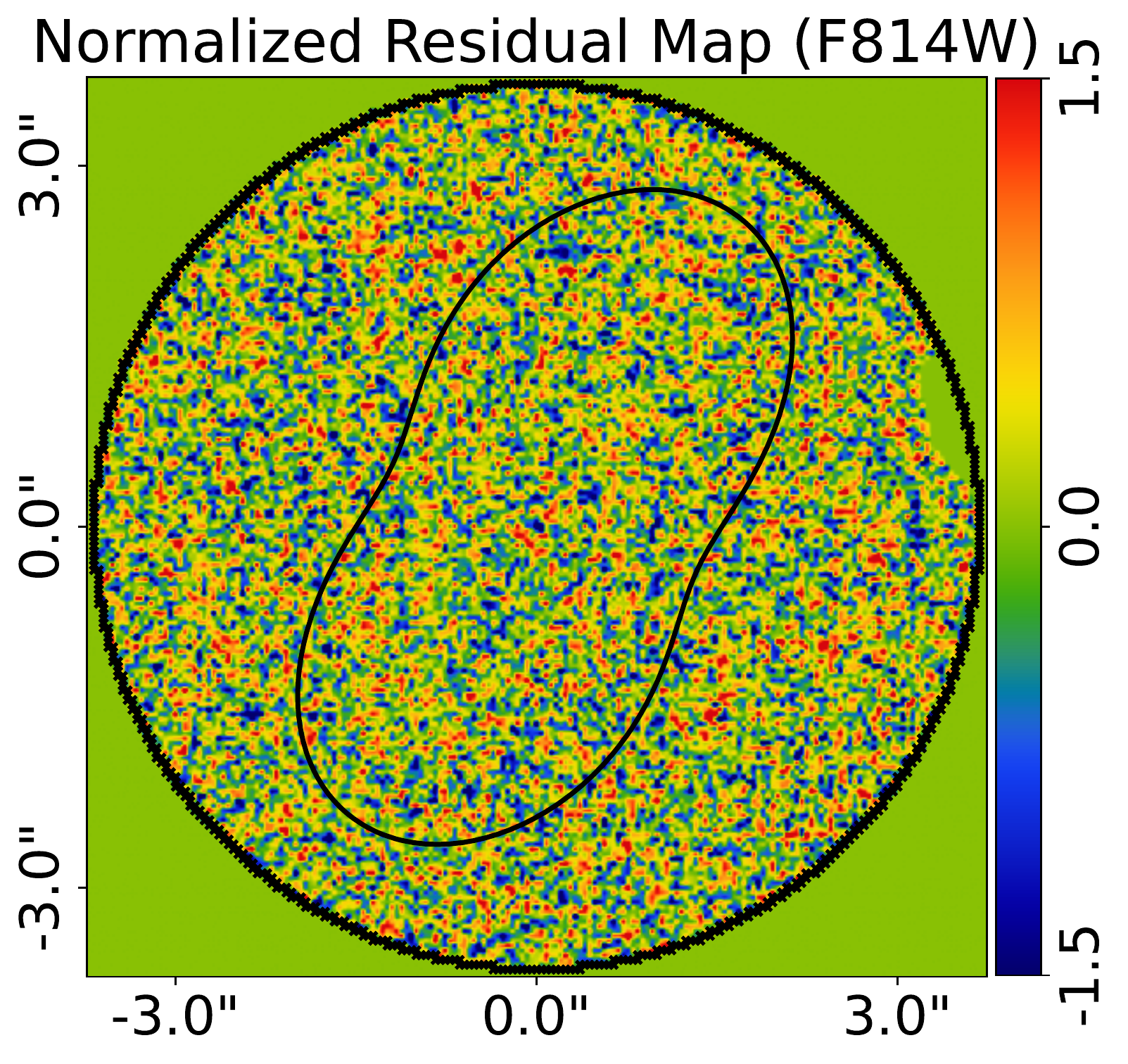}
\includegraphics[width=0.19\textwidth]{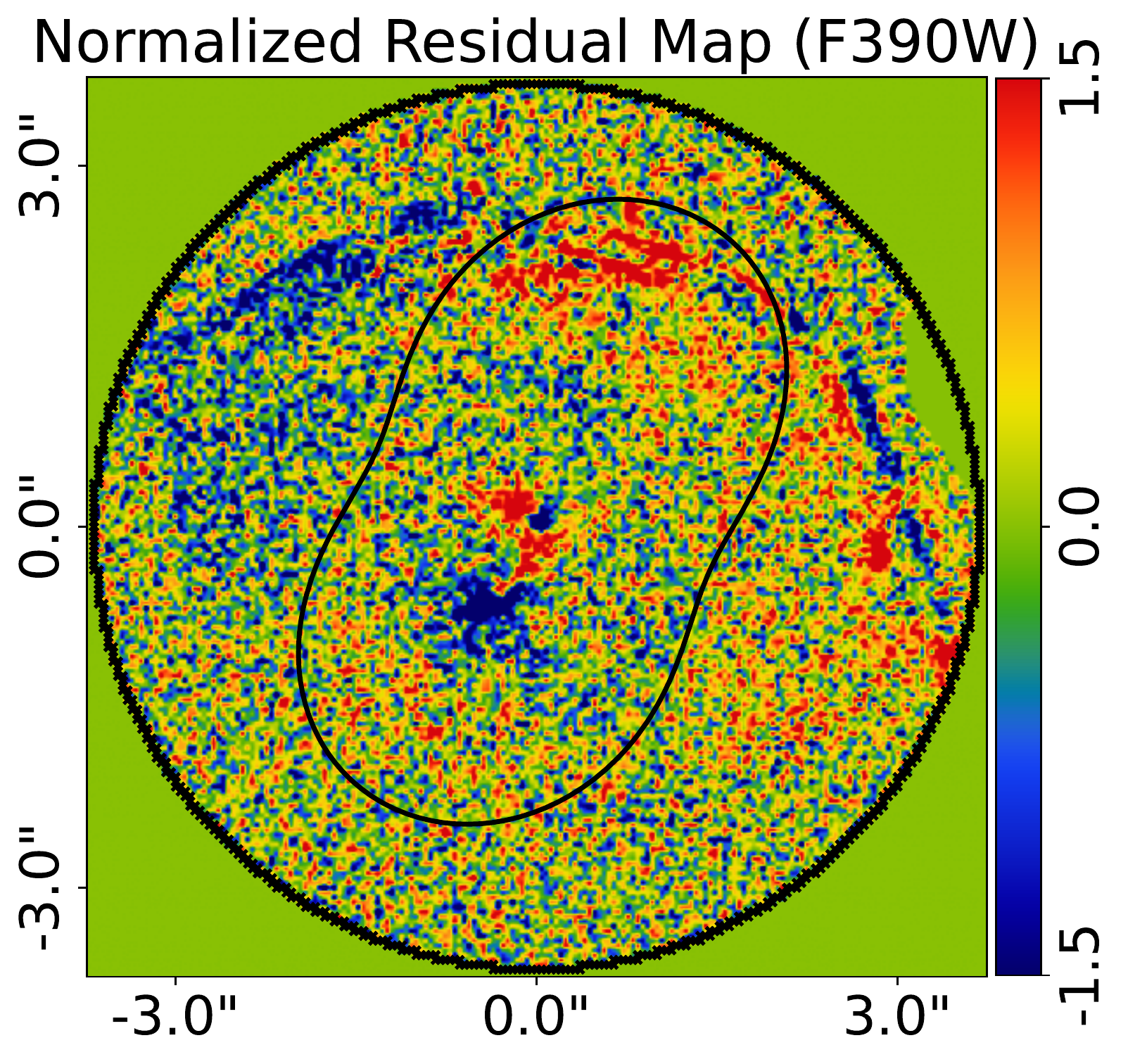}
\includegraphics[width=0.19\textwidth]{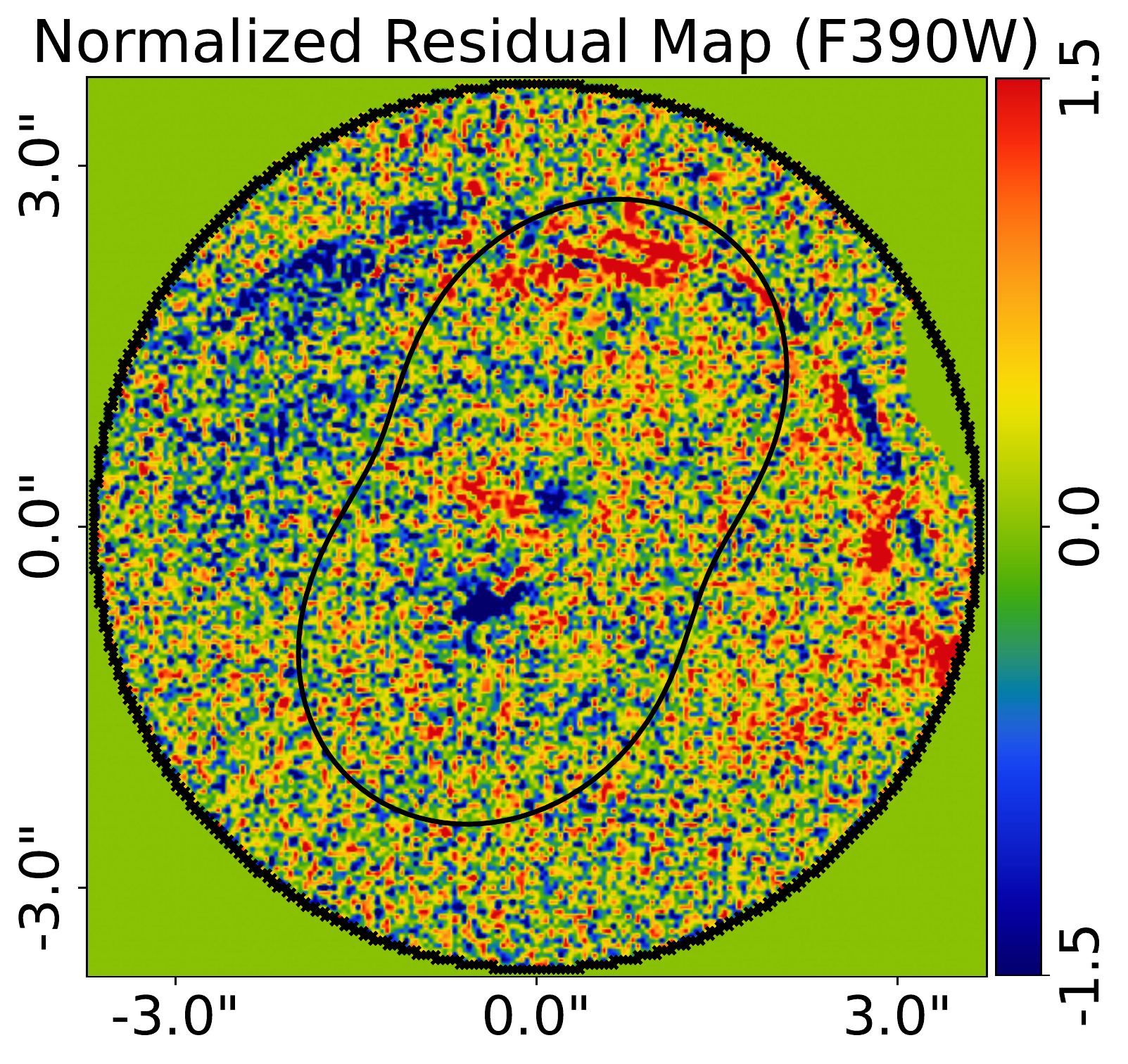}
\includegraphics[width=0.19\textwidth]{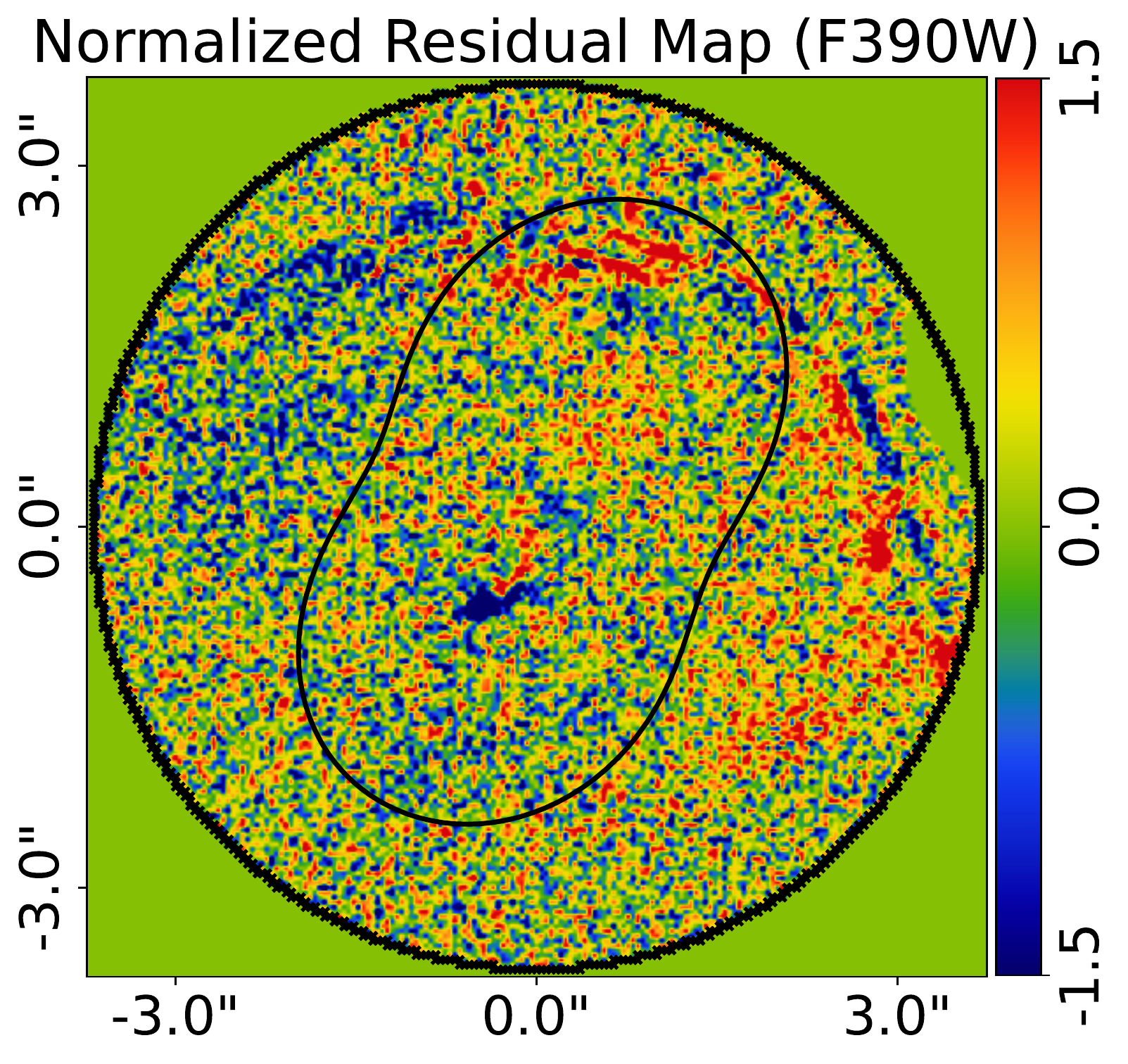}
\includegraphics[width=0.19\textwidth]{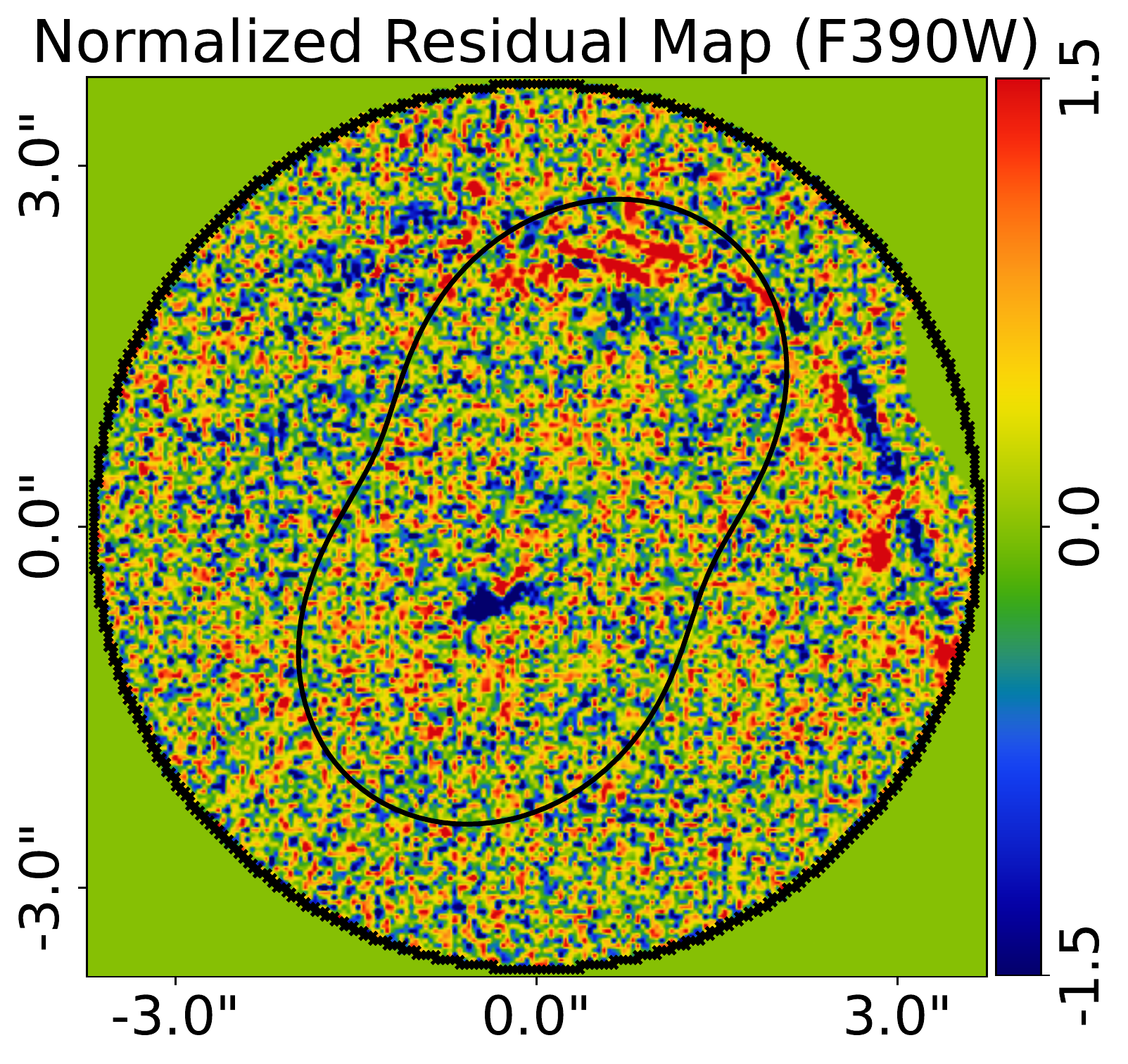}
\includegraphics[width=0.19\textwidth]{light_fit/f390w/light_fit_no_align_x2_norm.pdf}
\includegraphics[width=0.19\textwidth]{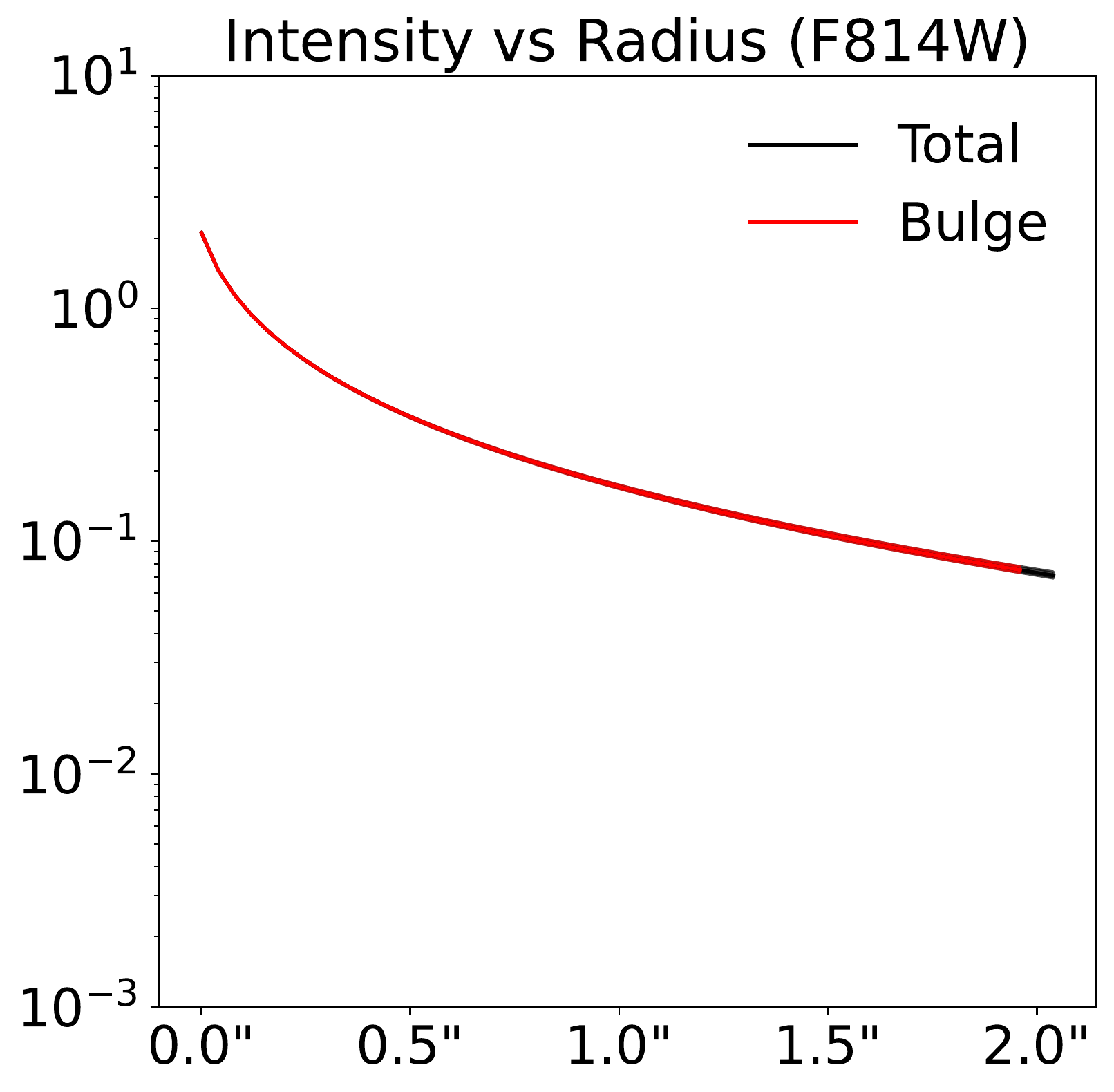}
\includegraphics[width=0.19\textwidth]{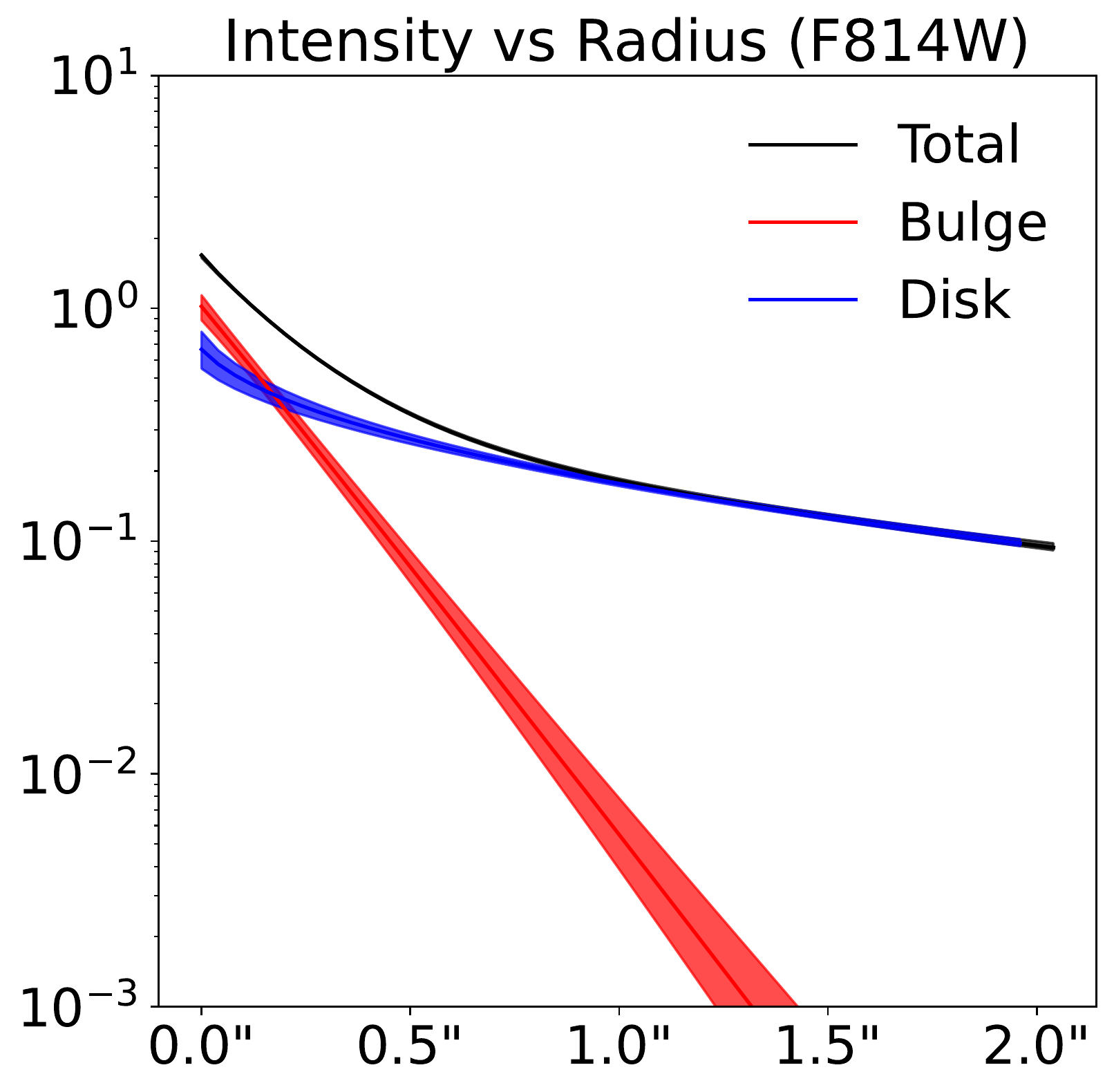}
\includegraphics[width=0.19\textwidth]{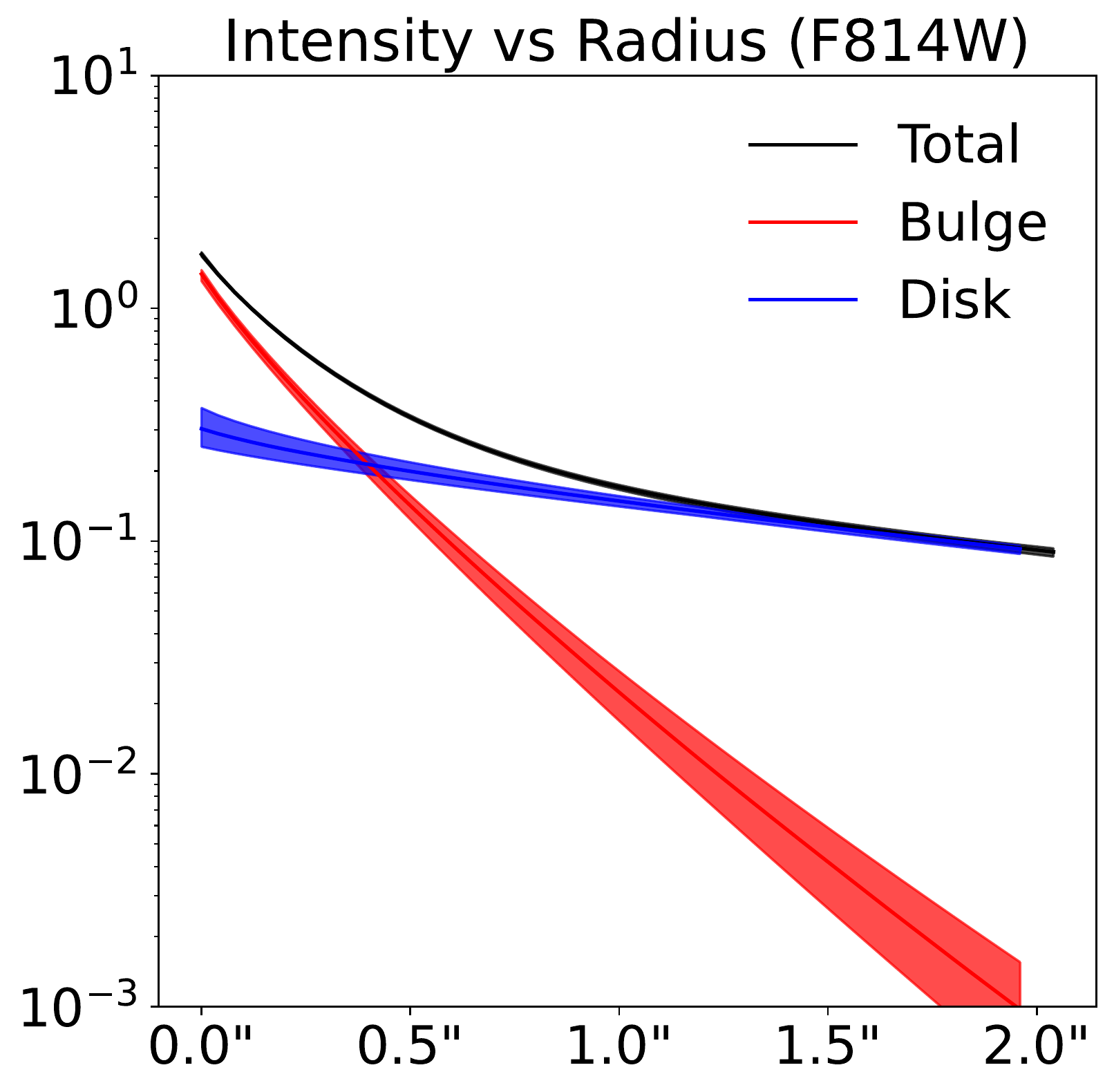}
\includegraphics[width=0.19\textwidth]{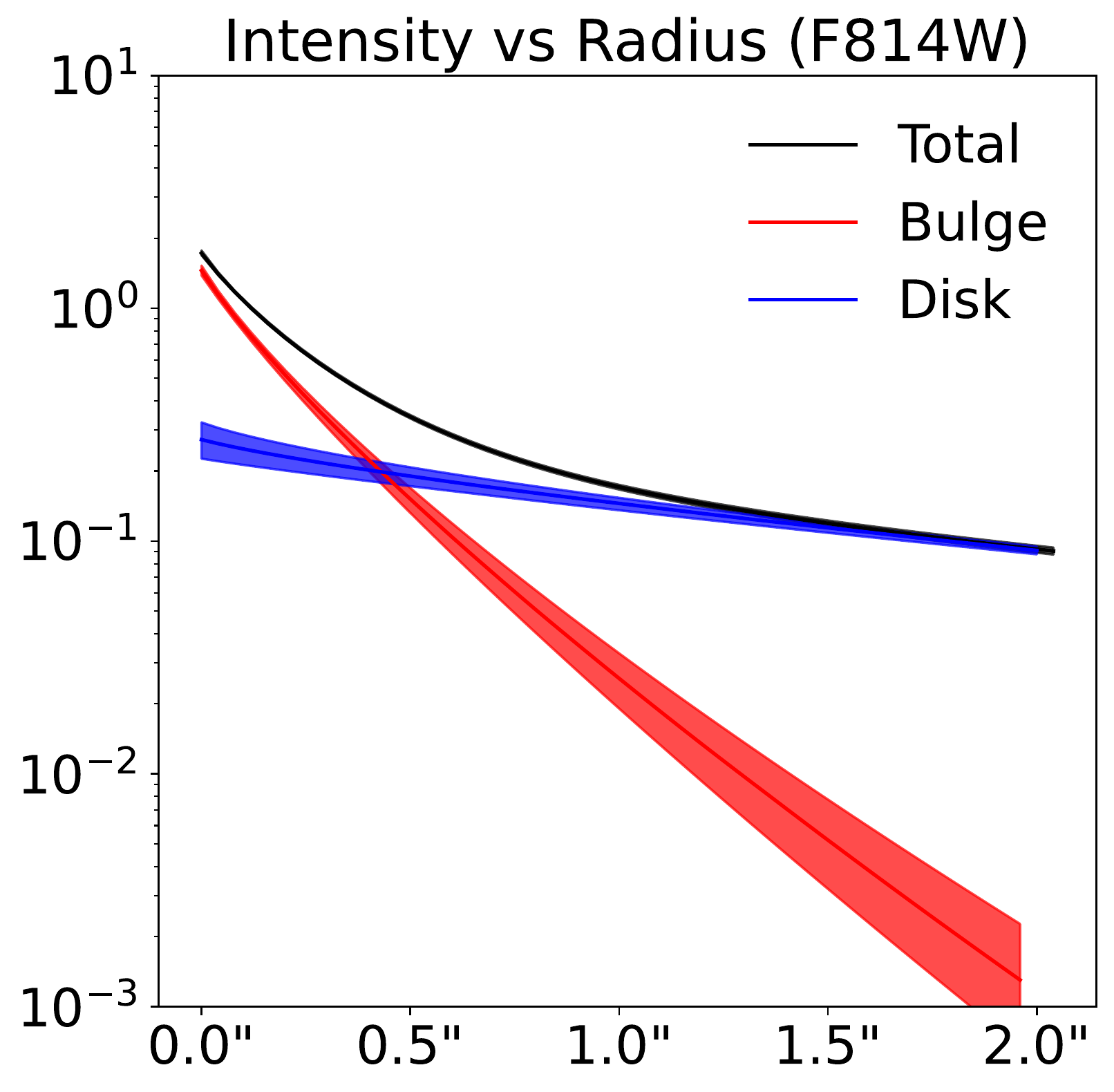}
\includegraphics[width=0.195\textwidth]{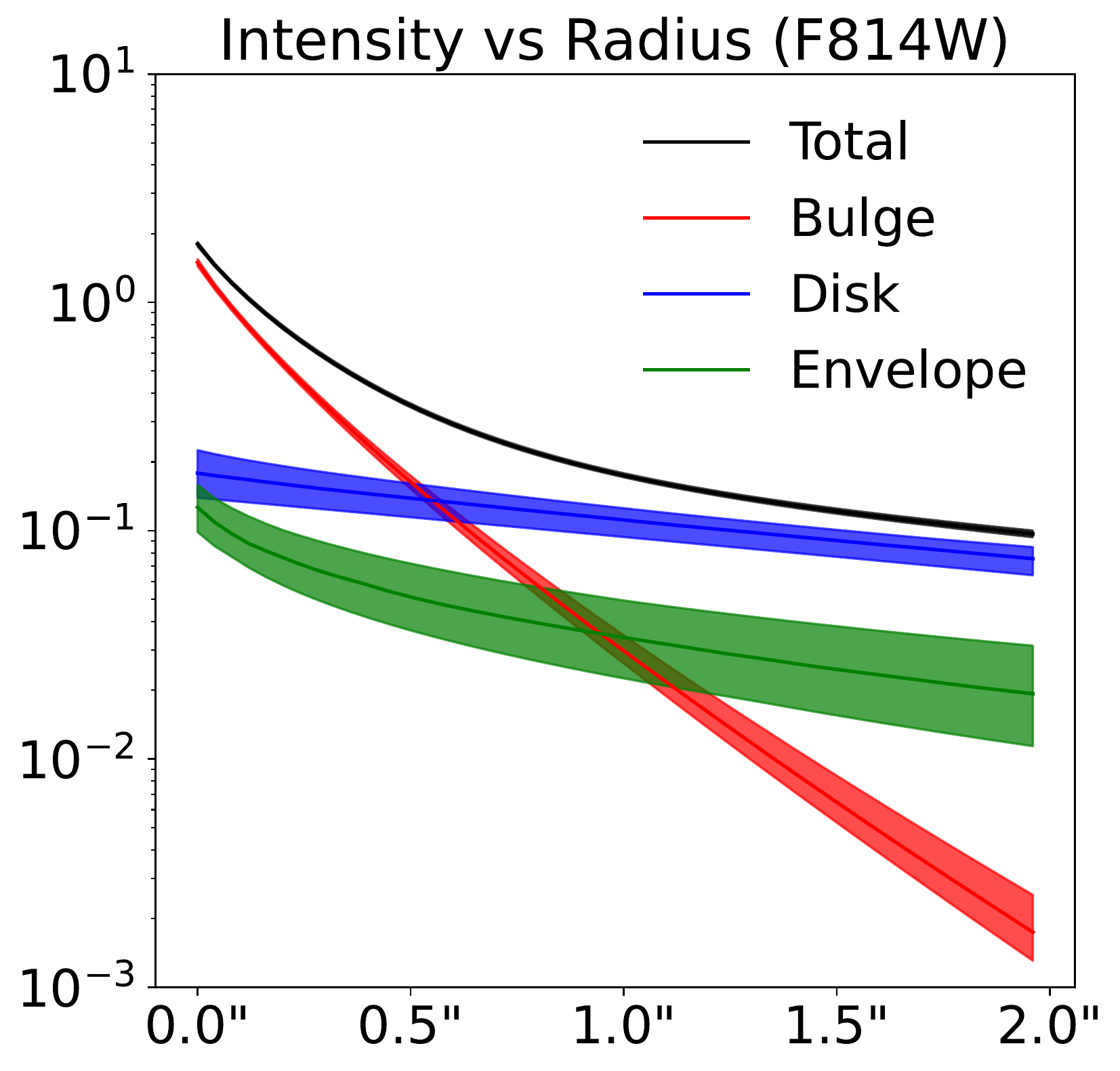}
\caption{The normalized residuals of fits to the F814W image (top row), F390W image (middle row) and 1D decomposed intensity profiles (bottom row) of the following five lens galaxy light models model-fits (from left to right): (i) a single Sersic profile; (ii) a double Sersic profile where the centres and elliptical components are aligned; (iii) where their centres are aligns but elliptical components are not; (iv) where neither components are aligned and; (v) a triple Sersic model where no geometric components are aligned. All models are fitted with a fixed isothermal mass profile with external shear and a pixelized source reconstruction which changes for every light profile fitted. 1D profiles are computed using coordinates that extend radially outwards from the centre of the light profile and are aligned with its major axis. Each plot corresponds to the maximum likelihood light model of a model-fit. The shaded regions show estimates of each light profile within $3\sigma$ confidence intervals. The black line shows the tangential critical curve of the mass model, the white line the radial critical curve, and the black cross(es) towards the centre of each figure the centre(s) of each light profile component.} 
\label{figure:LightFit}
\end{figure*}

This section presents the results of fitting the F390W and F814W images of Abell 1201 with different light models, which is performed in the Light pipeline. \cref{figure:Light2D} shows projected two-dimensional images of each model light profile, for the single Sersic model (top left panel), the double Sersic model assuming no geometric alignments (top-centre and top-right panels) and the triple Sersic model (bottom panels). The single Sersic model infers a compact central bulge with Sersic index $n^{\rm bulge} \simeq 4$, consistent with a massive elliptical galaxy. The double Sersic model decomposes the lens galaxy's light into two distinct components, consisting of a compact bulge similar to the single Sersic fit but with a much lower value of $n^{\rm bulge} \simeq 1.25$, surrounded by a more extended and elliptical component where $n^{\rm disk} \simeq 1.3$. The half-light radius of this extended component is $R_{\rm eff}^{\rm disk} \simeq 5.0"$, well beyond the strong lensing features. The triple Sersic model infers these two components, but includes a fainter additional inner structure.

\begin{table}
\resizebox{\linewidth}{!}{
\begin{tabular}{ l | l | l | l | l | l | l | l } 
\multicolumn{1}{p{1.1cm}|}{Filter} 
& \multicolumn{1}{p{1.3cm}|}{\centering Number of \\ Sersics} 
& \multicolumn{1}{p{1.5cm}|}{Aligned Elliptical Components} 
& \multicolumn{1}{p{1.6cm}|}{Aligned Centres} 
& \multicolumn{1}{p{1.5cm}|}{Evidence}  
\\ \hline
& & & & \\[-4pt]
F814W &  1 &       N/A &        N/A & 76664.15 \\[2pt]
F814W & 2 & \checkmark & \checkmark & 77616.09  \\[2pt]
F814W & 2 & \ding{55} & \checkmark & 78049.53 \\[2pt]
F814W & 2 & \ding{55} & \ding{55}  & 78181.48 \\[2pt]
F814W & 3 & \ding{55} & \ding{55}  & \textbf{78193.40} \\[2pt]
\hline
F390W & 1 &       N/A  &        N/A & 123604.22 \\[2pt]
F390W & 2 & \checkmark & \checkmark & 123962.79 \\[2pt] 
F390W & 2 & \ding{55}  & \checkmark & 124275.33 \\[2pt]
F390W & 2 & \ding{55}  & \ding{55}  & \textbf{124663.78} \\[2pt]
F390W & 3 & \ding{55}  & \ding{55}  & N/A \\[2pt]
\end{tabular}
}
\caption{The Bayesian Evidence, $\ln \mathcal{Z}$, of each model-fit performed by the Light pipeline, which compares models with one, two or three Sersic profiles. Fits to both the F814W and F390W images are shown. Models which make different assumptions for the alignment of the $(x,y)$ centre and $(\epsilon_{\rm 1}, \epsilon_{\rm 2}$) elliptical components of the bulge, disk and envelope are shown. A tick mark indicates that this assumption is used in the model, for example the second row is a model where both the elliptical components and centres are aligned. The triple Sersic model for the F390W is omitted because it went to unphysical solutions where one Sersic component was used to fit structure in the lensed source.}
\label{table:LightMC}
\end{table}

The Bayesian evidence values, $\ln \mathcal{Z}$, of the light models informs us which provides the best fit to the data. These are given for both F390W and F814W images in \cref{table:LightMC}. Models assuming a single Sersic profile give significantly worse fits than those using multiple profiles, indicating it does not capture the extended component. Three models assuming two Sersic profiles are compared, where: (i) their centre and elliptical components are aligned; (ii) their centres are aligned but elliptical components are not and; (iii) their centres are also free to vary. For both images model (iii) is preferred, with a value of $\Delta \ln \mathcal{Z} > 100$ the other models for the F814W data. For the F814W image a triple Sersic (with all geometric parameters free to vary) gives a value $\ln \mathcal{Z} = 11.55$ above that of the two Sersic model, indicating that it is the marginally favoured model. 

\cref{figure:LightFit} shows the normalized residuals of these fits. For the single Sersic model and models with geometric alignments residuals are evident around the lens galaxy's centre in both the F814W and F390W bands, consistent with the Bayesian evidences. In the F814W image the double Sersic model with free centres and the triple Sersic model gave a significant increase in $\ln \mathcal{Z}$. However, the improvements are not visible in the residuals, indicating they improve the light model fractionally over many pixels. 

The lower panels of this figure show 1D plots of the intensity as a function of radius for each component. The inner structure contributes to most of the stellar light within $\sim 1.0"$ where the counter image is observed, whereas at the location of the giant arc the extended component makes up over $95\%$ of the total emission. They also show that the outer component makes up the majority of the lens galaxy's total luminous emission, albeit most is beyond the $3.0"$ radius where the lensed source is constrained. 
\section{Double Sersic Models}\label{MassFits}

\begin{figure*}
\centering
\includegraphics[width=0.241\textwidth]{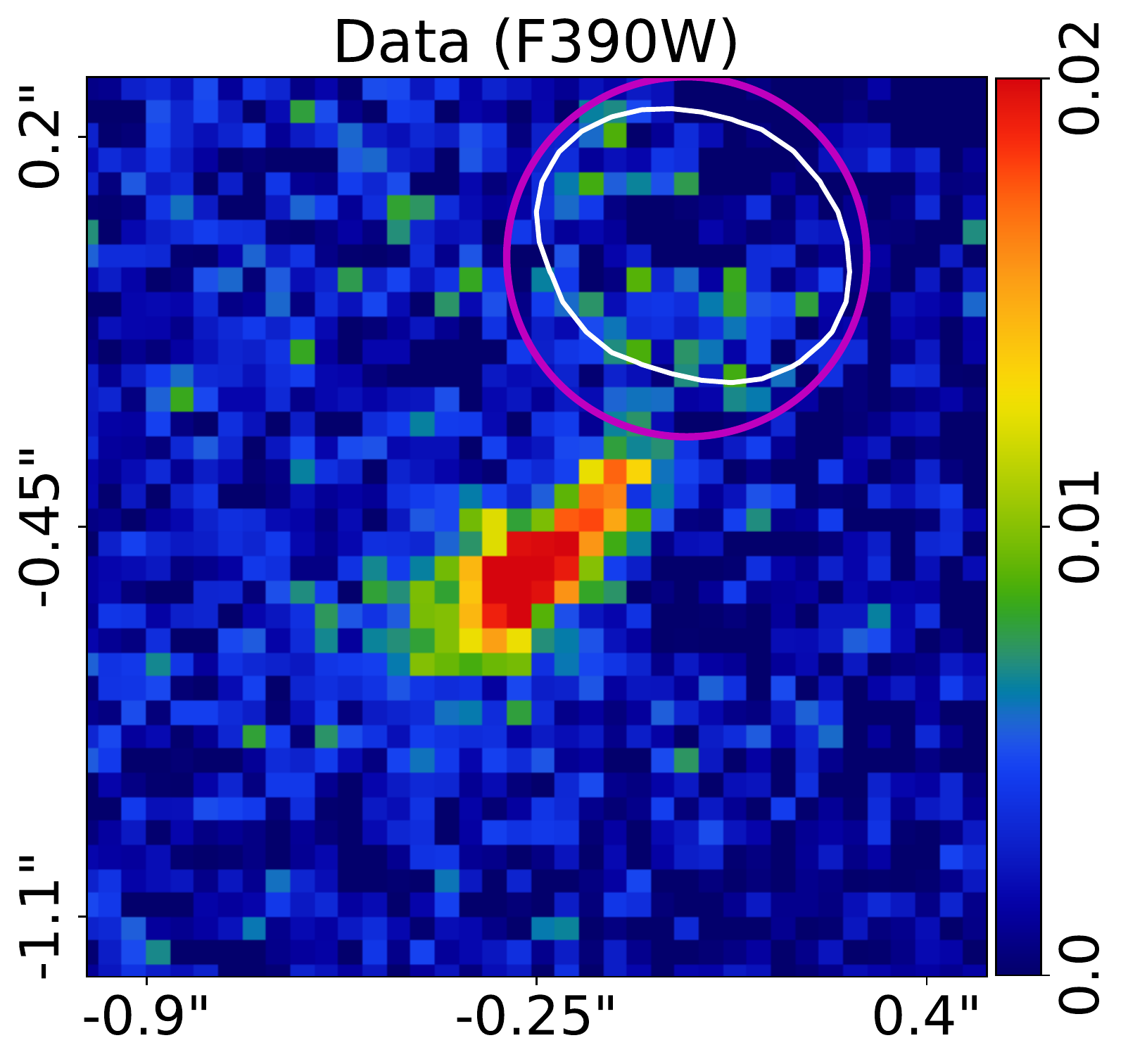}
\includegraphics[width=0.241\textwidth]{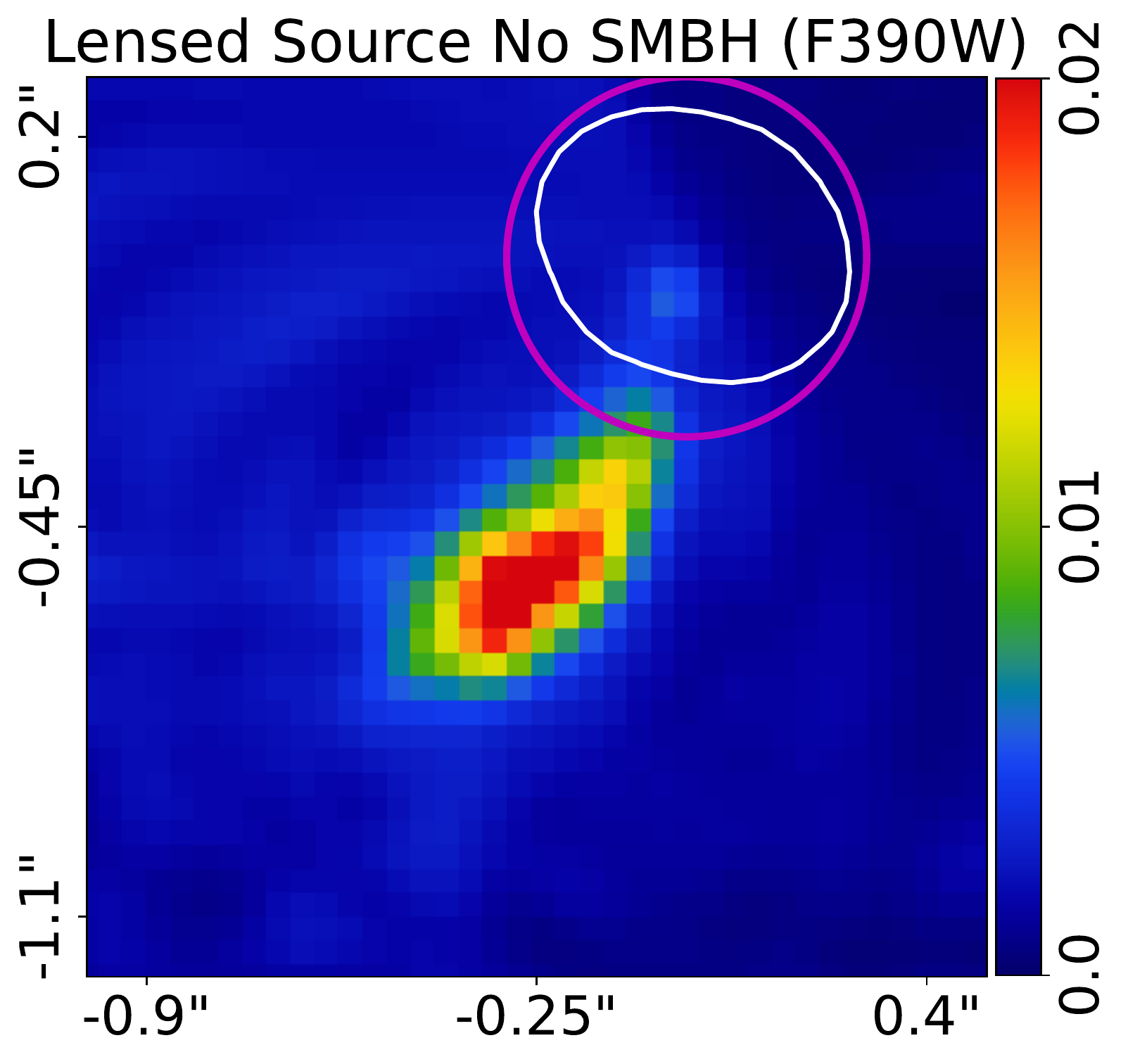}
\includegraphics[width=0.241\textwidth]{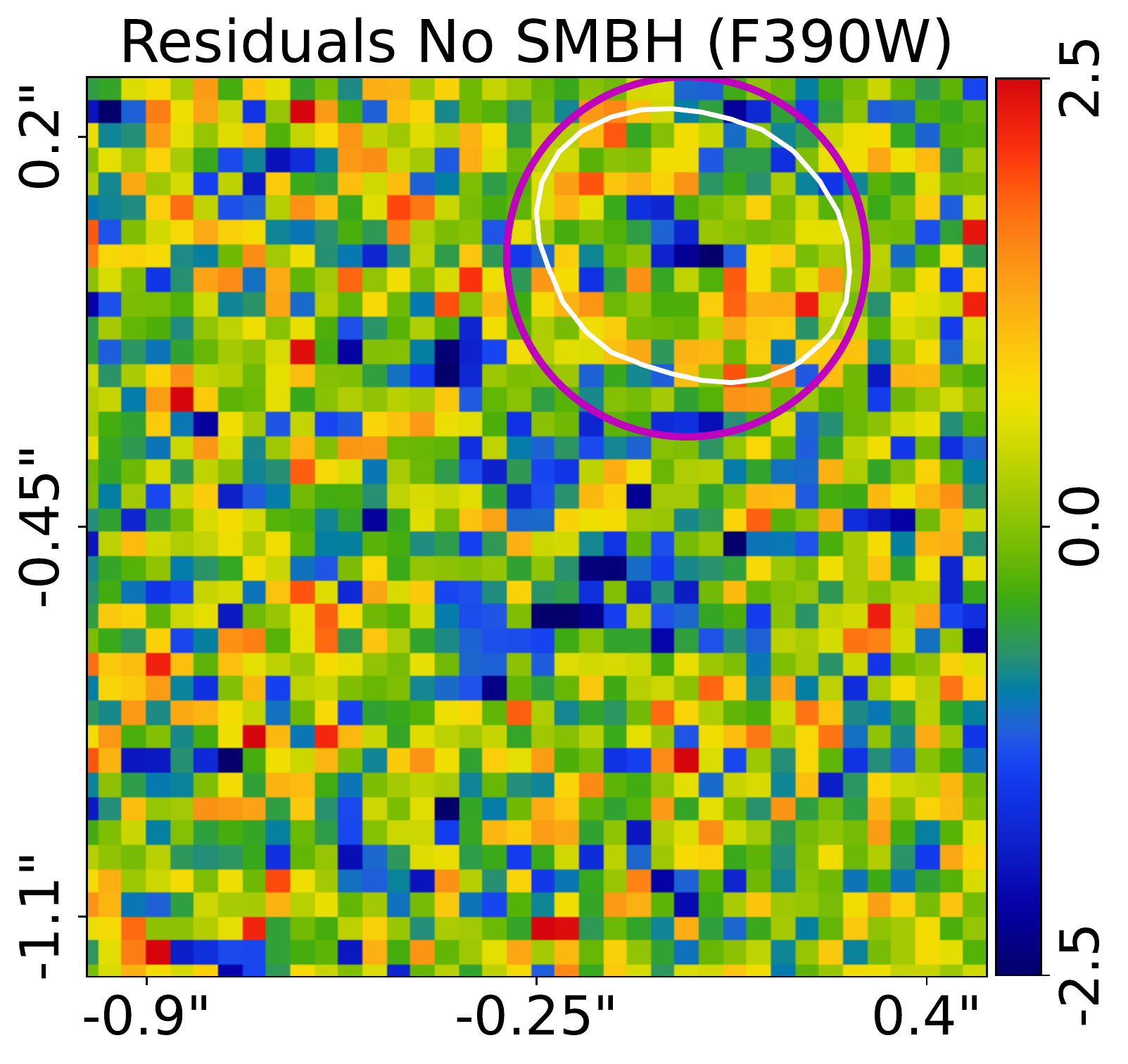}
\includegraphics[width=0.241\textwidth]{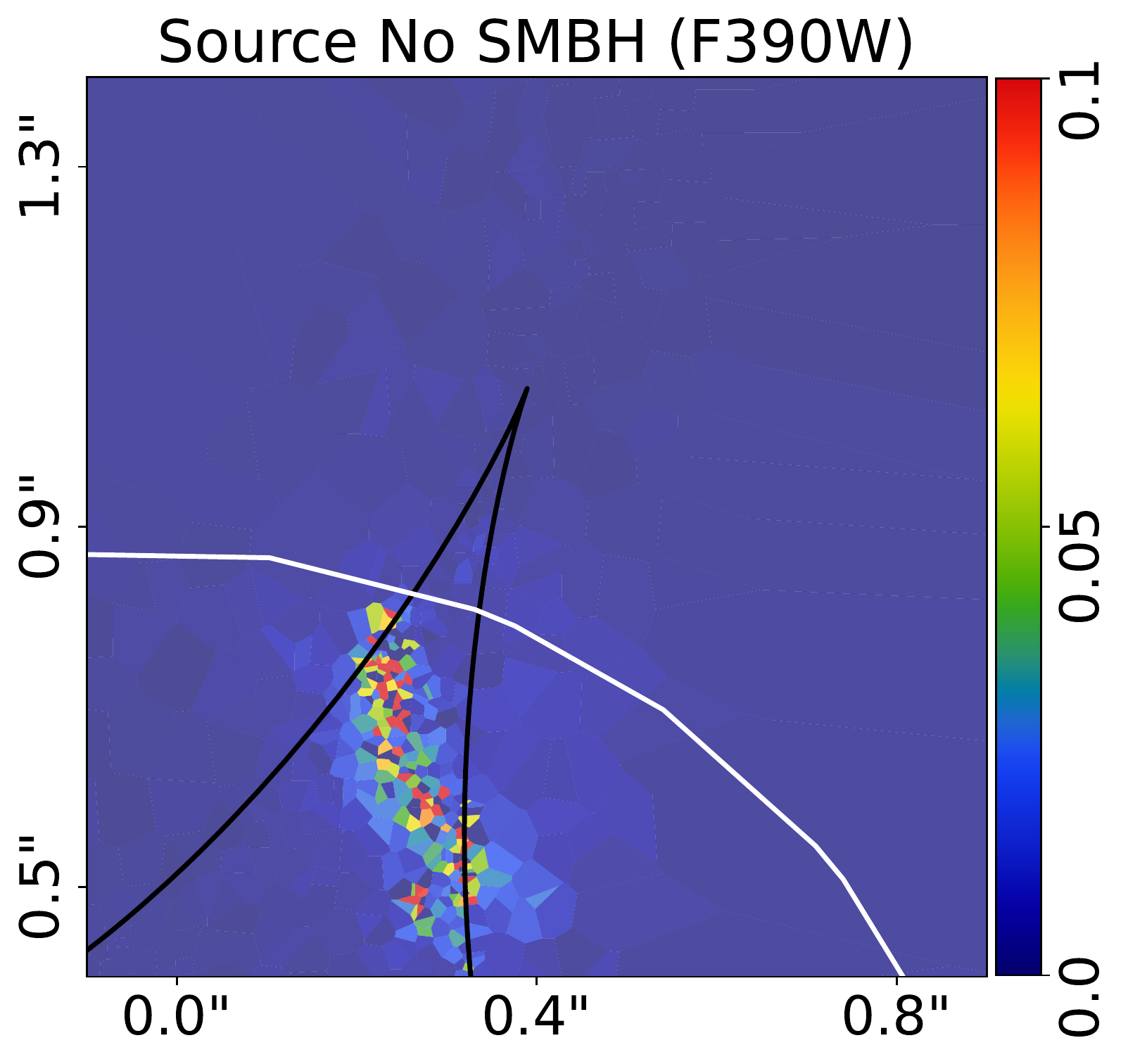}
\includegraphics[width=0.241\textwidth]{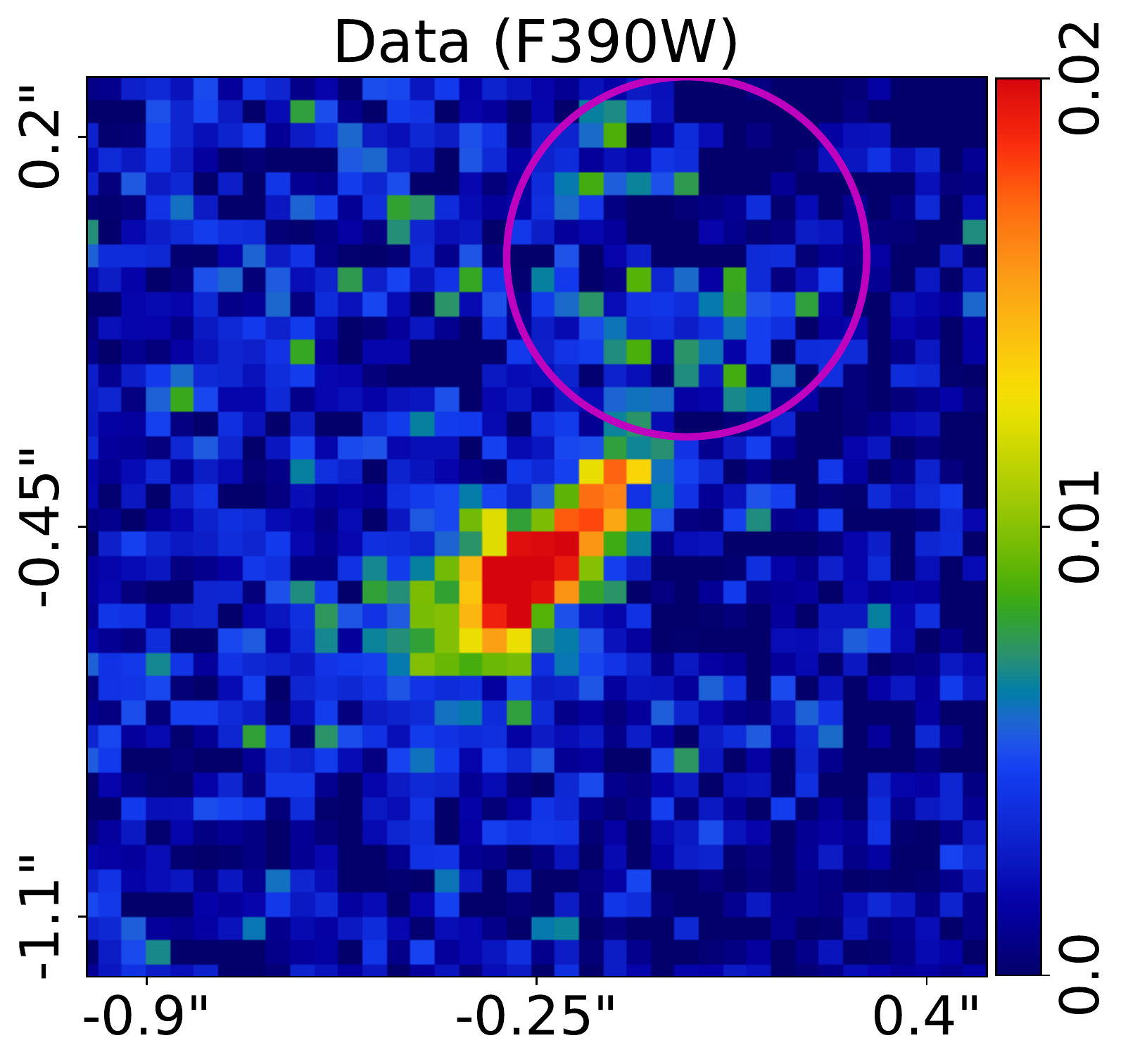}
\includegraphics[width=0.241\textwidth]{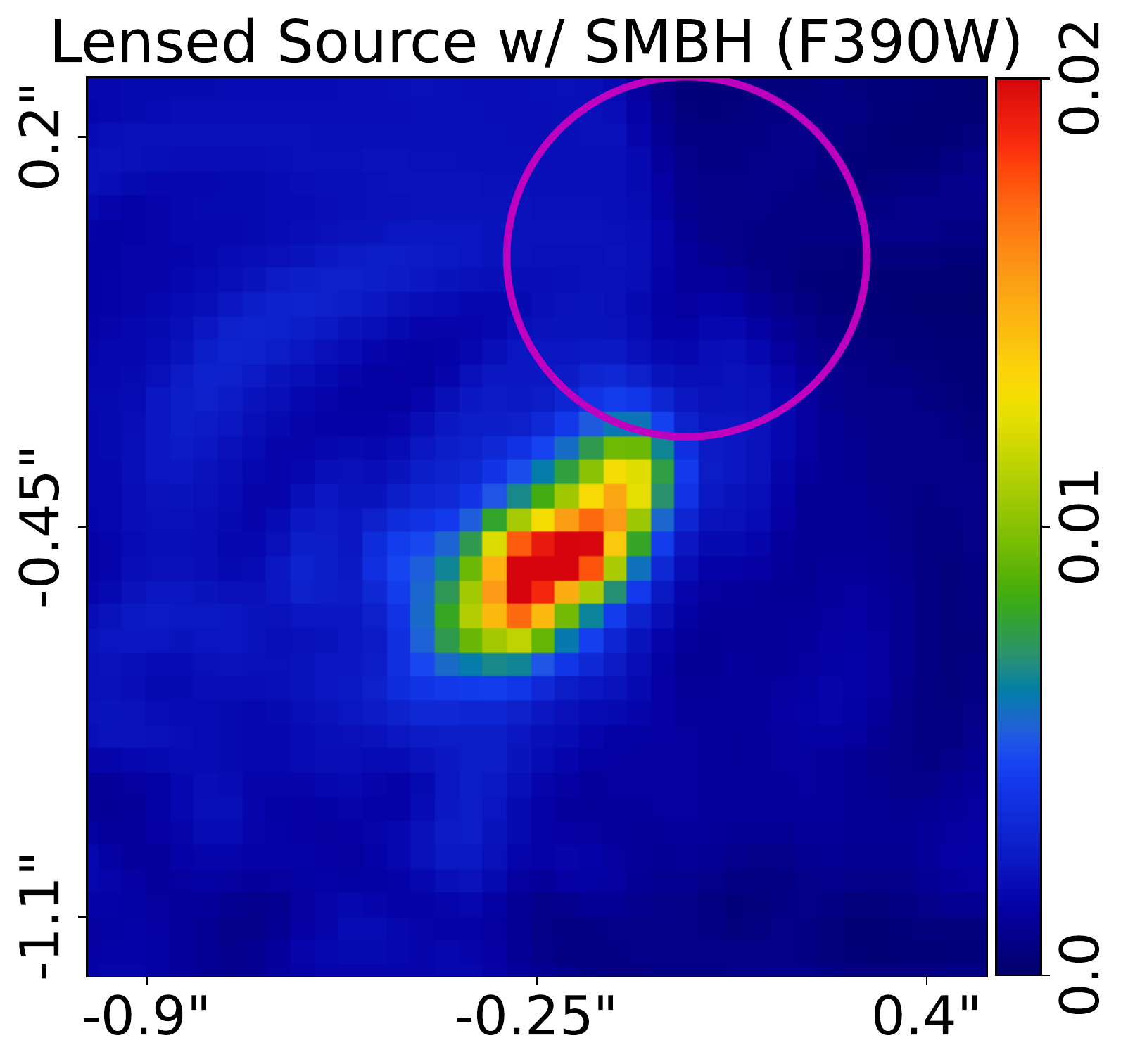}
\includegraphics[width=0.241\textwidth]{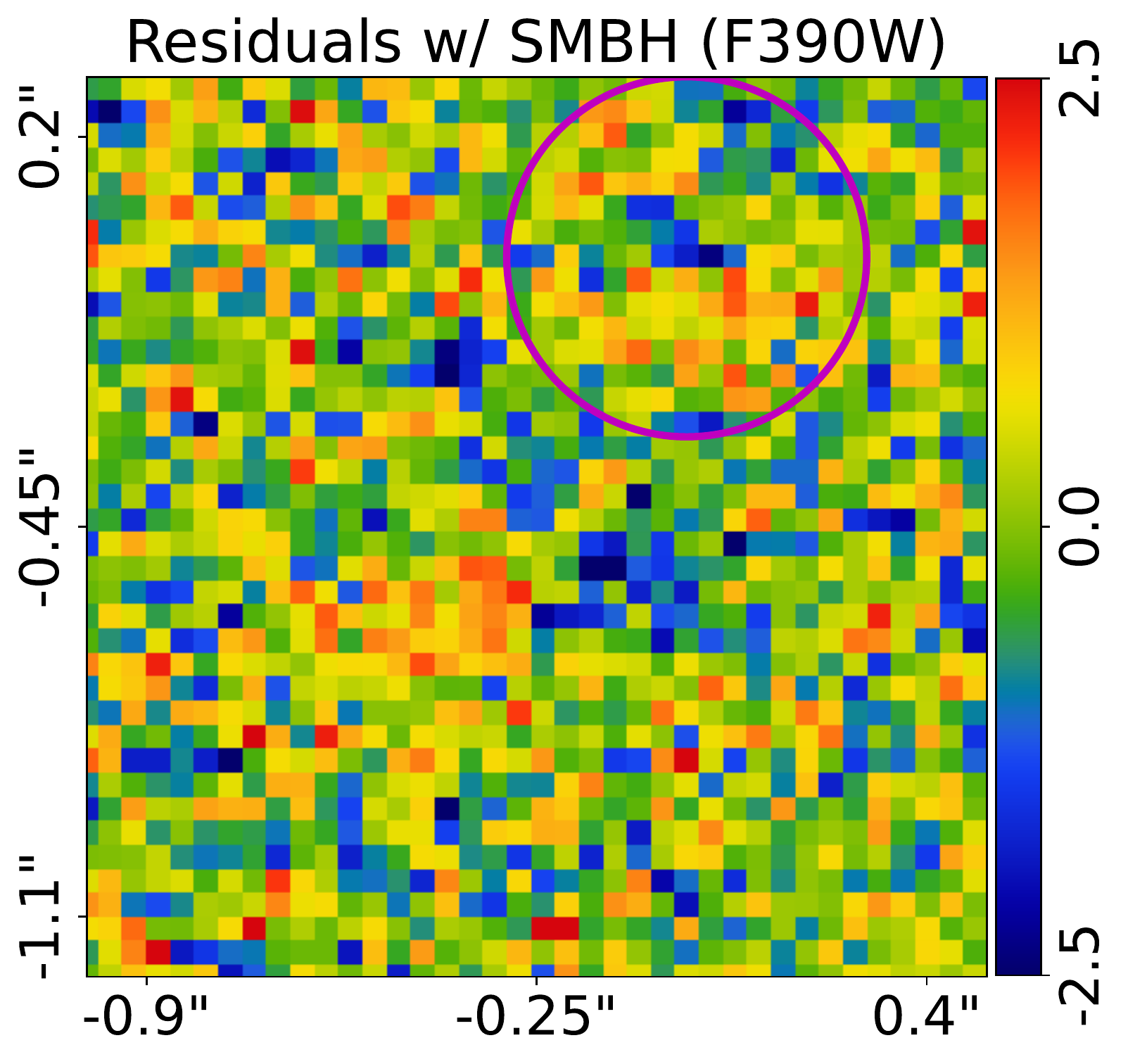}
\includegraphics[width=0.241\textwidth]{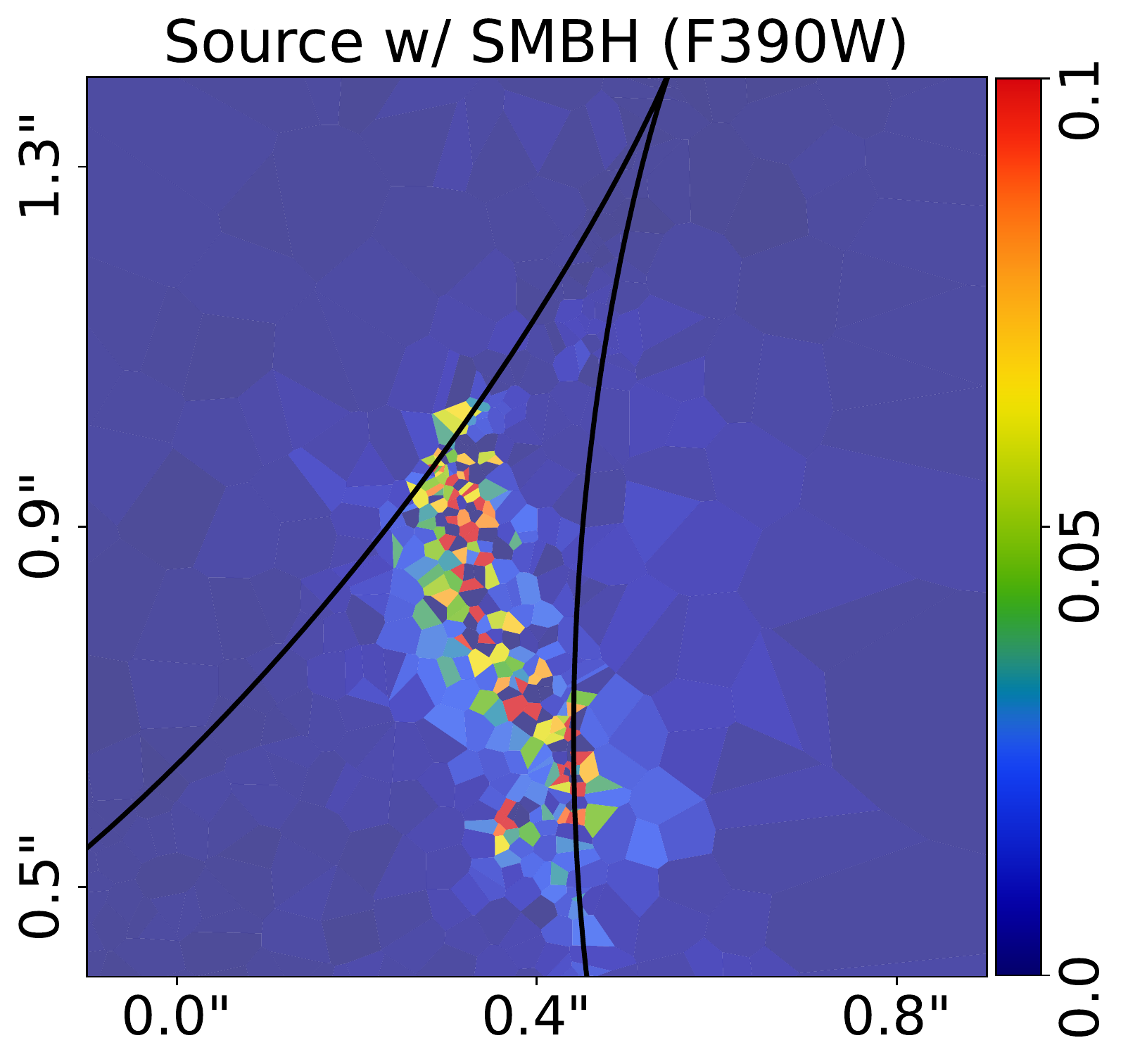}
\caption{
Zoom-ins of the observed counter image in the F390W data (left panel), the model lensed source (left-centre panel), the normalized residuals (right-centre panel) and the source reconstruction (right panel). The top and bottom rows shows double Sersic plus NFW decomposed model-fits without and with a SMBH respectively. All models include an external shear. Models which omit a SMBH form an additional clump of light in the counter image, which is not present in the data. The tangential caustic is shown by a black line and radial critical curve and caustic a white line; the latter does not form for models including a SMBH.
} 
\label{figure:LightDarkF390W}
\end{figure*}

\begin{figure*}
\centering
\includegraphics[width=0.241\textwidth]{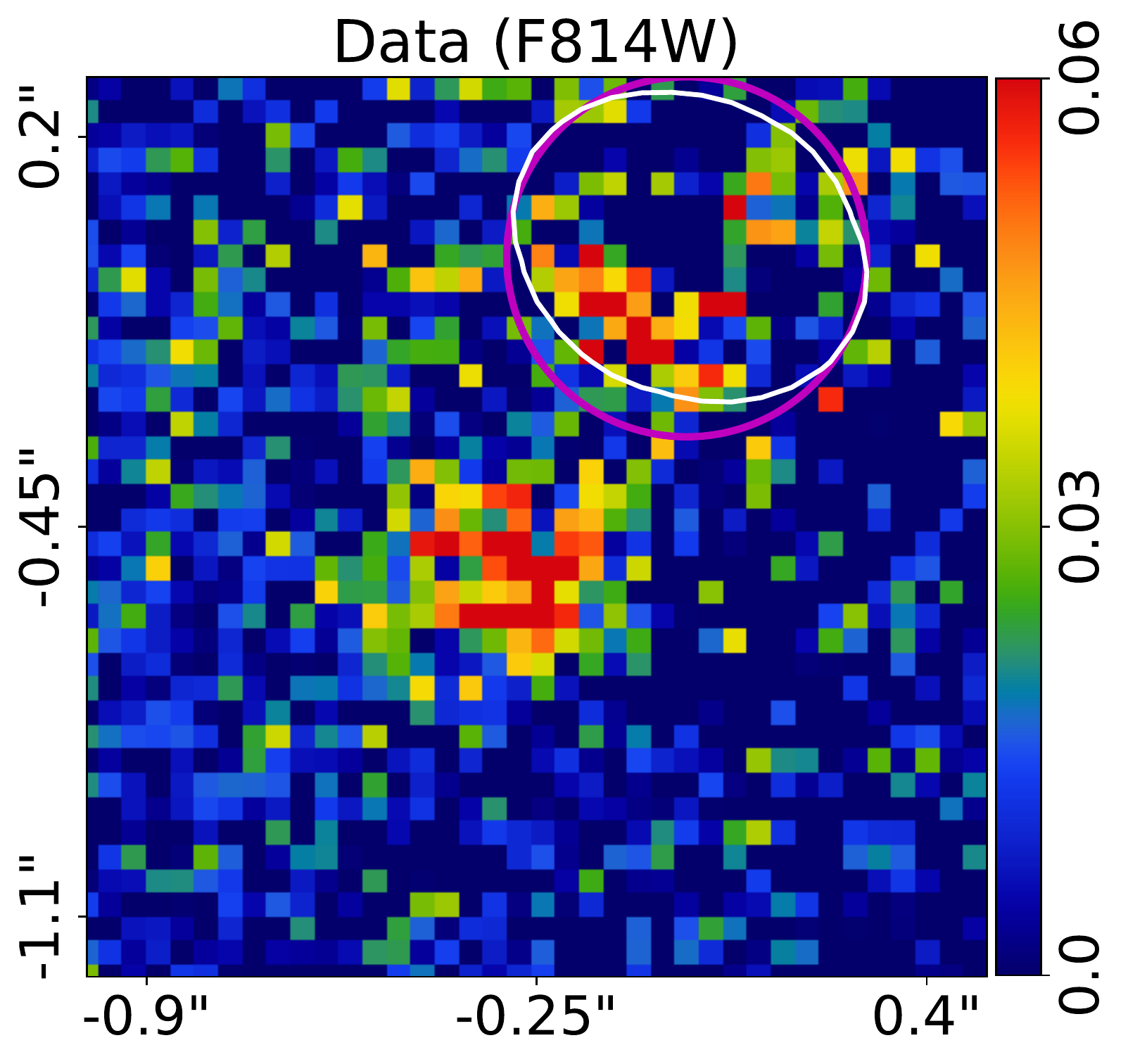}
\includegraphics[width=0.241\textwidth]{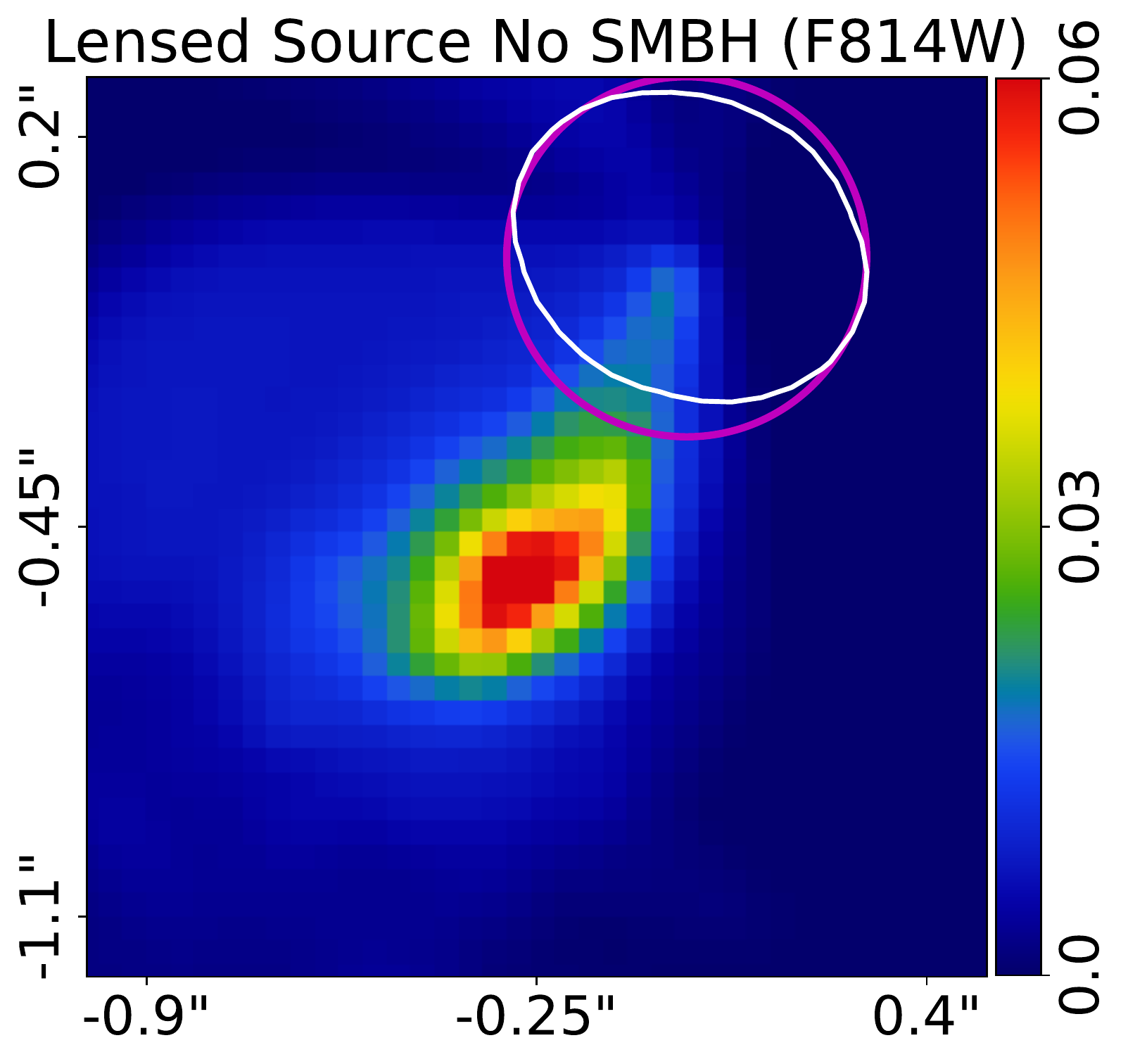}
\includegraphics[width=0.241\textwidth]{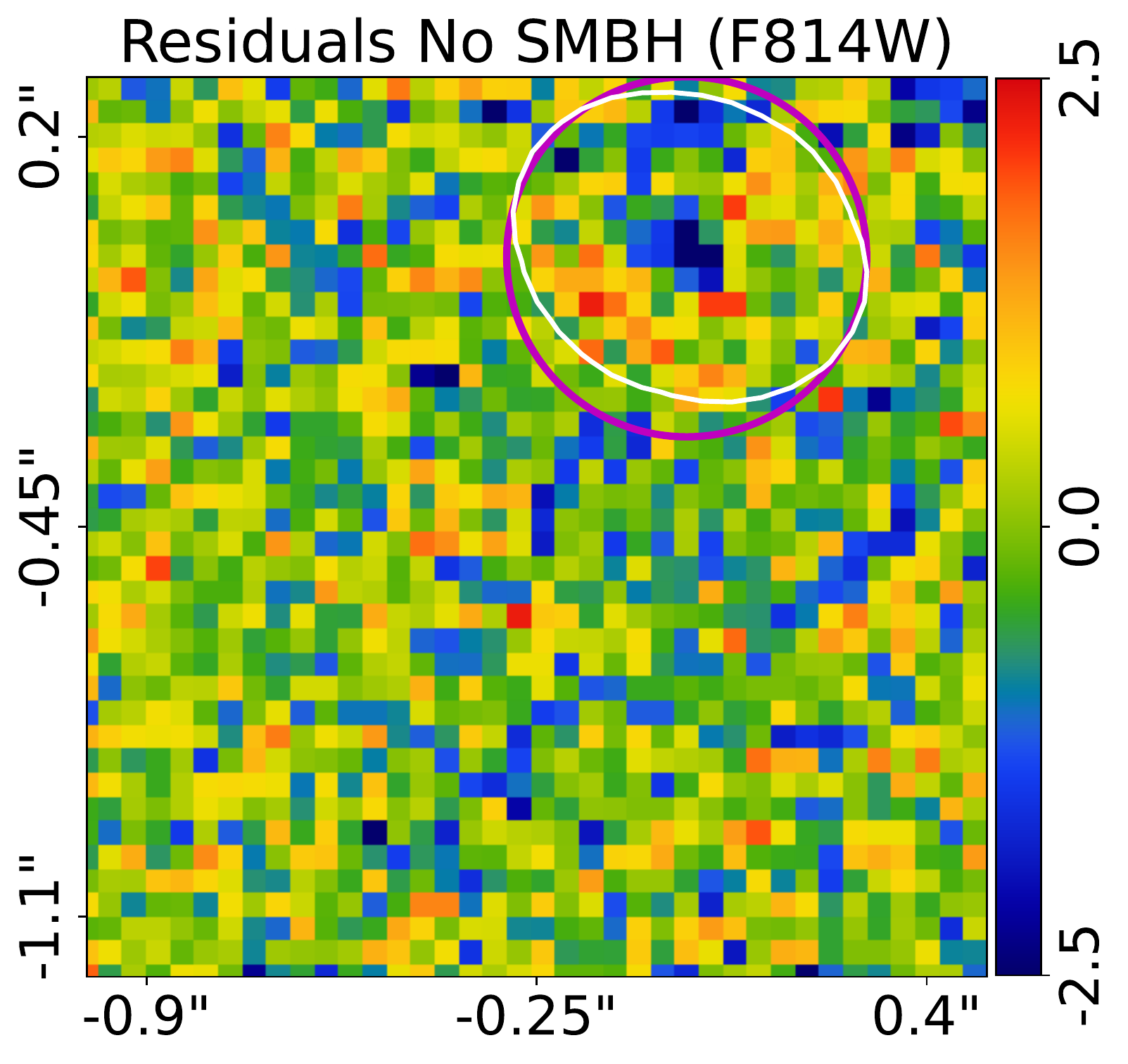}
\includegraphics[width=0.241\textwidth]{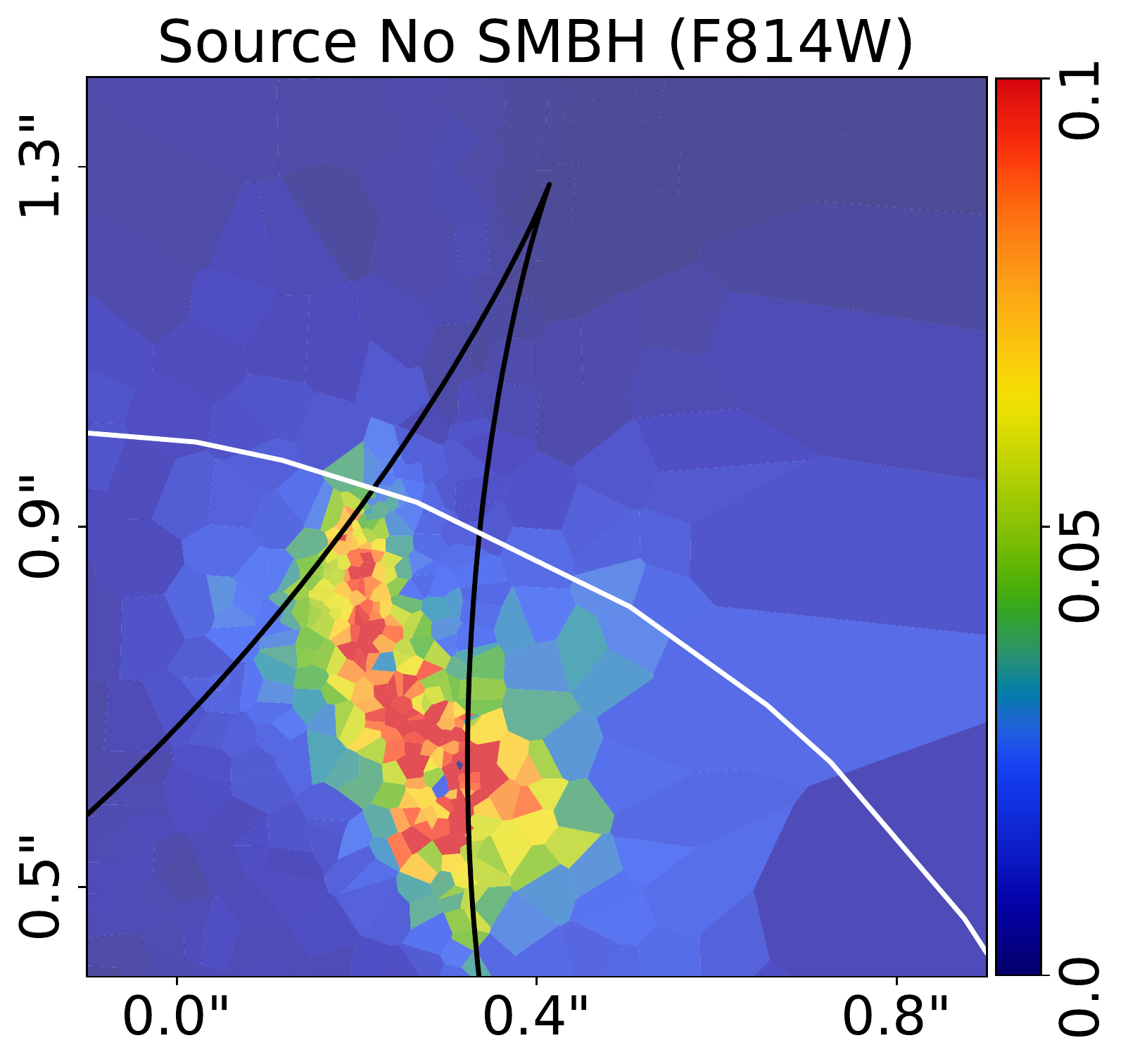}
\includegraphics[width=0.241\textwidth]{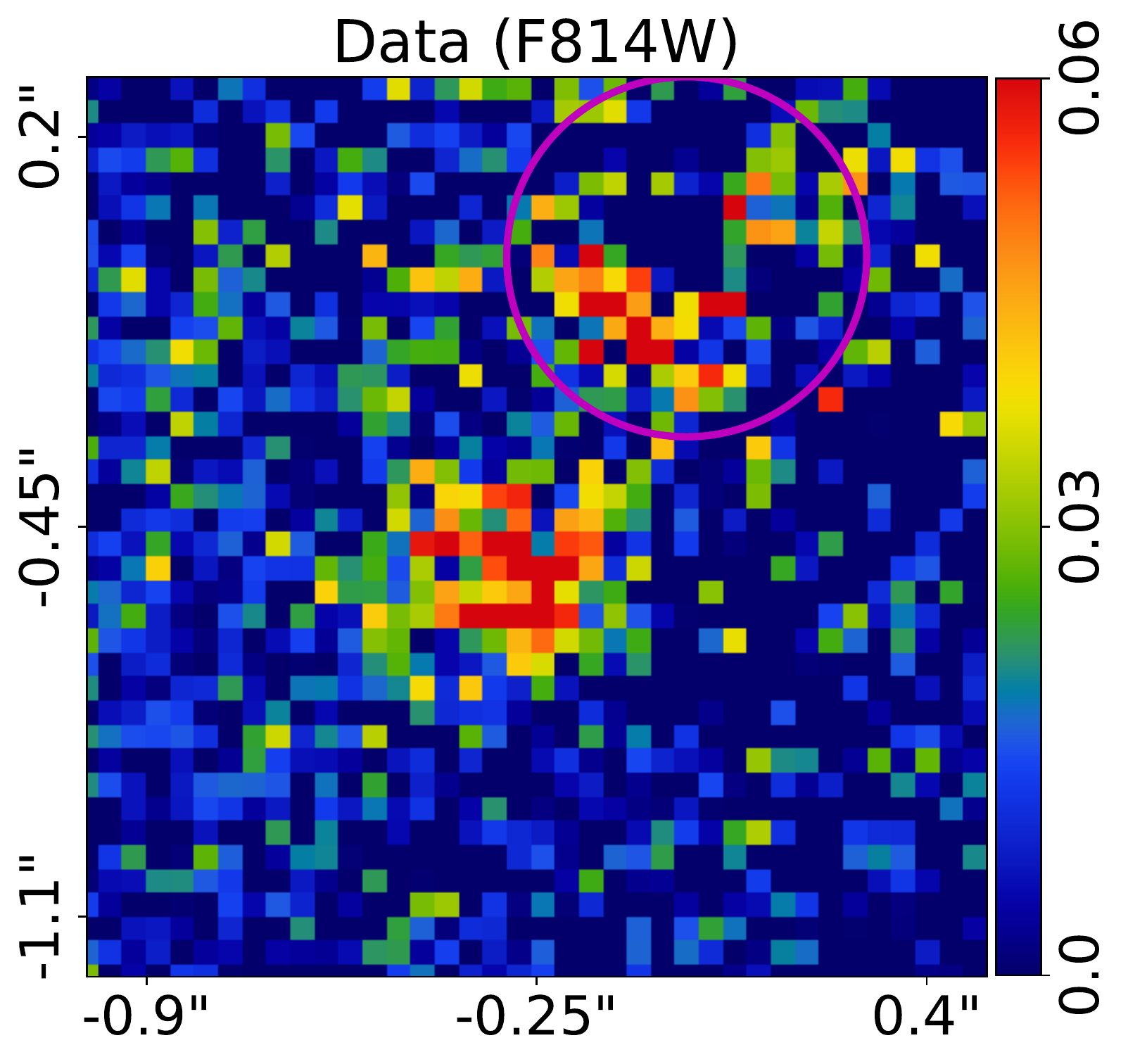}
\includegraphics[width=0.241\textwidth]{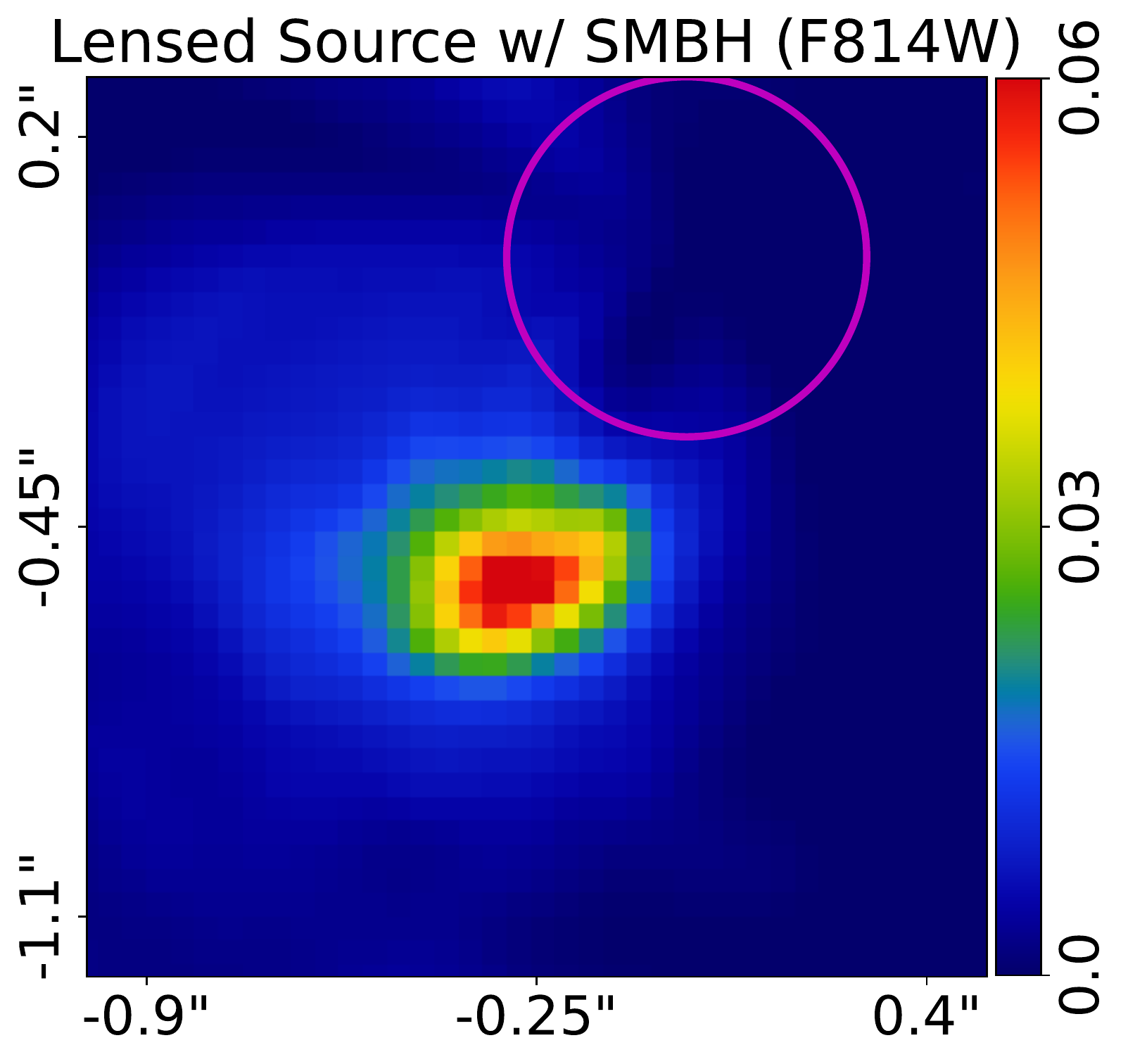}
\includegraphics[width=0.241\textwidth]{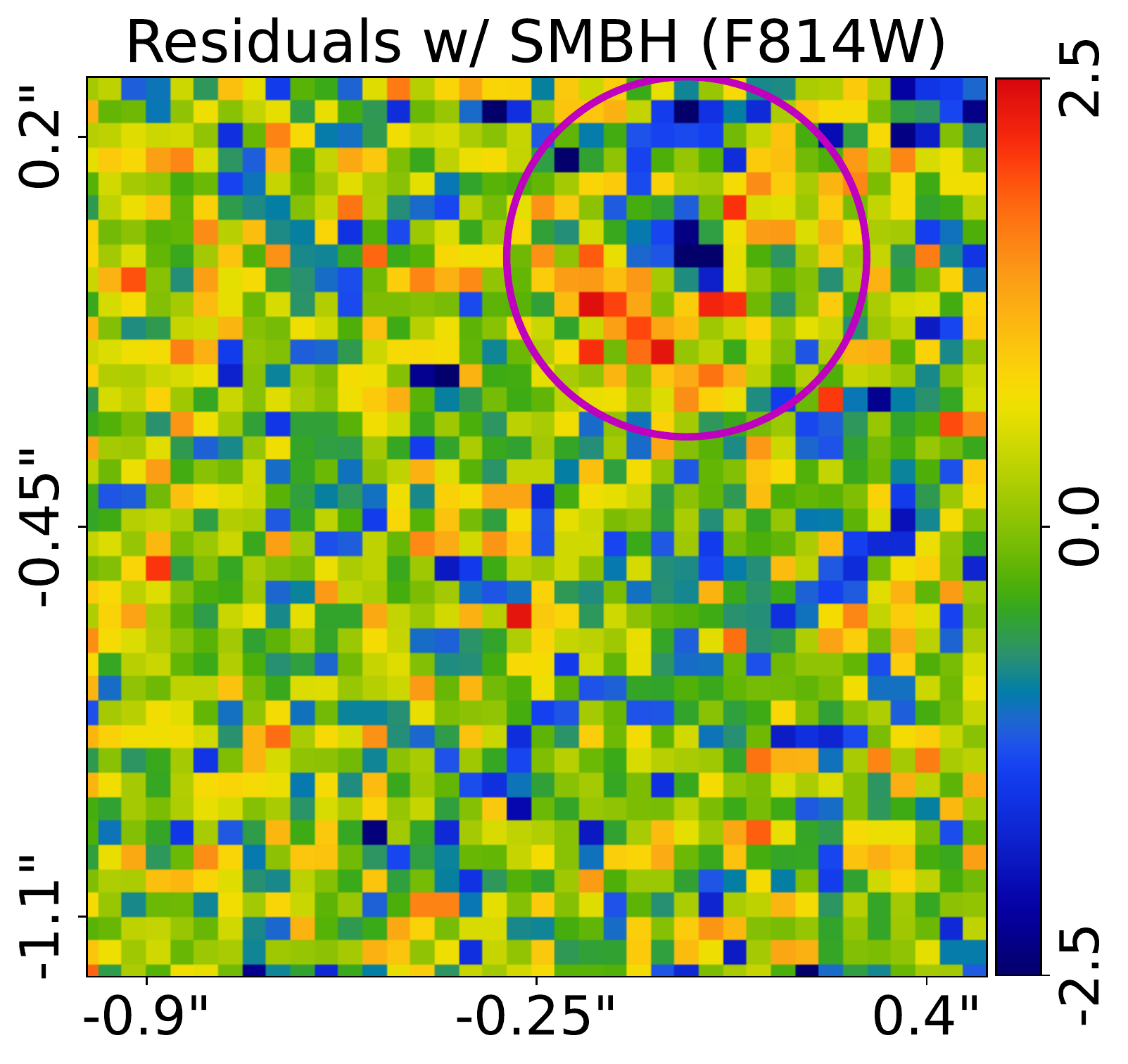}
\includegraphics[width=0.241\textwidth]{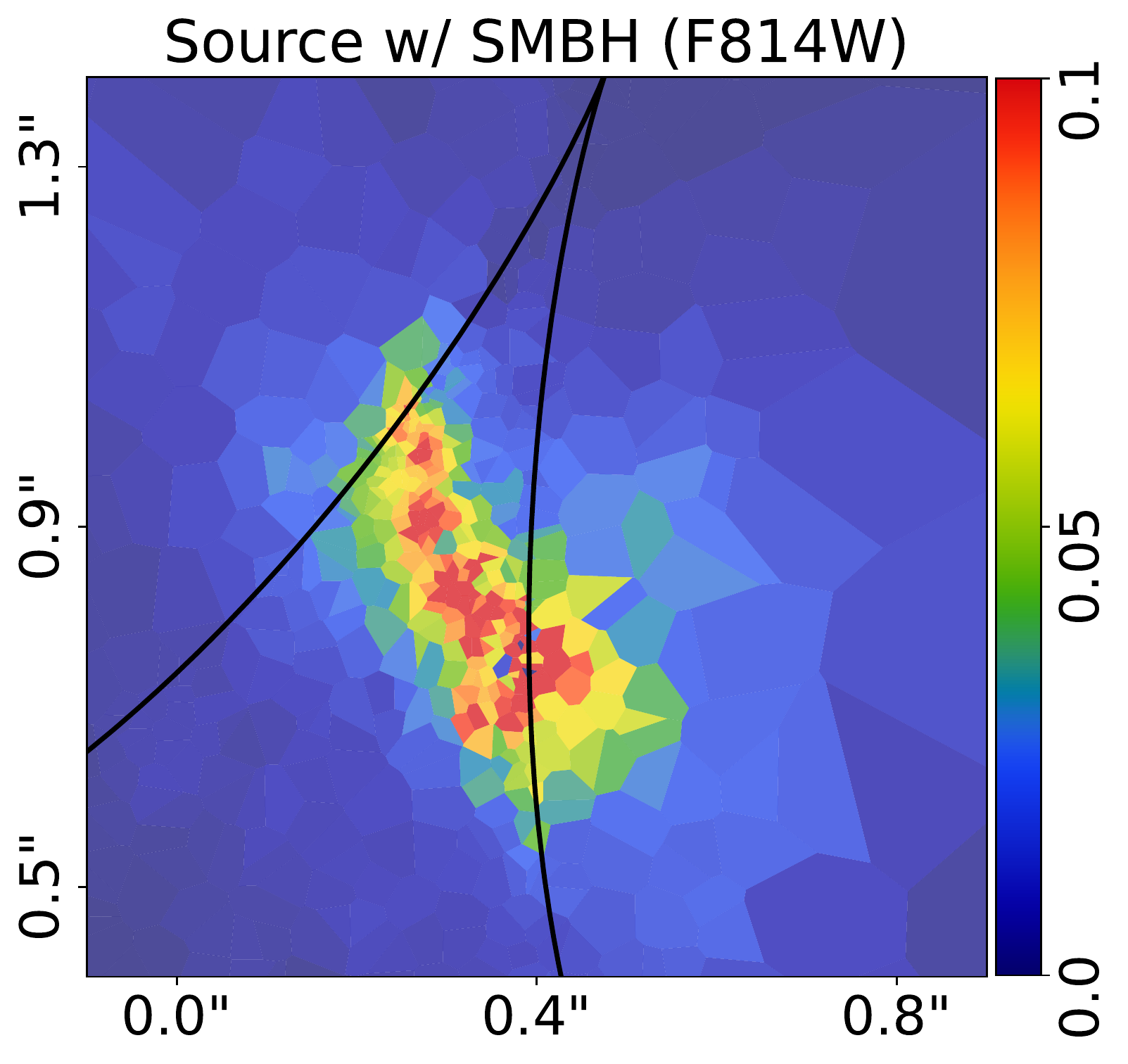}
\caption{
The same as \cref{figure:LightDarkF390W} but for the F814W data.
} 
\label{figure:LightDarkF814W}
\end{figure*}

The results of fitting the decomposed model with two Sersic profiles are shown in \cref{figure:LightDarkF390W} and \cref{figure:LightDarkF814W}. These figures follow the same layout as \cref{figure:LightDarkF390Wx3} and \cref{figure:LightDarkF814Wx3} in the main paper. Results show the same behaviour as the triple Sersic fitted in the main paper, including extraneous flux in the counter image reconstruction when the model omits a SMBH. The double Sersic fit with a SMBH has a $\ln \mathcal{Z}$ value $29.93$ below the triple Sersic with a SMBH. This suggests that the lensing effects of the faint inner structure the third Sersic represents plays a role in reconstructing the counter image.

\section{Total Mass Models}\label{ResultSIE}

\begin{table}
\resizebox{\linewidth}{!}{
\begin{tabular}{ l | l | l | l } 
\multicolumn{1}{p{1.5cm}|}{Filter} 
& \multicolumn{1}{p{1.3cm}|}{Model} 
& \multicolumn{1}{p{1.3cm}|}{Includes SMBH?} 
& \multicolumn{1}{p{1.5cm}|}{$\ln \mathcal{Z}$}  
\\ \hline
F390W & PL & \ding{55} & 125562.45 \\[1pt]
F390W & PL & \checkmark & \textbf{125707.20} \\[0pt]
\hline
F390W & BPL & \ding{55} & 125699.90 \\[1pt]
F390W & BPL & \checkmark & 125693.78 \\[0pt]
\hline
F814W & PL &  \ding{55}& 78301.58 \\[1pt]
F814W & PL & \checkmark & \textbf{78330.39} \\[0pt]
\hline
F814W & BPL & \ding{55} & 78331.17 \\[1pt]
F814W & BPL & \checkmark & 78329.28 \\[0pt]
\end{tabular}
}
\caption{
The Bayesian Evidence, $\ln \mathcal{Z}$, of each model-fit using total mass models that collectively represent the lens's stellar and dark matter, where all models also include an external shear. $\ln \mathcal{Z}$ values for both the F390W and F814W images are shown. The favoured models given our criteria of $\Delta \ln \mathcal{Z} > 10$ is shown in bold. The PL mass model without a SMBH produces lower values of $\ln \mathcal{Z}$ than the PL model with a SMBH and both BPL models. The BPL model without a SMBH produces a $\ln \mathcal{Z}$ comparable to all models including a SMBH. The PL models favoured by model comparison are shown in bold; no bold model is shown for the BPL because models with and without a SMBH are both within the threshold of $\Delta \ln \mathcal{Z} > 10$ of one another.
}
\label{table:SMBHMC}
\end{table}

\begin{figure*}
\centering
\includegraphics[width=0.241\textwidth]{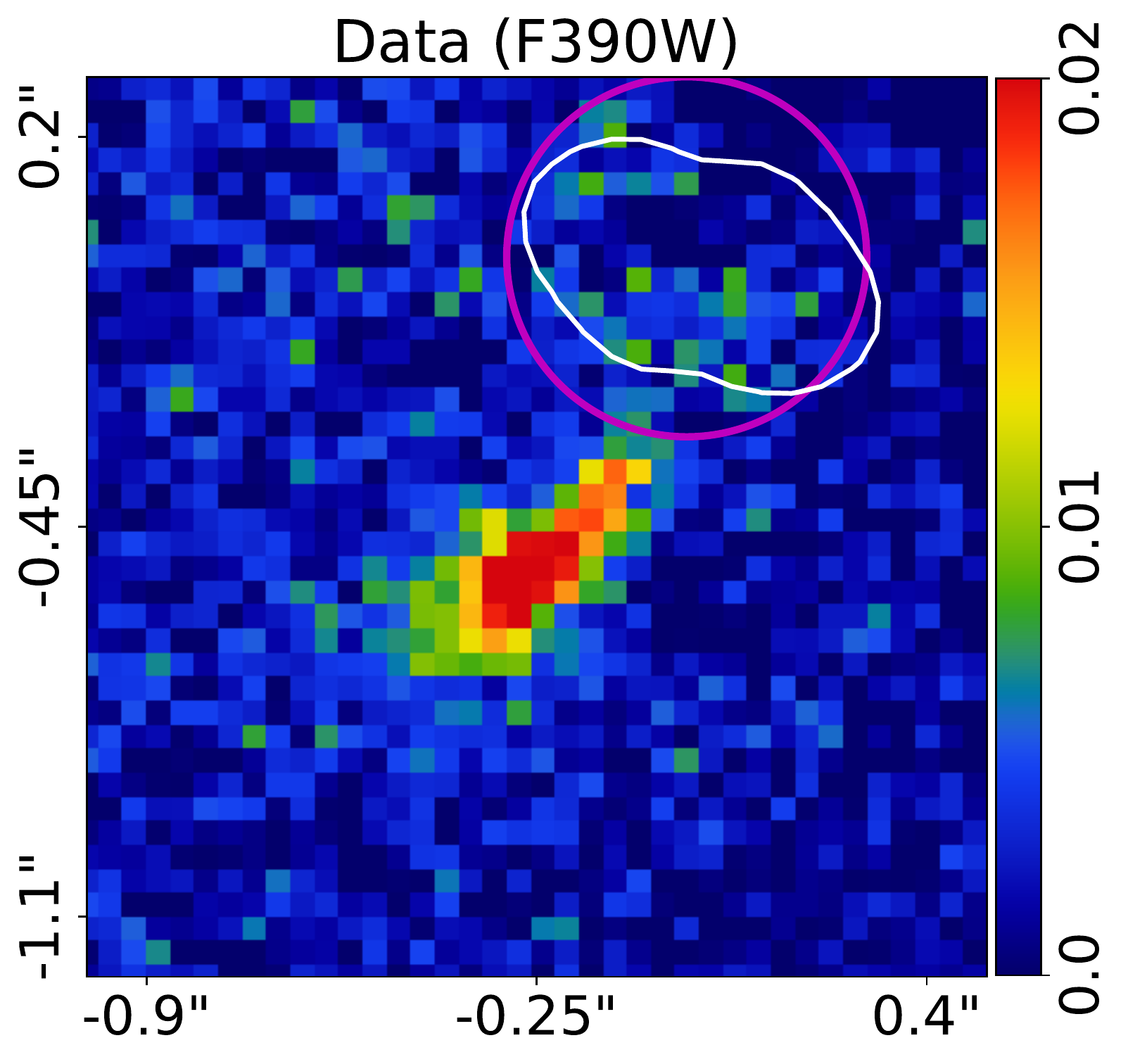}
\includegraphics[width=0.241\textwidth]{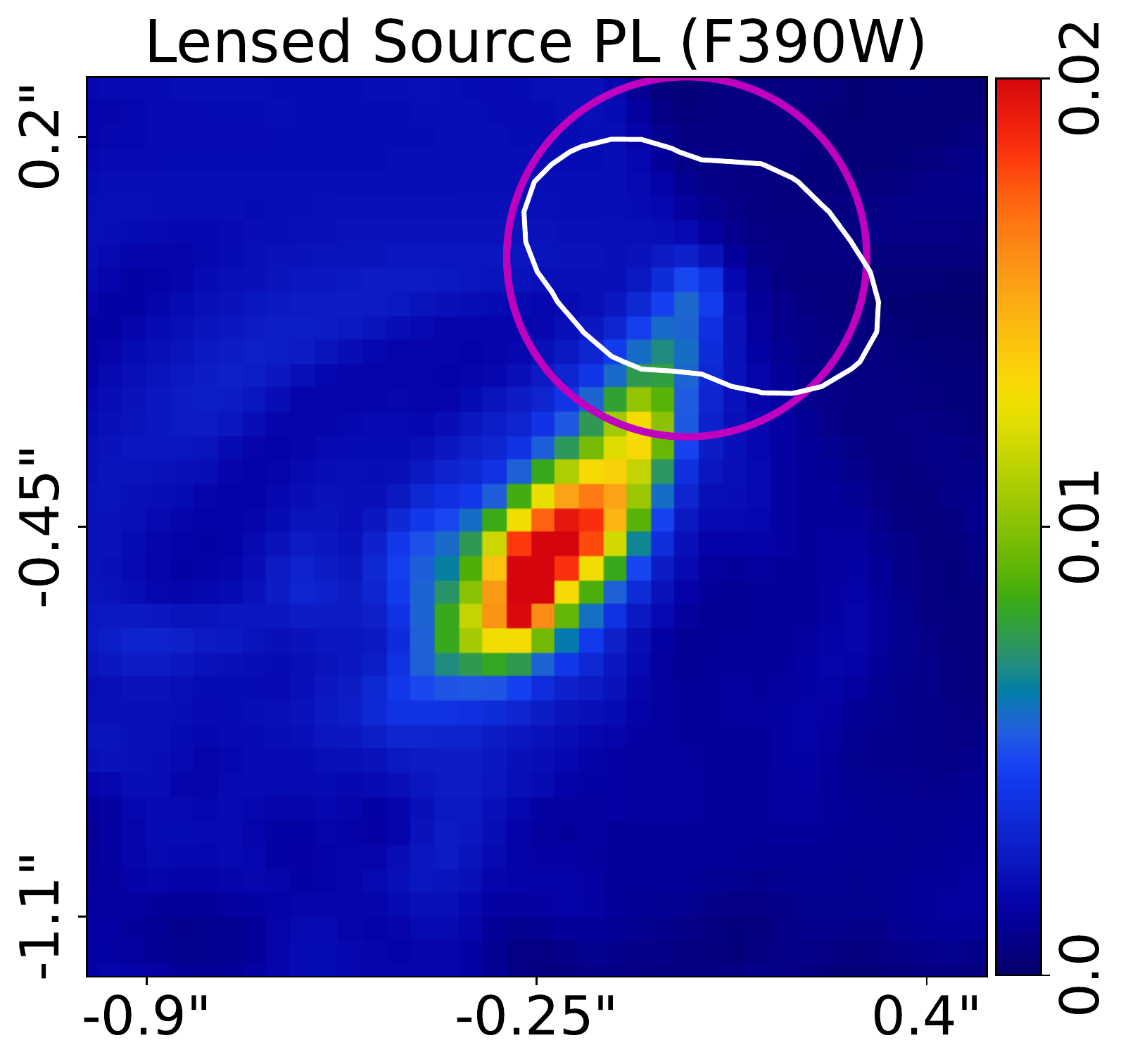}
\includegraphics[width=0.241\textwidth]{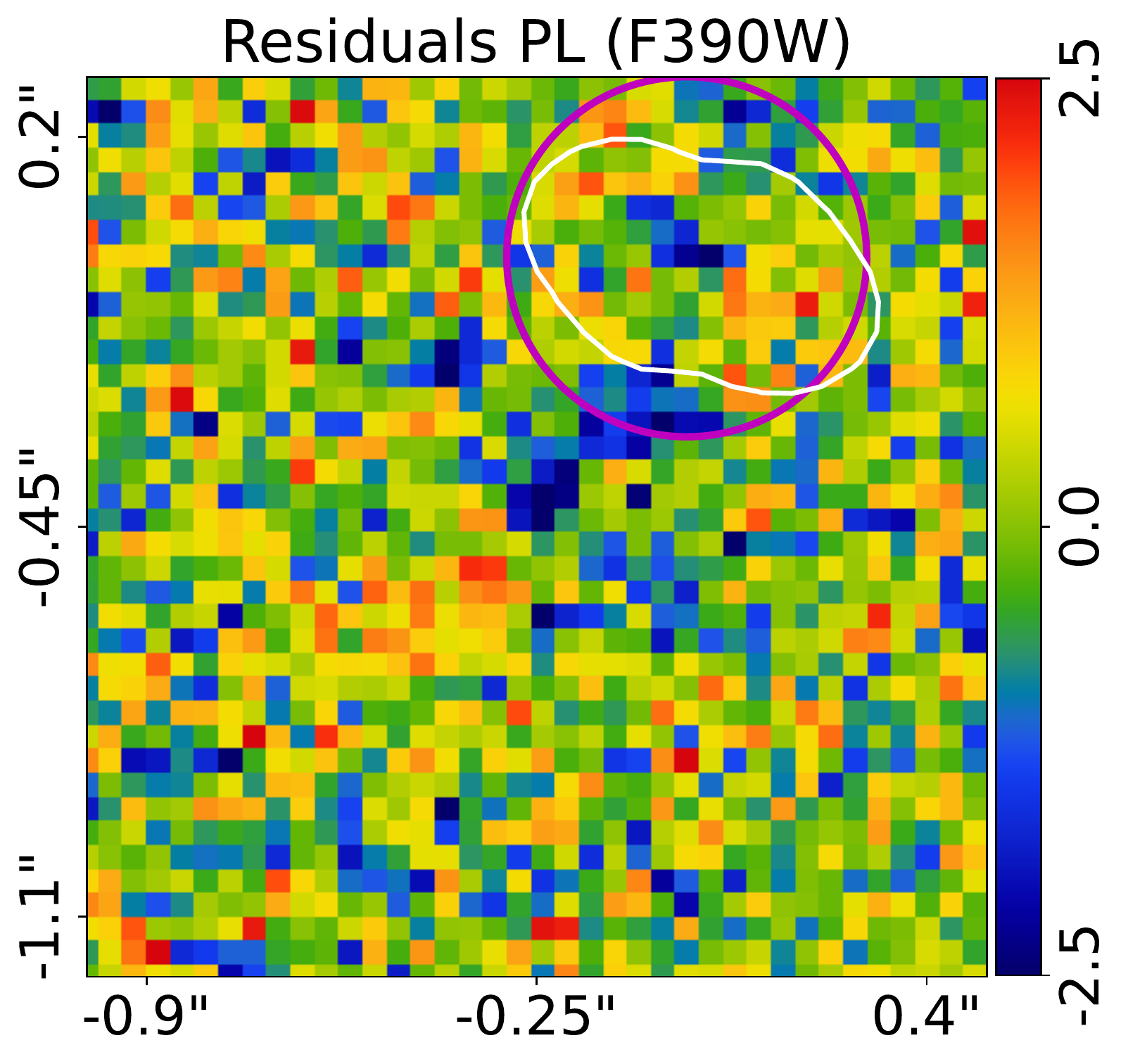}
\includegraphics[width=0.241\textwidth]{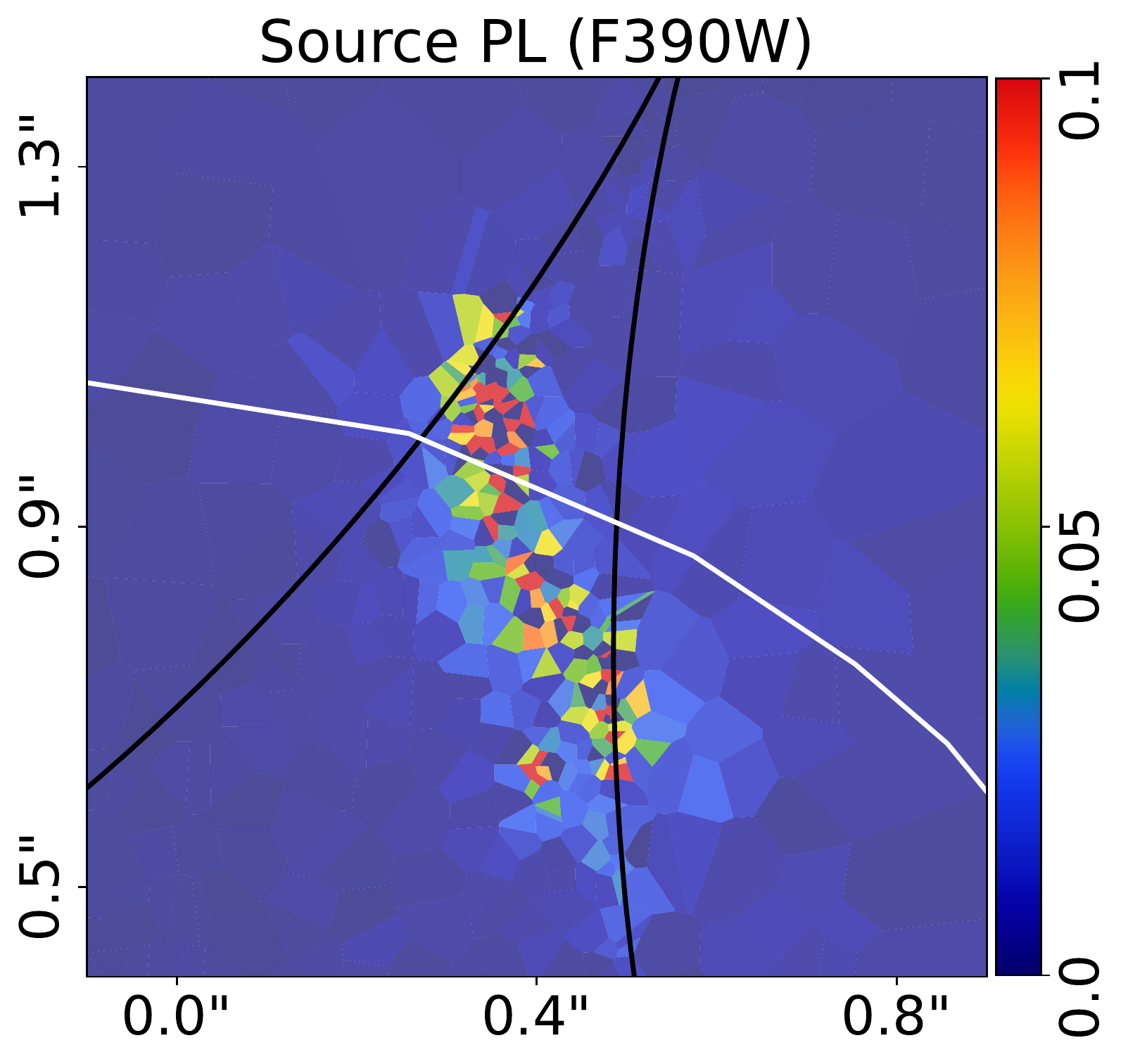}
\includegraphics[width=0.241\textwidth]{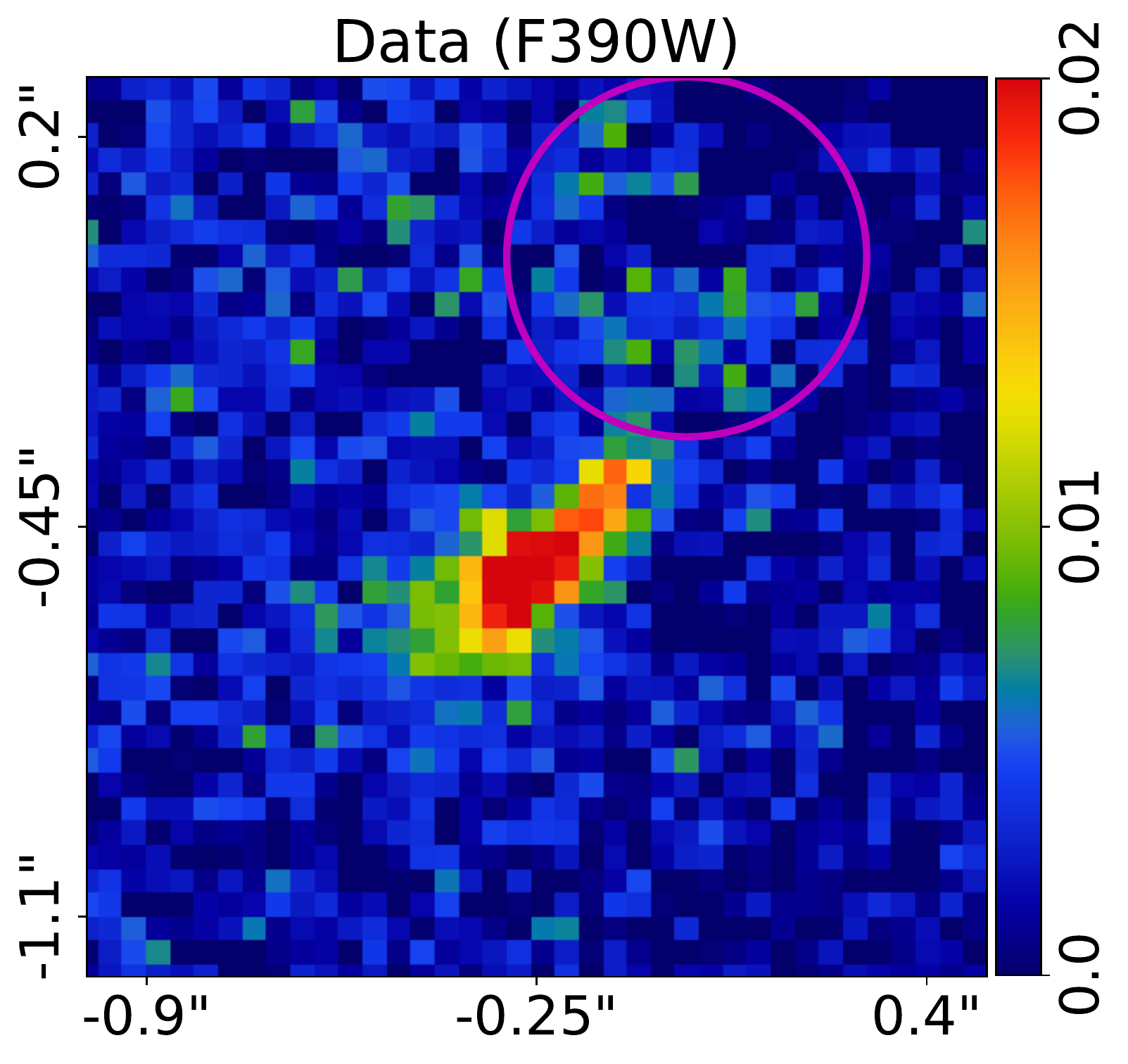}
\includegraphics[width=0.241\textwidth]{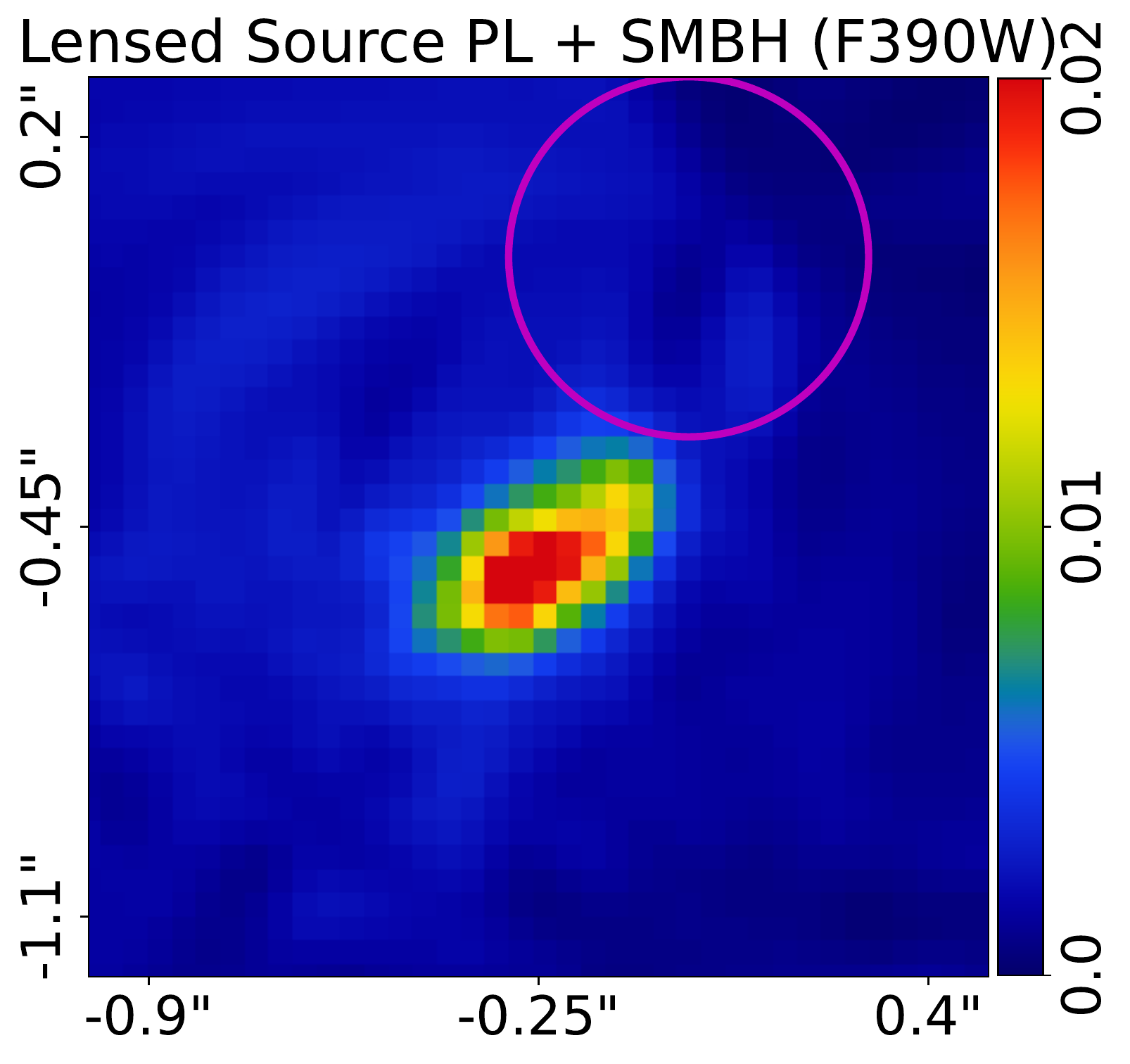}
\includegraphics[width=0.241\textwidth]{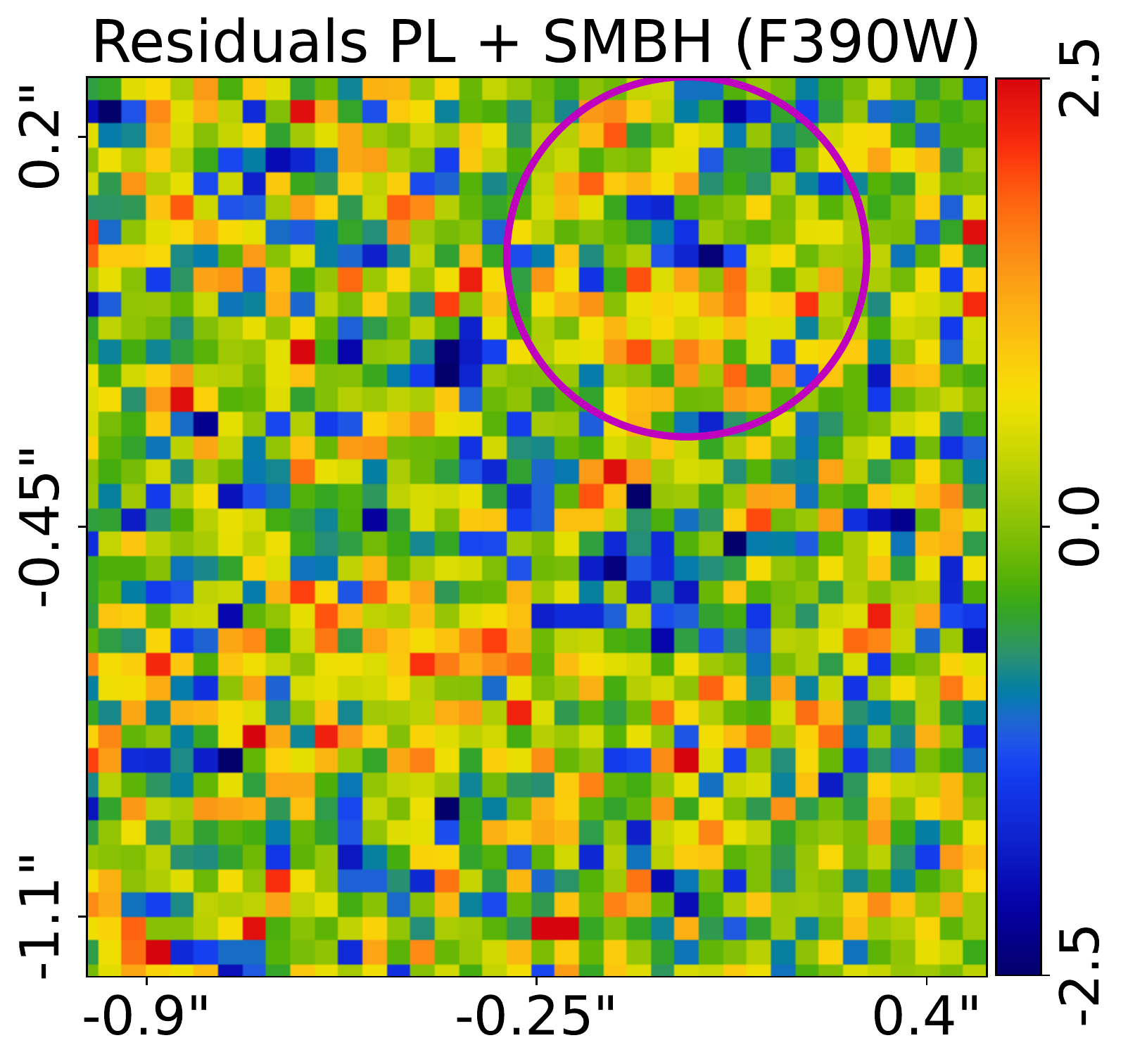}
\includegraphics[width=0.241\textwidth]{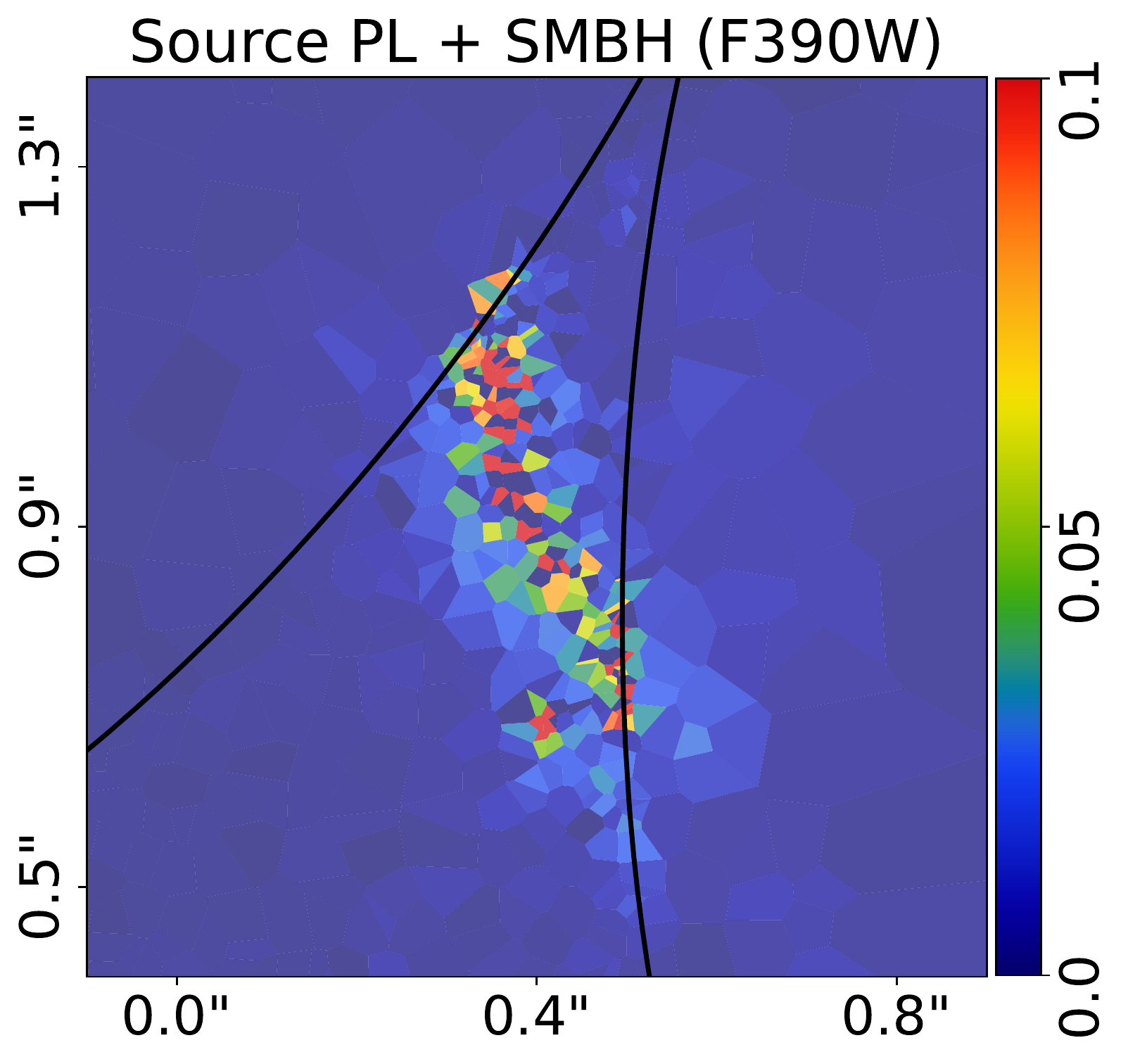}

\caption{
Zoom-ins of the observed counter image in the F390W data (left panel), the model lensed source (left-centre panel), the normalized residuals (right-centre panel) and the source reconstruction (right panel). The top and bottom rows show the power-law mass model without and with a SMBH respectively. All models include an external shear. Models which omit a SMBH form an additional clump of light in the counter image, which is not present in the data. The tangential caustic is shown by a black line and radial critical curve and caustic a white line; the latter does not form for models including a SMBH.
}
\label{figure:ModelsPLF390W}
\end{figure*}

\begin{figure*}
\centering
\includegraphics[width=0.241\textwidth]{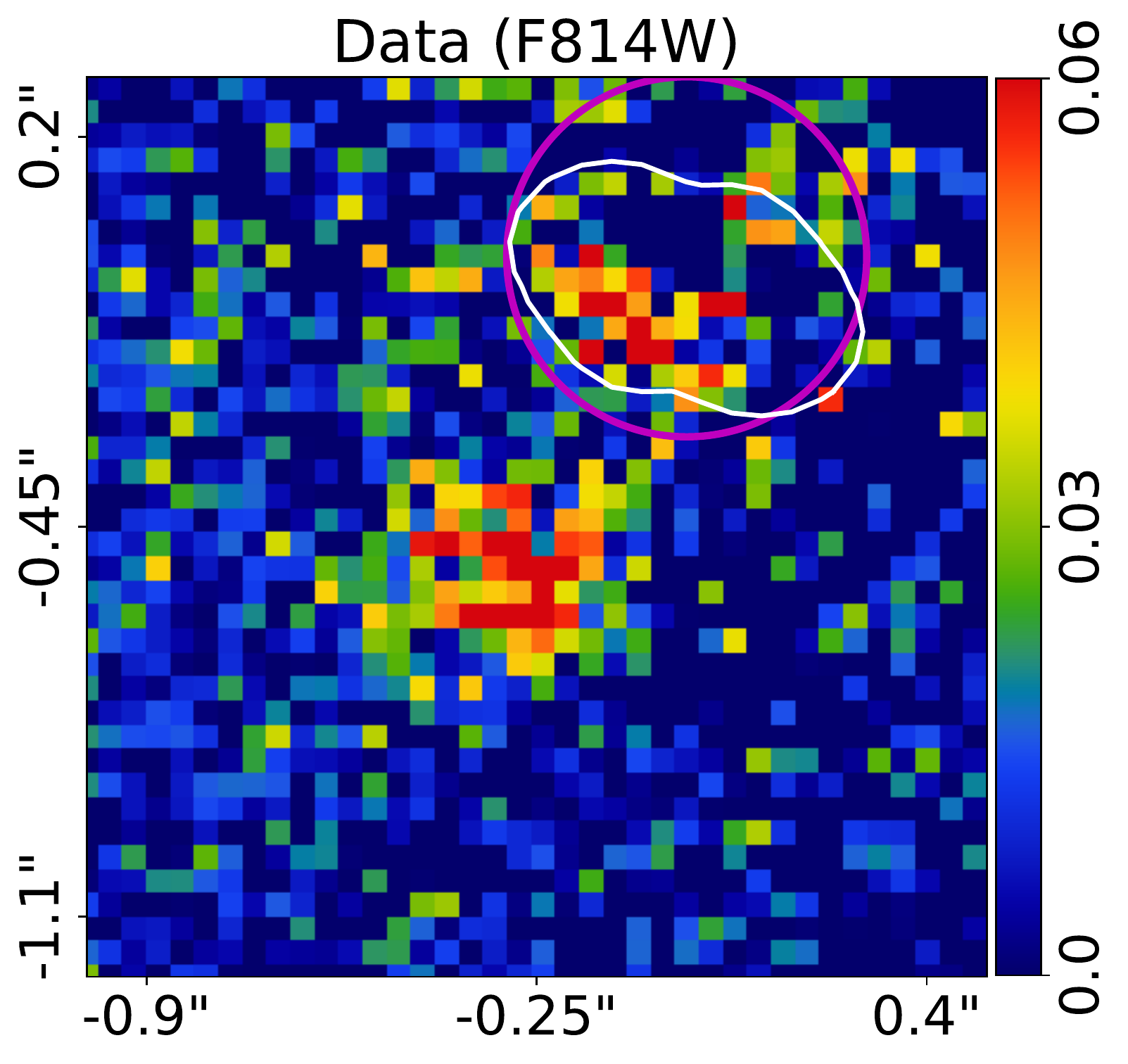}
\includegraphics[width=0.241\textwidth]{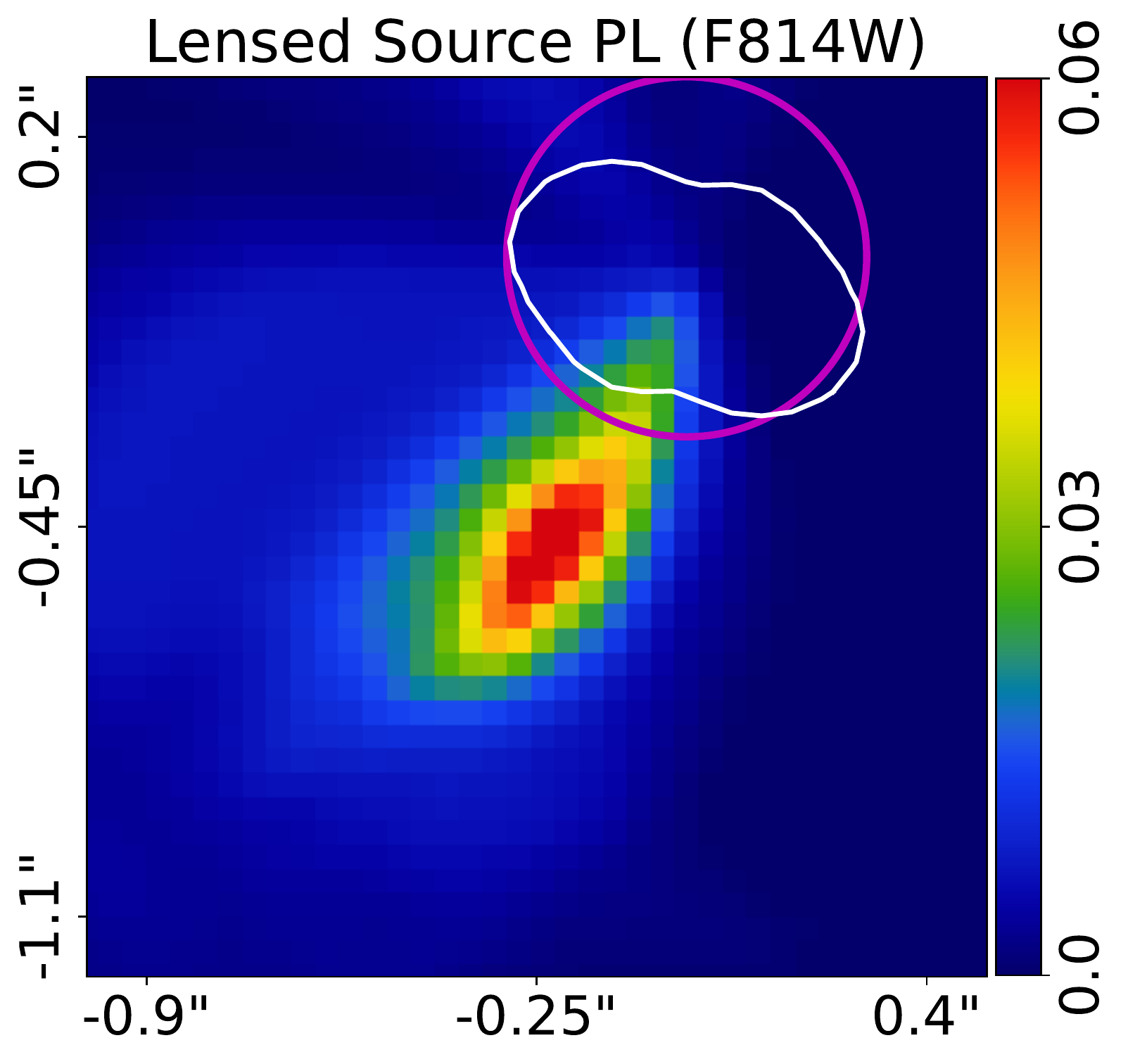}
\includegraphics[width=0.241\textwidth]{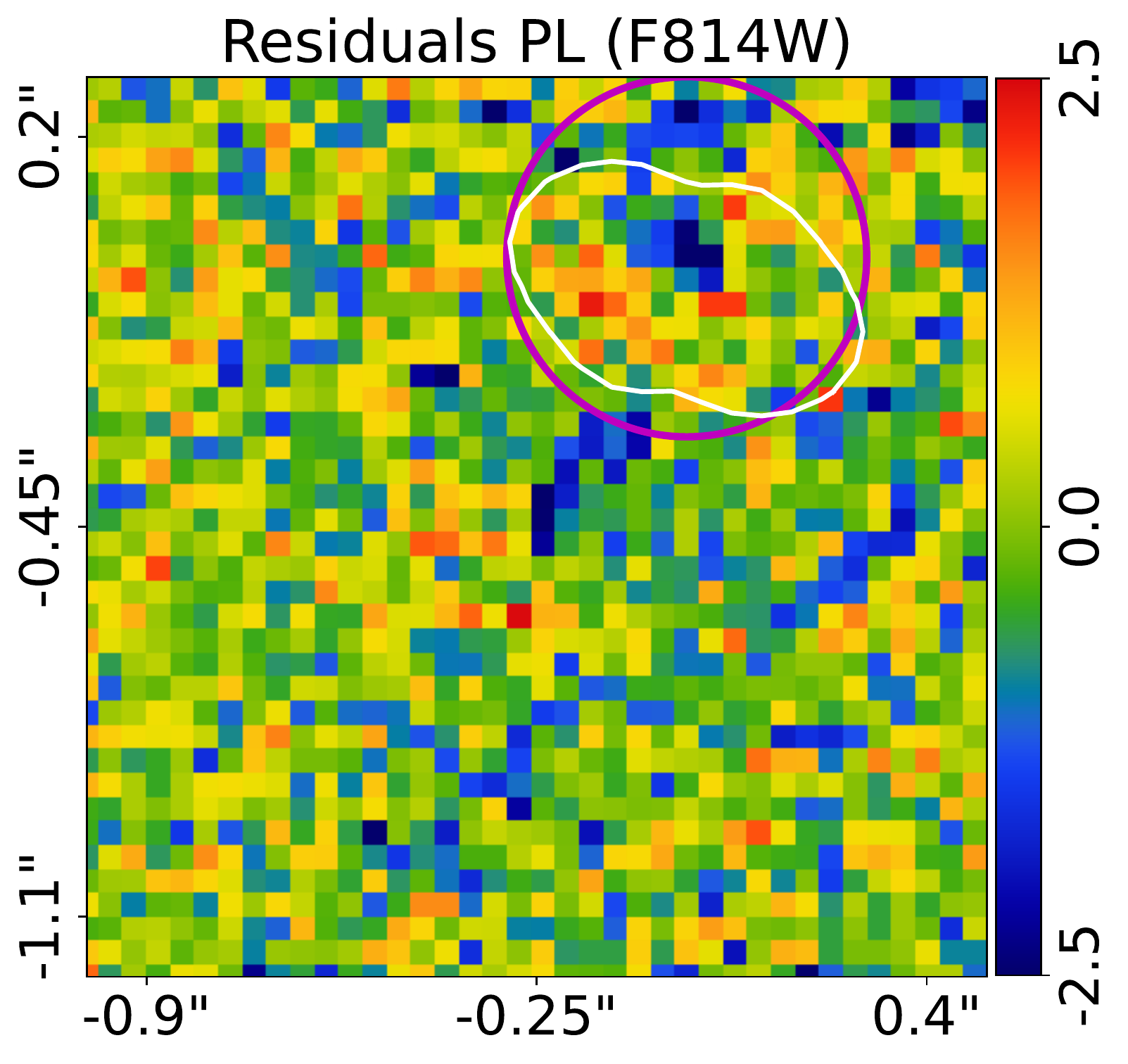}
\includegraphics[width=0.241\textwidth]{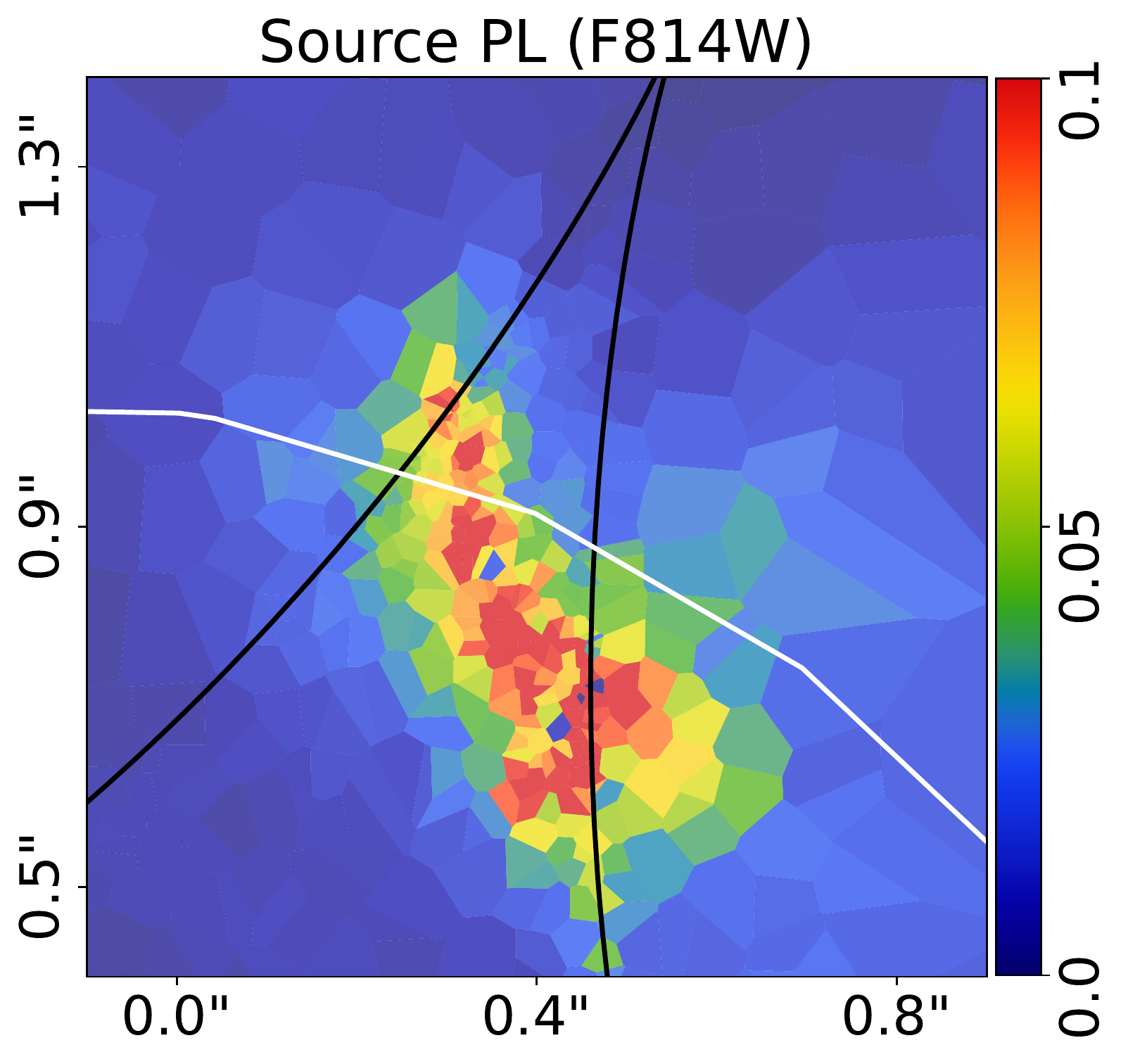}
\includegraphics[width=0.241\textwidth]{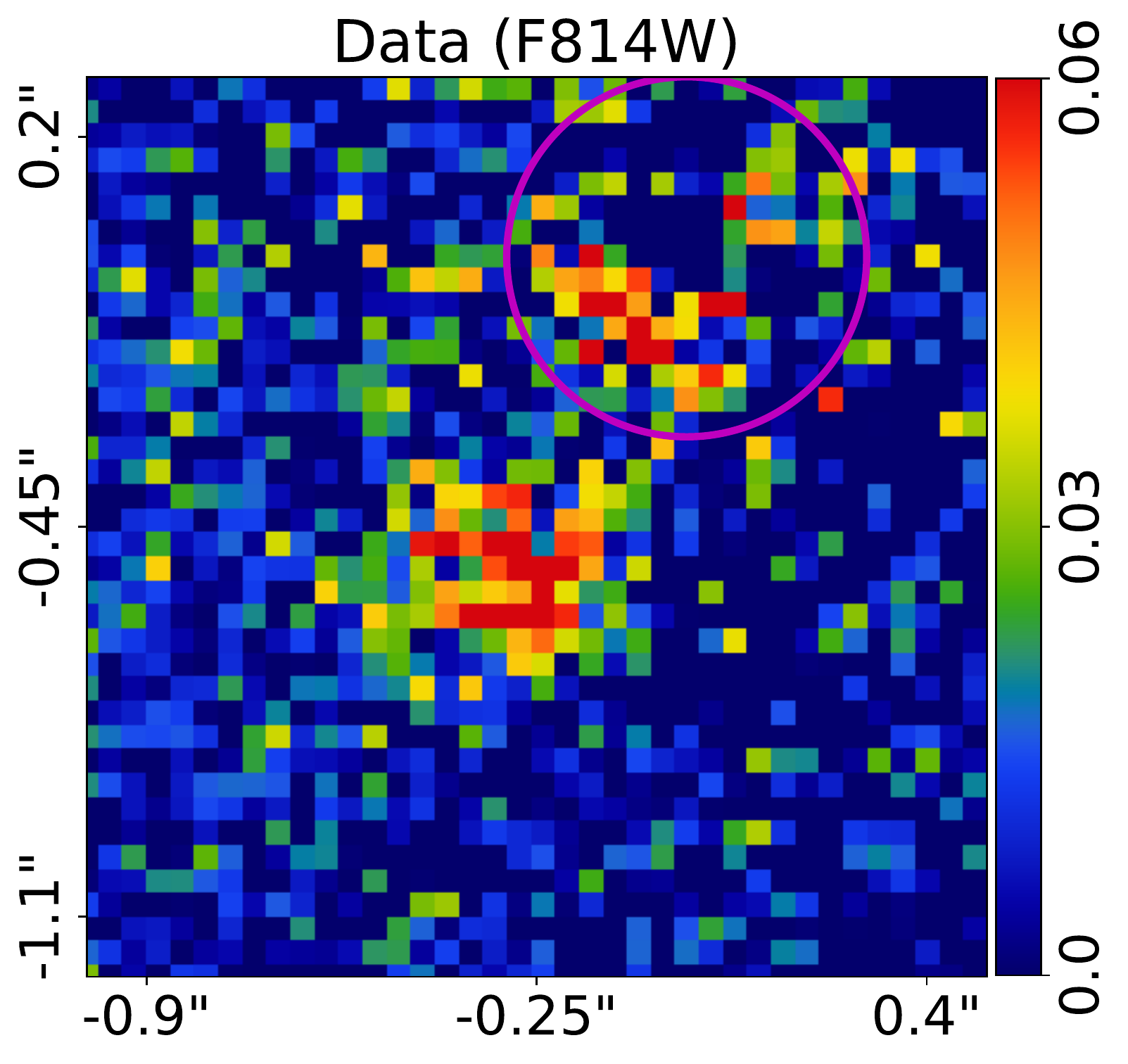}
\includegraphics[width=0.241\textwidth]{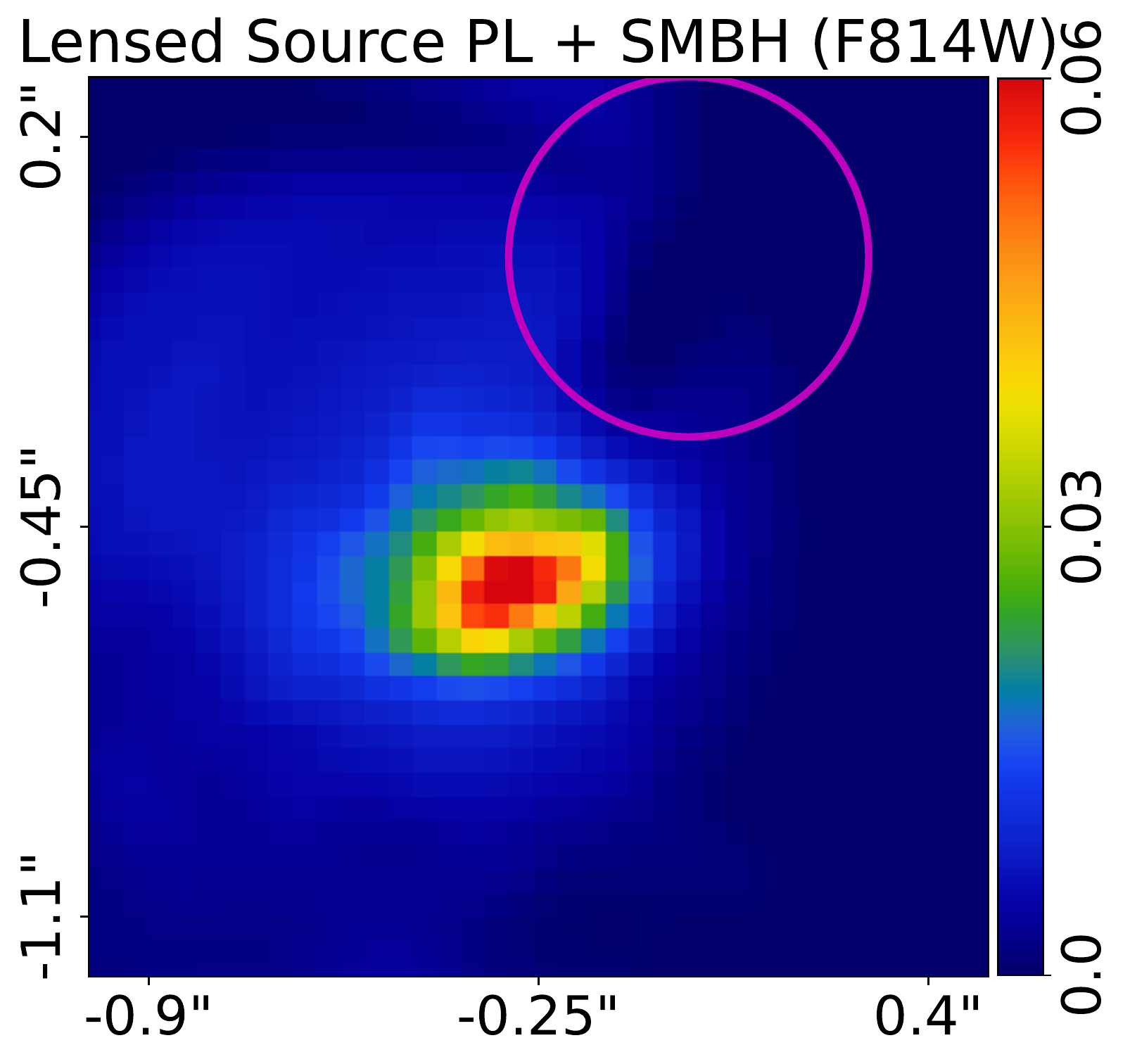}
\includegraphics[width=0.241\textwidth]{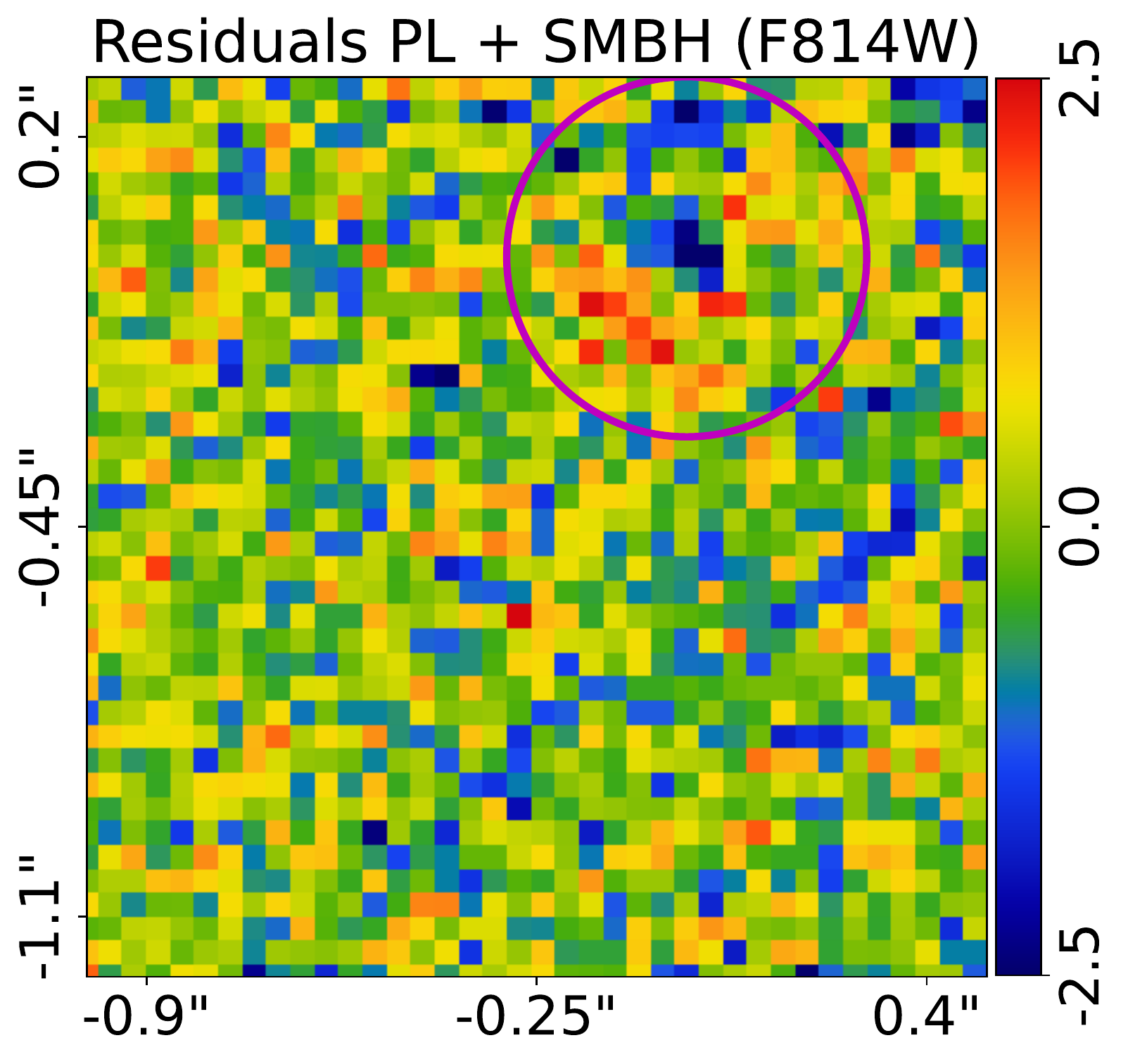}
\includegraphics[width=0.241\textwidth]{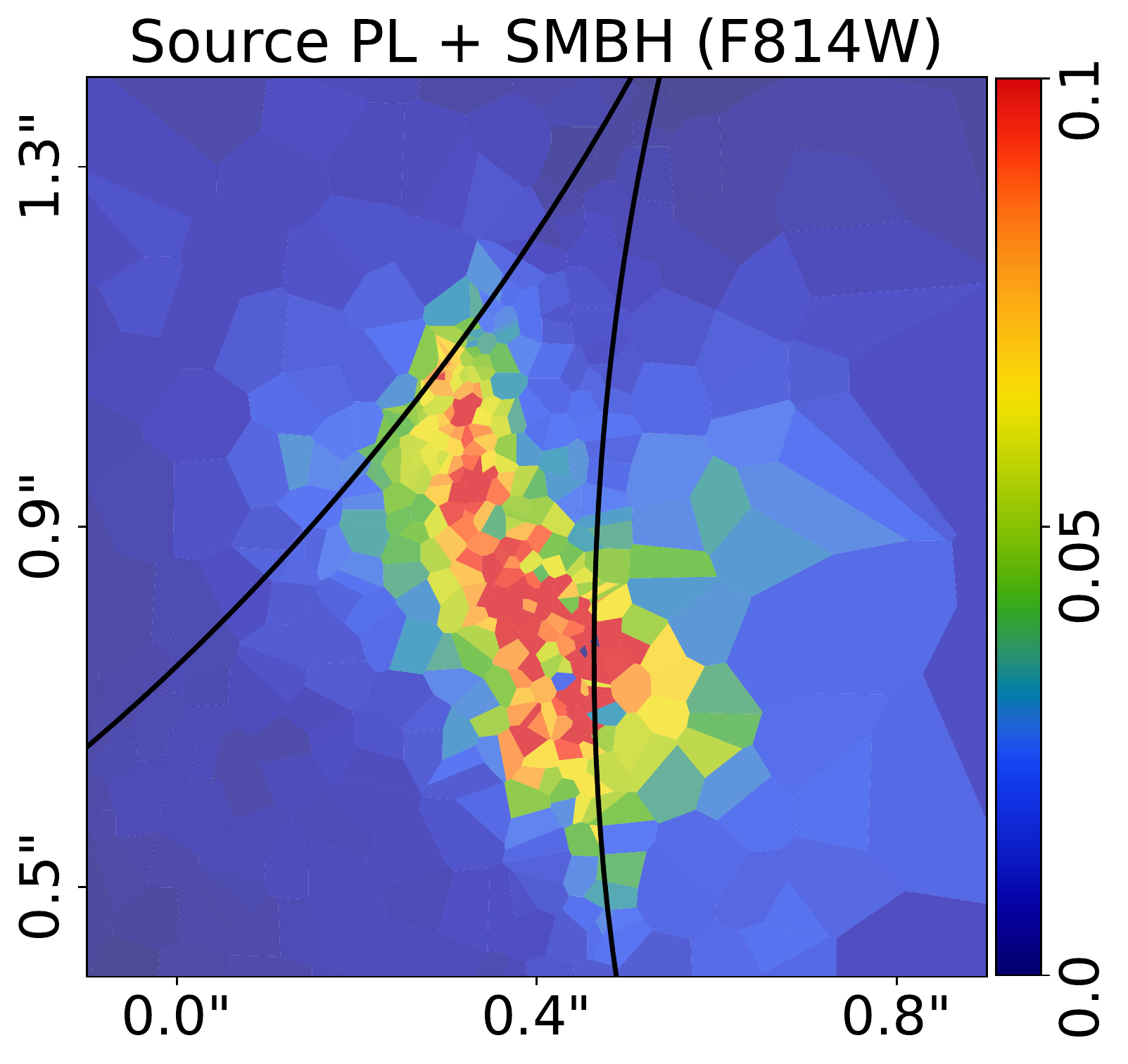}

\caption{
The same as \cref{figure:ModelsPLF390W} but for the F814W data.
}
\label{figure:ModelsPLF814W}
\end{figure*}

\begin{figure*}
\centering
\includegraphics[width=0.241\textwidth]{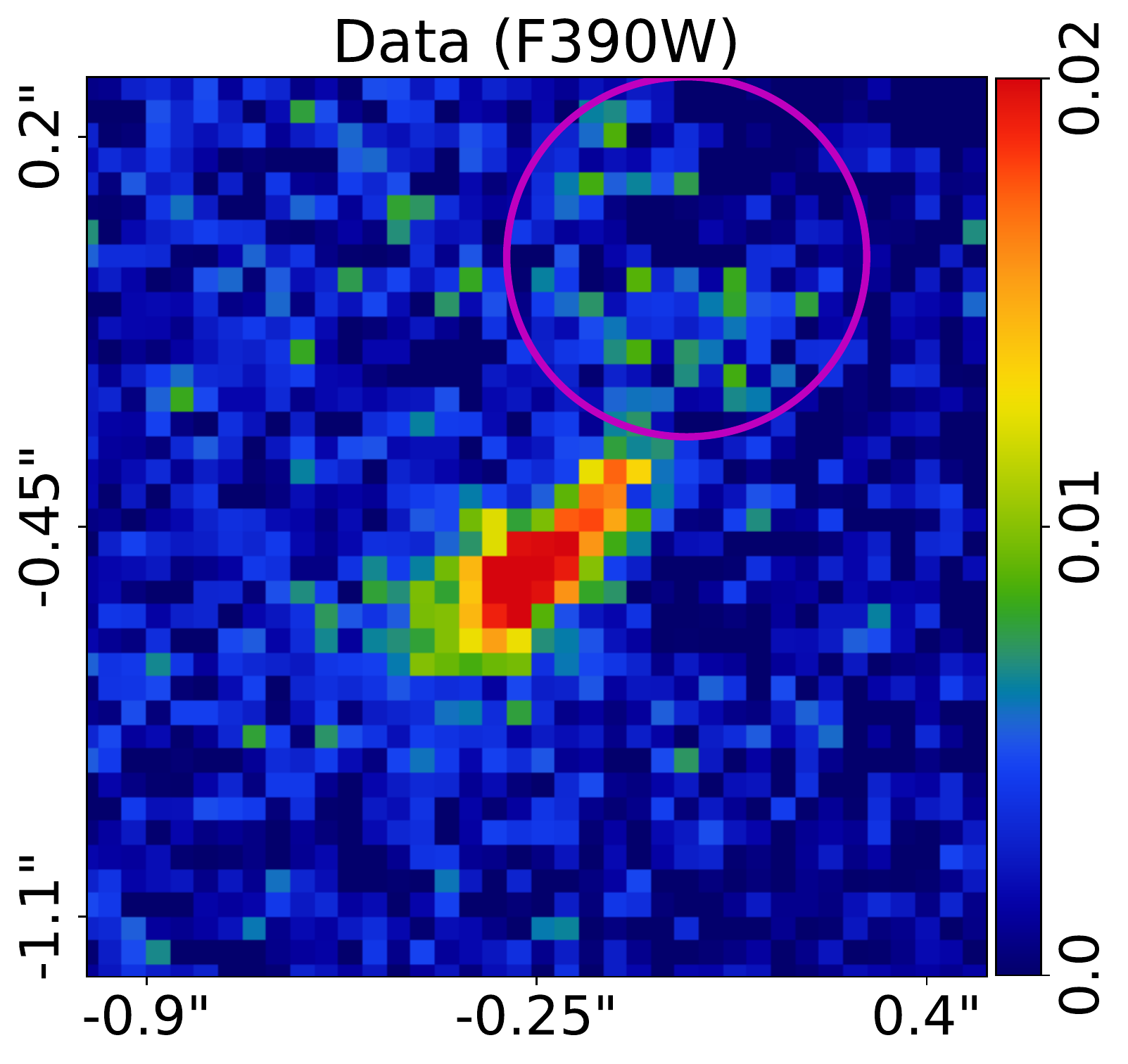}
\includegraphics[width=0.241\textwidth]{total_fit/f390w/bpl_image_zoomed.pdf}
\includegraphics[width=0.241\textwidth]{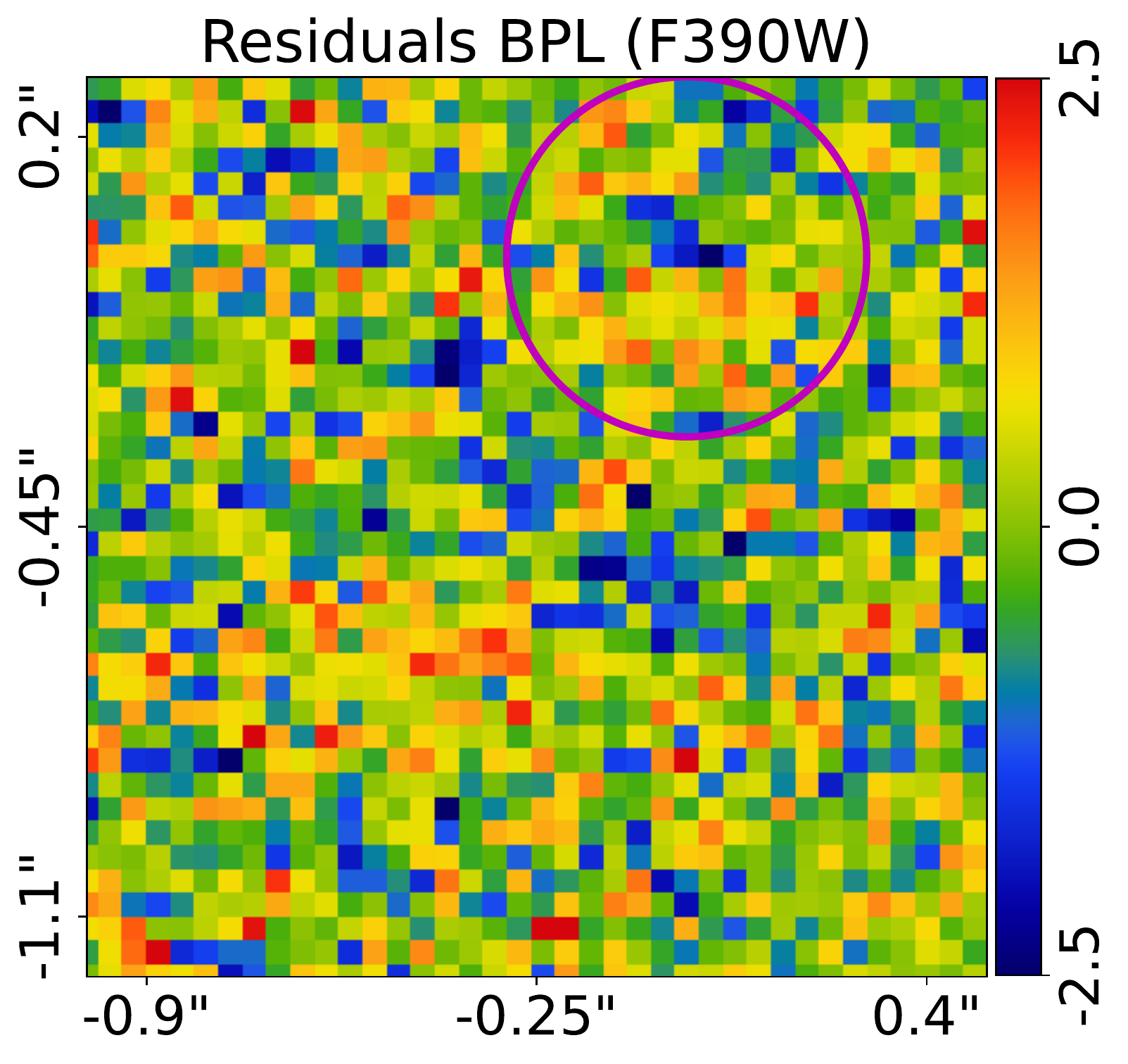}
\includegraphics[width=0.241\textwidth]{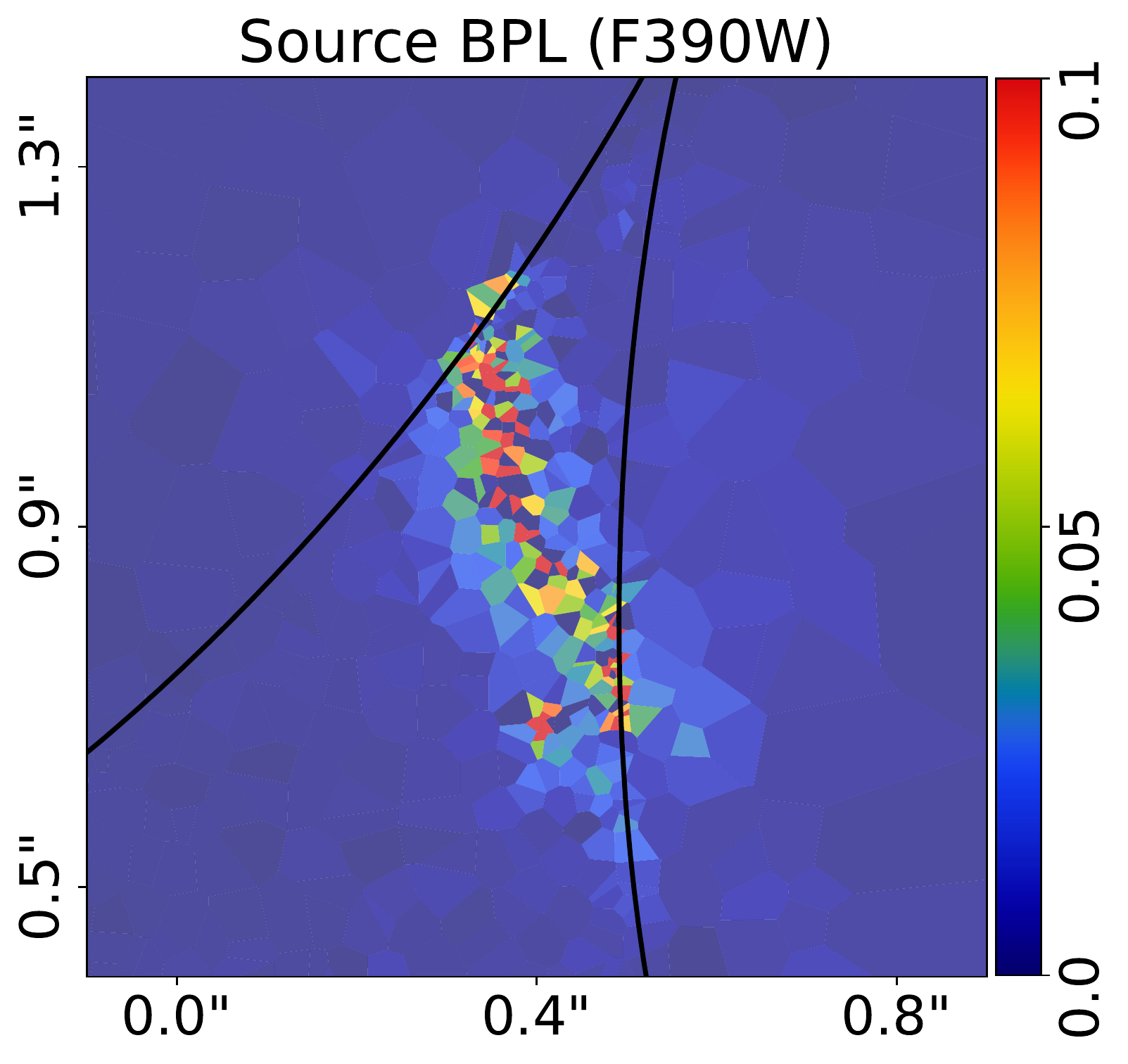}
\includegraphics[width=0.241\textwidth]{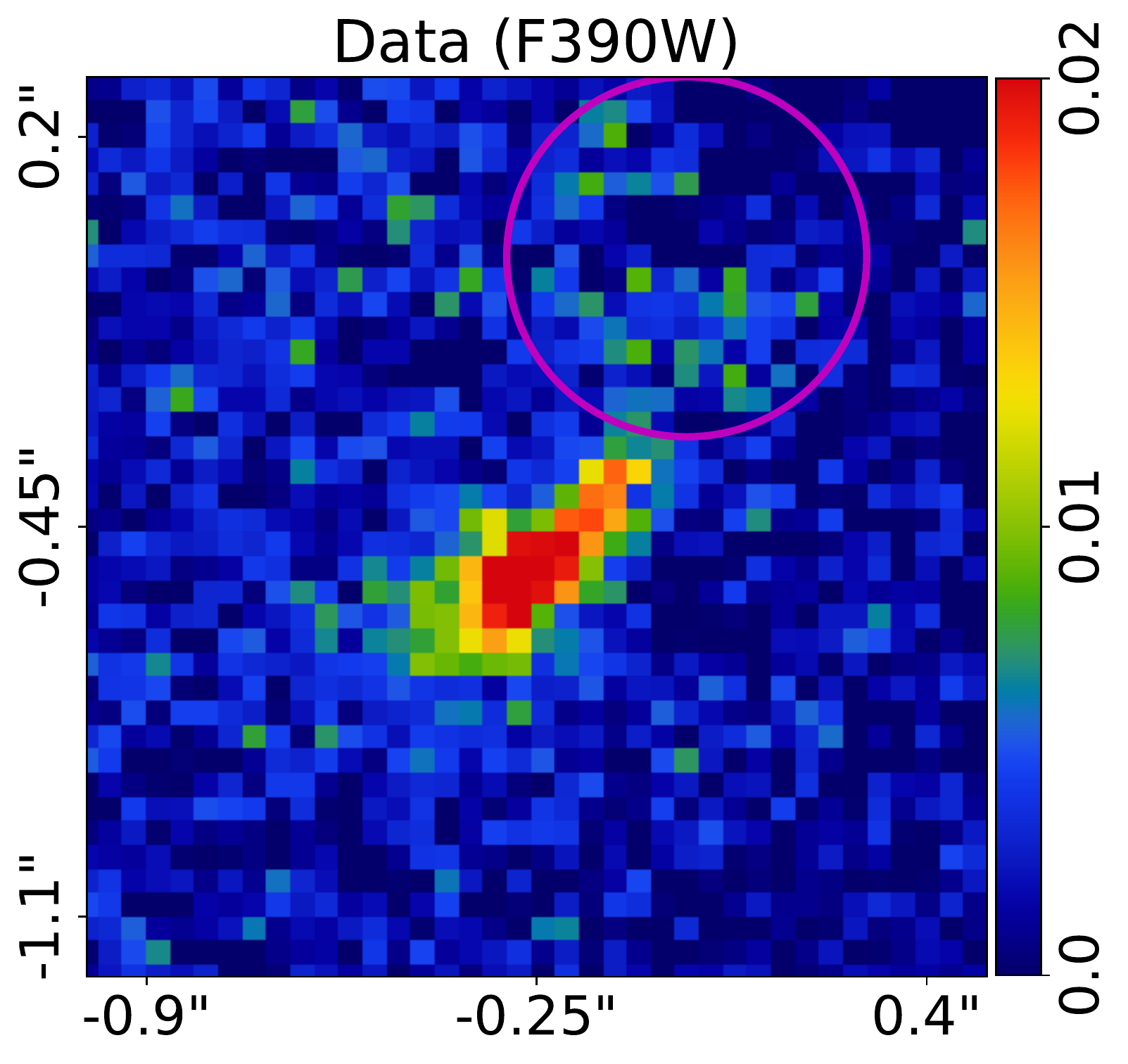}
\includegraphics[width=0.241\textwidth]{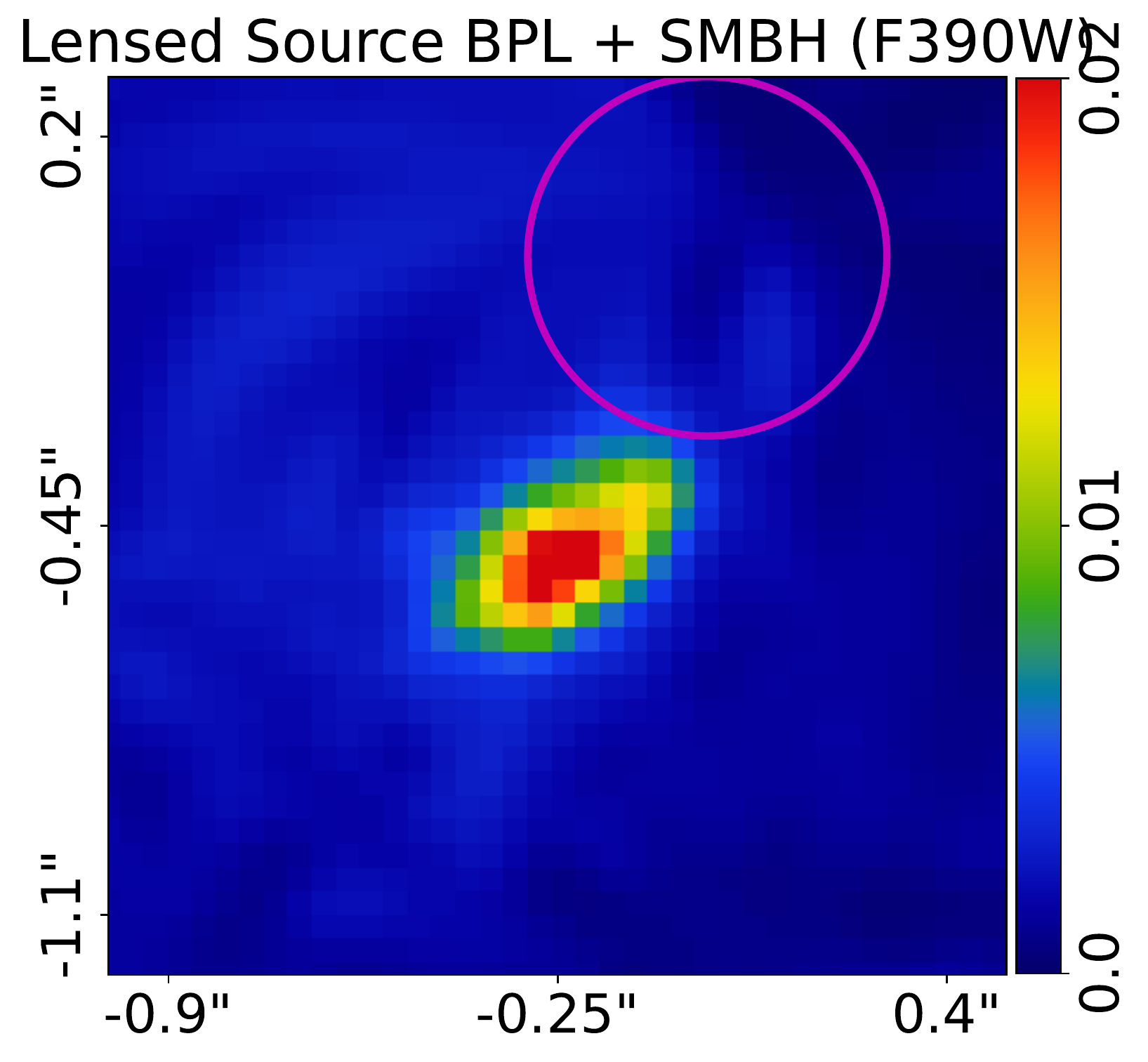}
\includegraphics[width=0.241\textwidth]{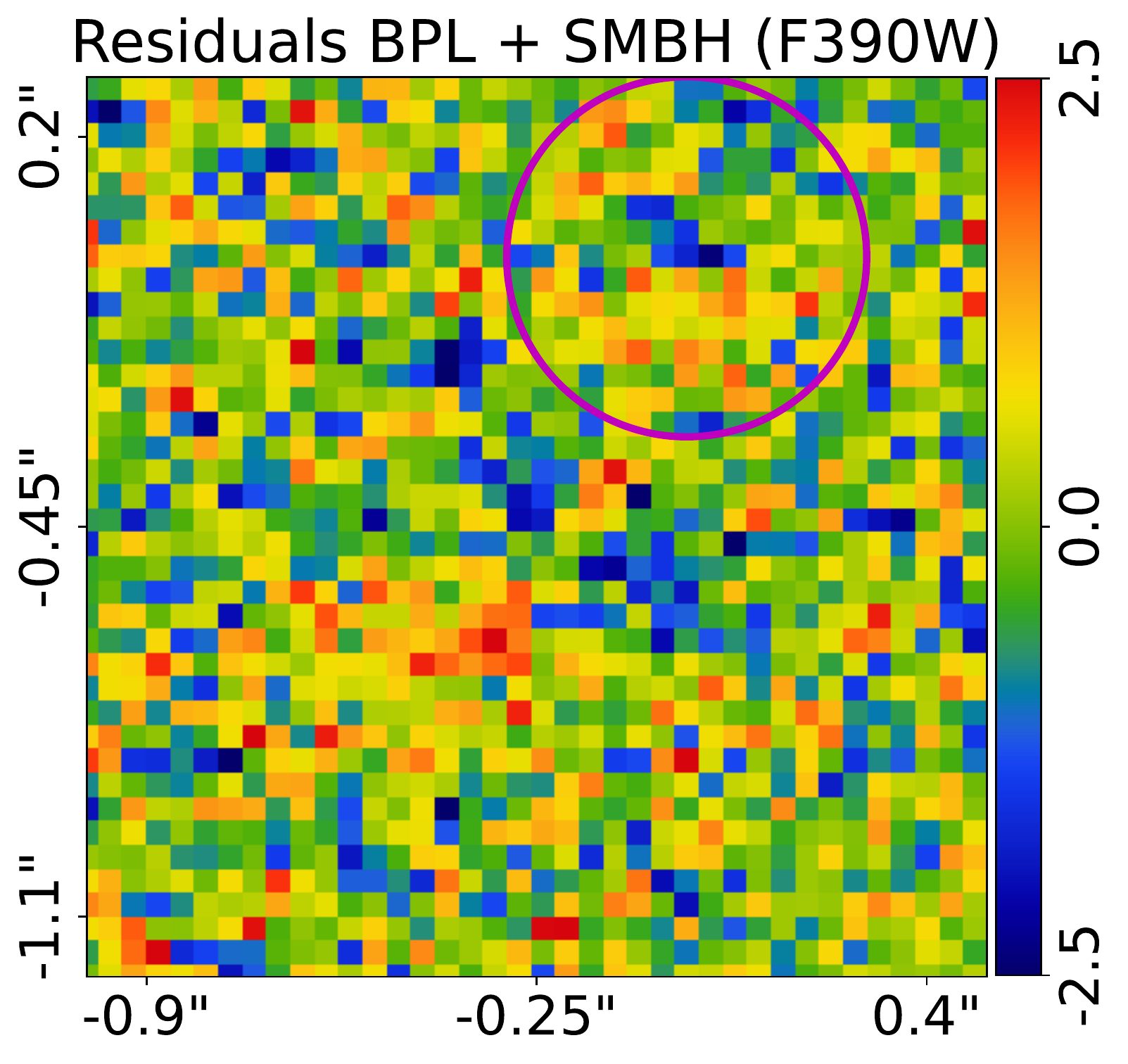}
\includegraphics[width=0.241\textwidth]{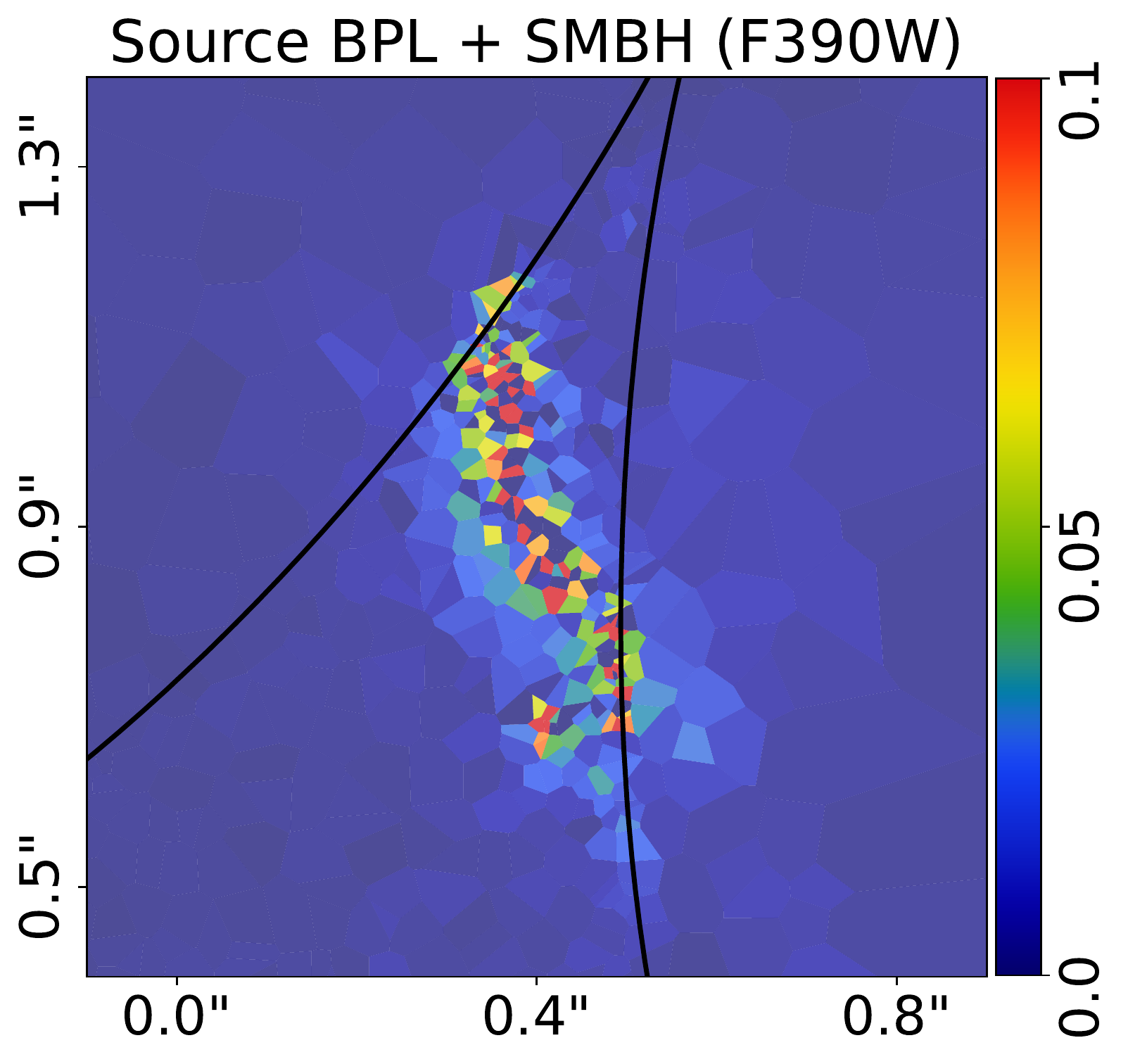}

\caption{
The same as \cref{figure:ModelsPLF390W} but for the broken power-law (BPL) model and F390W data.
}
\label{figure:ModelsBPLF390W}
\end{figure*}

\begin{figure*}
\centering
\includegraphics[width=0.241\textwidth]{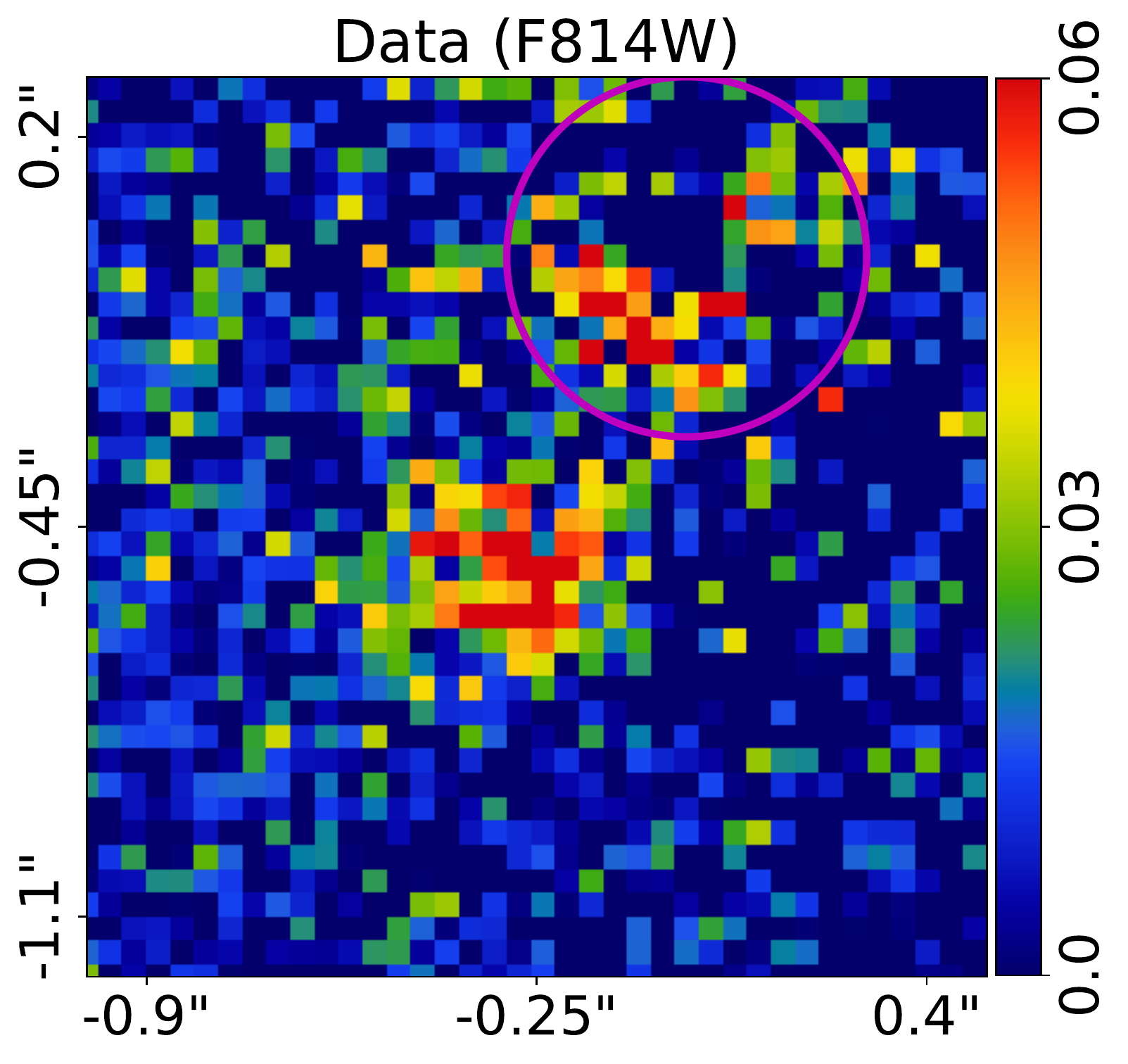}
\includegraphics[width=0.241\textwidth]{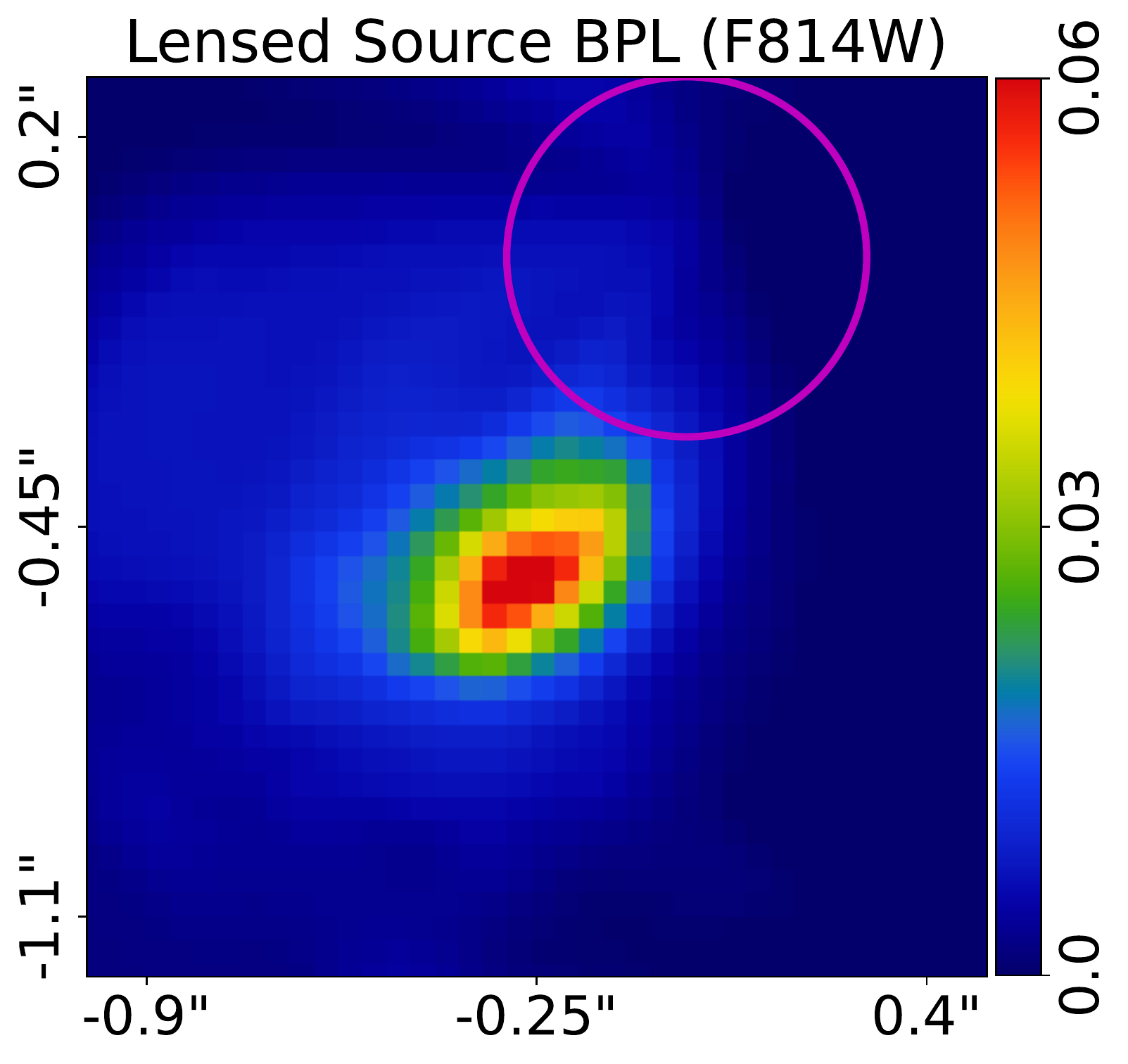}
\includegraphics[width=0.241\textwidth]{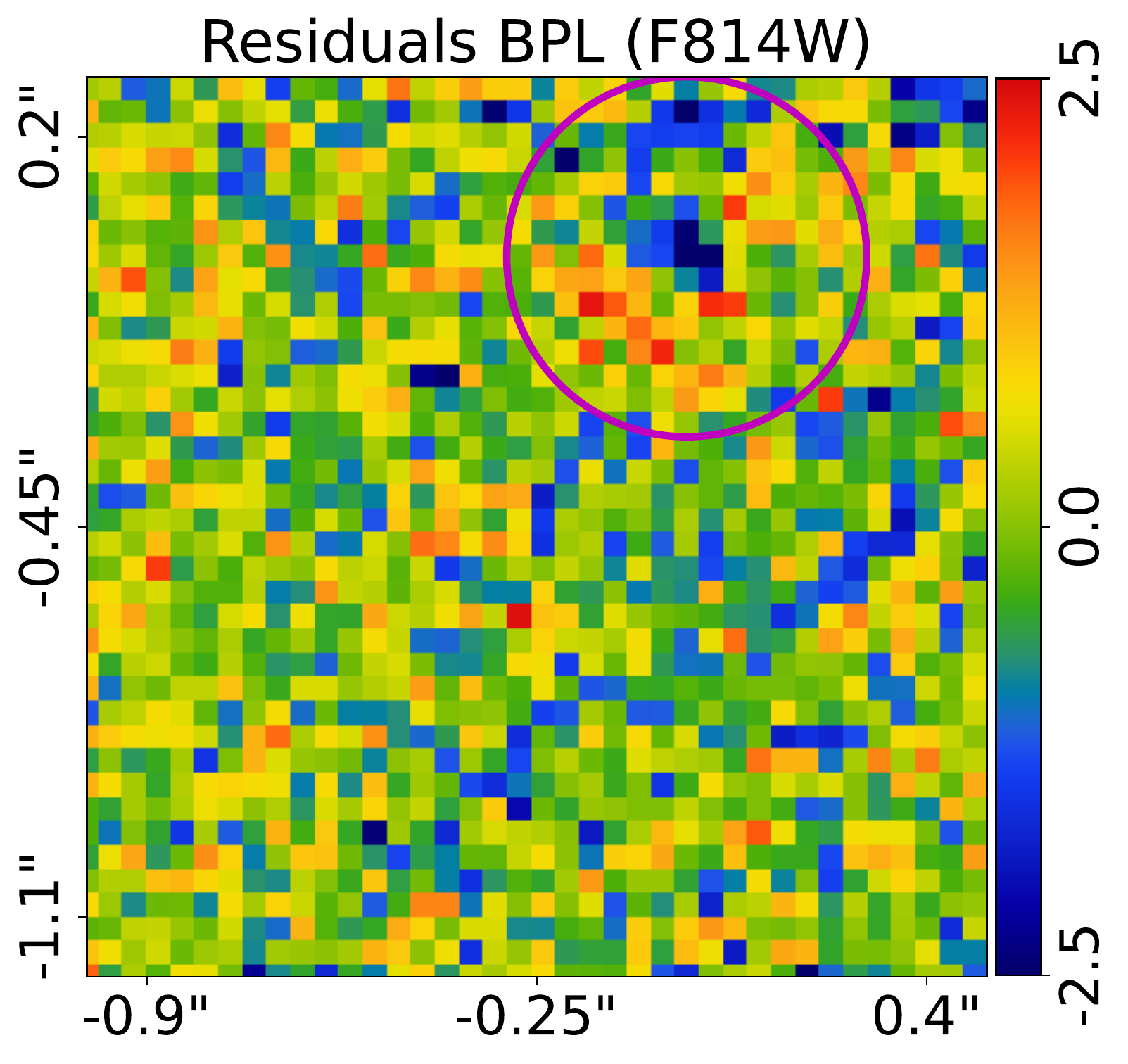}
\includegraphics[width=0.241\textwidth]{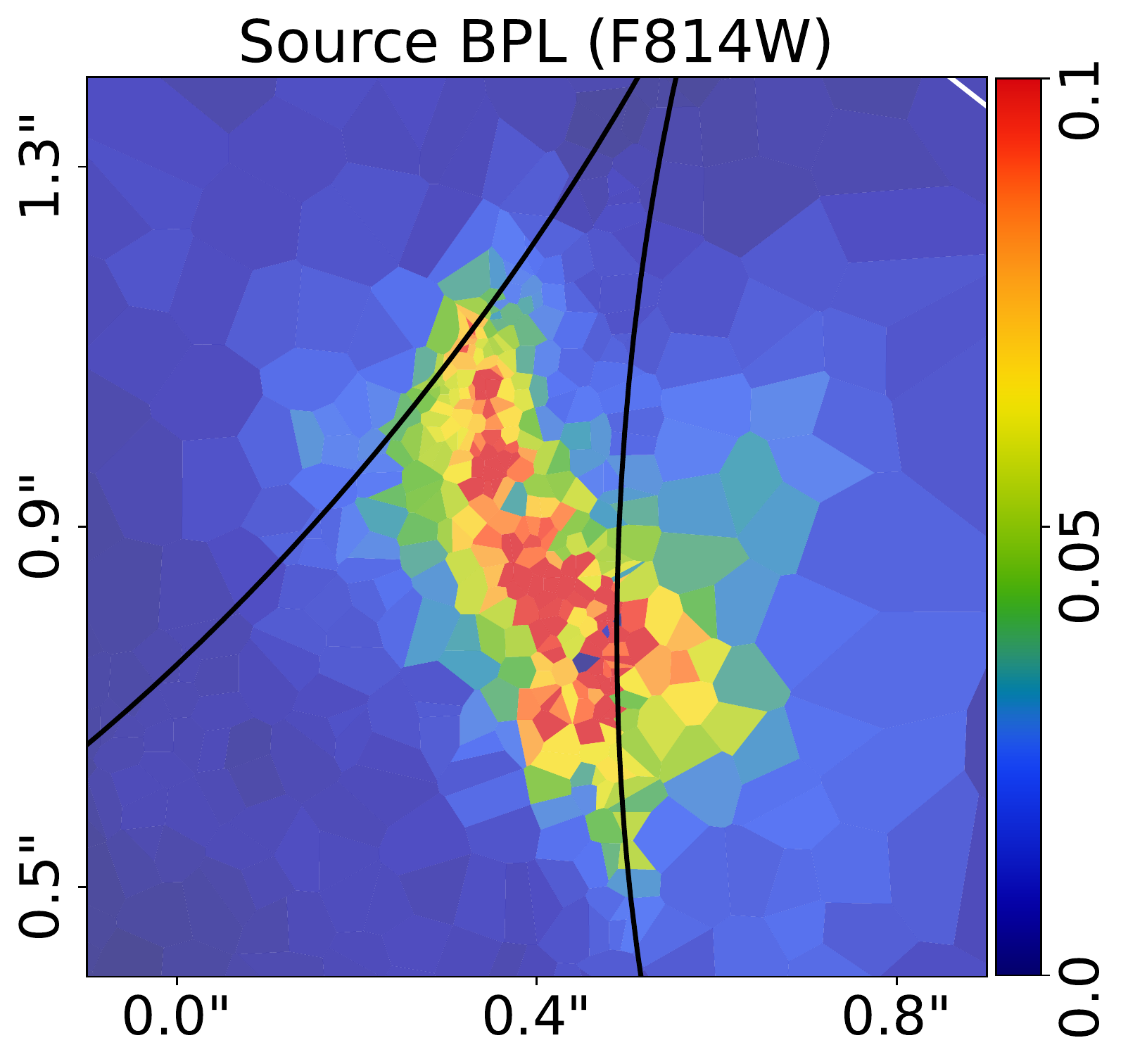}
\includegraphics[width=0.241\textwidth]{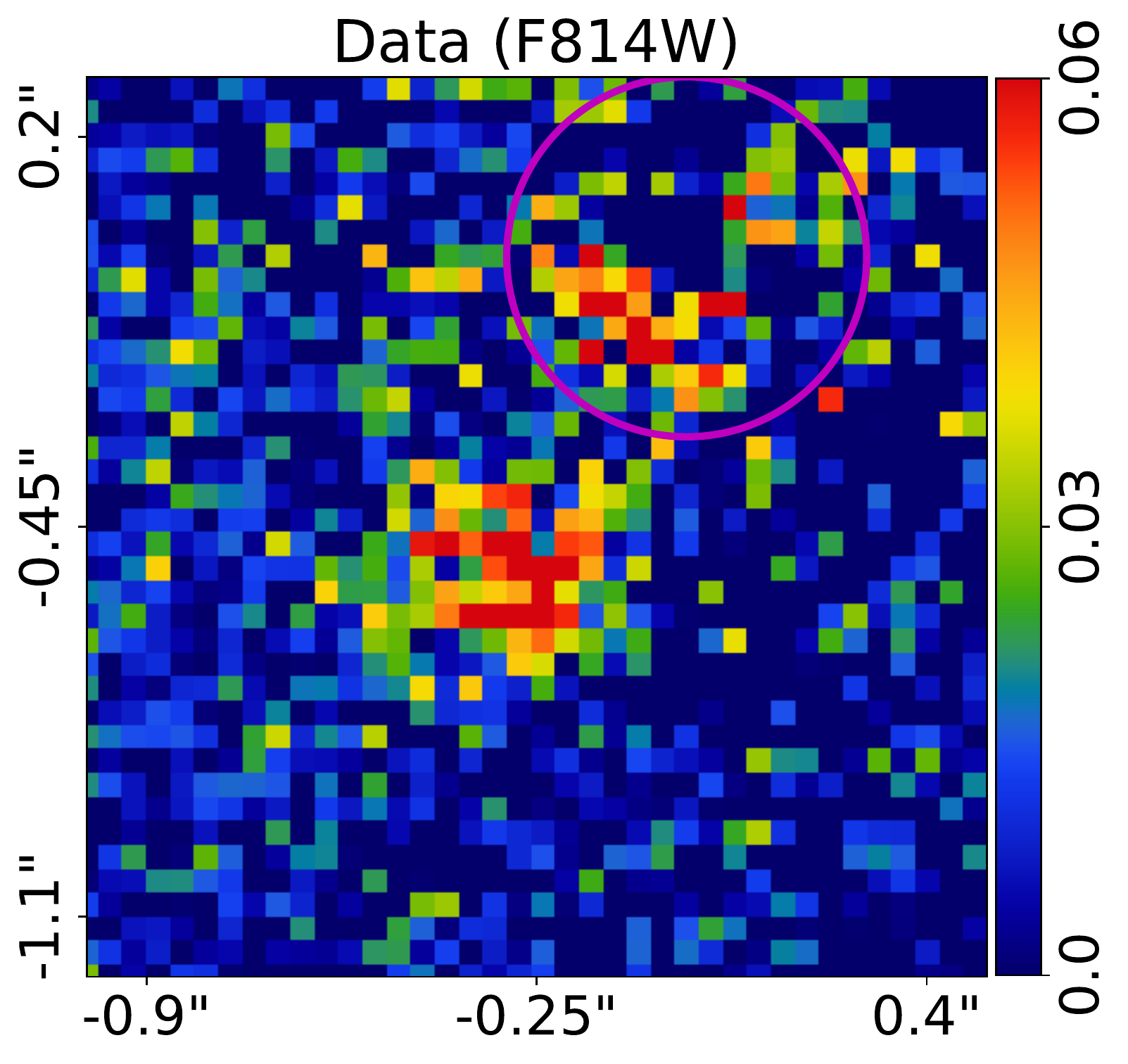}
\includegraphics[width=0.241\textwidth]{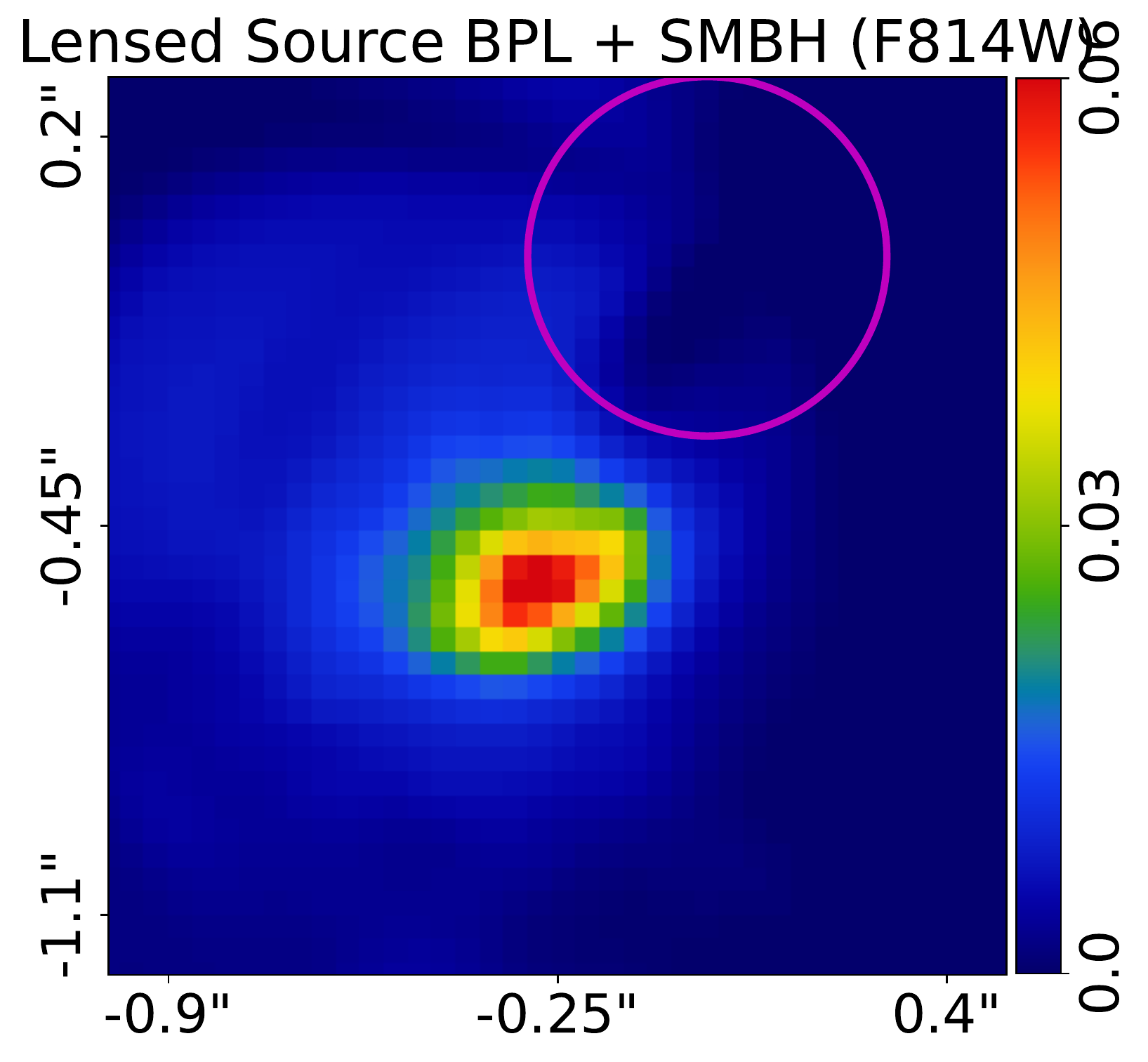}
\includegraphics[width=0.241\textwidth]{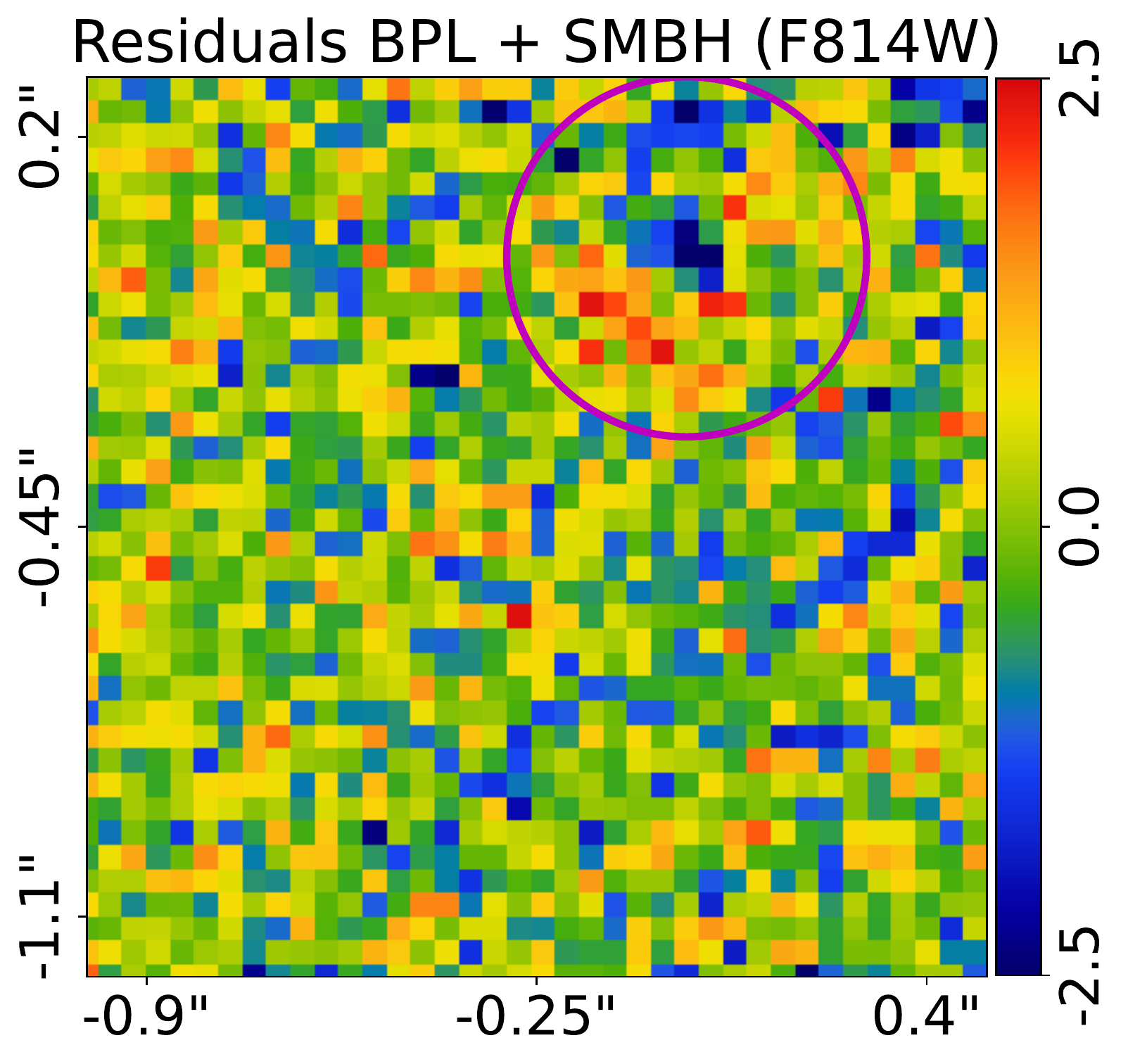}
\includegraphics[width=0.241\textwidth]{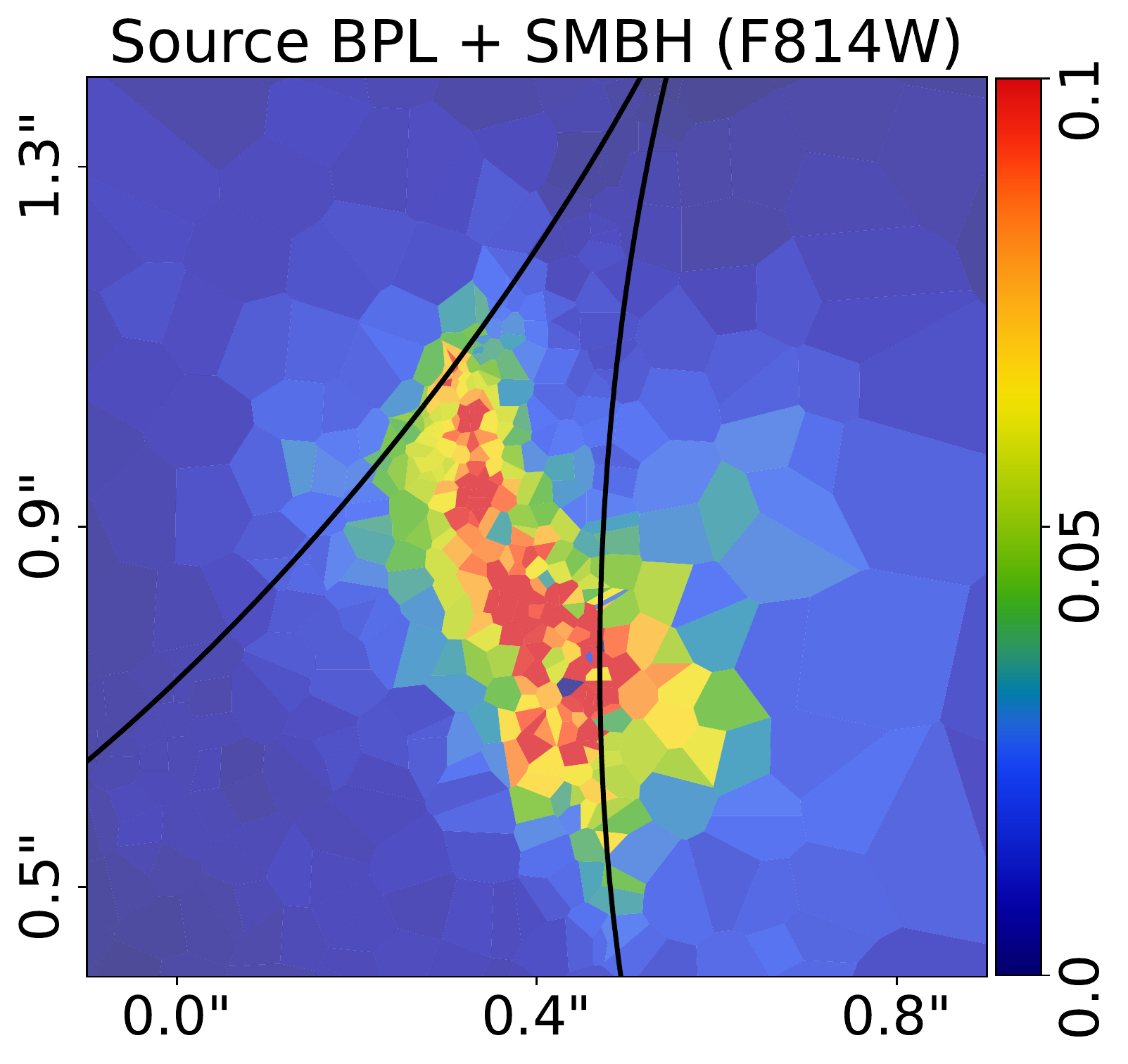}

\caption{
The same as \cref{figure:ModelsPLF390W} but for the broken power-law (BPL) model and F814W data.
}
\label{figure:ModelsBPLF814W}
\end{figure*}

\begin{figure}
\centering
\includegraphics[width=0.47\textwidth]{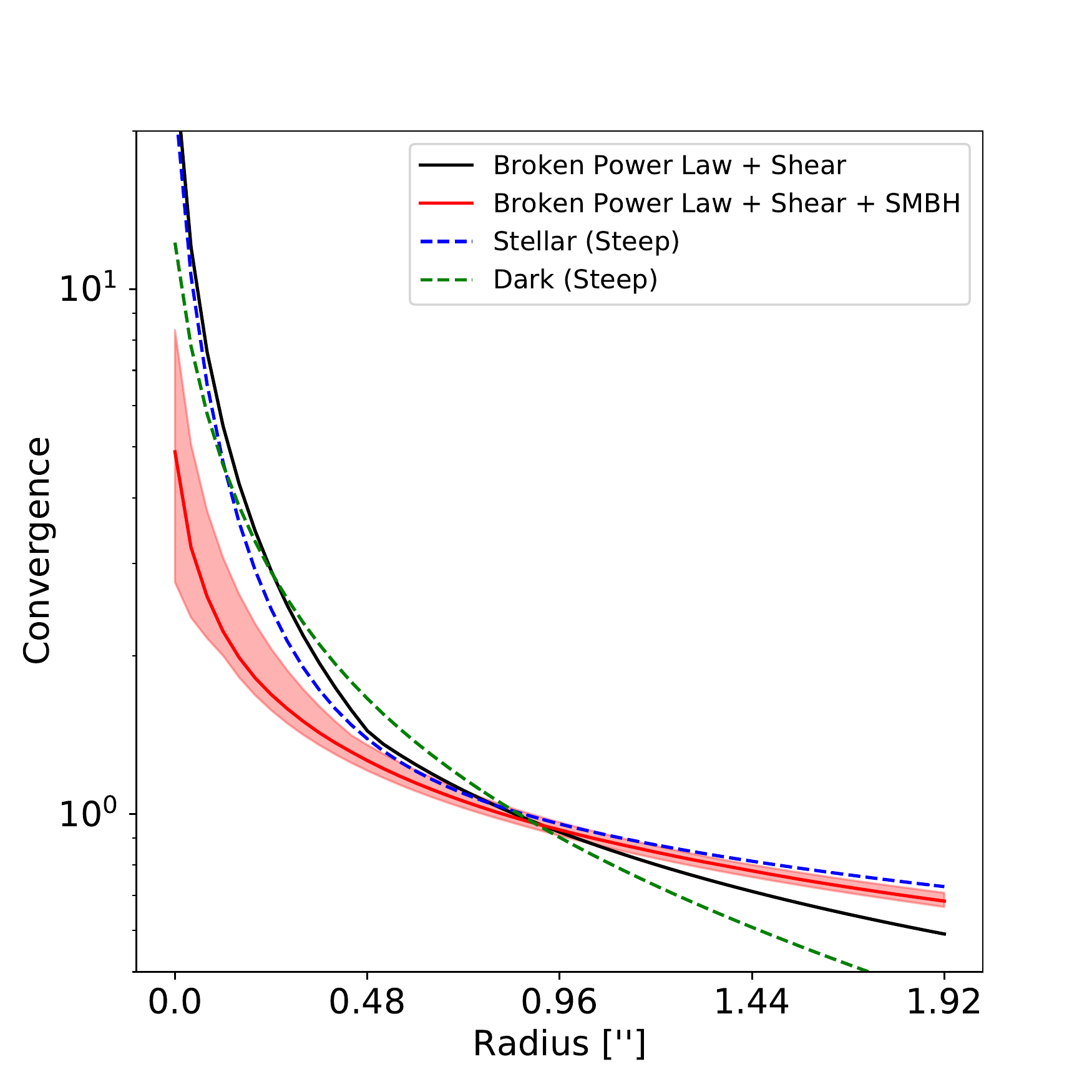}
\caption{The convergence as a function of radius inferred using the F390W image for the total mass models: (i) the power-law (black); (ii) the broken power-law (red); (iii) the power law and SMBH (blue) and; (iv) the broken power-law and SMBH (green), where all models include an external shear. Each line is computed using coordinates that extend radially outwards from the centre of the mass profile and are aligned with its major axis. Shaded regions for each mass model's convergence are shown, corresponding to the inferred $3\sigma$ confidence intervals. The BPL model places more mass centrally than all other models, consistent with its ability to reconstruct the counter image accurately.}
\label{figure:Total1D}
\end{figure}

This appendix shows the results of fitting two total mass-models: the power-law (PL) \citep{Tessore2015} and broken power-law (BPL) \citep{Oriordan2019, Oriordan2020, Oriordan2021}. Like in the main paper, we compare fits with and without a point-mass representing a SMBH. We focus on the Bayesian evidence, $\ln \mathcal{Z}$, and the reconstruction of the counter image. We investigate whether the extraneous flux removed by the SMBH for the decomposed models can be removed by either of these profiles without a SMBH. The inferred model parameters for the PL and BPL models are given in \cref{table:ModelsTotal1} and \cref{table:ModelsTotal2}.

\subsection{Profile Equations}

A softened power-law ellipsoid density profile of form
\begin{equation}
\label{eqn:SPLEkap}
\kappa^{\rm mass} (\xi) = \frac{3 - \gamma^{\rm mass}}{1 + q^{\rm mass}} \bigg( \frac{\theta^{\rm mass}_{\rm E}}{\xi} \bigg)^{\gamma^{\rm mass} - 1} ,
\end{equation}
is assumed, where $\theta^{\rm mass}_{\rm E}$ is the model Einstein radius in arcseconds. The power-law density slope is $\gamma^{\rm mass}$, and setting $\gamma^{\rm mass} = 2$ gives the singular isothermal ellipsoid (SIE) model. Deflection angles for the power-law are computed via an implemention of the method of \citep{Tessore2015} in {\tt PyAutoLens}.

We also use the elliptical broken power law (BPL) profile \citep{Oriordan2019, Oriordan2020, Oriordan2021} with convergence 
\begin{equation}\label{equ:power_law}
    \kappa^{\rm mass}{\left(r\right)}=\left\{
    \begin{array}{ll}
    \theta^{\rm mass}_{\rm E}\left(r^{\rm mass}_{\rm b} / r \right)^{t^{\rm mass}_1}, & r \leq r^{\rm mass}_{\rm b} \\
    \theta^{\rm mass}_{\rm E}\left(r^{\rm mass}_{\rm b} / r \right)^{t^{\rm mass}_2}, & r > r^{\rm mass}_{\rm b}
    \end{array}
    \right.,
\end{equation}
where $r^{\rm mass}_{\rm b}$ is the break radius, $\theta^{\rm mass}_{\rm E}$ is the convergence at the break radius, $t^{\rm mass}_1$ is the inner slope and $t^{\rm mass}_2$ is the outer slope. The isothermal case is given by $t^{\rm mass}_1 = t^{\rm mass}_2 = 1.0$.

\subsection{Power-Law Models}

We first investigate fits using the simpler PL mass model. The PL parameterization has less flexibility in adjusting its central density compared to the BPL. The top two rows of table \ref{table:SMBHMC} show the $\ln \mathcal{Z}$ values inferred for PL model-fits with and without a SMBH. Models including a SMBH are strongly favoured, giving $\Delta \ln \mathcal{Z} = 145$ for the F390W data and $\Delta \ln \mathcal{Z} = 29$ for the F814W. 

\cref{figure:ModelsPLF390W} shows zoom-ins of the PL model's reconstruction of the counter image. The figure shows the same behaviour seen for the decomposed model in the main paper, whereby the PL model without a SMBH produces central extraneous flux, which the inclusion of the SMBH removes. \cref{figure:ModelsPLF814W} shows this also occurs in the F814W image. The residuals of this extraneous flux are more significant than seen for the decomposed model fitted in the main paper, because of the PL model's reduced flexibility in adjusting its central density.    

When the PL mass model includes a SMBH a value of $M_{\rm BH} = 3.83^{+1.56}_{-1.72} \times 10^{10}$\,M$_{\rm \odot}$ is inferred, which is consistent with the $M_{\rm BH}$ values inferred for the decomposed models. The SMBH changes the ray-tracing such that the lens model can now reproduce the counter image's structure accurately. The PL also infers a shallower slope of $\gamma^{\rm mass} = 1.65^{+0.12}_{-0.12}$, compared to the value $\gamma^{\rm mass} = 1.82^{+0.05}_{-0.05}$ inferred without a SMBH. The model without a SMBH therefore tries (and fails) to better fit the counter image by placing more mass centrally.

Fits using the PL therefore support the inclusion of a SMBH is the lens model, and their reconstruction of the counter image produces the same behaviour seen for the decomposed model in the main paper.

\subsection{Broken Power-Law Models}

\begin{table*}
\tiny
\resizebox{\linewidth}{!}{
\begin{tabular}{ l l l l l l l} 
\multicolumn{1}{p{1.8cm}|}{\centering \textbf{Model}} 
& \multicolumn{1}{p{1.5cm}}{$x^{\rm{mass}}$ (\arcsec)} 
& \multicolumn{1}{p{1.5cm}}{$y^{\rm{mass}}$ (\arcsec)} 
& \multicolumn{1}{p{1.5cm}}{$\epsilon_{\rm 1}^{\rm{mass}}$} 
& \multicolumn{1}{p{1.5cm}}{$\epsilon_{\rm 2}^{\rm{mass}}$} 
& \multicolumn{1}{p{1.5cm}}{$\epsilon_{\rm 1}^{\rm{ext}}$} 
& \multicolumn{1}{p{1.5cm}}{$\epsilon_{\rm 2}^{\rm{ext}}$} 
\\ \hline
& & & & & & \\[-4pt]

PL & 
$0.014^{+0.035}_{-0.041}$ & 
$-0.054^{+0.053}_{-0.053}$ & 
$0.107^{+0.030}_{-0.028}$ & 
$-0.088^{+0.024}_{-0.003}$ & 
$-0.113^{+0.019}_{-0.018}$ & 
$0.152^{+0.018}_{-0.025}$ \\[-7pt]

\\ \hline
& & & & & & \\[-5pt]

BPL & 
$0.045^{+0.031}_{-0.033}$ & 
$0.063^{+0.065}_{-0.061}$ & 
$-0.105^{+0.026}_{-0.029}$ & 
$0.145^{+0.027}_{-0.032}$ & 
$-0.105^{+0.039}_{-0.024}$ & 
$0.145^{+0.025}_{-0.038}$ \\[-7pt]

\\ \hline
& & & & & & \\[-5pt]

PL + SMBH & 
$0.050^{+0.035}_{-0.040}$ & 
$0.063^{+0.045}_{-0.071}$ & 
$0.104^{+0.028}_{-0.026}$ & 
$-0.089^{+0.026}_{-0.037}$ & 
$-0.112^{+0.026}_{-0.021}$ & 
$0.147^{+0.019}_{-0.029}$ \\[-7pt]

\\ \hline
& & & & & & \\[-5pt]
BPL + SMBH & 
$0.052^{+0.001}_{-0.001}$ & 
$0.079^{+0.004}_{-0.009}$ & 
$0.101^{+0.002}_{-0.001}$ & 
$-0.083^{+0.003}_{-0.005}$ & 
$-0.117^{+0.002}_{-0.001}$ & 
$0.146^{+0.022}_{-0.041}$  \\[-2pt]
\end{tabular}
}
\caption{
The inferred geometric model parameters of the power-law (PL) and broken power-law (BPL) total mass models fitted to the F390W image in the Mass pipeline. Errors are given at 3$\sigma$ confidence intervals.
}
\label{table:ModelsTotal1}
\end{table*}

\begin{table*}
\tiny
\resizebox{\linewidth}{!}{
\begin{tabular}{ l l l l l l l} 
\multicolumn{1}{p{1.8cm}|}{\centering \textbf{Model}} 
& \multicolumn{1}{p{1.5cm}}{$\theta_{\rm Ein}^{\rm{mass}}$ (\arcsec)} 
& \multicolumn{1}{p{1.5cm}}{$\gamma^{\rm{mass}}$} 
& \multicolumn{1}{p{1.5cm}}{$t_{\rm 1}^{\rm{mass}}$} 
& \multicolumn{1}{p{1.5cm}}{$t_{\rm 2}^{\rm{mass}}$} 
& \multicolumn{1}{p{1.5cm}}{$\theta_{\rm B}^{\rm{mass}}$ (\arcsec)} 
& \multicolumn{1}{p{1.5cm}}{$\theta_{\rm Ein}^{\rm{smbh}}$ (\arcsec)} 
\\ \hline
& & & & & & \\[-4pt]

PL & 
$1.925^{+0.046}_{-0.040}$ & 
$1.818^{+0.042}_{-0.077}$ & 
& \\[-6pt]

\\ \hline
& & & & & & \\[-2pt]

BPL & 
$1.869^{+0.039}_{-0.041}$ & 
& 
$1.13^{+0.34}_{-0.20}$ & 
$0.64^{+0.09}_{-0.15}$ & 
$0.45^{+0.09}_{-0.22}$ & 
 \\[-6pt]

\\ \hline
& & & & & & \\[-2pt]

PL + SMBH & 
$1.56^{+0.18}_{-0.19}$ & 
$1.66^{+0.08}_{-0.10}$ & 
& 
&
&
$0.53^{+0.12}_{-0.14}$ \\[-6pt]

\\ \hline
& & & & & & \\[-2pt]
BPL + SMBH & 
$1.6165^{+0.0017}_{-0.0092}$ & 
&
$0.6920^{+0.0061}_{-0.0419}$ & 
$0.6637^{+0.0009}_{-0.0036}$ &
$0.2096^{+0.0320}_{-0.0030}$ & 
$0.5544^{+0.0107}_{-0.0013}$ \\[-2pt]
\end{tabular}
}
\caption{
The inferred model parameters of the power-law (PL) and broken power-law (BPL) total mass models fitted to the F390W image in the Mass pipeline. Errors are given at 3$\sigma$ confidence intervals.
}
\label{table:ModelsTotal2}
\end{table*}

We now inspect fits using the BPL, which has much greater flexibility than the PL in controlling its inner density. The bottom two rows of table \ref{table:SMBHMC} show the $\ln \mathcal{Z}$ values inferred for BPL model-fits with and without a SMBH. For the F390W image the $\ln \mathcal{Z}$ value for the BPL model without a SMBH is 125699.90; this is $6.22$ above the BPL model with a SMBH. This value is also within $\Delta \ln \mathcal{Z} \sim 1$ of the decomposed models including a SMBH fitted in the main paper (see table \ref{table:SMBHMCDecomp}). 

\cref{figure:ModelsBPLF390W} shows zoom-ins of the BPL model's reconstruction of the counter image. Irrespective of whether a SMBH is included in the model, the extraneous flux in the reconstructed counter image seen for decomposed models and the PL model without a SMBH is not produced. \cref{figure:ModelsBPLF814W} shows this is also true for fits to the F814W image. 

\cref{figure:Total1D} shows the 1D convergence profiles for the BPL mass models with and without a SMBH. Shaded regions shows $3\sigma$ confidence intervals for each profile. The inner density (e.g. within $0.3$") of the BPL without a SMBH is steeper than the decomposed models fitted in the main paper (and also the PL models). The BPL is therefore able to remove extraneous flux from the the reconstructed counter image because it places more mass centrally than any other mass model. The BPL model including a SMBH infers a shallower density profile, because the SMBH performs the ray-tracing which fits the counter image. 

Fits using the BPL model therefore raise the possibility that a SMBH is not required in the lens mass model.

\subsection{Decomposed Model Validation}

The BPL fits show that if the mass model has a sufficiently high inner density then it can reconstruct the counter image accurately. We therefore check whether the decomposed models fitted in the main paper can place as much mass centrally as the BPL without requiring a SMBH. The blue dashed line in \cref{figure:Total1D} shows that if the bulge of the triple Sersic model assumes a radial gradient parameter with the value $\Gamma^{\rm bulge} = 0.9$, its central density matches that of the BPL. The decomposed model parameterization therefore includes models with inner densities comparable to the BPL. We did not infer them because they correspond to lower likelihood solutions (our inferred value is $\Gamma^{\rm bulge} = 0.52^{+0.21}_{-0.32}$ at 3$\sigma$ confidence). We verify this by fitting decomposed models where a uniform prior on $\Gamma^{\rm bulge}$ for the bulge component is placed between 0.85 and 0.95. The $\ln \mathcal{Z}$ values of this model with three and two Sersics are $125560.70$ and $125630.27$ respectively, well below the value of $125699.06$ found for the triple Sersic decomposed model including a SMBH. 

We also investigate models which make the central dark matter density comparable to that of the BPL. The green dashed line in \cref{figure:Total1D} shows that an NFW profile with a concentration that is a $3.5\sigma$ positive outlier on the mass-concentration relation \citep{Ludlow2016} has a central density close to the BPL. We therefore fit a triple Sersic decomposed models which includes the scatter from the mass-concentration relation $\sigma^{\rm dark}$ as a free parameter with a uniform prior between 2.5 and 4.0. We infer $\ln \mathcal{Z} = 125329.01$, significantly below nearly all model fits, with or without a SMBH. 

We therefore conclude that decomposed models that place as much mass centrally as the BPL model cannot attain a comparable $\ln \mathcal{Z}$ without a SMBH for fits to the F390W data. They are also unable to prevent extraneous flux appearing in the counter image. 

\subsection{Mass Model Centering}\label{Projections}


The centre of the stellar mass component of the decomposed model is tied to that of the lens light, whereas the BPL has full freedom in choosing its center. We now inspect the centering of the decomposed and BPL models in more detail, to see if any model appears more or less realistic or physically plausible. This will allow us to argue in favour or against the need for a SMBH.

Upon inspection of the different mass model parameters, fits using the BPL model (with or without a SMBH) infer mass model centres in the range $0.03 < x^{\rm mass} < 0.06$ and $0.04 < y^{\rm mass} < 0.09$ for the F390W image and $0.0 < x^{\rm mass} < 0.03$ and $0.02 < y^{\rm mass} < 0.07$ for the F814W image. Inspecting the lens light model-fits, the inferred centre of the bulge at $3\sigma$ confidence is $x^{\rm bulge} = -0.008^{+0.003}_{-0.003}$ and $y^{\rm bulge} = 0.003^{+0.003}_{-0.003}$ for the F814W image and $x^{\rm bulge} = -0.013^{+0.007}_{-0.006}$ and $y^{\rm bulge} = 0.007^{+0.003}_{-0.003}$ for the F390W image. The BPL model is therefore shifting its centre $\geq 0.04$\," (a full pixel) away from the bulge centre, a shift which corresponds to $\geq 120$\,pc.

We now fit a BPL model without a SMBH where the centre is fixed to that of the bulge ($x^{\rm mass} = -0.008$ and $y^{\rm mass} = 0.003$). 
This model's fit to the F390W image infers $\ln \mathcal{Z} = 125317.26$, well below the value of $\ln \mathcal{Z} = 125699.90$ inferred for the BPL model with a free centre. When the BPL's centre is consistent with the luminous emission it therefore cannot reconstruct Abell 1201's source accurately. 

We can now explain why decomposed models without a SMBH but with a bulge radial gradient around $\Gamma^{\rm bulge} = 0.9$ or a very concentrated dark matter halo did not give as high $\ln \mathcal{Z}$ values or remove extraneous flux from the reconstructed counter image. Even though their central density is as steep as the BPL model, steepening the mass profile only improves the overall fit when its centre is offset from the bulge by $\geq 120$pc in the positive x and y directions. Thus, not only does the BPL show a nonphysical offset from the bulge, but its ability to reconstruct the counter image accurately is dependent on the existence of this offset. 

We therefore view the decomposed models with a SMBH fitted in the main paper as more reliable than the BPL model without a SMBH and discard the BPL model as nonphysical. 


\section{Line of sight Galaxy}\label{MassClump}

\begin{table}
\resizebox{\linewidth}{!}{
\begin{tabular}{ l | l | l | l } 
\multicolumn{1}{p{1.1cm}|}{Filter} 
& \multicolumn{1}{p{1.3cm}|}{Number of Sersics} 
& \multicolumn{1}{p{1.3cm}|}{Includes SMBH?} 
& \multicolumn{1}{p{1.5cm}|}{$\ln \mathcal{Z}$}  
\\ \hline
F390W & 2 & \ding{55} & 125559.11  \\[1pt]
F390W & 3 & \checkmark & 125608.05 \\[0pt]
\hline
F390W & 2 & \ding{55} &  125588.03 \\[1pt]
F390W & 3 &  \checkmark & 125596.66 \\[0pt]
\hline
F814W & 2 & \ding{55} & 78327.28   \\[1pt]
F814W & 3 & \checkmark & 78322.25  \\[0pt]
\hline
F814W & 2 & \ding{55} & 78318.61    \\[1pt]
F814W & 3 & \checkmark & 78316.00  \\[0pt]
\end{tabular}
}
\caption{
The same as \cref{table:SMBHMC} but with the $z = 0.273$ galaxy around $(4.0\,", \,1.0\,")$ included in the lens galaxy mass model. This table shows fits assuming a decomposed mass model with two and three Sersic profiles.
}
\label{table:SMBHMCDecompClump}
\end{table}

\begin{table}
\resizebox{\linewidth}{!}{
\begin{tabular}{ l | l | l | l } 
\multicolumn{1}{p{1.1cm}|}{Filter} 
& \multicolumn{1}{p{1.3cm}|}{Model} 
& \multicolumn{1}{p{1.3cm}|}{Includes SMBH?} 
& \multicolumn{1}{p{1.5cm}|}{$\ln \mathcal{Z}$}  
\\ \hline
F390W & PL & \ding{55} & 125434.63 \\[1pt]
F390W & PL &  \checkmark & 125589.40 \\[0pt]
\hline
F390W & BPL & \ding{55} & 125239.03 \\[1pt]
F390W & BPL & \checkmark & 125586.22 \\[0pt]
\hline
F814W & PL & \ding{55} & 78264.73 \\[1pt]
F814W & PL & \checkmark & 78323.10 \\[0pt]
\hline
F814W & BPL & \ding{55} & 78314.53 \\[1pt]
F814W & BPL & \checkmark & 78311.49 \\[0pt]
\end{tabular}
}
\caption{
The same as \cref{table:SMBHMC} but with the $z = 0.273$ galaxy around $(4.0\,", \,1.0\,")$ included in the lens galaxy mass model. This table shows fits assuming a PL and BPL lens model.
}
\label{table:SMBHMClump}
\end{table}

\cref{figure:Data} shows line-of-sight emission towards the right of the giant arc, around $(4.0\,", \,1.0\,")$. \citet{Smith2017a} show that this is a $z = 0.273$ galaxy, which is therefore located between the lens and source galaxies. The emission seen in the HST imaging appears as two (or more) distinct blobs. The [OIII] emission shows similar structure indicating this is likely a single galaxy. We fit additional lens models to Abell 1201 which include this galaxy in the lens model as a spherical isothermal mass profile (see \cref{eqn:SPLEkap}) where $\gamma^{\rm mass} = 2$), accounting for multi-plane ray-tracing effects \citep{Schneider2014c}. The centre of this model is fixed to $(3.6\,",\,0.95\,")$ in the image-plane, which is updated when performing multi-plane ray tracing. The $\ln \mathcal{Z}$ of these model fits are given in \cref{table:SMBHMCDecompClump} and \cref{table:SMBHMClump}. All models produce lower $\ln \mathcal{Z}$ values than those inferred in the main paper, indicating that including the galaxy does improve the lens model.
\section{SMBH with free centre}\label{MassCentreFree}

\begin{table}
\resizebox{\linewidth}{!}{
\begin{tabular}{ l | l | l | l } 
\multicolumn{1}{p{1.1cm}|}{Filter} 
& \multicolumn{1}{p{1.3cm}|}{Number of Sersics} 
& \multicolumn{1}{p{1.3cm}|}{Includes SMBH?}  
& \multicolumn{1}{p{1.5cm}|}{$\ln \mathcal{Z}$}  
\\ \hline
F390W & 2 & \ding{55} & 125637.18 \\[1pt]
F390W & 2 & \checkmark & 125665.66 \\[0pt]
\hline
F390W & 3 & \ding{55} & 125598.48  \\[1pt]
F390W & 3 & \checkmark & 125661.03  \\[0pt]
\hline
F814W & 2 & \ding{55} & 78330.51  \\[1pt]
F814W & 2 & \checkmark & 78327.26  \\[0pt]
\hline
F814W & 3 & \ding{55} & 78329.19 \\[1pt]
F814W & 3 & \checkmark & 78324.19  \\[0pt]
\end{tabular}
}
\caption{
The same as \cref{table:SMBHMC} but the SMBH centre is free to vary. This table shows fits assuming a decomposed mass model with two and three Sersic profiles.
}
\label{table:SMBHMCDecompCentreFree}
\end{table}

\begin{table}
\resizebox{\linewidth}{!}{
\begin{tabular}{ l | l | l | l } 
\multicolumn{1}{p{1.1cm}|}{Filter} 
& \multicolumn{1}{p{1.3cm}|}{Model} 
& \multicolumn{1}{p{1.3cm}|}{Includes SMBH?}  
& \multicolumn{1}{p{1.5cm}|}{$\ln \mathcal{Z}$}  
\\ \hline
F390W & PL & \ding{55} & 125562.45 \\[1pt]
F390W & PL & \checkmark & \textbf{125683.77} \\[0pt]
\hline
F390W & BPL & \ding{55} & 125699.90 \\[1pt]
F390W & BPL & \checkmark & 125557.48 \\[0pt]
\hline
F814W & PL & \ding{55} & 78301.58 \\[1pt]
F814W & PL & \checkmark & \textbf{78321.82} \\[0pt]
\hline
F814W & BPL & \ding{55} & 78331.17  \\[1pt]
F814W & BPL & \checkmark & 78323.26 \\[0pt]
\end{tabular}
}
\caption{
The same as \cref{table:SMBHMC} but the SMBH centre is free to vary. This table shows fits assuming a PL and BPL lens model.
}
\label{table:SMBHMCentreFree}
\end{table}

The $\ln \mathcal{Z}$ of model fits where the SMBH centre is free to vary are given in \cref{table:SMBHMCDecompCentreFree} and \cref{table:SMBHMCentreFree}. In agreement with the main paper's results, decomposed models including a SMBH with a free centre produce $\ln \mathcal{Z}$ increases at least $25$ above decomposed models without a SMBH. For the triple Sersic decomposed model, the model whose SMBH centre is free gives $\ln \mathcal{Z} = 125665.66$ compared to $\ln \mathcal{Z} = 125699.06$ when the SMBH centre is fixed to the bulge centre. We interpret this decrease as a consequence of Occam's Razor (see \cref{BayesEvi}), whereby the use of a too complex model that does not improve the fit to the data is being penalized. We find that the estimate of $M_{\rm BH}$ does not change when the SMBH center is free to vary. 
\section{Models With Shallow Inner Density}\label{Radial}

\begin{figure}
\centering
\includegraphics[width=0.235\textwidth]{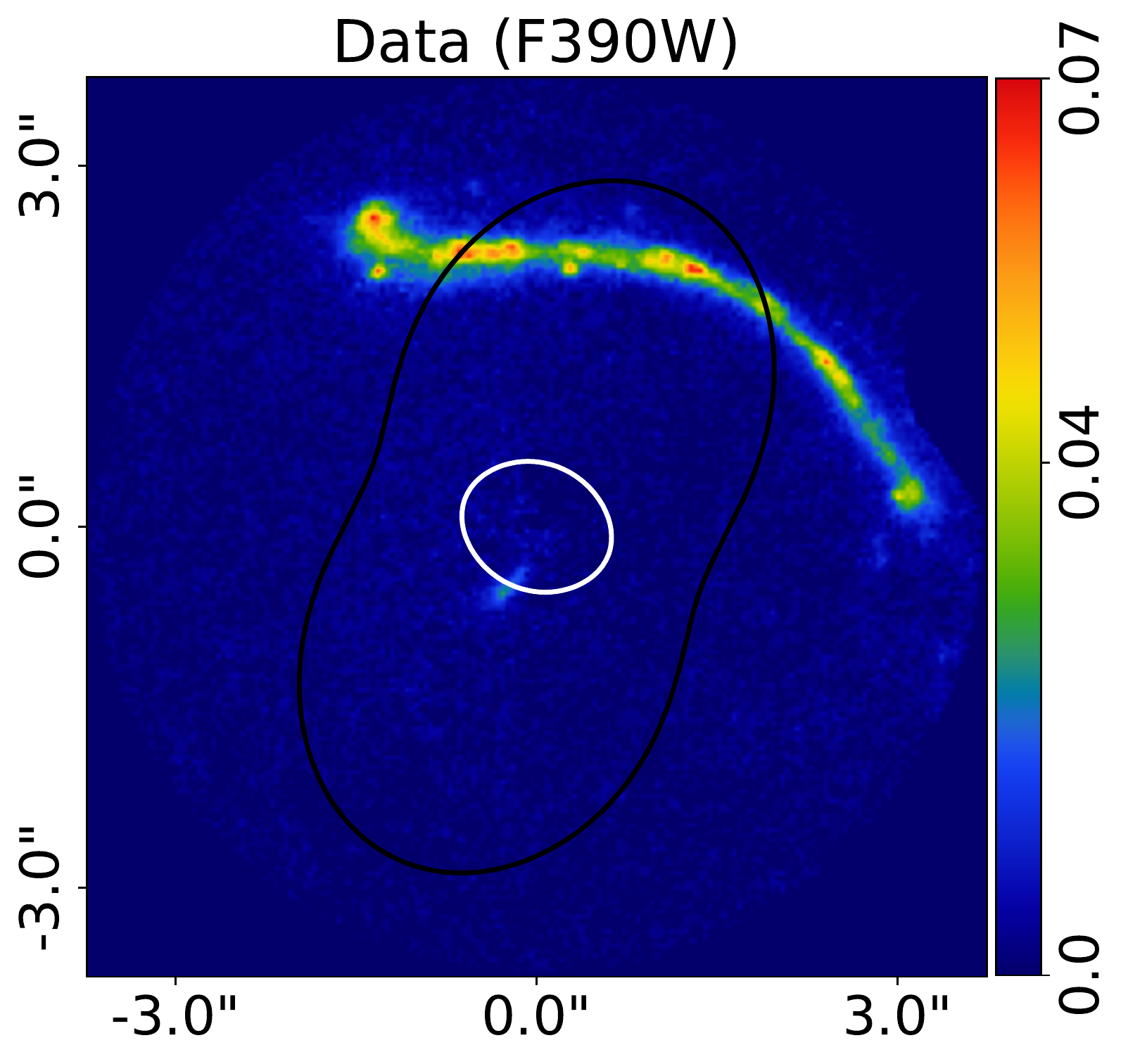}
\includegraphics[width=0.235\textwidth]{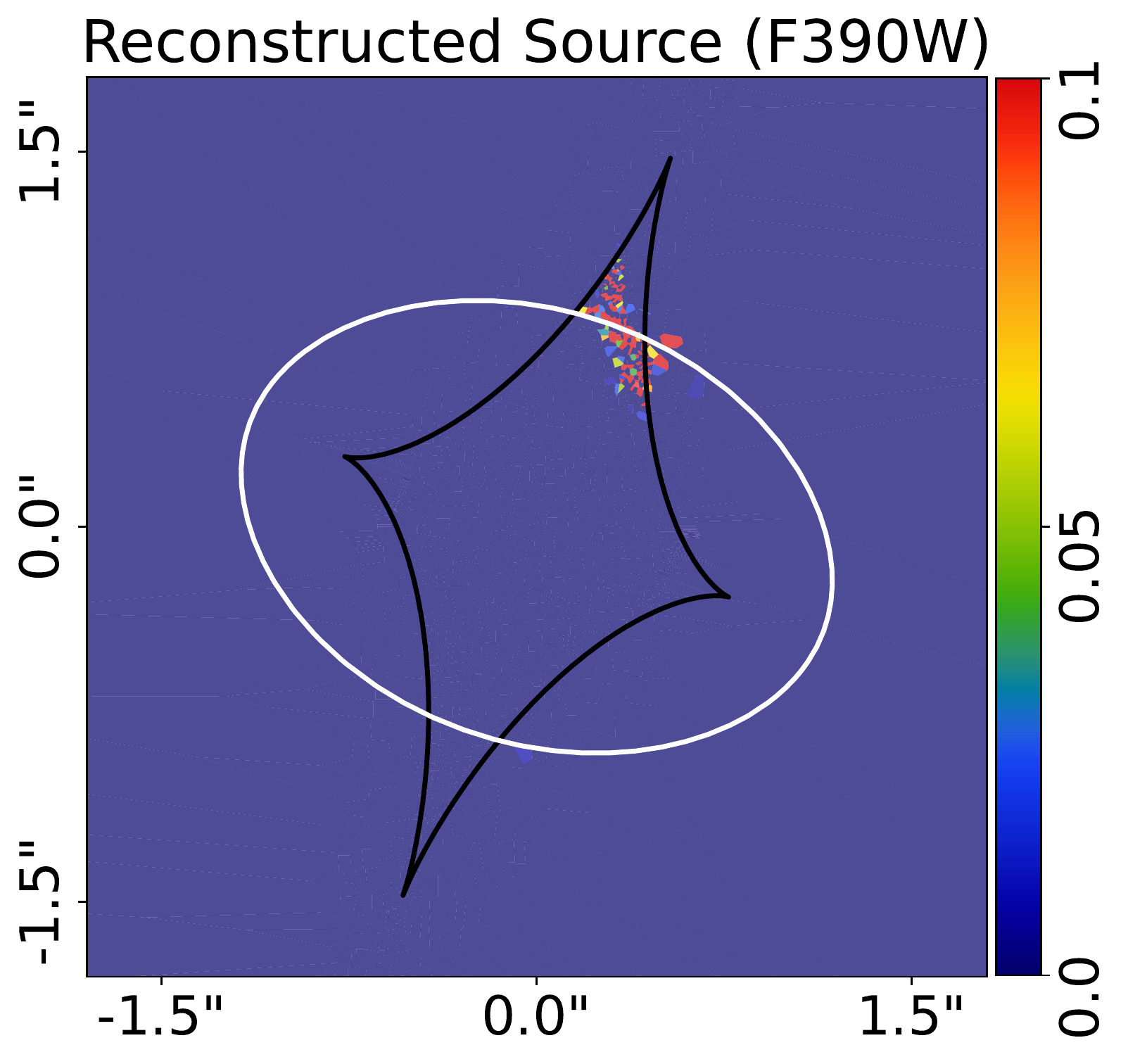}
\caption{
An example model where the lens mass model has a shallow inner density, which forms a larger radial critical curve than solutions presented in the main paper. The left panel shows the observed data, with radial and tangential critical curves (white and black respectively) overlaid. The right panel shows the corresponding source plane and source reconstruction, with the radial and tangential caustics (white and black respectively) overlaid.
} 
\label{figure:RadialDemo}
\end{figure}

We encountered an alternative family of solutions which are characterized by: (i) a shallow inner density profile that forms a larger radial critical curve than the solutions presented in the main paper, which cuts through the inner regions of the counter image and; (ii) the counter image reconstruction producing a pair of merging images (the models in the main paper reconstruct a single counter image). An example of such a model is shown in Fig.~\ref{figure:RadialDemo}.

For decomposed models, these solutions are found when the radial gradient parameters (e.g. $\Gamma^{\rm bulge}$) are below zero and there is less mass relative to light. The low Sersic indices of the lens galaxy’s light profiles ($n^{\rm bulge}$ = ~1.28 and $n^{\rm disk}$ = ~1.16) also help to produce a shallow inner density. For the BPL model, these correspond to solutions where the inner slope $t^{\rm mass}_{1} \sim 0.0$, the outer slope $t^{\rm mass}_{2} \sim 0.7$, and the break radius is $r^{\rm mass}_{\rm B} \sim 0.25"$. We verify that this family of models without a SMBH do not fit the data as well as models with a SMBH by performing \texttt{dynesty} fits, where the priors on certain mass-model parameters are constrained to uniform priors that restrict the analysis to these solutions. The priors can be found at \url{https://zenodo.org/record/7695438}.

\begin{figure*}
\centering
\includegraphics[width=0.241\textwidth]{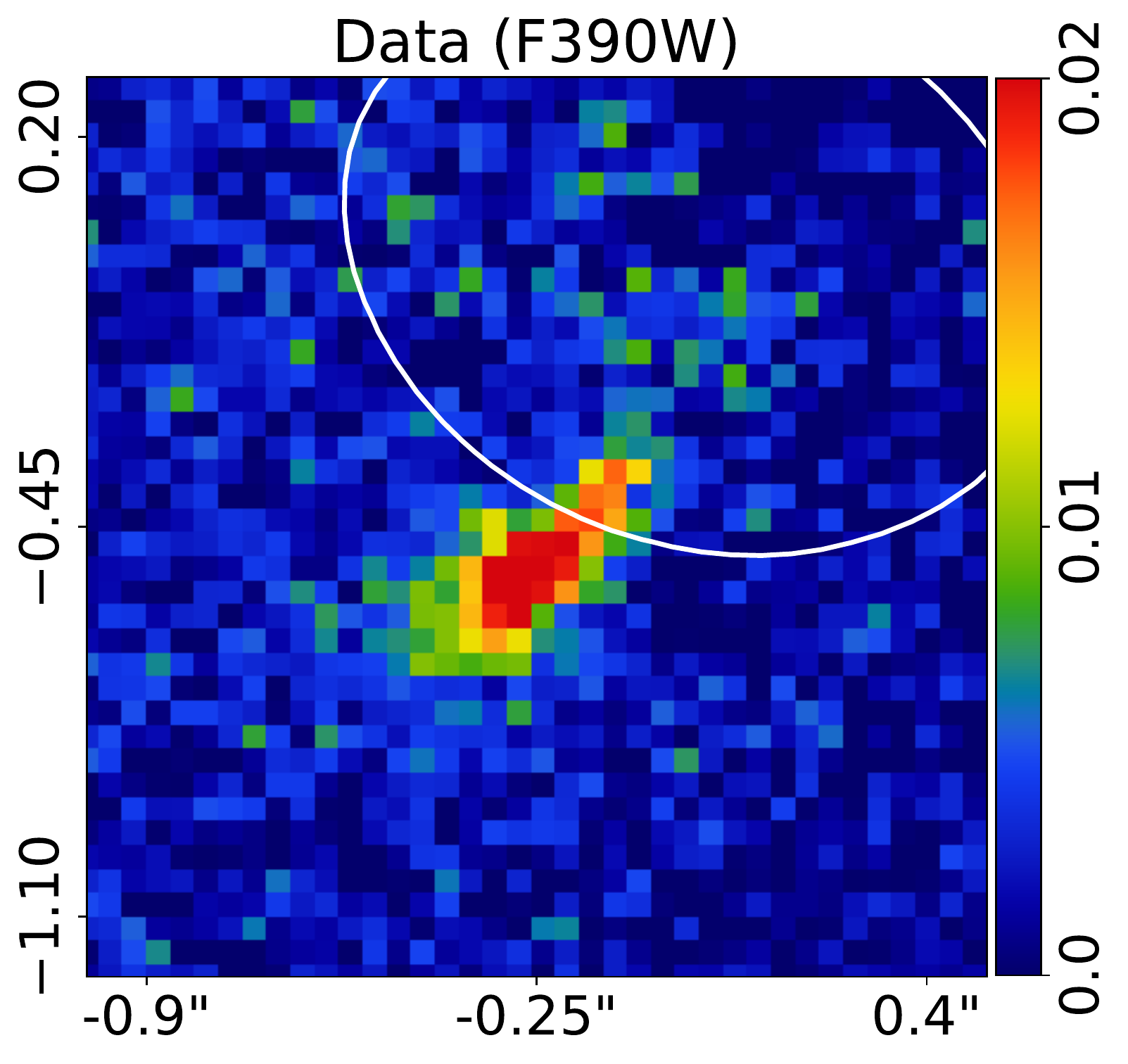}
\includegraphics[width=0.241\textwidth]{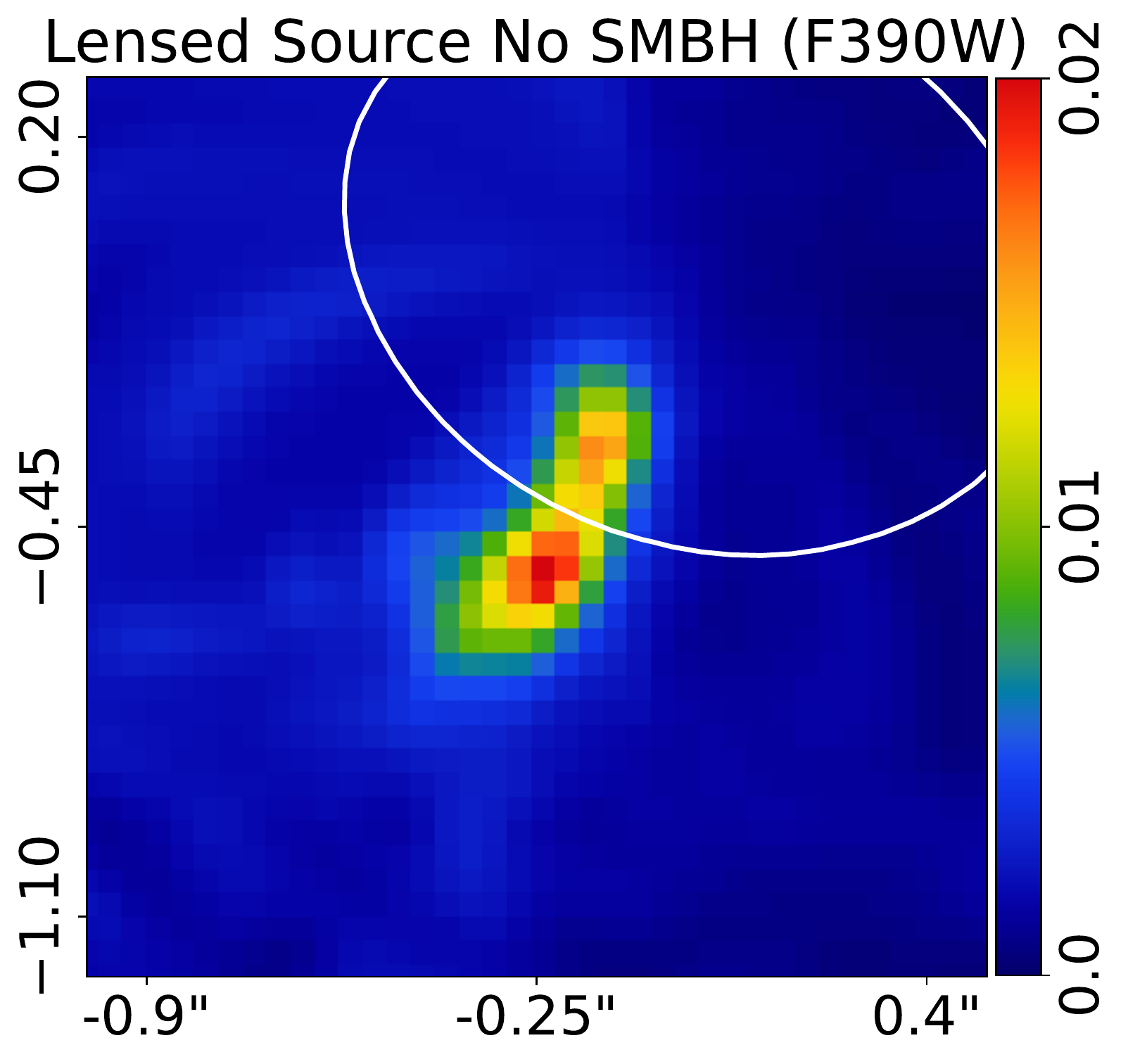}
\includegraphics[width=0.241\textwidth]{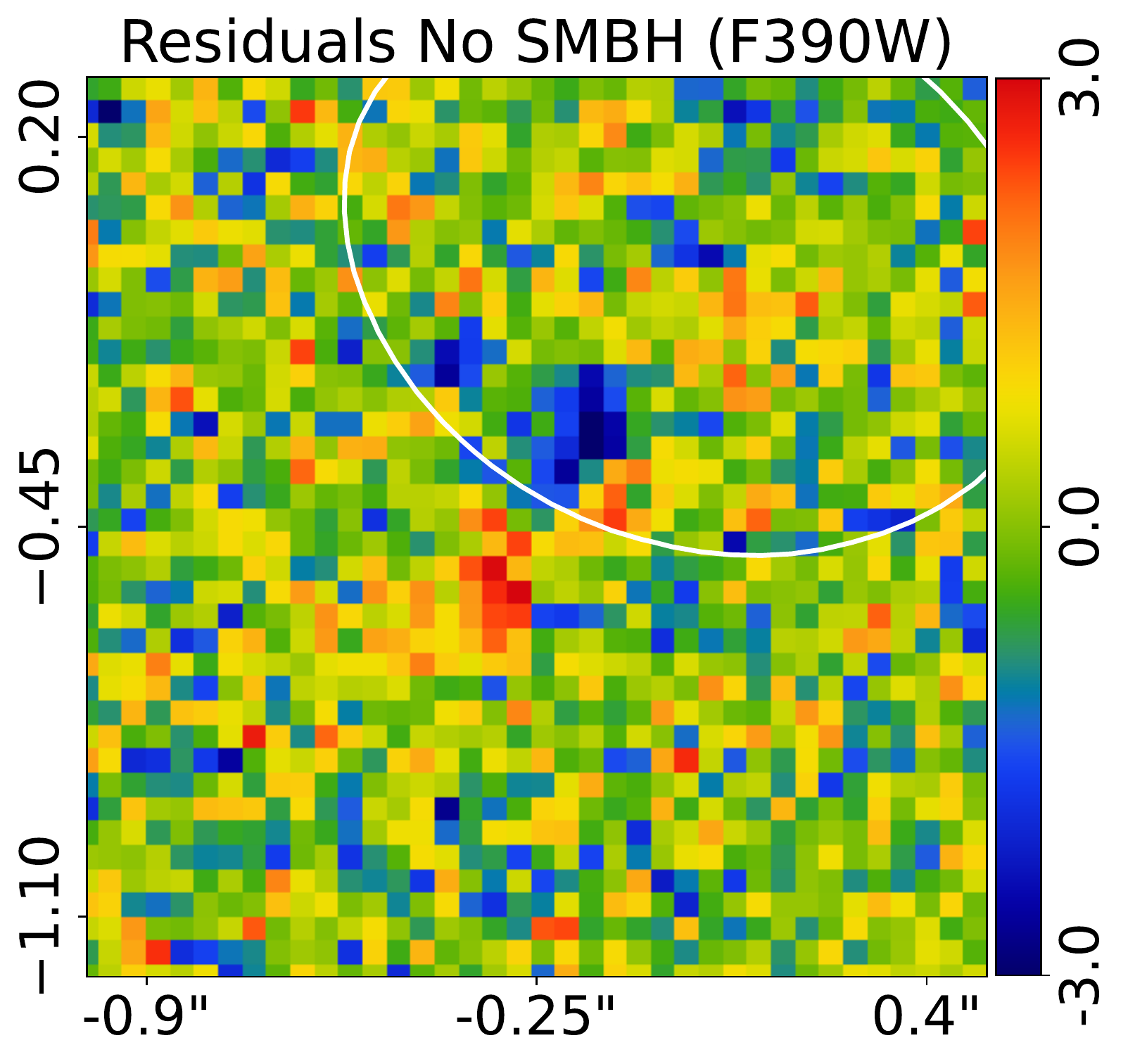}
\includegraphics[width=0.241\textwidth]{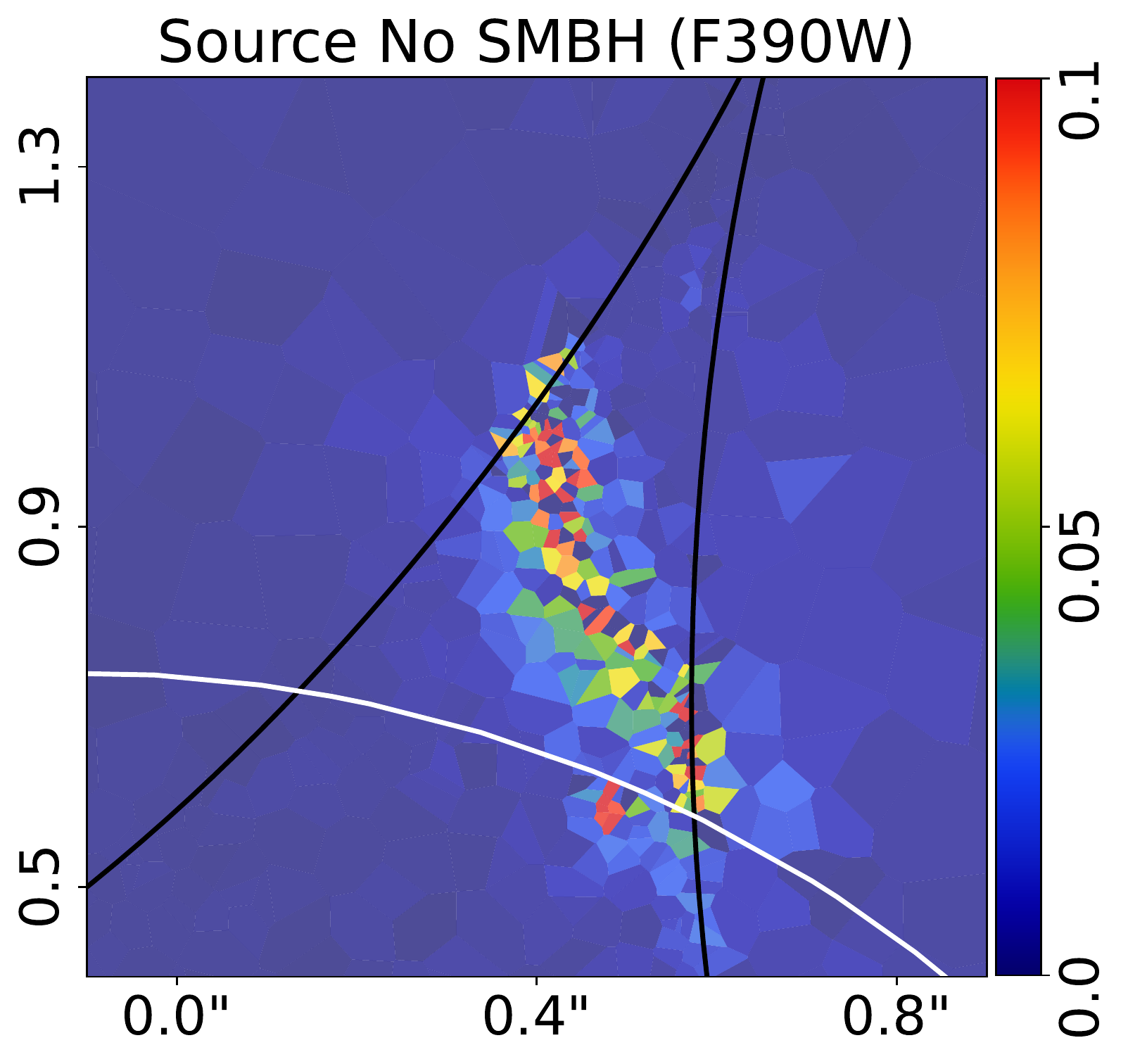}
\caption{
Zoom-ins of the observed counter image in the F390W data (left panel), the model lensed source (left-centre panel), the normalized residuals (right-centre panel) and the source reconstruction (right panel). These results are for the double Sersic plus NFW decomposed model-fits without a SMBH, where the parameter priors allow for solutions with a shallow inner density and large radial critical curve. The tangential caustic is shown by a black line and the radial critical curve and caustic are shown with a white line.
} 
\label{figure:RadialFit}
\end{figure*}

\begin{table}
\resizebox{\linewidth}{!}{
\begin{tabular}{ l | l | l | l} 
\multicolumn{1}{p{1.1cm}|}{Filter} 
& \multicolumn{1}{p{1.3cm}|}{Model} 
& \multicolumn{1}{p{1.3cm}|}{Shallow Density}  
& \multicolumn{1}{p{1.5cm}|}{SMBH}  
\\ \hline
F390W & Decomposed & 125649.72 & 125699.06  \\[1pt]
F390W & BPL & 125548.22 & 125693.78 \\[0pt]
\hline
F814W & Decomposed & 78289.00  & 78332.19  \\[1pt]
F814W & BPL & 78238.79 & 78329.28  \\[0pt]
\end{tabular}
}
\caption{
The Bayesian evidence, $\ln \mathcal{Z}$, of each model-fit performed by the Mass pipelines using: (i) a decomposed mass model assuming two Sersic profiles, an elliptical NFW and external shear or; (ii) a BPL mass model with external shear. Both models have the priors on various parameters adjusted such that they have a shallower inner density and can form a large radial critical curve. Log evidences are compared to the values found in the main paper, for models including a SMBH. Fits to both the F390W and F814W images are shown, where the F390W fits assume the Sersic parameters of the F814W image for the stellar mass. The favoured model is always that with a SMBH, because models with a shallow inner density fail to reconstruct the counter image's structure (see \cref{figure:RadialFit}).
}
\label{table:SMBHMCRadial}
\end{table}

The maximum likelihood solution for the double Sersic decomposed mass model are shown in \cref{figure:RadialFit}. The reconstructed counter image is split in two, and fails to capture the appearance of the counter image in the data. For this model, the log Bayesian evidence value is $\mathcal{Z} = \sim\,125649$, which is significantly below models with a SMBH which have a log evidence of $\mathcal{Z} =\sim\,125699$. \cref{table:SMBHMCRadial} compares the log Bayesian evidence values for the BPL model fits with a shallower inner density and also includes the values for the F814W. For both the F390W and F814W images, these solutions provide significantly worse fits to the data than models including a SMBH, confirming that they are ruled out by the data.

These fits also confirm that the central emission seen in the F814W data (\cref{figure:LightFit2}; within magenta circle) is not a central image. Lens model fits using cored mass profiles would reconstruct the counter image, if it were the physically correct solution. The fact these solutions are not inferred confirms it is not a central image.


\label{lastpage}

\end{document}